\documentclass{article}

\usepackage{amsmath}
\usepackage{amssymb}
\usepackage{authblk}
\usepackage{hyperref}
\usepackage{graphicx}
\usepackage{framed}
\usepackage{setspace}
\usepackage{fullpage}

\title{Computational General Relativity in the Wolfram Language using \textsc{Gravitas} II: ADM Formalism and Numerical Relativity}
\author[1]{Jonathan Gorard}
\affil[1]{Princeton University,\protect\\
Princeton, NJ, United States\footnote{\href{mailto:gorard@princeton.edu}{gorard@princeton.edu}}\footnote{Current affiliation. Research and development work was performed at, and funded by, the Wolfram Institute.}}

\begin{document}

\maketitle

\begin{abstract}
This is the second in a series of two articles introducing the \textsc{Gravitas} computational general relativity framework, in which we now focus upon the design and capabilities of \textsc{Gravitas}'s numerical subsystem, including its ability to perform general ${3 + 1}$ decompositions of spacetime via the ADM formalism, its support for the definition and construction of arbitrary Cauchy surfaces as initial data, its support for the definition and enforcement of arbitrary gauge and coordinate conditions, its various algorithms for ensuring the satisfaction of the ADM Hamiltonian and momentum constraints, and its unique (totally-unstructured) adaptive refinement algorithms based on hypergraph rewriting via Wolfram model evolution. Particular attention will be paid to the seamless integration between \textsc{Gravitas}'s symbolic and numerical subsystems, its ability to configure, run, analyze and visualize complex numerical relativity simulations and their outputs within a single notebook environment, and its capabilities for handling generic curvilinear coordinate systems and spacetimes with general (and often highly non-trivial) topologies using its specialized and highly efficient hypergraph-based numerical algorithms. We also provide illustrations of \textsc{Gravitas}'s functionality for the visualization of hypergraph geometries and spacetime embedding diagrams in two and three dimensions, the ability for \textsc{Gravitas}'s symbolic and numerical subsystems to be used in concert for the extraction of gravitational wave signals and other crucial simulation data, and \textsc{Gravitas}'s in-built library of standard initial data, matter distributions and gauge conditions. We conclude by demonstrating how the numerical subsystem can be used to set up, run, visualize and analyze a standard yet nevertheless reasonably challenging numerical relativity test case: a binary black hole collision and merger within a vacuum spacetime (including the extraction of its outgoing gravitational wave profile).
\end{abstract}

\section{Introduction}
\label{sec:Section1}

The first article in this series\cite{gorard} contained a more-or-less complete and self-contained introduction to the underlying philosophy of the \textsc{Gravitas} computational framework, a summary of its core design principles, and an incomplete but high-level survey of the field of computational and numerical relativity as a whole, along with a detailed explanation of how \textsc{Gravitas} compared to the many other existing open source tools that are presently available. In the interests of brevity, we do not seek to repeat those introductory remarks here, and we instead encourage the reader to consult the previous article before proceeding with this one. The previous article also provided a comprehensive overview of \textsc{Gravitas}'s symbolic and analytical capabilities, including its tensor calculus and differential geometry functionality, its methodology for handling general Riemannian/pseudo-Riemannian/Lorentzian metrics and coordinate systems, and its algorithms for solving the Einstein field equations both analytically and numerically, including particular support for electromagnetic fields and the Einstein-Maxwell equations, as well as support for arbitrary non-gravitational fields via the stress-energy tensor. The present article will place particular emphasis upon the seamless integration between the aforementioned symbolic and analytical functionality and \textsc{Gravitas}'s powerful numerical subsystem. One of the primary design features of \textsc{Gravitas} is its synthesis of the capabilities of an abstract tensor calculus package that is comparable in many respects to \href{http://www.xact.es}{\texttt{xAct}} (a Wolfram Language framework developed by Mart\'in-Garc\'ia et al., facilitating component calculus through \href{http://www.xact.es/xCoba/index.html}{\texttt{xCoba}}, manipulation of scalar invariants of the Riemann tensor through \href{http://www.xact.es/Invar/index.html}{\texttt{Invar}}\cite{martin-garcia}\cite{martin-garcia2} and canonicalization of tensor indices with respect to slot/permutation symmetries through \href{http://www.xact.es/xPerm/index.html}{\texttt{xPerm}}\cite{martin-garcia3}) or \href{https://einsteinpy.org}{\texttt{EinsteinPy}}\cite{bapat} (a Python package facilitating elementary tensor manipulations, as well as the visualization and computation of geodesics within a small number of in-built spacetime geometries), with the capabilities of a full-scale numerical relativity framework such as \href{http://www.cactuscode.org}{\texttt{Cactus}}\cite{goodale}/\href{https://einsteintoolkit.org/arrangementguide/McLachlan/documentation.html}{\texttt{McLachlan}}\cite{brown}\cite{reisswig} (a component of the popular \href{https://einsteintoolkit.org}{\texttt{Einstein}} toolkit\cite{zilhao}, facilitating the numerical solution of the Baumgarte-Shapiro-Shibata-Nakamura/BSSN formulation of the Einstein field equations\cite{nakamura}\cite{shibata}\cite{baumgarte} using finite-difference approximations with up to eighth-order accuracy and block-structured adaptive mesh refinement/AMR), \href{https://www.grchombo.org}{\texttt{GRChombo}}\cite{clough}\cite{andrade} (a similar block-structured AMR code for solving both the BSSN and conformally-covariant Z4/CCZ4 formulations\cite{alic}\cite{bona} of the Einstein field equations using fourth-order finite-difference methods) or \href{https://www.black-holes.org/code/SpEC.html}{\texttt{SpEC}}\cite{kidder}\cite{pfeiffer} (a code for solving the generalized harmonic formulation of the Einstein field equations using pseudospectral methods).

The principal advantage of this integration is the resulting ability of the user to configure initial conditions (i.e. initial spacelike hypersurface/Cauchy surface configurations), to define the desired evolution equations, to set up corresponding matter field configurations with appropriate coupling to the spacetime, to enforce appropriate gauge and coordinate constraints on the hyperbolic evolution, to run the resulting (automatically-optimized and automatically-parallelized) numerical simulations, and to analyze and visualize the raw outputs of those simulations (including the extraction of gravitational wave signals), all from within a single notebook environment. In particular, this eliminates the need to use separate computer algebra-based frameworks such as \href{http://kranccode.org}{\texttt{Kranc}}\cite{husa} in the Wolfram Language or \href{https://nrpyplus.net}{\texttt{SENR/NRPy+}}\cite{ruchlin} in Python for representing evolution equations and gauge conditions in an abstract tensorial form and then automatically synthesizing optimized low-level C or Fortran code from them, and indeed the need to use a separate tool such as \href{https://simulationtools.org}{\texttt{SimulationTools}} in the Wolfram Language or \href{https://sbozzolo.github.io/kuibit/}{\texttt{kuibit}}\cite{bozzola} in Python for performing standard post-processing tasks and the extraction or analysis of key simulation data. \textsc{Gravitas}'s use of next-generation adaptive refinement algorithms based upon totally-unstructured hypergraph rewriting/Wolfram model evolution\cite{gorard2}\cite{gorard3}\cite{gorard4} facilitates its handling of arbitrary (potentially overlapping) curvilinear coordinate systems and highly non-trivial spacetime topologies with ease, since one is not constrained to use a rigid Cartesian or spheroidal grid structure for the computational domain, and the hypergraph's topology can instead adapt organically to both the topology and the dynamics of the problem being solved. This hypergraph rewriting formalism\cite{gorard5}\cite{gorard6}\cite{gorard7} (including its efficient hypergraph canonicalization system\cite{gorard8}), and the resulting numerical algorithms that it enables, have previously been used to investigate a variety of physical problems in discrete spacetimes, including binary black hole mergers\cite{gorard9}, entanglement entropies in curved spacetime quantum field theories\cite{gorard10}, singularity theorems in idealized gravitational collapse scenarios\cite{gorard11}, amongst others.

In Section \ref{sec:Section2}, we begin by introducing the abstract formalism of the ADM (or ${3 + 1}$) metric decomposition of a Riemannian/pseudo-Riemannian/Lorentzian manifold into a sequence of immersed hypersurfaces, and demonstrate how \textsc{Gravitas} is able to use the \texttt{ADMDecomposition} object (the most fundamental object in \textsc{Gravitas}'s numerical subsystem, analogous to \texttt{MetricTensor} in the symbolic/analytial subsystem) to represent such decompositions and their corresponding geometrical properties using only the specification of the initial Cauchy surface of the manifold and a suitable gauge choice. In particular, we show an excerpt of \textsc{Gravitas}'s in-built library of Cauchy initial data (including initial conditions for both black hole and cosmological spacetimes), as well as functionality for calculating normal vectors, ``time vectors'', Gauss equations, Codazzi-Mainardi equations, ADM Hamiltonian constraints, ADM momentum constraints and (purely geometrical) ADM evolution equations from the underlying \texttt{ADMDecomposition} object. We also highlight functions such as \texttt{ExtrinsicCurvatureTensor} for computing the second fundamental form/extrinsic curvature of immersed hypersurfaces within the decomposition, and \texttt{CottonTensor} for determining conformal-flatness of immersed three-dimensional submanifolds. In Section \ref{sec:Section3}, we proceed to demonstrate \textsc{Gravitas}'s library of supported ADM gauge conditions, including the geodesic, maximal, harmonic and ${1 + log}$ slicing conditions on the lapse function, and the normal, harmonic, minimal distortion and pseudo-minimal distortion coordinate conditions on the shift vector. We also show how the (vacuum) ADM evolution equations and (vacuum) ADM Hamiltonian and momentum constraint equations can be solved, both analytically and numerically, using the \texttt{SolveVacuumADMEquations} function, with such solutions being represented symbolically as \texttt{VacuumADMSolution} objects. In Section \ref{sec:Section4}, we illustrate how one can introduce a generic matter field into such a solution (and evolve it either numerically or analytically) using the \texttt{ADMStressEnergyDecomposition} function, which produces an ADM/${3 + 1}$ decomposition of an arbitrary \texttt{StressEnergyTensor} object, thus allowing one to compute energy, momentum and stress projections of the tensor, as well as timelike (energy conservation) and spacelike (momentum conservation) projections of the stress-energy continuity equations, within any ADM decomposition and with any choice of gauge. We show how this can be combined with the aforementioned functionality in order to solve the full ADM evolution equations and full ADM Hamiltonian and momentum constraint equations, again both analytically and numerically, using the \texttt{SolveADMEquations} function, with such solutions being represented symbolically as \texttt{ADMSolution} objects. Finally, in Section \ref{sec:Section5}, we demonstrate how the \texttt{DiscreteHypersurfaceDecomposition} function can be used to expose and visualize some of the internal details of how \textsc{Gravitas} obtains its numerical solutions, namely by means of a fourth-order Runge-Kutta time integration algorithm defined over hypergraphs of arbitrary topology, with a generalized local adaptive refinement algorithm that automatically refines and coarsens the hypergraph (using hypergraph rewriting methods) based upon the dynamics of the problem being solved. We show how all of these methods can be combined in order to simulate a non-trivial numerical relativistic problem in vacuum spacetime: the head-on collision of a binary black hole system, including the extraction of the resulting outgoing gravitational radiation. We conclude in Section \ref{sec:Section6} with some directions for future research and development.

Note that, as with the previous article\cite{gorard} in this series, much of the \textsc{Gravitas} functionality that is demonstrated and discussed here is fully-documented and available via the \textit{Wolfram Function Repository}, including \href{https://resources.wolframcloud.com/FunctionRepository/resources/ADMDecomposition/}{ADMDecomposition}, \href{https://resources.wolframcloud.com/FunctionRepository/resources/SolveVacuumADMEquations/}{SolveVacuumADMEquations} and \href{https://resources.wolframcloud.com/FunctionRepository/resources/DiscreteHypersurfaceDecomposition/}{DiscreteHypersurfaceDecomposition}. However, there are many more functions that have not yet been documented in this way, or which have received further development beyond their documented versions; as always, an up-to-date version of the \textsc{Gravitas} framework may be obtained from its \href{https://github.com/JonathanGorard/Gravitas/}{GitHub repository} (currently approximately 30,000 lines of Wolfram Language code, including experimental and research functionality). Note moreover that this article will employ the same notational and terminological conventions as the previous one\cite{gorard}; for instance, we use geometric units with ${c = G = \hbar = 1}$, we assume a metric signature of ${\left( -, +, +, + \right)}$ in all relevant cases, we adopt the Einstein summation convention of summing over all repeated tensor indices throughout, and \textsc{Gravitas} functions use the keywords ``\textit{Reduced}'' and ``\textit{Symbolic}'' within certain property names in order to signify that \textsc{Gravitas} should apply all known algebraic equivalences and return a given tensor expression in canonical form, or that \textsc{Gravitas} should leave all partial derivatives as purely symbolic/unevaluated operators, respectively.

\section{ADM Formalism and Spacetime Foliations}
\label{sec:Section2}

At the core of the \textsc{Gravitas} framework's numerical subsystem is the \texttt{ADMDecomposition} object, which plays a role analogous to the role played by the \texttt{MetricTensor} object within \textsc{Gravitas}'s symbolic subsystem\cite{gorard}, and which allows one to represent a decomposition or ``foliation'' of a given differentiable Riemannian or pseudo-Riemannian manifold ${\left( \mathcal{M}, g_{\mathcal{M}} \right)}$ into an ordered sequence of differentiable submanifolds ${\left( \mathcal{N}, g_{\mathcal{N}} \right)}$, where ${\mathcal{N} \subset \mathcal{M}}$, each of codimension-1\cite{candel}\cite{candel2}. In the context of general relativity, the ``ambient'' manifold ${\left( \mathcal{M}, g_{\mathcal{M}} \right)}$ is typically a pseudo-Riemannian/Lorentzian spacetime, and the submanifolds ${\left( \mathcal{N}, g_{\mathcal{N}} \right)}$ are typically Riemannian spacelike hypersurfaces, ordered by a gauge-dependent ``time'' parameter\cite{gourgoulhon}. Mathematically, if ${i : \mathcal{N} \to \mathcal{M}}$ denotes the embedding of the submanifold ${\left( \mathcal{N}, g_{\mathcal{N}} \right)}$ into the ambient manifold ${\left( \mathcal{M}, g_{\mathcal{M}} \right)}$, then the tangent bundle ${\bigsqcup\limits_{\mathbf{x} \in \mathcal{N}} T_{\mathbf{x}}^{\top} \mathcal{N}}$ on the submanifold ${\left( \mathcal{N}, g_{\mathcal{N}} \right)}$ is included in the tangent bundle ${\bigsqcup\limits_{\mathbf{x} \in \mathcal{M}} T_{\mathbf{x}}^{\top} \mathcal{M}}$ on the ambient manifold ${\left( \mathcal{M}, g_{\mathcal{M}} \right)}$ by means of the pushforward construction, whose cokernel/quotient is then the normal bundle ${\bigsqcup\limits_{\mathbf{x} \in \mathcal{N}} T_{\mathbf{x}}^{\bot} \mathcal{N}}$ on the submanifold ${\left( \mathcal{N}, g_{\mathcal{N}} \right)}$, i.e. one has the following short exact sequence\cite{kobayashi}\cite{kobayashi2}:

\begin{equation}
0 \longrightarrow \left( \bigsqcup\limits_{\mathbf{x} \in \mathcal{N}} T_{\mathbf{x}}^{\top} \mathcal{N} \right) \longrightarrow \left. \left( \bigsqcup\limits_{\mathbf{x} \in \mathcal{M}} T_{\mathbf{x}}^{\top} \mathcal{M} \right) \right\vert_{i \left( \mathcal{N} \right)} \longrightarrow \left( \bigsqcup\limits_{\mathbf{x} \in \mathcal{N}} T_{\mathbf{x}}^{\bot} \mathcal{N} \right) \longrightarrow 0.
\end{equation}
In the above, ${T_{\mathbf{x}}^{\top} \mathcal{N}}$ and ${T_{\mathbf{x}}^{\top} \mathcal{M}}$ denote the tangent spaces of the manifolds ${\left( \mathcal{N}, g_{\mathcal{N}} \right)}$ and ${\left( \mathcal{M}, g_{\mathcal{M}} \right)}$ at points ${\mathbf{x} \in \mathcal{N}}$ and ${\mathbf{x} \in \mathcal{M}}$, respectively; ${\left. \left( \bigsqcup\limits_{\mathbf{x} \in \mathcal{M}} T_{\mathbf{x}}^{\top} \mathcal{M} \right) \right\vert_{i \left( \mathcal{N} \right)}}$ designates the restriction of the tangent bundle ${\bigsqcup\limits_{\mathbf{x} \in \mathcal{M}} T_{\mathbf{x}}^{\top} \mathcal{M}}$ on the ambient manifold ${\left( \mathcal{M}, g_{\mathcal{M}} \right)}$ to the submanifold ${\left( \mathcal{N}, g_{\mathcal{N}} \right)}$, i.e., more formally, it designates the pullback ${i^{*} \left( \bigsqcup\limits_{\mathbf{x} \in \mathcal{M}} T_{\mathbf{x}}^{\top} \mathcal{M} \right)}$ of the tangent bundle ${\bigsqcup\limits_{\mathbf{x} \in \mathcal{M}} T_{\mathbf{x}}^{\top} \mathcal{M}}$ on the ambient manifold ${\left( \mathcal{M}, g_{\mathcal{M}} \right)}$ to a corresponding vector bundle on the submanifold ${\left( \mathcal{N}, g_{\mathcal{N}} \right)}$; ${T_{\mathbf{x}}^{\bot} \mathcal{N}}$ denotes the normal space of the manifold ${\left( \mathcal{N}, g_{\mathcal{N}} \right)}$ at the point ${\mathbf{x} \in \mathcal{N}}$, i.e. the set of all vectors in the tangent space ${T_{\mathbf{x}}^{\top} \mathcal{M}}$ of the ambient manifold ${\left( \mathcal{M}, g_{\mathcal{M}} \right)}$ that are normal to every vector in the tangent space ${T_{\mathbf{x}}^{\top} \mathcal{N}}$ of the submanifold ${\left( \mathcal{N}, g_{\mathcal{N}} \right)}$\cite{lee}:

\begin{equation}
\forall \mathbf{x} \in \mathcal{N}, \qquad T_{\mathbf{x}}^{\bot} \mathcal{N} = \left\lbrace \left. \mathbf{n} \in T_{\mathbf{x}}^{\top} \mathcal{M} \right\vert g_{\mathbf{x}} \left( \mathbf{n}, \mathbf{v} \right) = 0, \text{ for every } \mathbf{v} \in T_{\mathbf{x}}^{\top} \mathcal{N} \right\rbrace,
\end{equation}
such that the normal bundle ${\bigsqcup\limits_{\mathbf{x} \in \mathcal{N}} T_{\mathbf{x}}^{\bot} \mathcal{N}}$ is simply given by the quotient bundle:

\begin{equation}
\left( \bigsqcup\limits_{\mathbf{x} \in \mathcal{N}} T_{\mathbf{x}}^{\bot} \mathcal{N} \right) = \left. \left( \bigsqcup\limits_{\mathbf{x} \in \mathcal{M}} T_{\mathbf{x}}^{\top} \mathcal{M} \right) \right\vert_{i \left( \mathcal{N} \right)} \Bigg/ \left( \bigsqcup\limits_{\mathbf{x} \in \mathcal{N}} T_{\mathbf{x}}^{\top} \mathcal{N} \right);
\end{equation}
and $0$ designates the trivial (rank-0) vector bundle.

The metric tensor ${g_{\mathcal{M}}}$ on the ambient manifold ${\mathcal{M}}$ has the effect of splitting this exact sequence, thus guaranteeing that the restriction of the tangent bundle on the ambient manifold ${\left( \mathcal{M}, g_{\mathcal{M}} \right)}$ to the submanifold ${\left( \mathcal{N}, g_{\mathcal{N}} \right)}$ can be written as the following direct sum of tangent and normal bundles on the submanifold ${\left( \mathcal{N}, g_{\mathcal{N}} \right)}$:

\begin{equation}
\left. \left( \bigsqcup\limits_{\mathbf{x} \in \mathcal{M}} T_{\mathbf{x}}^{\top} \mathcal{M} \right) \right\vert_{i \left( \mathcal{N} \right)} = \left( \bigsqcup\limits_{\mathbf{x} \in \mathcal{N}} T_{\mathbf{x}}^{\top} \mathcal{N} \right) \oplus \left( \bigsqcup\limits_{\mathbf{x} \in \mathcal{N}} T_{\mathbf{x}}^{\bot} \mathcal{N} \right).
\end{equation}
This splitting allows one, in turn, to write the Levi-Civita connection ${\nabla^{\left( \mathcal{M} \right)}}$ on the ambient manifold ${\left( \mathcal{M}, g_{\mathcal{M}} \right)}$ as a sum of tangential and normal components, i.e. one has:

\begin{equation}
\forall \mathbf{x} \in \mathcal{M}, \qquad \forall \mathbf{X} \in T_{\mathbf{x}} \mathcal{M}, \qquad \forall \mathbf{Y} \in \Gamma \left( \bigsqcup\limits_{\mathbf{x} \in \mathcal{M}} T_{\mathbf{x}} \mathcal{M} \right), \qquad \nabla_{\mathbf{X}}^{\left( \mathcal{M} \right)} \mathbf{Y} = \top \left( \nabla_{\mathbf{X}}^{\left( \mathcal{M} \right)} \mathbf{Y} \right) + \bot \left( \nabla_{\mathbf{X}}^{\left( \mathcal{M} \right)} \mathbf{Y} \right),
\end{equation}
where ${\mathbf{X} \in T_{\mathbf{x}} \mathcal{M}}$ denotes an arbitrary tangent vector at the point ${\mathbf{x} \in \mathcal{M}}$, ${\mathbf{Y} \in \Gamma \left( \bigsqcup\limits_{\mathbf{x} \in \mathcal{M}} T_{\mathbf{x}} \mathcal{M} \right)}$ denotes an arbitrary vector field on the ambient manifold ${\left( \mathcal{M}, g_{\mathcal{M}} \right)}$,  ${\Gamma \left( \bigsqcup\limits_{\mathbf{x} \in \mathcal{M}} T_{\mathbf{x}} \mathcal{M} \right)}$ denotes the space of smooth sections of the tangent bundle ${\bigsqcup\limits_{\mathbf{x} \in \mathcal{M}} T_{\mathbf{x}} \mathcal{M}}$ (i.e. the space of vector fields definable on the ambient manifold ${\left( \mathcal{M}, g_{\mathcal{M}} \right)}$), the tangential component ${\top \left( \nabla_{\mathbf{X}}^{\left( \mathcal{M} \right)} \mathbf{Y} \right)}$ denotes the orthogonal projection of the covariant derivative ${\nabla_{\mathbf{X}}^{\left( \mathcal{M} \right)} \mathbf{Y}}$ onto the tangent bundle ${\bigsqcup\limits_{\mathbf{x} \in \mathcal{N}} T_{\mathbf{x}}^{\top} \mathcal{N}}$, and the normal component ${\bot \left( \nabla_{\mathbf{X}}^{\left( \mathcal{M} \right)} \mathbf{Y} \right)}$ denotes the orthogonal projection of the covariant derivative ${\nabla_{\mathbf{X}}^{\left( \mathcal{M} \right)} \mathbf{Y}}$ onto the normal bundle ${\bigsqcup\limits_{\mathbf{x} \in \mathcal{N}} T_{\mathbf{x}}^{\bot} \mathcal{N}}$. The induced Levi-Civita connection ${\nabla^{\left( \mathcal{N} \right)}}$ on the submanifold ${\left( \mathcal{N}, g_{\mathcal{N}} \right)}$ is then given by the orthogonal projection of the ambient Levi-Civita connection ${\nabla^{\left( \mathcal{M} \right)}}$ onto the tangent bundle:

\begin{equation}
\forall \mathbf{x} \in \mathcal{M}, \qquad \forall \mathbf{X} \in T_{\mathbf{x}} \mathcal{M}, \qquad \forall \mathbf{Y} \in \Gamma \left( \bigsqcup\limits_{\mathbf{x} \in \mathcal{M}} T_{\mathbf{x}} \mathcal{M} \right), \qquad \nabla_{\mathbf{X}}^{\left( \mathcal{N} \right)} \mathbf{Y} = \top \left( \nabla_{\mathbf{X}}^{\left( \mathcal{M} \right)} \mathbf{Y} \right),
\end{equation}
and the second fundamental form/extrinsic curvature tensor $K$ (a symmetric, vector-valued differential form taking values in the normal bundle ${\bigsqcup\limits_{\mathbf{x} \in \mathcal{N}} T_{\mathbf{x}}^{\bot} \mathcal{N}}$) is given by the orthogonal projection of the ambient Levi-Civita connection ${\nabla^{\left( \mathcal{M} \right)}}$ onto the normal bundle\cite{kobayashi2}:

\begin{equation}
\forall \mathbf{x} \in \mathcal{M}, \qquad \forall \mathbf{X} \in T_{\mathbf{x}} \mathcal{M}, \qquad \forall \mathbf{Y} \in \Gamma \left( \bigsqcup\limits_{\mathbf{x} \in \mathcal{M}} T_{\mathbf{x}} \mathcal{M} \right), \qquad K \left( \mathbf{X}, \mathbf{Y} \right) = \bot \left( \nabla_{\mathbf{X}}^{\left( \mathcal{M} \right)} \mathbf{Y} \right),
\end{equation}
i.e. one has, overall:

\begin{equation}
\forall \mathbf{x} \in \mathcal{M}, \qquad \forall \mathbf{X} \in T_{\mathbf{x}} \mathcal{M}, \qquad \forall \mathbf{Y} \in \Gamma \left( \bigsqcup\limits_{\mathbf{x} \in \mathcal{M}} T_{\mathbf{x}} \mathcal{M} \right), \qquad \nabla_{\mathbf{X}}^{\left( \mathcal{M} \right)} \mathbf{Y} = \nabla_{\mathbf{X}}^{\left( \mathcal{N} \right)} \mathbf{Y} + K \left( \mathbf{X}, \mathbf{Y} \right).
\end{equation}

\begin{figure}[ht]
\centering
\begin{framed}
\includegraphics[width=0.595\textwidth]{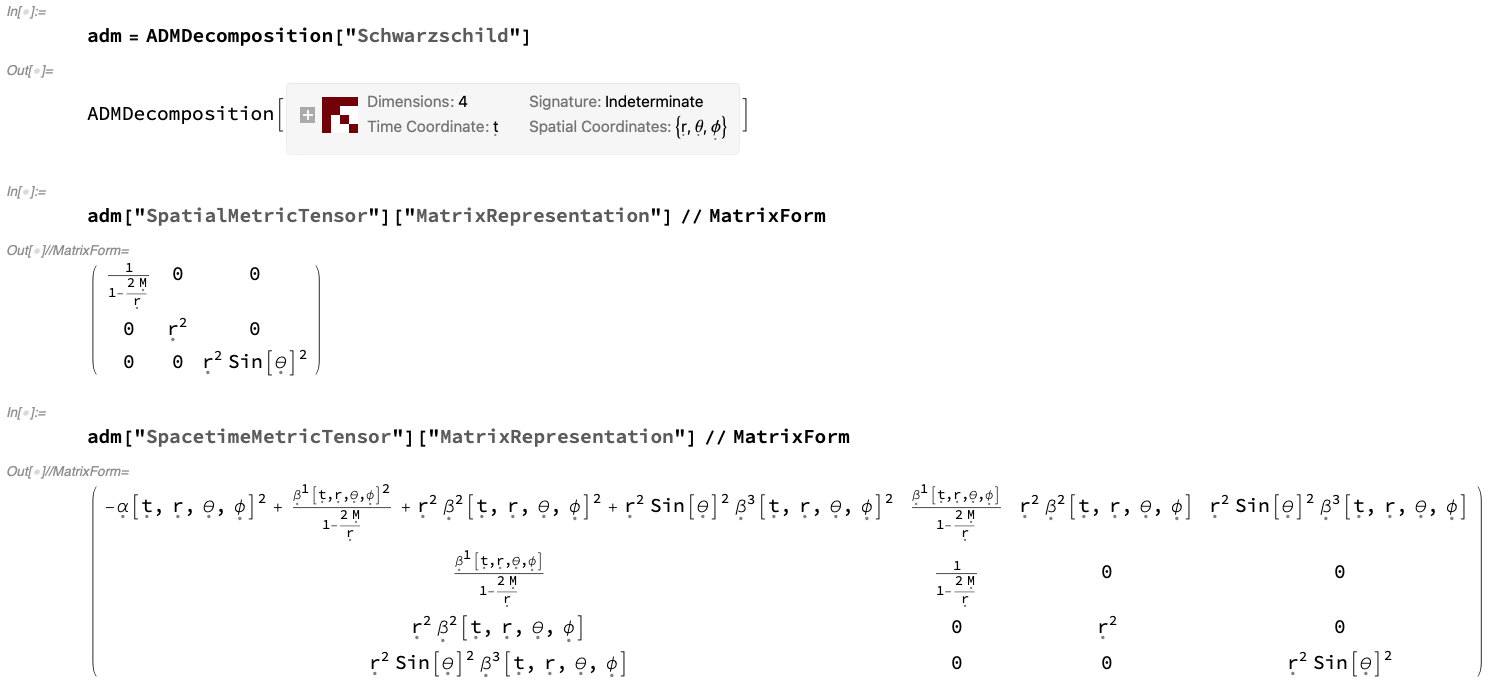}
\vrule
\includegraphics[width=0.395\textwidth]{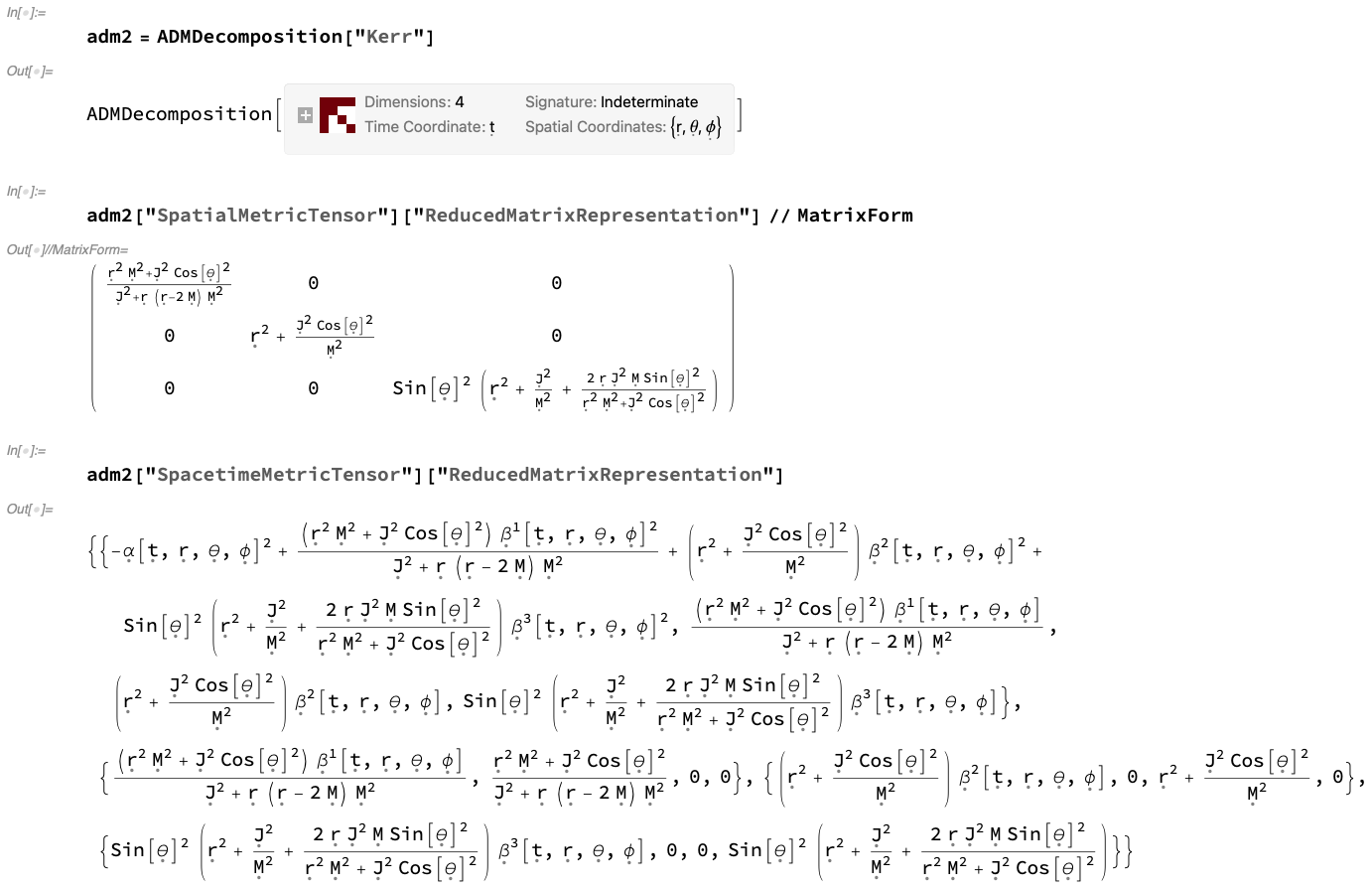}
\end{framed}
\caption{On the left, the \texttt{ADMDecomposition} object for a Schwarzschild geometry (representing, for instance, an uncharged, non-rotating black hole with mass $M$ in Schwarzschild or spherical polar coordinates ${\left( t, r, \theta, \phi \right)}$) with lapse function ${\alpha \left( t, r, \theta, \phi \right)}$ and shift vector ${\left( \beta^1 \left( t, r, \theta, \phi \right), \beta^2 \left( t, r, \theta, \phi \right), \beta^3 \left( t, r, \theta, \phi \right) \right)}$, represented by its induced (spatial) and ambient (spacetime) \texttt{MetricTensor} objects in explicit covariant matrix form. On the right, the \texttt{ADMDecomposition} object for a Kerr geometry (representing, for instance, an uncharged, spinning black hole with mass $M$ and angular momentum $J$ in Boyer-Lindquist or oblate spheroidal coordinates ${\left( t, r, \theta, \phi \right)}$) with lapse function ${\alpha \left( t, r, \theta, \phi \right)}$ and shift vector ${\left( \beta^1 \left( t, r, \theta, \phi \right), \beta^2 \left( t, r, \theta, \phi \right), \beta^3 \left( t, r, \theta, \phi \right) \right)}$, represented by its induced (spatial) and ambient (spacetime) \texttt{MetricTensor} objects in explicit covariant matrix form.}
\label{fig:Figure1}
\end{figure}

\begin{figure}[ht]
\centering
\begin{framed}
\includegraphics[width=0.595\textwidth]{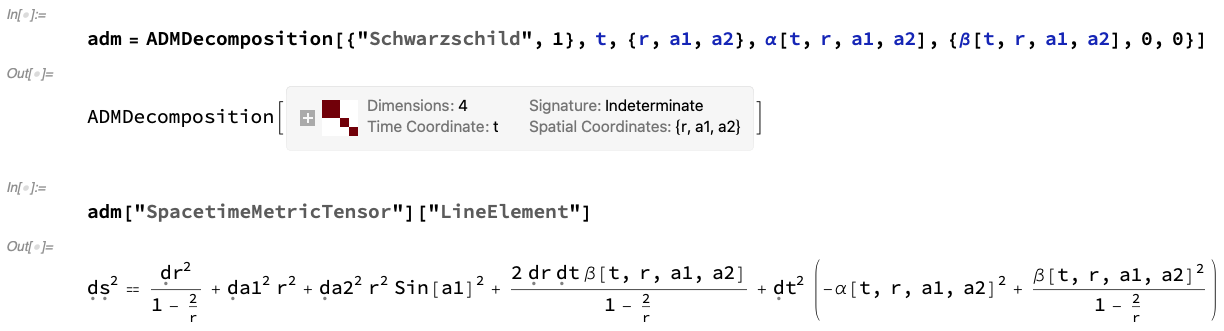}
\vrule
\includegraphics[width=0.395\textwidth]{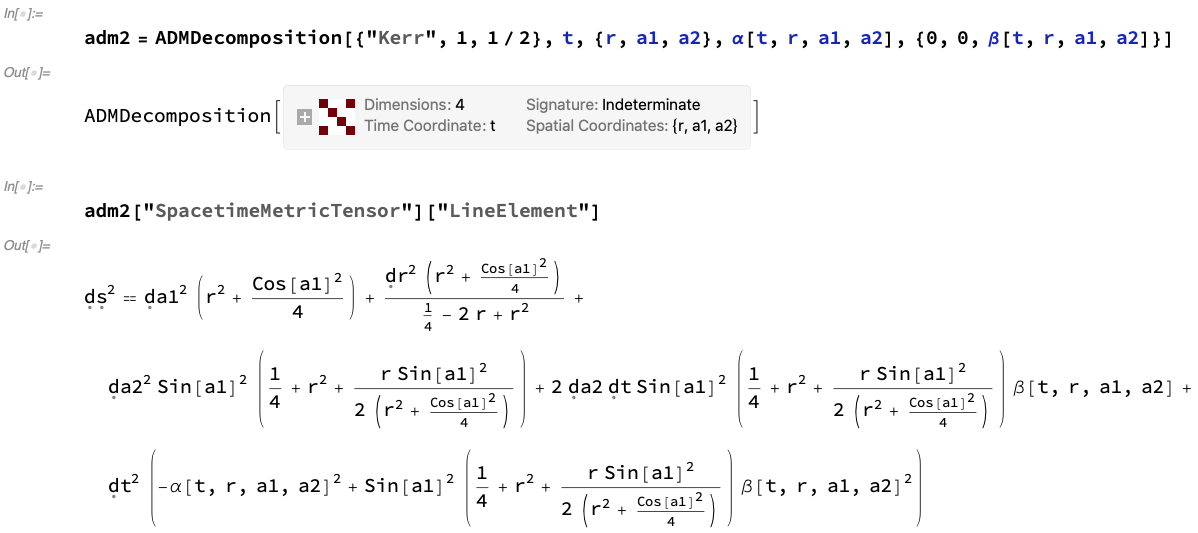}
\end{framed}
\caption{On the left, the \texttt{ADMDecomposition} object for a Schwarzschild geometry (representing, for instance, an uncharged, non-rotating black hole with numerical mass 1) in modified Schwarzschild or spherical polar coordinates ${\left( t, r, a1, a2 \right)}$, with lapse function ${\alpha \left( t, r, a1, a2 \right)}$ and modified shift vector ${\left( \beta \left( t, r, a1, a2 \right), 0, 0 \right)}$. On the right, the \texttt{ADMDecomposition} object for a Kerr geometry (representing, for instance, an uncharged, spinning black hole with numerical mass 1 and numerical angular momentum ${\frac{1}{2}}$) in modified Boyer-Lindquist or oblate spheroidal coordinates ${\left( t, r, a1, a2 \right)}$, with lapse function ${\alpha \left( t, r, a1, a2 \right)}$ and modified shift vector ${\left( 0, 0, \beta \left( t, r, a1, a2 \right) \right)}$.}
\label{fig:Figure2}
\end{figure}

Returning now to the specific case of general relativity, in which our ambient manifold ${\left( \mathcal{M}, g_{\mathcal{M}} \right)}$ is a Lorentzian spacetime, we proceed to define a universal ``time function'' ${t : \mathcal{M} \to \mathbb{R}}$ mapping each point/event to its corresponding coordinate time value; the submanifolds ${\left( N, g_{\mathcal{N}} \right)}$ are then given by the level surfaces of this universal time function, i.e:

\begin{equation}
\forall t_0 \in \mathbb{R}, \qquad \mathcal{N}_{t_0} = \left\lbrace \left. \mathbf{x} \in \mathcal{M} \right\vert t \left( \mathbf{x} \right) = t_0 \right\rbrace \subset \mathcal{M}.
\end{equation}
We thus obtain a time-ordered sequence of purely spacelike hypersurfaces (each of codimension-1, with purely Riemannian signature) ${\left( \mathcal{N}_{t_0}, g_{\mathcal{N}_{t_0}} \right)}$ which covers the entire ambient spacetime ${\left( \mathcal{M}, g_{\mathcal{M}} \right)}$:

\begin{equation}
\mathcal{M} = \bigcup_{t_0 \in \mathbb{R}} \mathcal{N}_{t_0}.
\end{equation}
If our ambient spacetime ${\left( \mathcal{M}, g_{\mathcal{M}} \right)}$ is, moreover, globally hyperbolic, then these spacelike hypersurfaces are also strictly non-intersecting, i.e:

\begin{equation}
\forall t_1, t_2 \in \mathbb{R}, \qquad \left( \mathcal{N}_{t_1} \cap \mathcal{N}_{t_2} \right) \neq \varnothing \qquad \Longleftrightarrow \qquad t_1 = t_2.
\end{equation}
Henceforth, we shall denote the overall metric tensor on the ambient spacetime manifold ${\mathcal{M}}$ (assumed to be of dimension $n$) by $g$, and the ``induced'' metric tensor on the spacelike hypersurfaces ${\mathcal{N}_{t_0}}$ (assumed to be of dimension ${n - 1}$) by ${\gamma}$. Concretely, \texttt{ADMDecomposition} represents such spacetime ``foliations'' by means of the ADM formalism first proposed by Arnowitt, Deser and Misner\cite{arnowitt}\cite{arnowitt2} (although the particular mathematical formulation of the ${3 + 1}$ decomposition that \textsc{Gravitas} uses is due to York\cite{york}), wherein one defines a metric tensor ${\gamma}$ on some ``initial'' spacelike hypersurface ${\left( \mathcal{N}_{t_0}, \gamma \right)}$, as well as the Lagrange multipliers ${\alpha}$ (a scalar field known as the \textit{lapse function}) and ${\beta^{\mu}}$ (an ${\left( n - 1 \right)}$-dimensional vector field known as the \textit{shift vector}, with ${\mu \in \left\lbrace 0, \dots, n - 2 \right\rbrace}$, yet in which every such ${\beta^{\mu}}$ depends upon all $n$ spacetime coordinates), which play the role of the gauge variables and consequently define how neighboring hypersurfaces are related to one another geometrically. More precisely, the lapse function ${\alpha}$ determines the proper time distance ${d \tau}$ between corresponding points on the spacelike hypersurfaces at coordinate time values ${t = t_0}$ and ${t = t_0 + d t}$, as measured along the unit normal direction ${\mathbf{n}}$ to the hypersurface at ${t = t_0}$ (i.e. the direction followed by normal observers):

\begin{equation}
d \tau \left( t_0, t_0 + d t \right) = \alpha d t,
\end{equation}
whereas the shift vector (field) ${\beta^{\mu}}$ determines how the spatial coordinates ${x^{\mu} \left( t_0 \right)}$ get relabeled as one moves from the spacelike hypersurface at coordinate time value ${t = t_0}$ to the spacelike hypersurface at coordinate time value ${t = t_0 + d t}$, i.e:

\begin{equation}
x^{\mu} \left( t_0 + d t \right) = x^{\mu} \left( t_0 \right) - \beta^{\mu} d t.
\end{equation}
The overall line element/first fundamental form ${d s^2}$ for the ambient (spacetime) \texttt{MetricTensor} object ${g_{\mu \nu}}$ is therefore given in terms of the components of the induced (spatial) \texttt{MetricTensor} object ${\gamma_{\mu \nu}}$, the lapse function ${\alpha}$ and the components of the shift vector (field) ${\beta^{\mu}}$ as:

\begin{multline}
d s^2 = - \alpha^2 d t^2 + \gamma_{\mu \nu} \left( d x^{\mu} + \beta^{\mu} d t \right) \left( d x^{\nu} + \beta^{\nu} d t \right)\\
= \left( - \alpha^2 + \gamma_{\mu \sigma} \beta^{\sigma} \beta^{\mu} \right) d t^2 + 2 \gamma_{\mu \sigma} \beta^{\sigma} d t d x^{\mu} + \gamma_{\mu \nu} d x^{\mu} d x^{\nu}\\
= \left( - \alpha^2 + \beta_{\mu} \beta^{\mu} \right) d t^2 + 2 \beta_{\mu} d t d x^{\mu} + \gamma_{\mu \nu} d x^{\mu} d x^{\nu},
\end{multline}
with ${\mu, \nu, \sigma}$ ranging across all ${\left\lbrace 0, \dots, n - 2 \right\rbrace}$ (i.e across spatial coordinate indices only). Note that the spatial line element/first fundamental form ${d l^2}$ for the induced (spatial) \texttt{MetricTensor} object ${\gamma_{\mu \nu}}$ constituting the initial spacelike hypersurface of the Schwarzschild metric\cite{schwarzschild}, as developed independently by Droste\cite{droste} (representing, for instance, an uncharged, non-rotating black hole with mass $M$ in Schwarzschild or spherical polar coordinates ${\left( t, r, \theta, \phi \right)}$) is given by:

\begin{equation}
d l^2 = \gamma_{\mu \nu} d x^{\mu} d x^{\nu} = \left( 1 - \frac{2 M}{r} \right)^{-1} d r ^2 + r^2 \left( d \theta^2 + \sin^2 \left( \theta \right) d \phi^2 \right),
\end{equation}
whereas, for the initial spacelike hypersurface of the Kerr metric\cite{kerr} (representing, for instance, an uncharged, spinning black hole with mass $M$ and angular momentum $J$ in Boyer-Lindquist\cite{boyer} or oblate spheroidal coordinates ${\left( t, r, \theta, \phi \right)}$), the spatial line element is given instead by:

\begin{multline}
d l^2 = \gamma_{\mu \nu} d x^{\mu} d x^{\nu} = \left( \frac{r^2 + \left( \frac{J}{M} \right)^2 \cos^2 \left( \theta \right)}{r^2 - 2 M r + \left( \frac{J}{M} \right)^2} \right) d r^2 + \left( r^2 + \left( \frac{J}{M} \right)^2 \cos^2 \left( \theta \right) \right) d \theta^2\\
+ \left( r^2 + \left( \frac{J}{M} \right)^2 + \frac{2 J^2 \sin^2 \left( \theta \right)}{M \left( r^2 + \left( \frac{J}{M} \right)^2 \cos^2 \left( \theta \right) \right)} \right) \sin^2 \left( \theta \right) d \phi^2,
\end{multline}
with ${\mu, \nu}$ ranging across all ${\left\lbrace 0, \dots, n - 2 \right\rbrace}$ (i.e. across spatial coordinate indices only) in both cases. The corresponding \texttt{ADMDecomposition} objects, including representations of both the induced/spatial and ambient/spacetime \texttt{MetricTensor} objects, for the Schwarzschild and Kerr metrics are shown in Figure \ref{fig:Figure1}, assuming the most general choice of gauge consisting of the lapse function ${\alpha \left( t,r, \theta, \phi \right)}$ and the shift vector (field) ${\left( \beta^1 \left( t, r, \theta, \phi \right), \beta^2 \left( t, r, \theta, \phi \right), \beta^3 \left( t, r, \theta, \phi \right) \right)}$. By default, the most general forms of the ADM gauge variables are always assumed, and appropriate formal symbols are chosen for the various parameters of the decomposition (such as mass $M$ and angular momentum $J$), as well as for the coordinates (such as time coordinate $t$ and radial coordinate $r$), although these defaults may always be overridden by passing additional arguments to \texttt{ADMDecomposition}, as illustrated in Figure \ref{fig:Figure2} for the case of modified shift vectors ${\left( \beta \left( t, r, \theta, \phi \right), 0, 0 \right)}$ (for Schwarzschild) and ${\left( 0, 0, \beta \left( t, r, \theta, \phi \right) \right)}$ (for Kerr).

\begin{figure}[ht]
\centering
\begin{framed}
\includegraphics[width=0.595\textwidth]{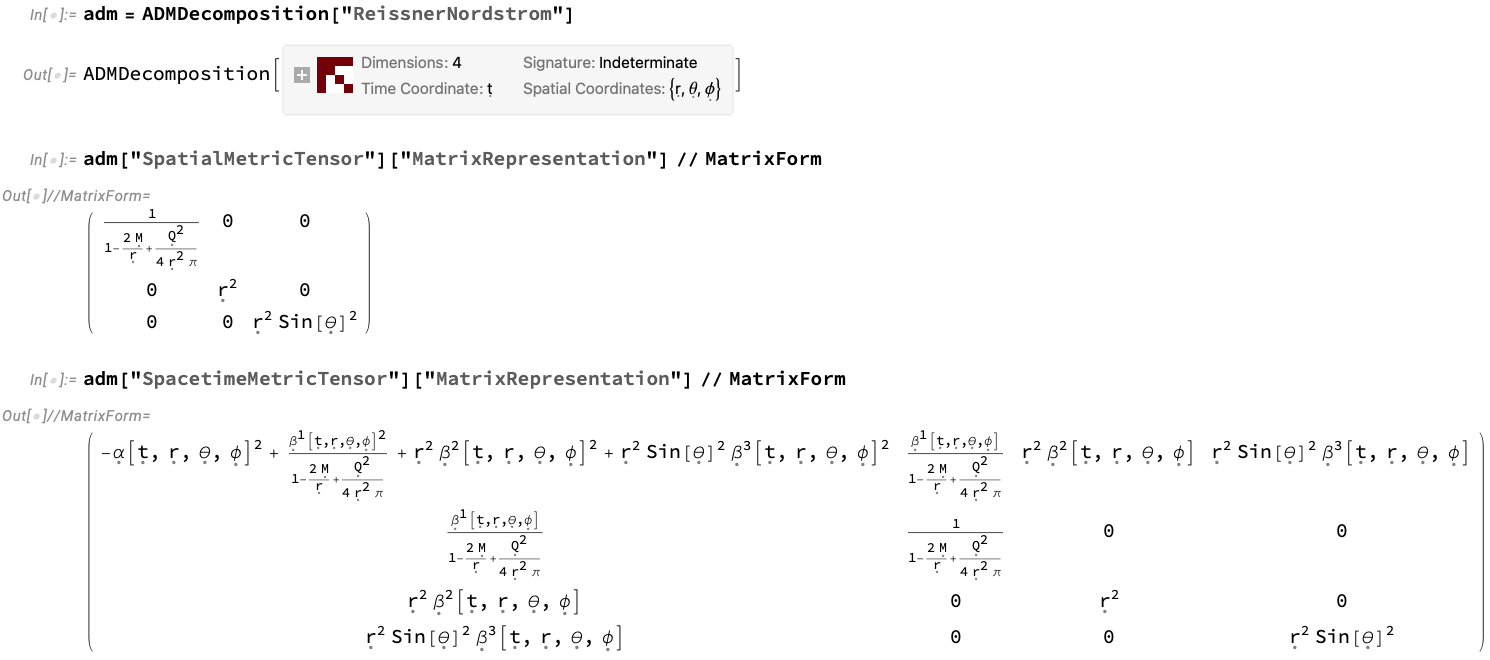}
\vrule
\includegraphics[width=0.395\textwidth]{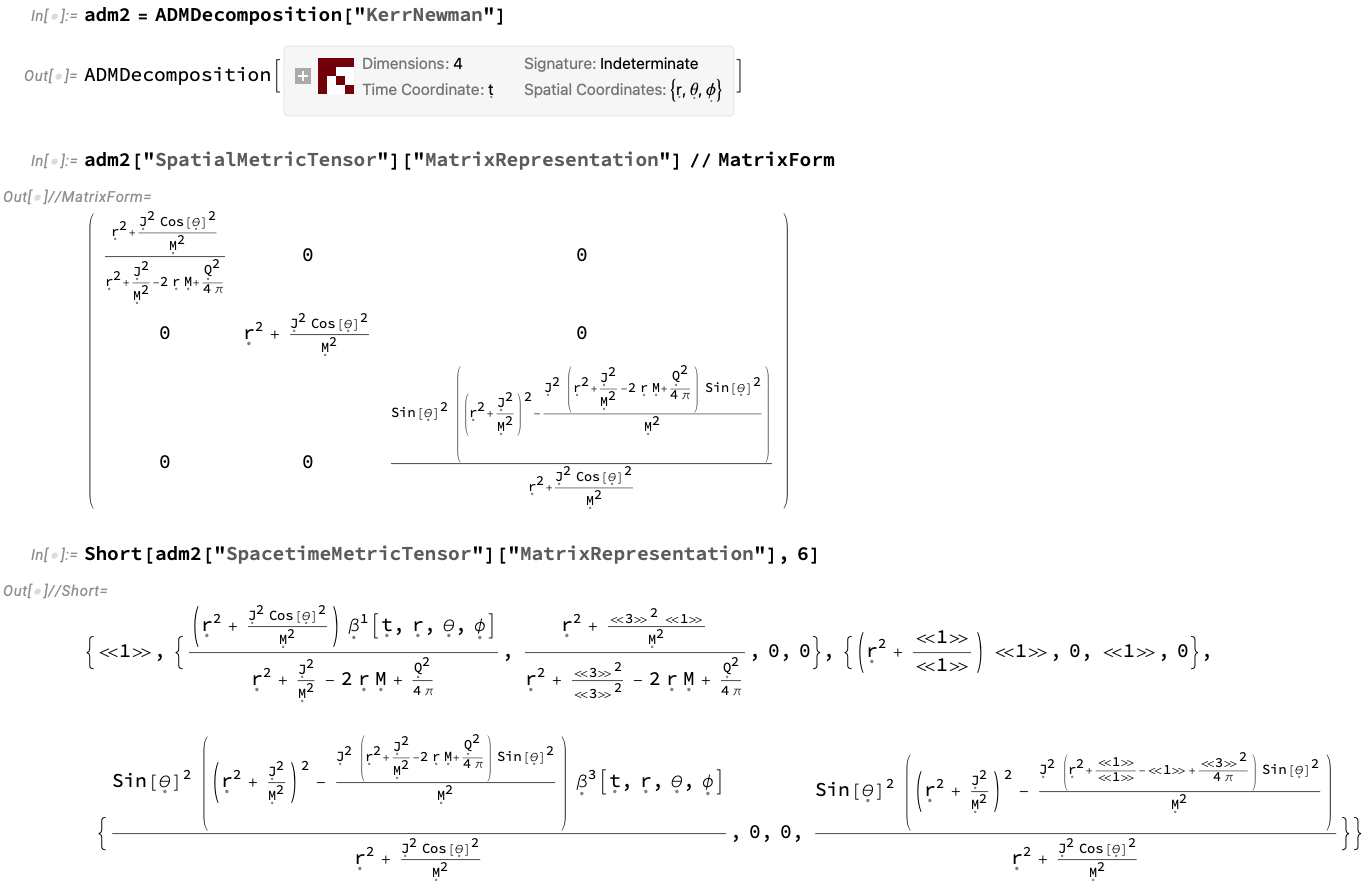}
\end{framed}
\caption{On the left, the \texttt{ADMDecomposition} object for a Reissner-Nordstr\"om geometry (representing, for instance, a charged, non-rotating black hole with mass $M$ and electric charge $Q$ in Schwarzschild or spherical polar coordinates ${\left( t, r, \theta, \phi \right)}$) with lapse function ${\alpha \left( t, r, \theta, \phi \right)}$ and shift vector ${\left( \beta^1 \left( t, r, \theta, \phi \right), \beta^2 \left( t, r, \theta, \phi \right), \beta^3 \left( t, r, \theta, \phi \right) \right)}$. On the right, the \texttt{ADMDecomposition} object for a Kerr-Newman geometry (representing, for instance, a charged, spinning black hole with mass $M$, angular momentum $J$ and electric charge $Q$ in Boyer-Lindquist or oblate spheroidal coordinates ${\left( t, r, \theta, \phi \right)}$) with lapse function ${\alpha \left( t, r, \theta, \phi \right)}$ and shift vector ${\left( \beta^1 \left( t, r, \theta, \phi \right), \beta^2 \left( t, r, \theta, \phi \right), \beta^3 \left( t, r, \theta, \phi \right) \right)}$.}
\label{fig:Figure44}
\end{figure}

\begin{figure}[ht]
\centering
\begin{framed}
\includegraphics[width=0.595\textwidth]{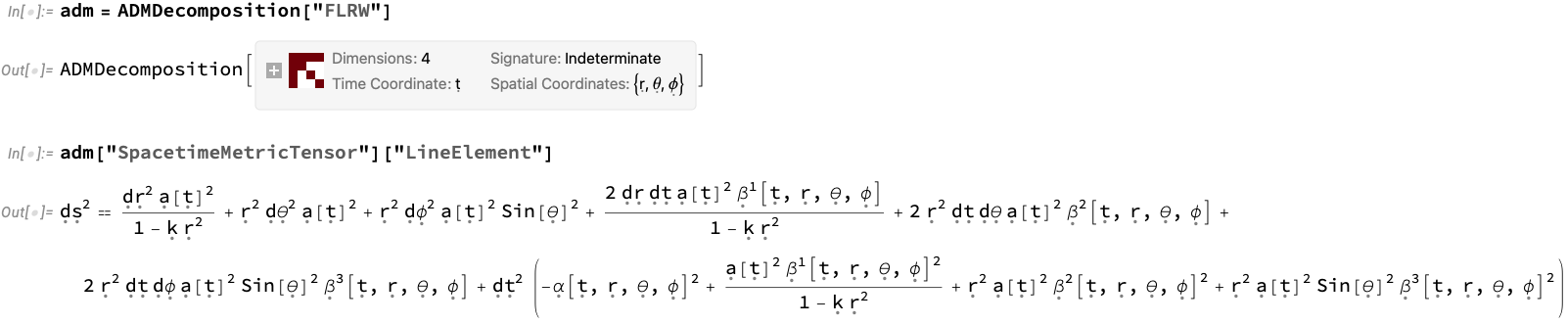}
\vrule
\includegraphics[width=0.395\textwidth]{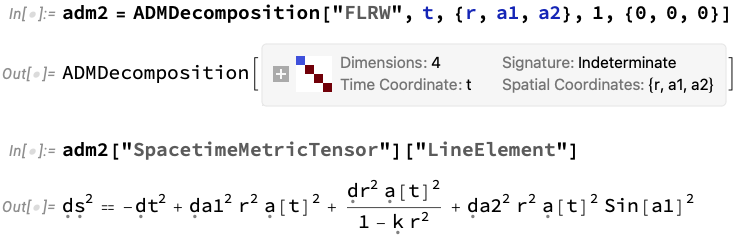}
\end{framed}
\caption{On the left, the \texttt{ADMDecomposition} object for a Friedmann-Lema\^itre-Robertson-Walker or FLRW geometry (representing, for instance, a perfectly homogeneous and isotropic universe with scale factor ${a \left( t \right)}$ and curvature $k$ in spherical polar coordinates ${\left( t, r, \theta, \phi \right)}$) with lapse function ${\alpha \left( t, r, \theta, \phi \right)}$ and shift vector ${\left( \beta^1 \left( t, r, \theta, \phi \right), \beta^2 \left( t, r, \theta, \phi \right), \beta^3 \left( t, r, \theta, \phi \right) \right)}$. On the right, the \texttt{ADMDecomposition} object for a Friedmann-Lema\^itre-Robertson-Walker or FLRW geometry (representing, for instance, a perfectly homogeneous and isotropic universe with scale factor ${a \left( t \right)}$ and curvature $k$ in modified spherical polar coordinates ${\left( t, r, a1, a2 \right)}$) with trivial lapse function $1$ and trivial shift vector ${\left( 0, 0, 0 \right)}$.}
\label{fig:Figure45}
\end{figure}

Note that several other Cauchy surface geometries are built into \texttt{ADMDecomposition}'s small (though continually growing) library of initial spacelike hypersurface configurations, including Cauchy initial data for the Reissner-Nordstr\"om metric\cite{reissner}\cite{nordstrom}, as developed independently by Weyl\cite{weyl} and Jeffery\cite{jeffery} (representing, for instance, a charged, non-rotating black hole with mass $M$ and electric charge $Q$ in Schwarzschild or spherical polar coordinates ${\left( t, r, \theta, \phi \right)})$ and the Kerr-Newman metric\cite{newman}\cite{newman2} (representing, for instance, a charged, spinning black hole with mass $M$, angular momentum $J$ and electric charge $Q$ in Boyer-Lindquist or oblate spheroidal coordinates ${\left( t, r, \theta, \phi \right)}$), with spatial line elements given by:

\begin{equation}
d l^2 = \gamma_{\mu \nu} d x^{\mu} d x^{\nu} = \left( 1 - \frac{2 M}{r} + \frac{Q^2}{4 \pi r^2} \right)^{-1} d r^2 + r^2 \left( d \theta^2 + \sin^2 \left( \theta \right) d \phi^2 \right),
\end{equation}
and:

\begin{multline}
d l^2 = \gamma_{\mu \nu} d x^{\mu} d x^{\nu} = - \left( \frac{ d r^2}{r^2 - 2 M + \left( \frac{J}{M} \right)^2 + \frac{Q^2}{4 \pi}} + d \theta^2 \right) \left( r^2 + \left( \frac{J}{M} \right)^2 \cos^2 \left( \theta \right) \right)\\
+ \left( \frac{J}{M} \right)^2 \sin^4 \left( \theta \right) d \phi^2 \left( \frac{r^2 - 2 M + \left( \frac{J}{M} \right)^2 + \frac{Q^2}{4 \pi}}{r^2 + \left( \frac{J}{M} \right)^2 \cos^2 \left( \theta \right)} \right) - \left( r^2 + \left( \frac{J}{M} \right)^2 \right)^2 d \phi^2 \left( \frac{\sin^2 \left( \theta \right)}{r^2 + \left( \frac{J}{M} \right)^2 \cos^2 \left( \theta \right)} \right),
\end{multline}
respectively, as shown in Figure \ref{fig:Figure44}. Cosmological Cauchy initial data is also supported, such as for the Friedmann-Lema\^itre-Robertson-Walker or FLRW metric\cite{friedmann}\cite{lemaitre}\cite{robertson}\cite{walker} (representing, for instance, a perfectly homogeneous and isotropic universe with scale factor ${a \left( t \right)}$ and curvature $k$ in spherical polar coordinates ${\left( t, r, \theta, \phi \right)}$), with spatial line element given by:

\begin{equation}
d l^2 = \gamma_{\mu \nu} d x^{\mu} d x^{\nu} = a^2 \left( t \right) \left( \frac{d r^2}{1 - k r^2} + r^2 \left( d \theta^2 + \sin^2 \left( \theta \right) d \phi^2 \right) \right),
\end{equation}
as shown in Figure \ref{fig:Figure45}, both for the default choice of gauge with general lapse function ${\alpha \left( t, r, \theta, \phi \right)}$ and general shift vector (field) ${\left( \beta^1 \left( t, r, \theta, \phi \right), \beta^2 \left( t, r, \theta, \phi \right), \beta^3 \left( t, r, \theta, \phi \right) \right)}$, and for the trivial choice of gauge with lapse function identically equal to 1 (i.e. ${\alpha \left( t, r, \theta, \phi \right) = 1}$) and shift vector (field) identically equal to zero (i.e. ${\boldsymbol \beta \left( t, r, \theta, \phi \right) = \left( 0, 0, 0 \right)}$). In the above, as before, ${\mu}$ and ${\nu}$ represent spatial indices, and therefore range across ${\left\lbrace 0, \dots, n - 2 \right\rbrace}$ only.

\begin{figure}[ht]
\centering
\begin{framed}
\includegraphics[width=0.645\textwidth]{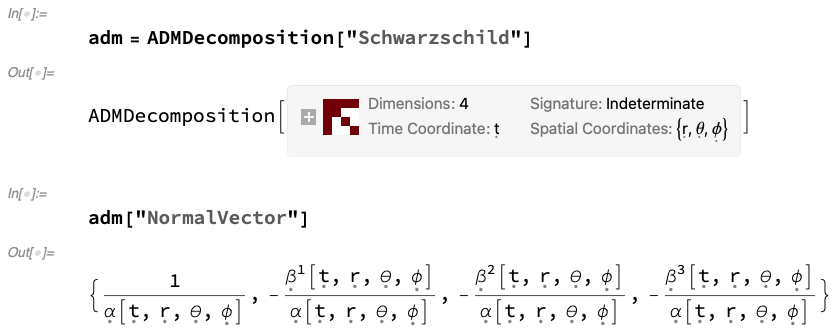}
\vrule
\includegraphics[width=0.345\textwidth]{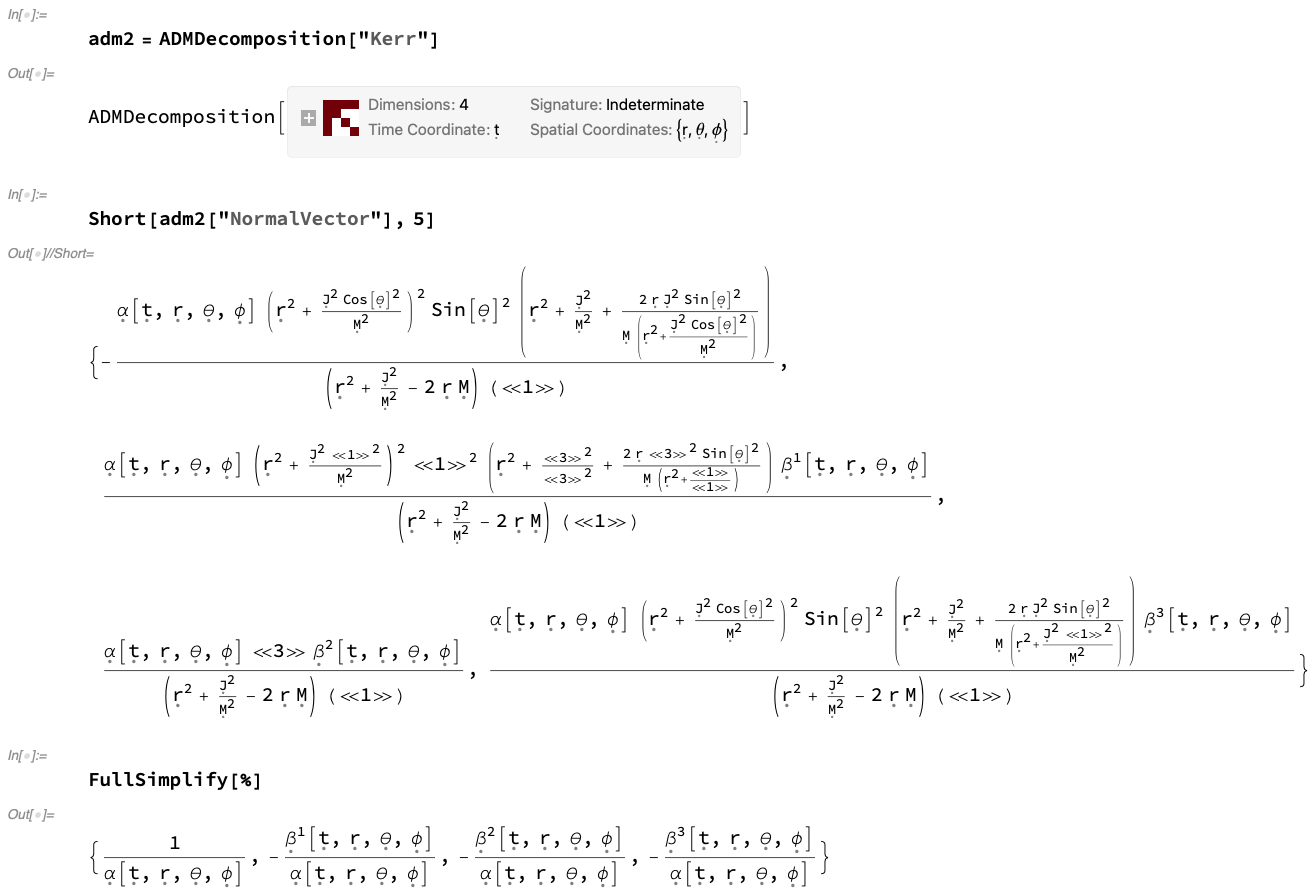}
\end{framed}
\caption{On the left, the future-pointing, timelike unit vector normal to each spacelike hypersurface of the \texttt{ADMDecomposition} object for a Schwarzschild geometry (representing, for instance, an uncharged, non-rotating black hole with mass $M$ in Schwarzschild or spherical polar coordinates ${\left( t, r, \theta, \phi \right)}$) with lapse function ${\alpha \left( t, r, \theta, \phi \right)}$ and shift vector ${\left( \beta^1 \left( t, r, \theta, \phi \right), \beta^2 \left( t, r, \theta, \phi \right), \beta^3 \left( t, r, \theta, \phi \right) \right)}$. On the right, the future-pointing, timelike unit vector normal to each spacelike hypersurface of the \texttt{ADMDecomposition} object for a Kerr geometry (representing, for instance, an uncharged, spinning black hole with mass $M$ and angular momentum $J$ in Boyer-Lindquist or oblate spheroidal coordinates ${\left( t, r, \theta, \phi \right)}$) with lapse function ${\alpha \left( t, r, \theta, \phi \right)}$ and shift vector ${\left( \beta^1 \left( t, r, \theta, \phi \right), \beta^2 \left( t, r, \theta, \phi \right), \beta^3 \left( t, r, \theta, \phi \right) \right)}$.}
\label{fig:Figure3}
\end{figure}

\begin{figure}[ht]
\centering
\begin{framed}
\includegraphics[width=0.645\textwidth]{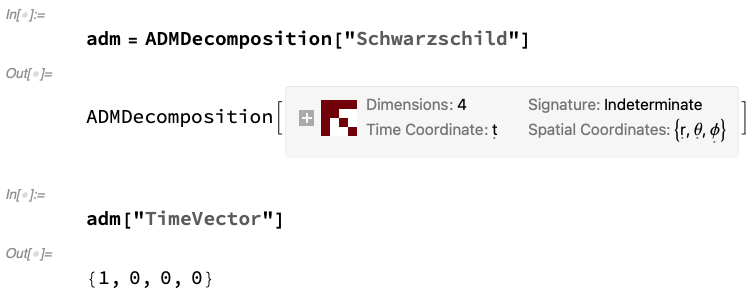}
\vrule
\includegraphics[width=0.345\textwidth]{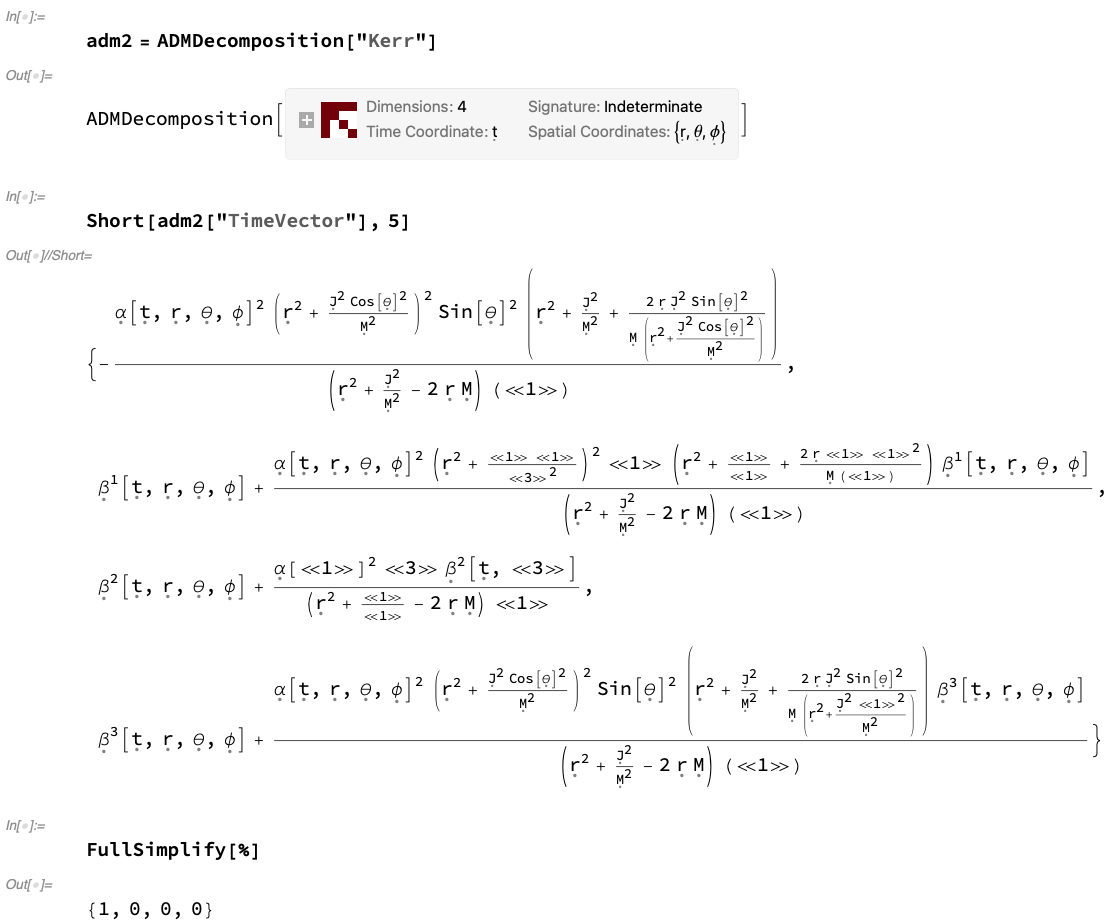}
\end{framed}
\caption{On the left, the future-pointing unit ``time vector'' for each spacelike hypersurface of the \texttt{ADMDecomposition} object for a Schwarzschild geometry (representing, for instance, an uncharged, non-rotating black hole with mass $M$ in Schwarzschild or spherical polar coordinates ${\left( t, r, \theta, \phi \right)}$) with lapse function ${\alpha \left( t, r, \theta, \phi \right)}$ and shift vector ${\left( \beta^1 \left( t, r, \theta, \phi \right), \beta^2 \left( t, r, \theta, \phi \right), \beta^3 \left( t, r, \theta, \phi \right) \right)}$. On the right, the future-pointing unit ``time vector'' for each spacelike hypersurface of the \texttt{ADMDecomposition} object for a Kerr geometry (representing, for instance, an uncharged, spinning black hole with mass $M$ and angular momentum $J$ in Boyer-Lindquist or oblate spheroidal coordinates ${\left( t, r, \theta, \phi \right)}$) with lapse function ${\alpha \left( t, r, \theta, \phi \right)}$ and shift vector ${\left( \beta^1 \left( t, r, \theta, \phi \right), \beta^2 \left( t, r, \theta, \phi \right), \beta^3 \left( t, r, \theta, \phi \right) \right)}$.}
\label{fig:Figure4}
\end{figure}

\begin{figure}[ht]
\centering
\begin{framed}
\includegraphics[width=0.545\textwidth]{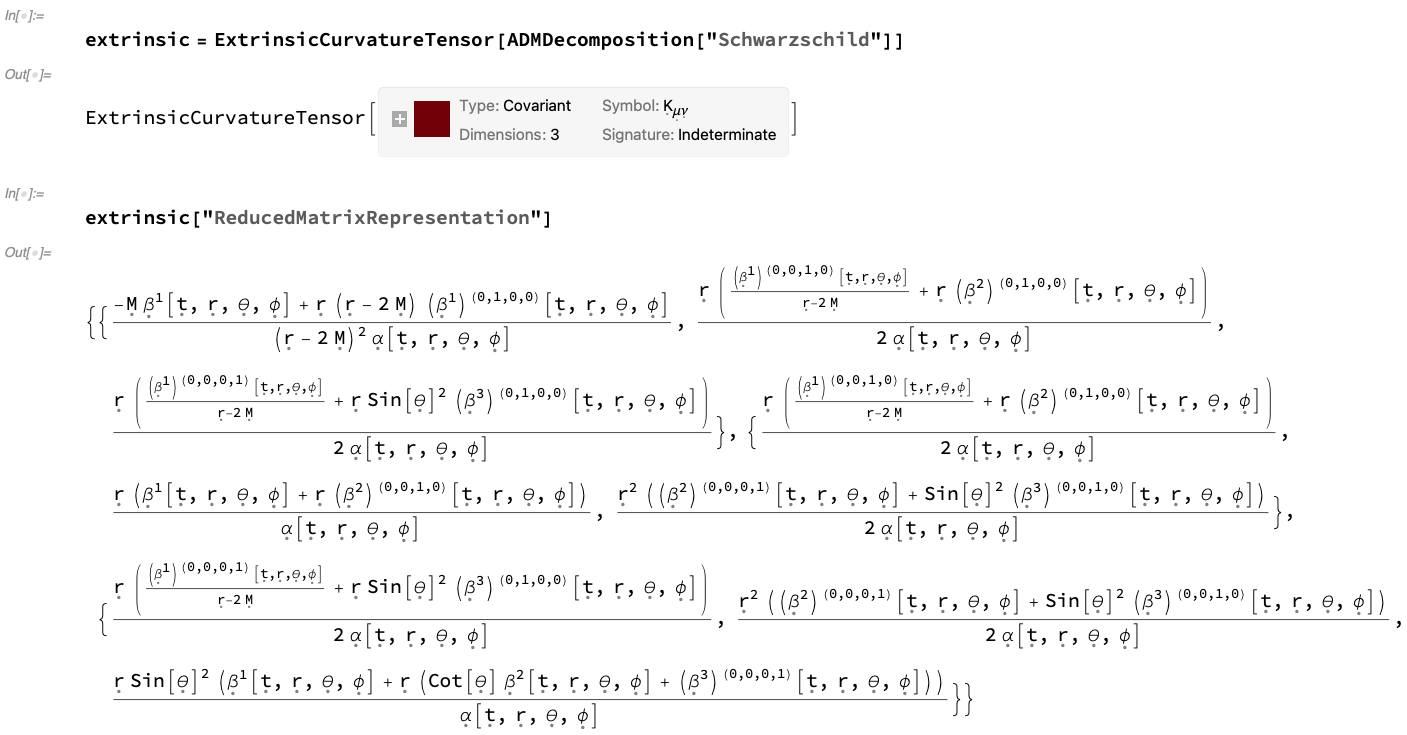}
\vrule
\includegraphics[width=0.445\textwidth]{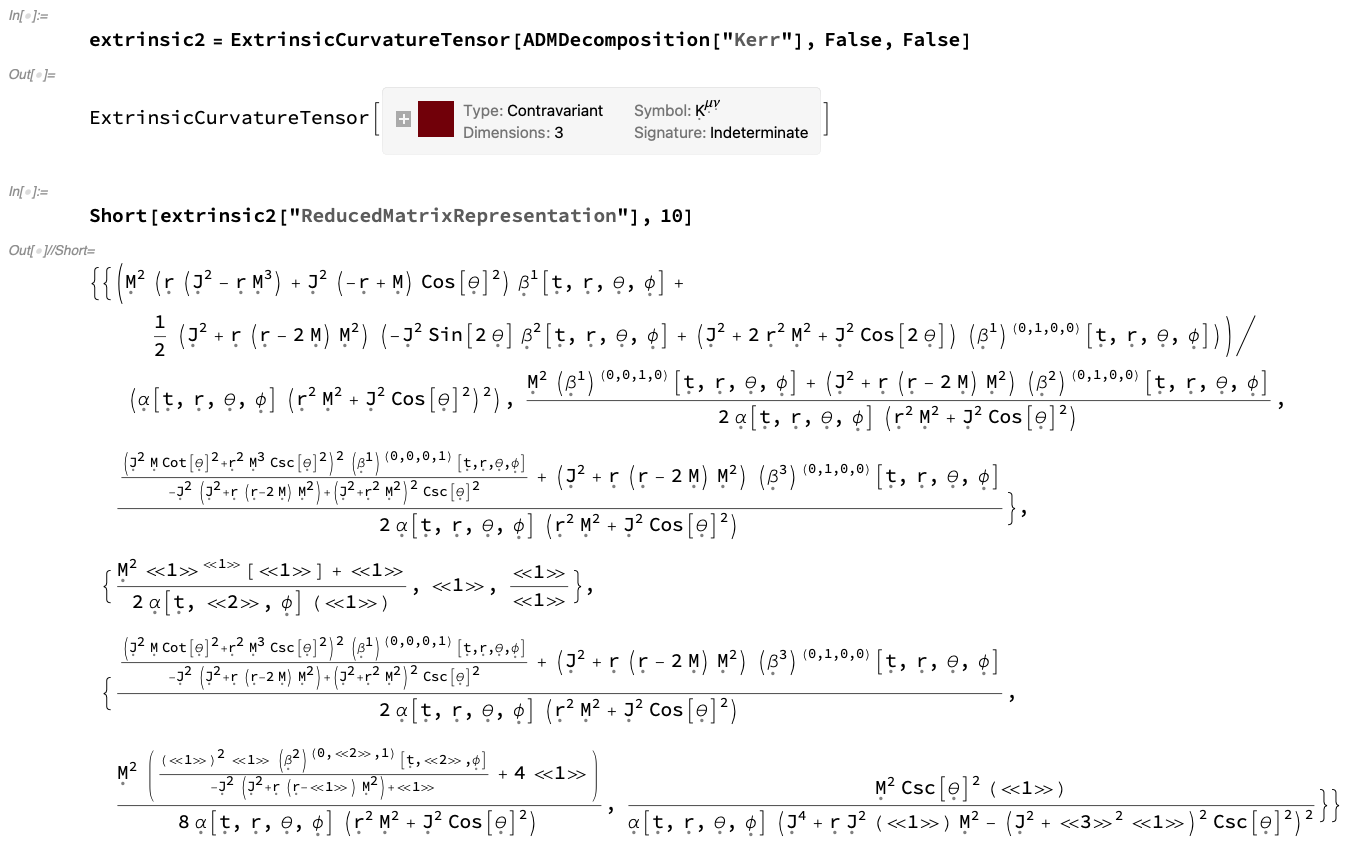}
\end{framed}
\caption{On the left, the \texttt{ExtrinsicCurvatureTensor} object computed from the \texttt{ADMDecomposition} object for a Schwarzschild geometry (representing, for instance, an uncharged, non-rotating black hole with mass $M$ in Schwarzschild or spherical polar coordinates ${\left( t, r, \theta, \phi \right)}$) with lapse function ${\alpha \left( t, r, \theta, \phi \right)}$ and shift vector ${\left( \beta^1 \left( t, r, \theta, \phi \right), \beta^2 \left( t, r, \theta, \phi \right), \beta^3 \left( t, r, \theta, \phi \right) \right)}$, in explicit covariant matrix form, with both indices lowered/covariant (default). On the right, the \texttt{ExtrinsicCurvatureTensor} object computed from the \texttt{ADMDecomposition} object for a Kerr geometry (representing, for instance, an uncharged, spinning black hole with mass $M$ and angular momentum $J$ in Boyer-Lindquist or oblate spheroidal coordinates ${\left( t, r, \theta, \phi \right)}$) with lapse function ${\alpha \left( t, r, \theta, \phi \right)}$ and shift vector ${\left( \beta^1 \left( t, r, \theta, \phi \right), \beta^2 \left( t, r, \theta, \phi \right), \beta^3 \left( t, r, \theta, \phi \right) \right)}$, in explicit contravariant matrix form, with both indices raised/contravariant.}
\label{fig:Figure5}
\end{figure}

\begin{figure}[ht]
\centering
\begin{framed}
\includegraphics[width=0.595\textwidth]{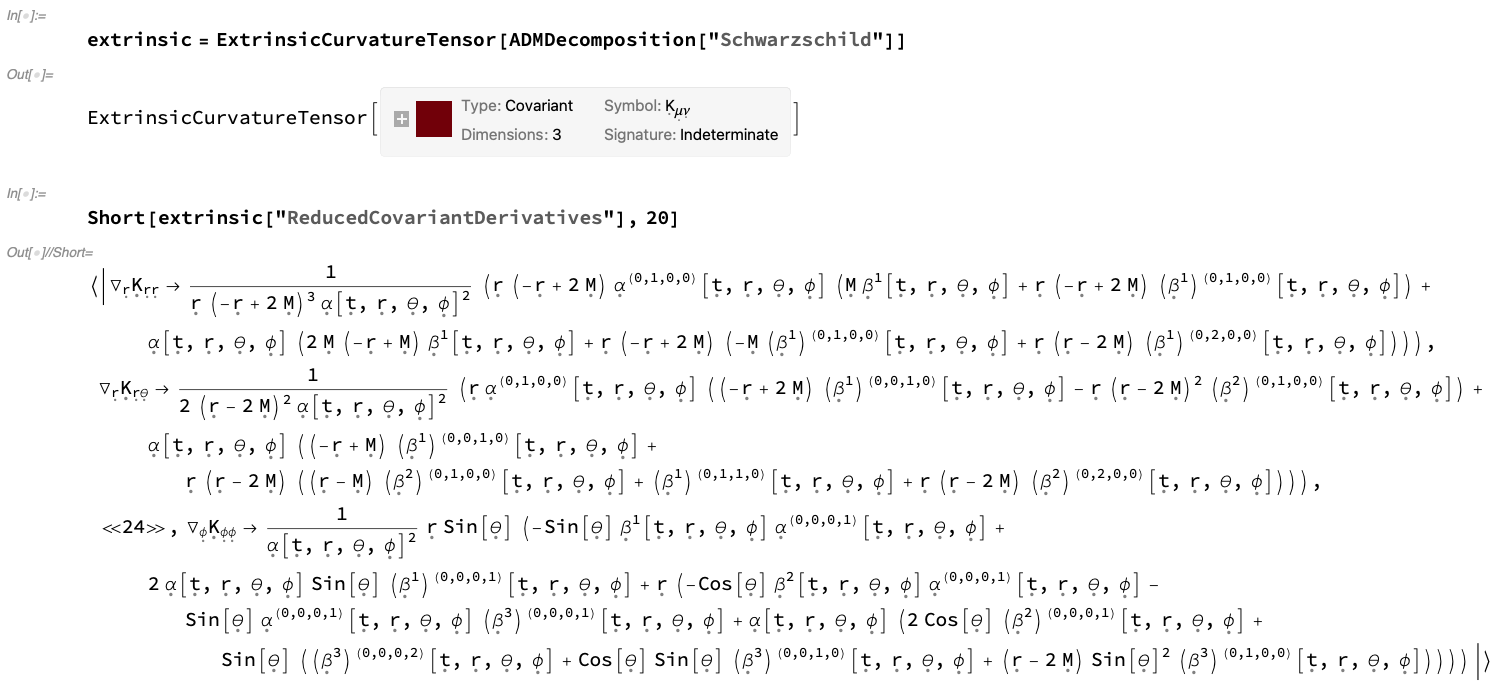}
\vrule
\includegraphics[width=0.395\textwidth]{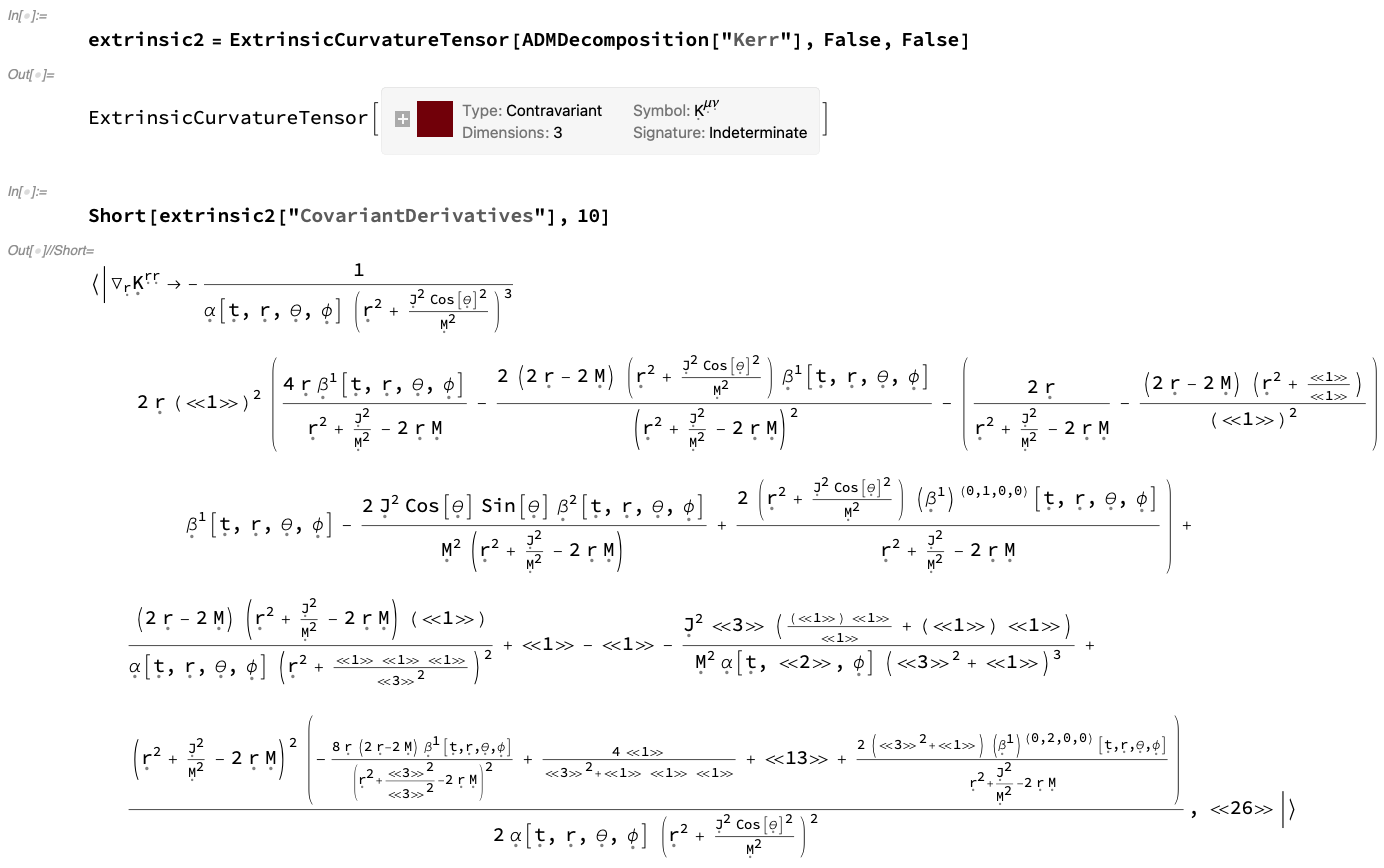}
\end{framed}
\caption{On the left, the association of all covariant derivatives of the \texttt{ExtrinsicCurvatureTensor} object computed from the \texttt{ADMDecomposition} object for a Schwarzschild geometry (representing, for instance, an uncharged, non-rotating black hole with mass $M$ in Schwarzschild or spherical polar coordinates ${\left( t, r, \theta, \phi \right)}$) with lapse function ${\alpha \left( t, r, \theta, \phi \right)}$ and shift vector ${\left( \beta^1 \left( t, r, \theta, \phi \right), \beta^2 \left( t, r, \theta, \phi \right), \beta^3 \left( t, r, \theta, \phi \right) \right)}$, with both indices lowered/covariant (default). On the right, the association of all covariant derivatives of the \texttt{ExtrinsicCurvatureTensor} object computed from the \texttt{ADMDecomposition} object for a Kerr geometry (representing, for instance, an uncharged, spinning black hole with mass $M$ and angular momentum $J$ in Boyer-Lindquist or oblate spheroidal coordinates ${\left( t, r, \theta, \phi \right)}$) with lapse function ${\alpha \left( t, r, \theta, \phi \right)}$ and shift vector ${\left( \beta^1 \left( t, r, \theta, \phi \right), \beta^2 \left( t, r, \theta, \phi \right), \beta^3 \left( t, r, \theta, \phi \right) \right)}$, with both indices raised/contravariant.}
\label{fig:Figure6}
\end{figure}

In all that follows, we will use a bracketed ``3'' to indicate that a given object is restricted to spacelike hypersurfaces, with a bracketed ``4'' indicating instead that a given object is defined over the entire ambient spacetime (for instance, ${{}^{\left( 3 \right)} \nabla_{\mu}}$ denotes the covariant derivative operator on spacelike hypersurfaces, whose corresponding Christoffel symbols ${{}^{\left( 3 \right)} \Gamma_{\mu \nu}^{\rho}}$ are defined in terms of derivatives of the induced/spatial metric tensor ${\gamma_{\mu \nu}}$, while ${{}^{\left( 4 \right)} \nabla_{\mu}}$ denotes the covariant derivative operator on the ambient spacetime, whose corresponding Christoffel symbols ${{}^{\left( 4 \right)} \Gamma_{\mu \nu}^{\rho}}$ are defined in terms of derivatives of the overall spacetime metric tensor ${g_{\mu \nu}}$); note that we use these particular numbers purely for illustrative purposes (since the formalism is often referred to as ``${3 + 1}$ formalism'') - the objects themselves are defined within \textsc{Gravitas} for arbitrary $n$-dimensional spacetimes, which are then foliated into ${\left( n - 1 \right)}$-dimensional spacelike hypersurfaces. The future-pointing, timelike unit vector ${\mathbf{n}}$ that is normal to each spacelike hypersurface can now be computed as the spacetime contravariant derivative ${{}^{\left( 4 \right)} \nabla^{\mu}}$ of the distinguished time coordinate $t$\cite{gourgoulhon}:

\begin{equation}
n^{\mu} = - \alpha {}^{\left( 4 \right)} \nabla^{\mu} t = - \alpha g^{\mu \sigma} {}^{\left( 4 \right)} \nabla_{\sigma} t = - \alpha g^{\mu \sigma} \frac{\partial}{\partial x^{\sigma}} \left( t \right),
\end{equation}
whereas the ``time vector'' ${\mathbf{t}}$ determines how corresponding points on neighboring hypersurfaces are related (and so, in particular, will differ from the normal vector ${\mathbf{n}}$ if the corresponding observer does not move in a normal direction to the hypersurfaces, which in turn corresponds to the case of a non-vanishing shift vector ${\boldsymbol\beta \neq \mathbf{0}}$):
\begin{equation}
t^{\mu} = \alpha n^{\mu} + \beta^{\mu} = - \alpha^2 g^{\mu \sigma} \frac{\partial}{\partial x^{\sigma}} \left( t \right) + \beta^{\mu}.
\end{equation}
In both of the above, ${\mu, \sigma}$ range across all ${\left\lbrace 0, \dots, n - 1 \right\rbrace}$ (i.e. across all spacetime coordinate indices). Figures \ref{fig:Figure3} and \ref{fig:Figure4} show the components of the future-pointing unit normal vector ${n^{\mu}}$ and of the future-pointing unit ``time vector'' ${t^{\mu}}$, computed directly from the respective \texttt{ADMDecomposition} objects for the Schwarzschild and Kerr metrics, assuming a fully generic choice of gauge with lapse function ${\alpha \left( t, r, \theta, \phi \right)}$ and shift vector ${\left( \beta^1 \left( t, r, \theta, \phi \right), \beta^2 \left( t, r, \theta, \phi \right), \beta^3 \left( t, r, \theta, \phi \right) \right)}$ in both cases. Recall from the discussion above that the second fundamental form/extrinsic curvature tensor $K$ is simply the symmetric, vector-valued differential form given by the orthogonal projection ${\bot}$ of the ambient/spacetime Levi-Civita connection ${{}^{\left( 4 \right)} \nabla}$ onto the normal bundle; we can express this in component form by taking the Lie derivative ${\mathcal{L}}$ of the spatial metric tensor ${\gamma_{\mu \nu}}$ in the normal direction ${\mathbf{n}}$\cite{alcubierre}:

\begin{equation}
K_{\mu \nu} = - \frac{1}{2} \mathcal{L}_{\mathbf{n}} \gamma_{\mu \nu},
\end{equation}
i.e., in expanded form:

\begin{multline}
K_{\mu \nu} = \frac{1}{2 \alpha} \left( {}^{\left( 3 \right)} \nabla_{\nu} \beta_{\mu} + {}^{\left( 3 \right)} \nabla_{\mu} \beta_{\nu} - \frac{\partial}{\partial t} \left( \gamma_{\mu \nu} \right) \right)\\
= \frac{1}{2 \alpha} \left( \frac{\partial}{\partial x^{\nu}} \left( \beta_{\mu} \right) - {}^{\left( 3 \right)} \Gamma_{\nu \mu}^{\sigma} \beta_{\sigma} + \frac{\partial}{\partial x^{\mu}} \left( \beta_{\nu} \right) - {}^{\left( 3 \right)} \Gamma_{\mu \nu}^{\sigma} \beta_{\sigma} - \frac{\partial}{\partial t} \left( \gamma_{\mu \nu} \right) \right),
\end{multline}
where ${\mu, \nu, \sigma}$ range across all ${\left\lbrace 0, \dots, n - 2 \right\rbrace}$ (i.e. across spatial coordinate indices only), as shown in Figure \ref{fig:Figure5} for the case of \texttt{ExtrinsicCurvatureTensor} objects computed directly from the \texttt{ADMDecomposition} objects for the Schwarzschild and Kerr metrics, assuming again a fully generic choice of gauge with lapse function ${\alpha \left( t, r, \theta, \phi \right)}$ and shift vector ${\left( \beta^1 \left( t, r, \theta, \phi \right), \beta^2 \left( t, r, \theta, \phi \right), \beta^3 \left( t, r, \theta, \phi \right) \right)}$. Note that both the shift vector ${\beta^{\mu}}$ and the extrinsic curvature tensor ${K_{\mu \nu}}$ are purely spatial objects, and so their indices are raised and lowered using the spatial metric tensor ${\gamma_{\mu \nu}}$ rather than full spacetime metric tensor ${g_{\mu \nu}}$, i.e:

\begin{equation}
\beta^{\mu} = \gamma^{\mu \sigma} \beta_{\sigma}, \qquad \text{ and } \qquad \beta_{\mu} = \gamma_{\mu \sigma} \beta^{\sigma},
\end{equation}
for the case of the rank-1 shift vector/covector (field) ${\boldsymbol\beta}$, and:

\begin{equation}
K^{\mu \nu} = \gamma^{\mu \sigma} K_{\sigma}^{\nu} = \gamma^{\sigma \nu} K_{\sigma}^{\mu} = \gamma^{\mu \sigma} \gamma^{\lambda \nu} K_{\sigma \lambda}, \qquad K_{\mu}^{\nu} = \gamma_{\mu \sigma} K^{\sigma \nu} = \gamma_{\mu \sigma} \gamma^{\lambda \nu} K_{\lambda}^{\sigma} = \gamma^{\sigma \nu} K_{\mu \sigma},
\end{equation}
\begin{equation}
K_{\nu}^{\mu} = \gamma_{\sigma \nu} K^{\mu \sigma} = \gamma^{\mu \sigma} \gamma_{\lambda \nu} K_{\sigma}^{\lambda} = \gamma^{\mu \sigma} K_{\sigma \nu}, \qquad \text{ and } \qquad K_{\mu \nu} = \gamma_{\mu \sigma} \gamma_{\lambda \nu} K^{\sigma \lambda} = \gamma_{\sigma \nu} K_{\mu}^{\sigma} = \gamma_{\mu \sigma} K_{\nu}^{\sigma}
\end{equation}
for the case of the rank-2 extrinsic curvature tensor (field) $K$. Their covariant derivatives ${{}^{\left( 3 \right)} \nabla_{\mu}}$ (i.e. their derivatives along tangent vectors) are therefore represented in terms of the coefficients of the induced/spatial Levi-Civita connection ${{}^{\left( 3 \right)} \nabla}$, i.e. the induced/spatial Christoffel symbols ${{}^{\left( 3 \right)} \Gamma_{\mu \nu}^{\rho}}$, which are themselves represented in terms of partial derivatives of the spatial metric tensor ${\gamma_{\mu \nu}}$:

\begin{equation}
{}^{\left( 3 \right)} \Gamma_{\mu \nu}^{\rho} = \frac{1}{2} \gamma^{\rho \sigma} \left( \frac{\partial}{\partial x^{\mu}} \left( \gamma_{\sigma \nu} \right) + \frac{\partial}{\partial x^{\nu}} \left( \gamma_{\mu \sigma} \right) - \frac{\partial}{\partial x^{\sigma}} \left( \gamma_{\mu \nu} \right) \right),
\end{equation}
namely:

\begin{equation}
{}^{\left( 3 \right)} \nabla_{\rho} \beta^{\mu} = \frac{\partial}{\partial x^{\rho}} \left( \beta^{\mu} \right) + {}^{\left( 3 \right)} \Gamma_{\rho \sigma}^{\mu} \beta^{\sigma}, \qquad \text{ and } \qquad {}^{\left( 3 \right)} \nabla_{\rho} \beta_{\mu} = \frac{\partial}{\partial x^{\rho}} \left( \beta_{\mu} \right) - {}^{\left( 3 \right)} \Gamma_{\rho \mu}^{\sigma} \beta_{\sigma},
\end{equation}
for the case of the rank-1 shift vector/covector (field) ${\boldsymbol\beta}$, and:

\begin{equation}
{}^{\left( 3 \right)} \nabla_{\rho} K^{\mu \nu} = \frac{\partial}{\partial x^{\rho}} \left( K^{\mu \nu} \right) + {}^{\left( 3 \right)} \Gamma_{\rho \sigma}^{\mu} K^{\sigma \nu} + {}^{\left( 3 \right)} \Gamma_{\rho \sigma}^{\nu} K^{\mu \sigma}, \qquad {}^{\left( 3 \right)} \nabla_{\rho} K_{\mu}^{\nu} = \frac{\partial}{\partial x^{\rho}} \left( K_{\mu}^{\nu} \right) + {}^{\left( 3 \right)} \Gamma_{\rho \sigma}^{\nu} K_{\mu}^{\sigma} - {}^{\left( 3 \right)} \Gamma_{\rho \mu}^{\sigma} K_{\sigma}^{\nu},
\end{equation}
\begin{equation}
{}^{\left( 3 \right)} \nabla_{\rho} K_{\nu}^{\mu} = \frac{\partial}{\partial x^{\rho}} \left( K_{\nu}^{\mu} \right) + {}^{\left( 3 \right)} \Gamma_{\rho \sigma}^{\mu} K_{\nu}^{\sigma} - {}^{\left( 3 \right)} \Gamma_{\rho \nu}^{\sigma} K_{\sigma}^{\mu}, \qquad \text{ and } \qquad {}^{\left( 3 \right)} \nabla_{\rho} K_{\mu \nu} = \frac{\partial}{\partial x^{\rho}} \left( K_{\mu \nu} \right) - {}^{\left( 3 \right)} \Gamma_{\rho \mu}^{\sigma} K_{\sigma \nu} - {}^{\left( 3 \right)} \Gamma_{\rho \nu}^{\sigma} K_{\mu \sigma},
\end{equation}
for the case of the rank-2 extrinsic curvature tensor (field) $K$. This is demonstrated in Figure \ref{fig:Figure6}, in which all covariant derivatives of the \texttt{ExtrinsicCurvatureTensor} object in lowered-index/covariant form (i.e. ${K_{\mu \nu}}$, the default case), as well as all covariant derivatives of the \texttt{ExtrinsicCurvatureTensor} object in raised-index/contravariant form (i.e. ${K^{\mu \nu}}$) are computed for the \texttt{ADMDecomposition} objects for the Schwarzschild and Kerr metrics, respectively, assuming generic gauge choice.

\begin{figure}[ht]
\centering
\begin{framed}
\includegraphics[width=0.495\textwidth]{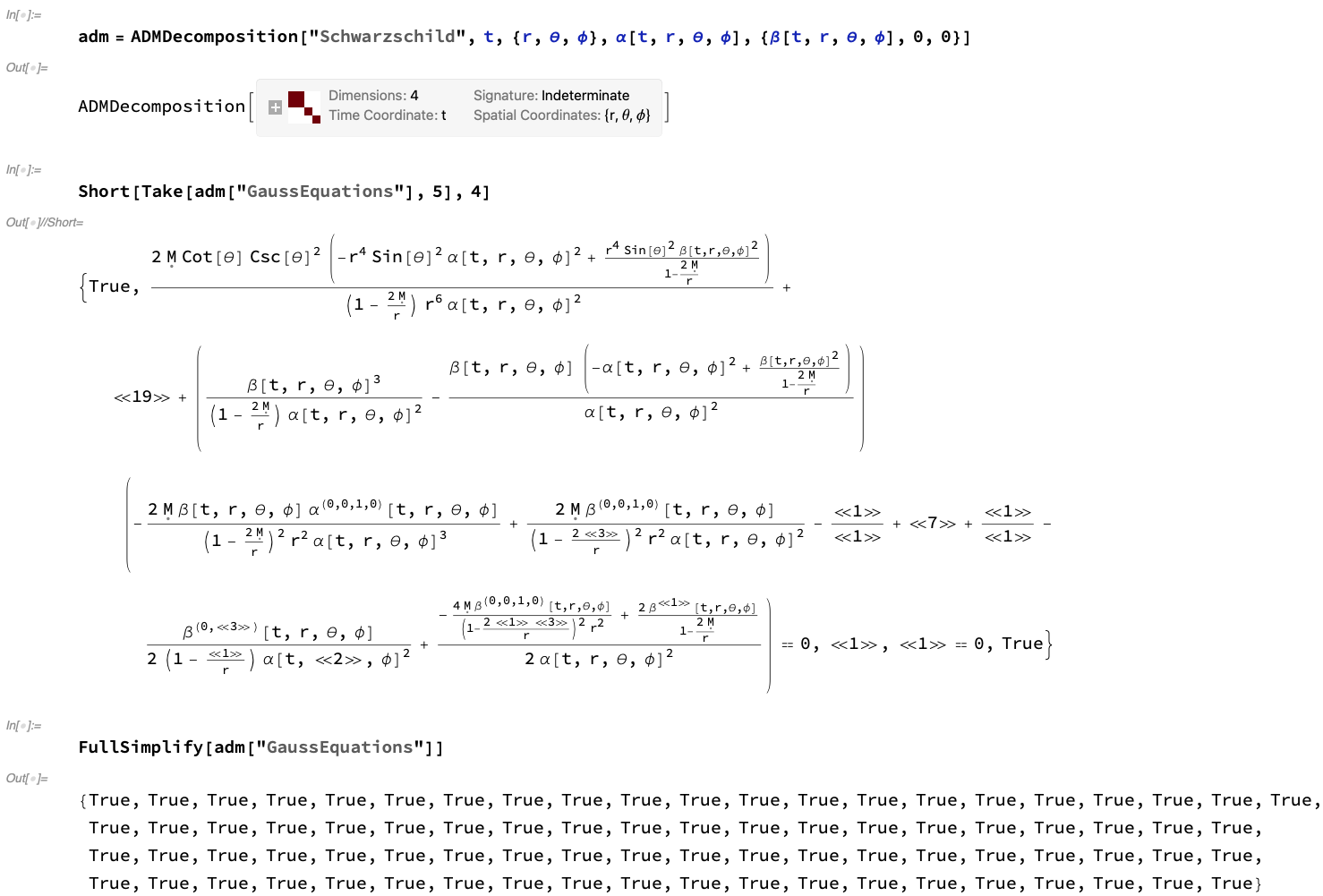}
\vrule
\includegraphics[width=0.495\textwidth]{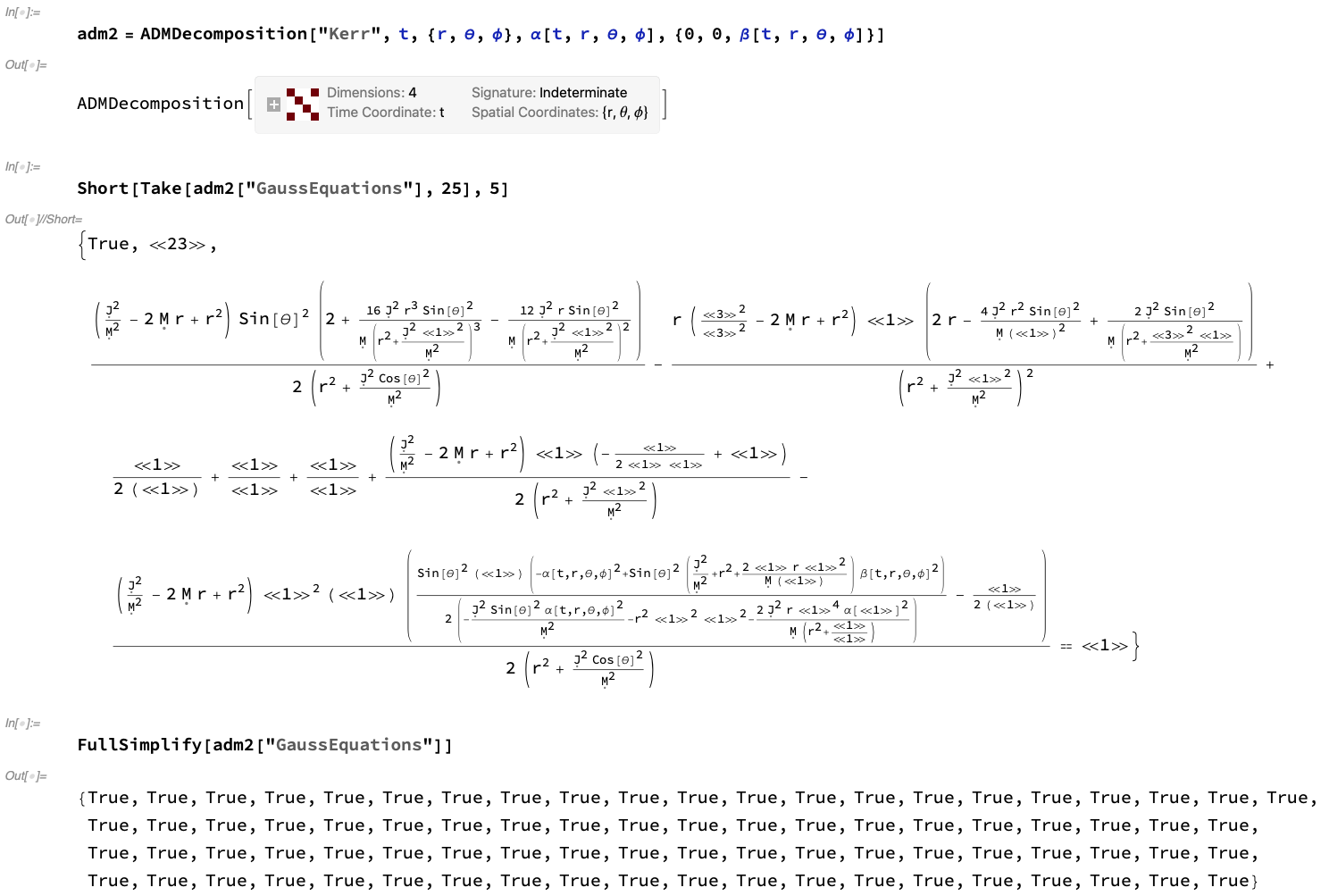}
\end{framed}
\caption{On the left, the list of Gauss equations asserting the relationship between the components of the ambient/spacetime Riemann tensor and the components of the induced/spatial Riemann tensor for the \texttt{ADMDecomposition} object for a Schwarzschild geometry (representing, for instance, an uncharged, non-rotating black hole with mass $M$ in Schwarzschild or spherical polar coordinates ${\left( t, r, \theta, \phi \right)}$) with lapse function ${\alpha \left( t, r, \theta, \phi \right)}$ and modified shift vector ${\left( \beta \left( t, r, \theta, \phi \right), 0, 0 \right)}$, together with a verification that they all hold identically. On the right, the list of Gauss equations asserting the relationship between the components of the ambient/spacetime Riemann tensor and the components of the induced/spatial Riemann tensor for the \texttt{ADMDecomposition} object for a Kerr geometry (representing, for instance, an uncharged, spinning black hole with mass $M$ and angular momentum $J$ in Boyer-Lindquist or oblate spheroidal coordinates ${\left( t, r, \theta, \phi \right)}$) with lapse function ${\alpha \left( t, r, \theta, \phi \right)}$ and modified shift vector ${\left( 0, 0, \beta \left( t, r, \theta, \phi \right) \right)}$, together with a verification that they all hold identically.}
\label{fig:Figure7}
\end{figure}

\begin{figure}[ht]
\centering
\begin{framed}
\includegraphics[width=0.445\textwidth]{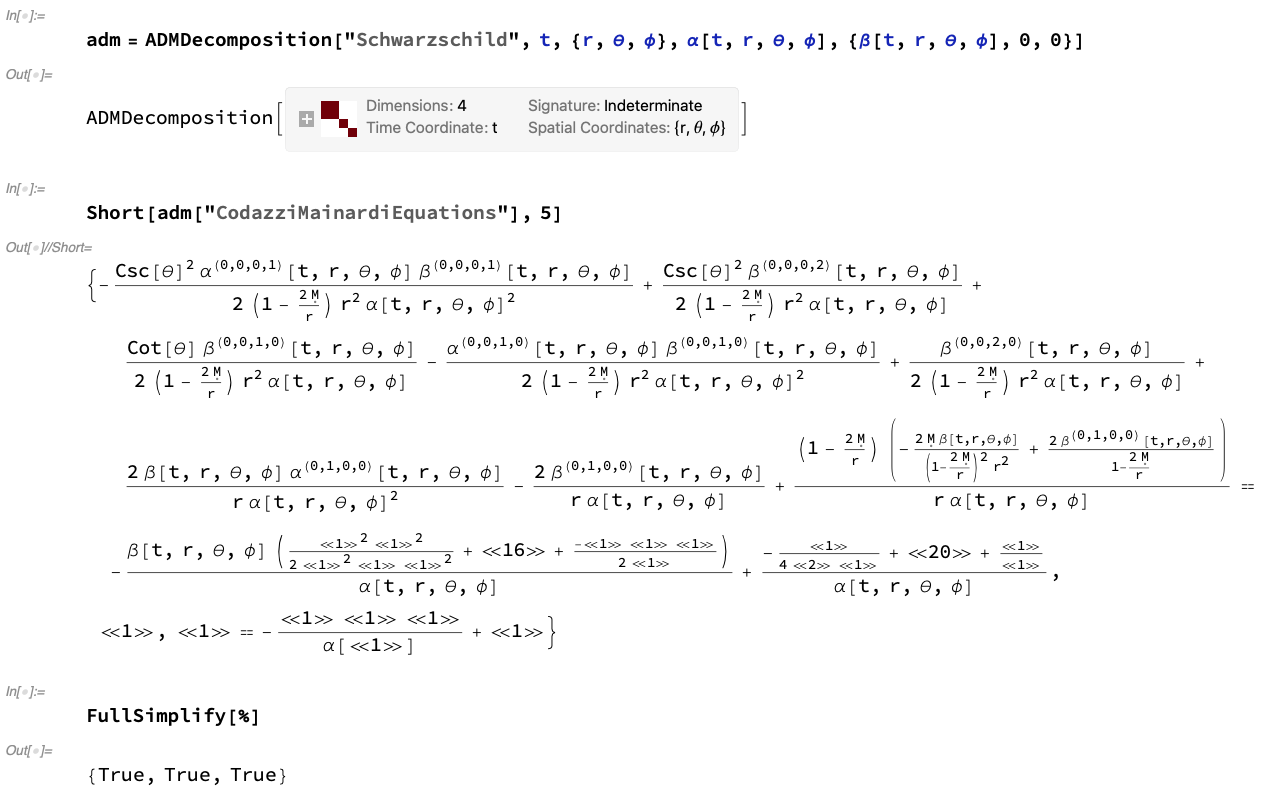}
\vrule
\includegraphics[width=0.545\textwidth]{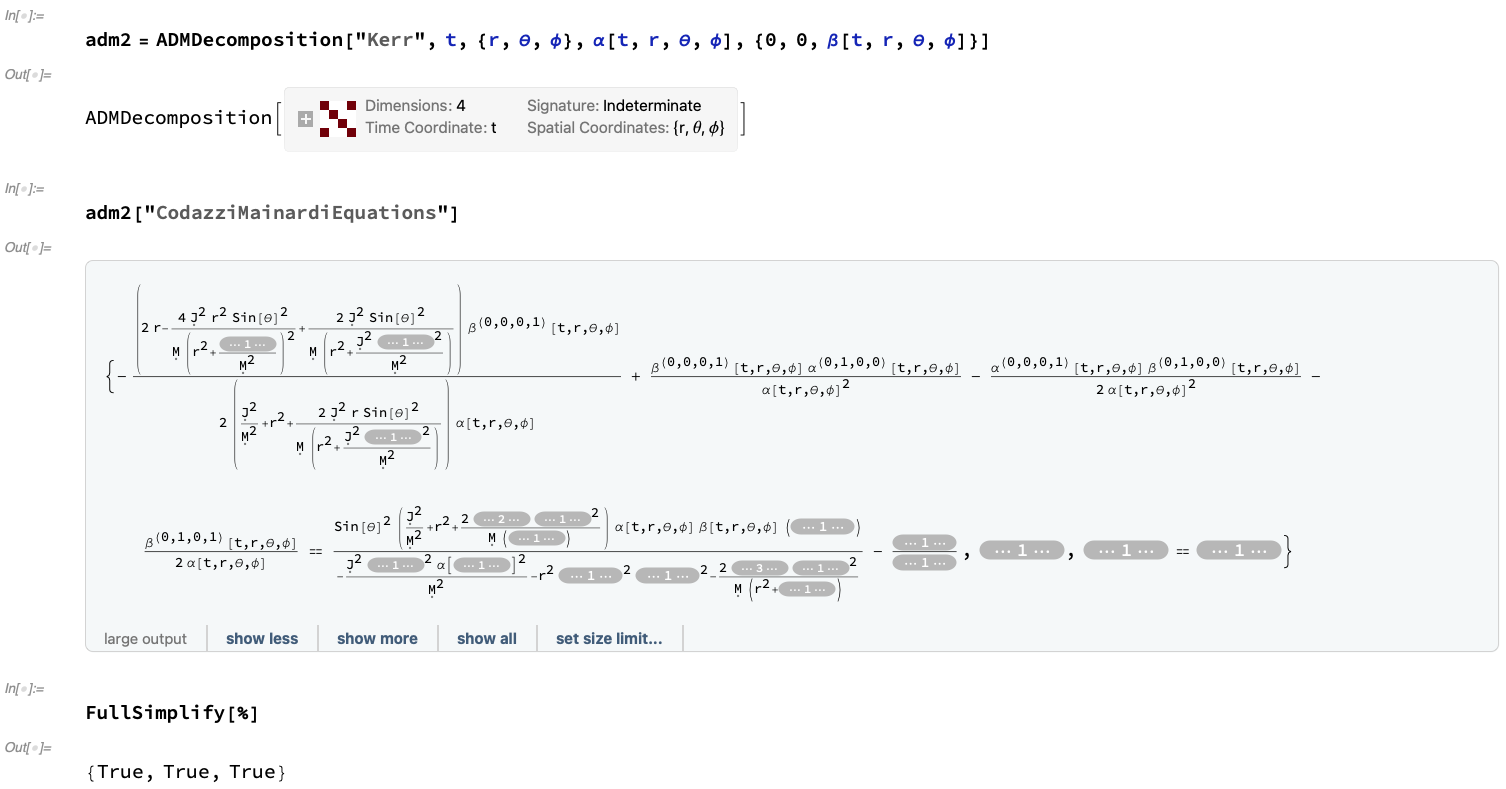}
\end{framed}
\caption{On the left, the list of Codazzi-Mainardi equations asserting the relationship between the components of the ambient/spacetime Ricci tensor and the covariant derivatives of the extrinsic curvature tensor for the \texttt{ADMDecomposition} object for a Schwarzschild geometry (representing, for instance, an uncharged, non-rotating black hole with mass $M$ in Schwarzschild or spherical polar coordinates ${\left( t, r, \theta, \phi \right)}$) with lapse function ${\alpha \left( t, r, \theta, \phi \right)}$ and modified shift vector ${\left( \beta \left( t, r, \theta, \phi \right), 0, 0 \right)}$, together with a verification that they all hold identically. On the right, the list of Codazzi-Mainardi equations asserting the relationship between the components of the ambient/spacetime Ricci tensor and the covariant derivatives of the extrinsic curvature tensor for the \texttt{ADMDecomposition} object for a Kerr geometry (representing, for instance, an uncharged, spinning black hole with mass $M$ and angular momentum $J$ in Boyer-Lindquist or oblate spheroidal coordinates ${\left( t, r, \theta, \phi \right)}$) with lapse function ${\alpha \left( t, r, \theta, \phi \right)}$ and modified shift vector ${\left( 0, 0, \beta \left( t, r, \theta, \phi \right) \right)}$, together with a verification that they all hold identically.}
\label{fig:Figure8}
\end{figure}

As stated previously, the orthogonal projection ${\top}$ of the ambient/spacetime Levi-Civita connection ${{}^{\left( 4 \right)} \nabla}$ onto the tangent bundle yields the induced/spatial Levi-Civita connection ${{}^{\left( 3 \right)} \nabla}$, while the orthogonal projection ${\bot}$ of the ambient/spacetime Levi-Civita connection ${{}^{\left( 4 \right)} \nabla}$ onto the normal bundle yields the symmetric (vector-valued) second fundamental form/extrinsic curvature tensor $K$. The statement that this is necessarily the case is often referred to as the Gauss relation\cite{oneill}, which asserts the following general relationship between (projections of) the ambient/spacetime Riemann tensor ${{}^{\left( 4 \right)} R}$ and the induced/spatial Riemann tensor ${{}^{\left( 3 \right)} R}$ (assuming that, as before, that ${\left( \mathcal{M}, g_{\mathcal{M}} \right)}$ denotes the ambient/spacetime manifold and ${\left( \mathcal{N}, g_{\mathcal{N}} \right)}$ denotes the induced/spatial submanifold):

\begin{multline}
\forall \mathbf{x} \in \mathcal{M}, \qquad \forall \mathbf{X}, \mathbf{Y}, \mathbf{Z}, \mathbf{W} \in T_{\mathbf{x}} \mathcal{M},\\
g_{\mathcal{M}} \left( {}^{\left( 4 \right)} R \left( \mathbf{X}, \mathbf{Y} \right) \mathbf{Z}, \mathbf{W} \right) = g_{\mathcal{M}} \left( {}^{\left( 3 \right)} R \left( \mathbf{X}, \mathbf{Y} \right) \mathbf{Z}, \mathbf{W} \right) + g_{\mathcal{M}} \left( K \left( \mathbf{X}, \mathbf{Z} \right), K \left( \mathbf{Y}, \mathbf{W} \right) \right) - g_{\mathcal{M}} \left( K \left( \mathbf{Y}, \mathbf{Z} \right), K \left( \mathbf{X}, \mathbf{W} \right) \right),
\end{multline}
or, in explicit (component-based) form:

\begin{equation}
\bot_{\rho}^{\alpha} \bot_{\beta}^{\sigma} \bot_{\gamma}^{\mu} \bot_{\delta}^{\nu} {}^{\left( 4 \right)} R_{\sigma \mu \nu}^{\rho} = {}^{\left( 3 \right)} R_{\beta \gamma \delta}^{\alpha} + K_{\gamma}^{\alpha} K_{\beta \delta} - K_{\delta}^{\alpha} K_{\beta \gamma},
\end{equation}
where ${\bot_{\mu}^{\nu}}$ designates the orthogonal projector, i.e. the projection operator in the normal direction ${\mathbf{n}}$:

\begin{equation}
\bot_{\mu}^{\nu} = \delta_{\mu}^{\nu} + n_{\mu} n^{\nu},
\end{equation}
with ${\delta_{\mu}^{\nu}}$ being the identity tensor (i.e. the Kronecker delta function). In the above, ${\rho, \sigma, \mu, \nu}$ range across all ${\left\lbrace 0, \dots, n - 1 \right\rbrace}$ (i.e. across all spacetime coordinate indices), while ${\alpha, \beta, \gamma, \delta}$ range across all ${\left\lbrace 0, \dots, n - 2 \right\rbrace}$ (i.e. across spatial coordinate indices only), with ${{}^{\left( 4 \right)} R_{\sigma \mu \nu}^{\rho}}$ being the components of the ambient/spacetime Riemann tensor:

\begin{equation}
{}^{\left( 4 \right)} R_{\sigma \mu \nu}^{\rho} = \frac{\partial}{\partial x^{\mu}} \left( {}^{\left( 4 \right)} \Gamma_{\sigma \nu}^{\rho} \right) - \frac{\partial}{\partial x^{\nu}} \left( {}^{\left( 4 \right)} \Gamma_{\mu \sigma}^{\rho} \right) + {}^{\left( 4 \right)} \Gamma_{\mu \lambda}^{\rho} {}^{\left( 4 \right)} \Gamma_{\sigma \nu}^{\lambda} - {}^{\left( 4 \right)} \Gamma_{\lambda \nu}^{\rho} {}^{\left( 4 \right)} \Gamma_{\mu \sigma}^{\lambda},
\end{equation}
where ${\lambda}$ here ranges across all spacetime coordinate indices ${\left\lbrace 0, \dots, n - 1 \right\rbrace}$, ${{}^{\left( 3 \right)} R_{\beta \gamma \delta}^{\alpha}}$ being the components of the induced/spatial Riemann tensor:

\begin{equation}
{}^{\left( 3 \right)} R_{\beta \gamma \delta}^{\alpha} = \frac{\partial}{\partial x^{\gamma}} \left( {}^{\left( 3 \right)} \Gamma_{\beta \delta}^{\alpha} \right) - \frac{\partial}{\partial x^{\delta}} \left( {}^{\left( 3 \right)} \Gamma_{\gamma \beta}^{\alpha} \right) + {}^{\left( 3 \right)} \Gamma_{\gamma \lambda}^{\alpha} {}^{\left( 3 \right)} \Gamma_{\beta \delta}^{\lambda} - {}^{\left( 3 \right)} \Gamma_{\lambda \delta}^{\alpha} {}^{\left( 3 \right)} \Gamma_{\gamma \beta}^{\lambda},
\end{equation}
where ${\lambda}$ here ranges across spatial coordinate indices ${\left\lbrace 0, \dots, n - 2 \right\rbrace}$ only, and with ${{}^{\left( 4 \right)} \Gamma_{\mu \nu}^{\rho}}$ being the ambient/spacetime Christoffel symbols, represented in terms of partial derivatives of the overall spacetime metric tensor ${g_{\mu \nu}}$:

\begin{equation}
{}^{\left( 4 \right)} \Gamma_{\mu \nu}^{\rho} = \frac{1}{2} g^{\rho \sigma} \left( \frac{\partial}{\partial x^{\mu}} \left( g_{\sigma \nu} \right) + \frac{\partial}{\partial x^{\nu}} \left( g_{\mu \sigma} \right) - \frac{\partial}{\partial x^{\sigma}} \left( g_{\mu \nu} \right) \right).
\end{equation}
The Gauss equations necessarily hold identically for any \texttt{ADMDecomposition} object computed directly from an (initial) spatial \texttt{MetricTensor} object in the manner described previously, as illustrated in Figure \ref{fig:Figure7} for the cases of the Schwarzschild metric (representing, for instance, an uncharged, non-rotating black hole with mass $M$ in Schwarzschild or spherical polar coordinates ${\left( t, r, \theta, \phi \right)}$) and the Kerr metric (representing, for instance, an uncharged, spinning black hole with mass $M$ and angular momentum $J$ in Boyer-Lindquist or oblate spheroidal coordinates ${\left( t, r, \theta, \phi \right)}$), assuming a restricted choice of gauge consisting of the lapse function ${\alpha \left( t, r, \theta, \phi \right)}$ and the modified shift vectors ${\left( \beta \left( t, r, \theta, \phi \right), 0, 0 \right)}$ (for Schwarzschild) and ${\left( 0, 0, \beta \left( t, r, \theta, \phi \right) \right)}$ (for Kerr). On the other hand, the Codazzi-Mainardi relation\cite{kobayashi2}\cite{oneill} asserts the following general relationship between (orthogonal projections of) the ambient/spacetime Ricci tensor and the covariant derivatives of the second fundamental form/extrinsic curvature tensor:

\begin{equation}
\forall \mathbf{x} \in \mathcal{M}, \qquad \forall \mathbf{X}, \mathbf{Y}, \mathbf{Z} \in T_{\mathbf{x}} \mathcal{M}, \qquad \bot \left( {}^{\left( 4 \right)} R \left( \mathbf{X}, \mathbf{Y} \right) \mathbf{Z} \right) = \left( {}^{\left( 3 \right)} \nabla_{\mathbf{X}} K \right) \left( \mathbf{Y}, \mathbf{Z} \right) - \left( {}^{\left( 3 \right)} \nabla_{\mathbf{Y}} K \right) \left( \mathbf{X}, \mathbf{Z} \right),
\end{equation}
or, in explicit (component-based) form:

\begin{equation}
- {}^{\left( 4 \right)} R_{\rho \sigma} n^{\sigma} \bot_{\mu}^{\rho} = \nabla_{\nu} K_{\mu \nu} - \nabla_{\mu} K,
\end{equation}
i.e., in expanded form:

\begin{equation}
- {}^{\left( 4 \right)} R_{\rho \sigma} n^{\sigma} \bot_{\mu}^{\rho} = \frac{\partial}{\partial x^{\nu}} \left( K_{\mu}^{\nu} \right) + {}^{\left( 3 \right)} \Gamma_{\nu \lambda}^{\nu} K_{\mu}^{\lambda} - {}^{\left( 3 \right)} \Gamma_{\nu \mu}^{\lambda} K_{\lambda}^{\nu} - \frac{\partial}{\partial x^{\mu}} \left( K \right),
\end{equation}
with ${\rho, \sigma}$ ranging across all ${\left\lbrace 0, \dots, n - 1 \right\rbrace}$ (i.e. across all spacetime coordinate indices), while ${\mu, \nu, \lambda}$ range across all ${\left\lbrace 0, \dots, n - 2 \right\rbrace}$ (i.e. across spatial coordinate indices only). In the above, ${{}^{\left( 4 \right)} R_{\mu \nu}}$ designates the ambient/spacetime Ricci tensor, obtained through appropriate contraction of the ambient/spacetime Riemann tensor ${{}^{\left( 4 \right)} R_{\sigma \mu \nu}^{\rho}}$ (i.e. ${{}^{\left( 4 \right)} R_{\mu \nu} = {}^{\left( 4 \right)} R_{\mu \sigma \nu}^{\sigma}}$), and $K$ designates the trace of the extrinsic curvature tensor ${K_{\mu \nu}}$ (i.e. ${K = K_{\mu}^{\mu} = \gamma^{\mu \nu} K_{\mu \nu}}$). Just as for the Gauss equations, the Codazzi-Mainardi equations also necessarily hold identically for any \texttt{ADMDecomposition} object computed directly from an (initial) spatial \texttt{MetricTensor} object in this way, as demonstrated in Figure \ref{fig:Figure8} for the cases of the Schwarzschild and Kerr metrics with a restricted choice of gauge consisting, as before, of the lapse function ${\alpha \left( t, r, \theta, \phi \right)}$ and the modified shift vectors ${\left( \beta \left( t, r, \theta, \phi \right), 0, 0 \right)}$ (for Schwarzschild) and ${\left( 0, 0, \beta \left( t, r, \theta, \phi \right) \right)}$ (for Kerr). In some sense, the Gauss and Codazzi-Mainardi equations therefore represent a set of geometrical consistency conditions that must necessarily be satisfied in order for the hypersurfaces in the foliation to ``plumb together'' in some appropriate way.

\begin{figure}[ht]
\centering
\begin{framed}
\includegraphics[width=0.495\textwidth]{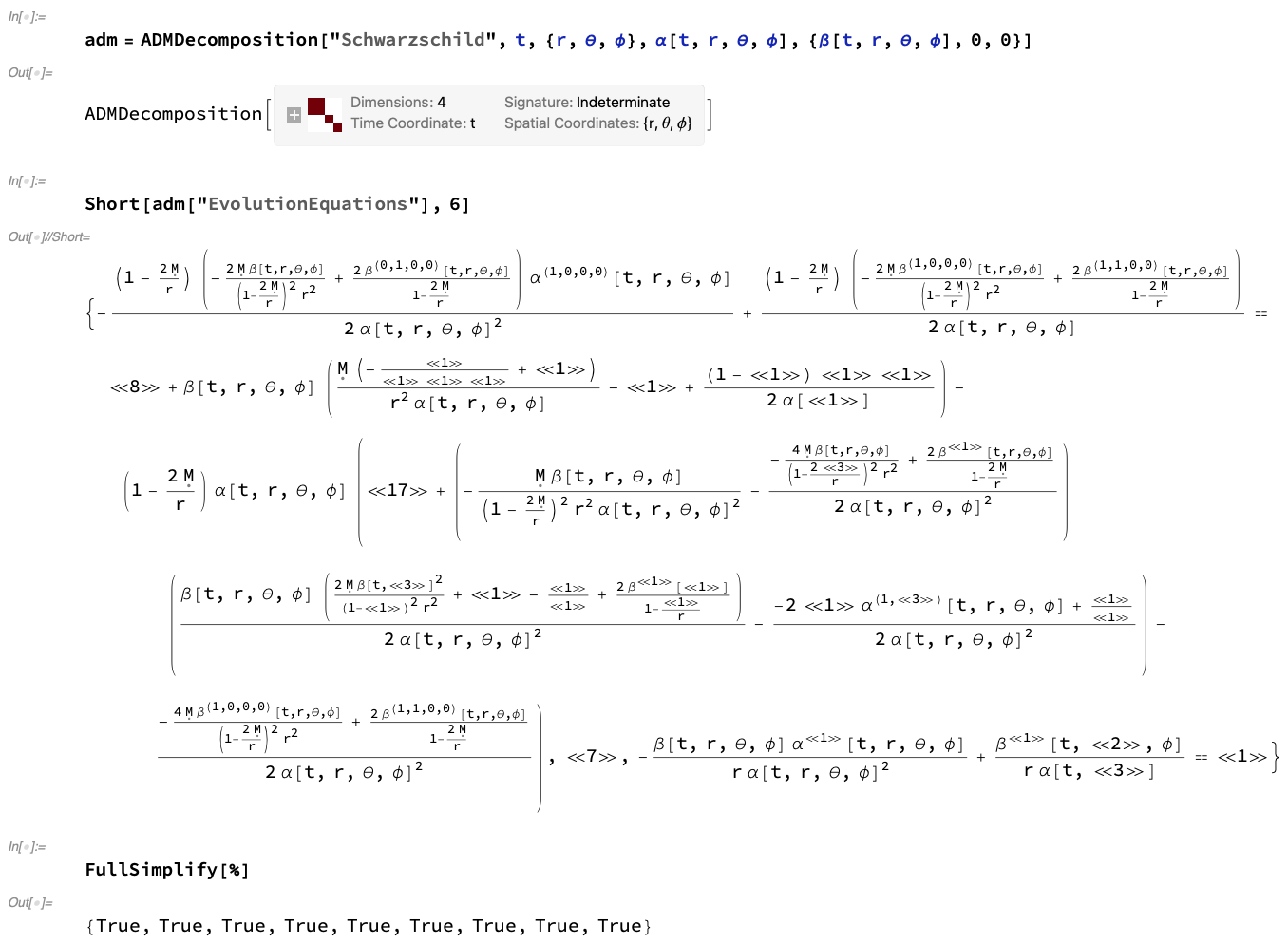}
\vrule
\includegraphics[width=0.495\textwidth]{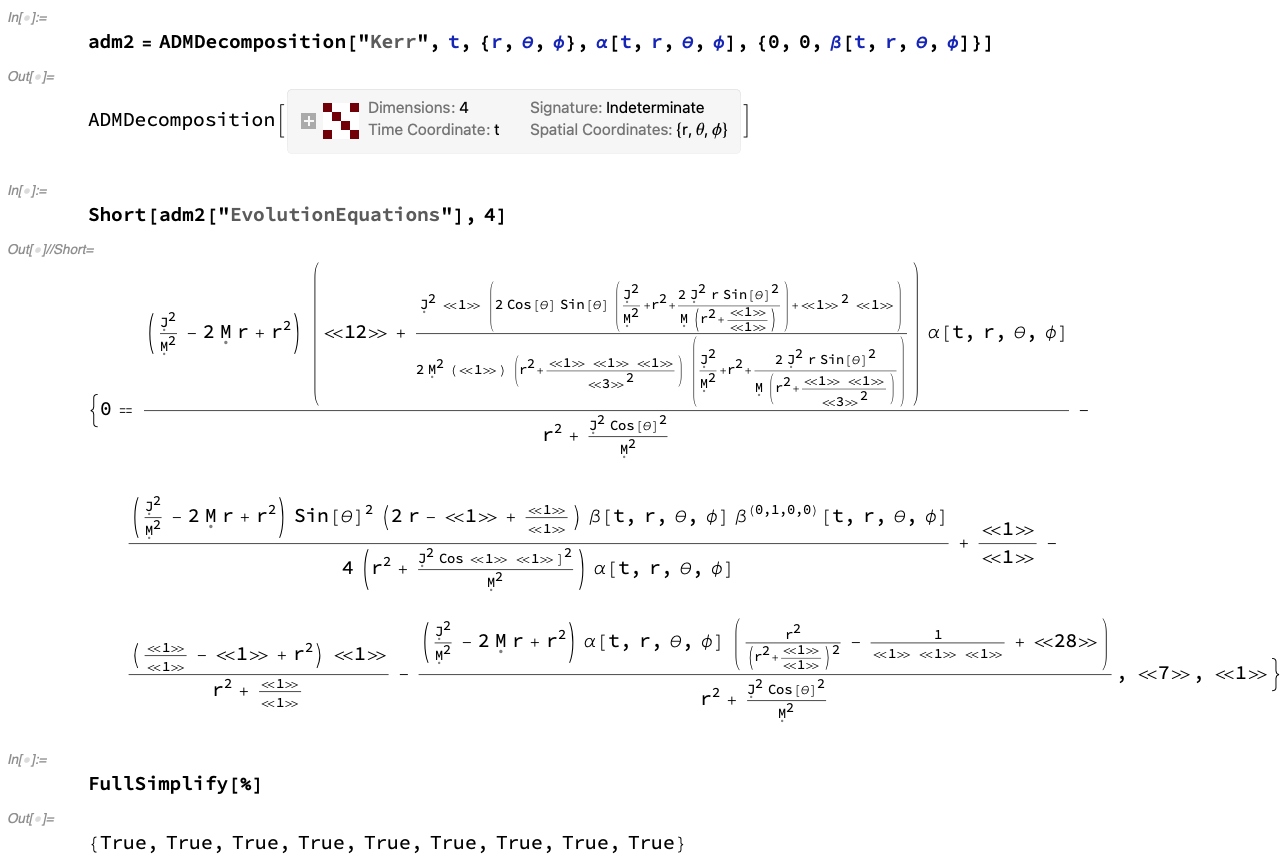}
\end{framed}
\caption{On the left, the list of evolution equations asserting the relationship between the components of the ambient/spacetime Ricci tensor and the components of the induced/spatial Ricci tensor for the \texttt{ADMDecomposition} object for a Schwarzschild geometry (representing, for instance, an uncharged, non-rotating black hole with mass $M$ in Schwarzschild or spherical polar coordinates ${\left( t, r, \theta, \phi \right)}$) with lapse function ${\alpha \left( t, r, \theta, \phi \right)}$ and modified shift vector ${\left( \beta \left( t, r, \theta, \phi \right), 0, 0 \right)}$, together with a verification that they all hold identically. On the right, the list of evolution equations asserting the relationship between the components of the ambient/spacetime Ricci tensor and the components of the induced/spatial Ricci tensor for the \texttt{ADMDecomposition} object for a Kerr geometry (representing, for instance, an uncharged, spinning black hole with mass $M$ and angular momentum $J$ in Boyer-Lindquist or oblate spheroidal coordinates ${\left( t, r, \theta, \phi \right)}$) with lapse function ${\alpha \left( t, r, \theta, \phi \right)}$ and modified shift vector ${\left( 0, 0, \beta \left( t, r, \theta, \phi \right) \right)}$, together with a verification that they all hold identically.}
\label{fig:Figure9}
\end{figure}

\begin{figure}[ht]
\centering
\begin{framed}
\includegraphics[width=0.545\textwidth]{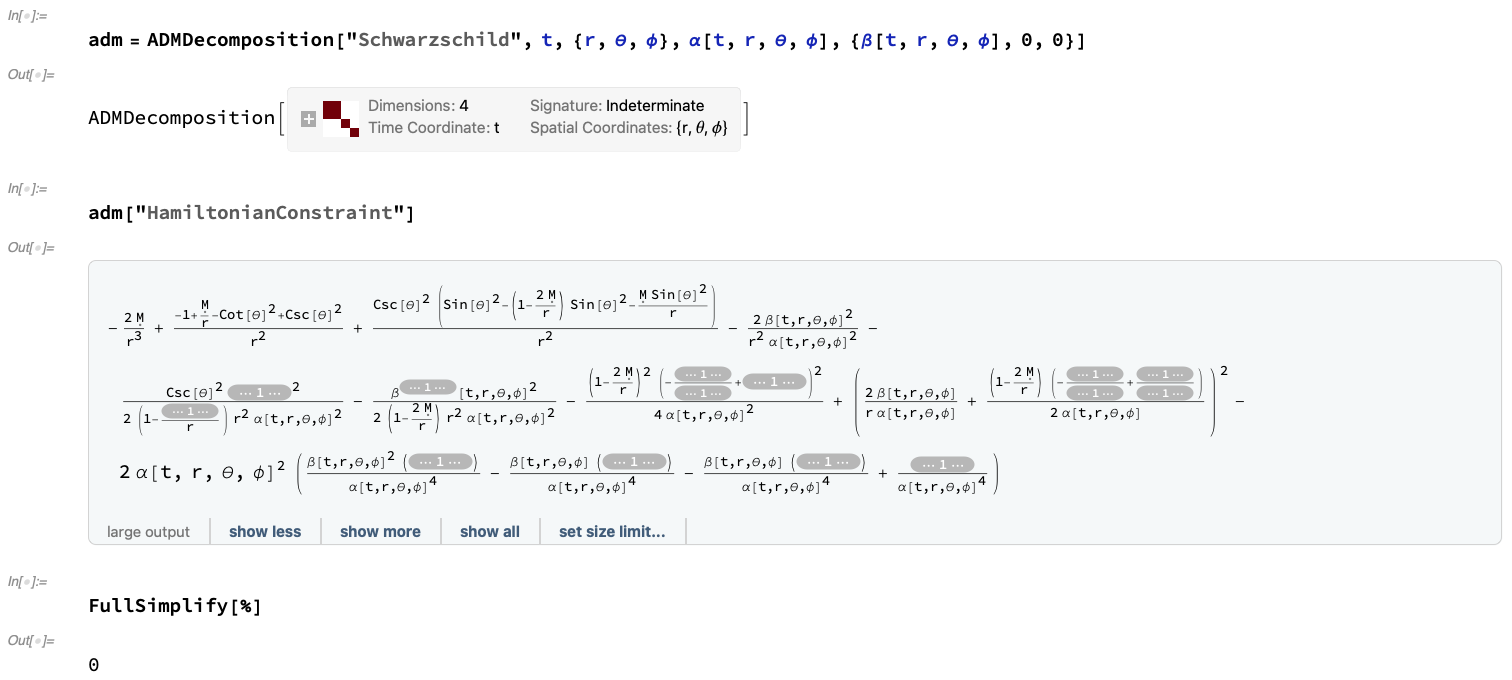}
\vrule
\includegraphics[width=0.445\textwidth]{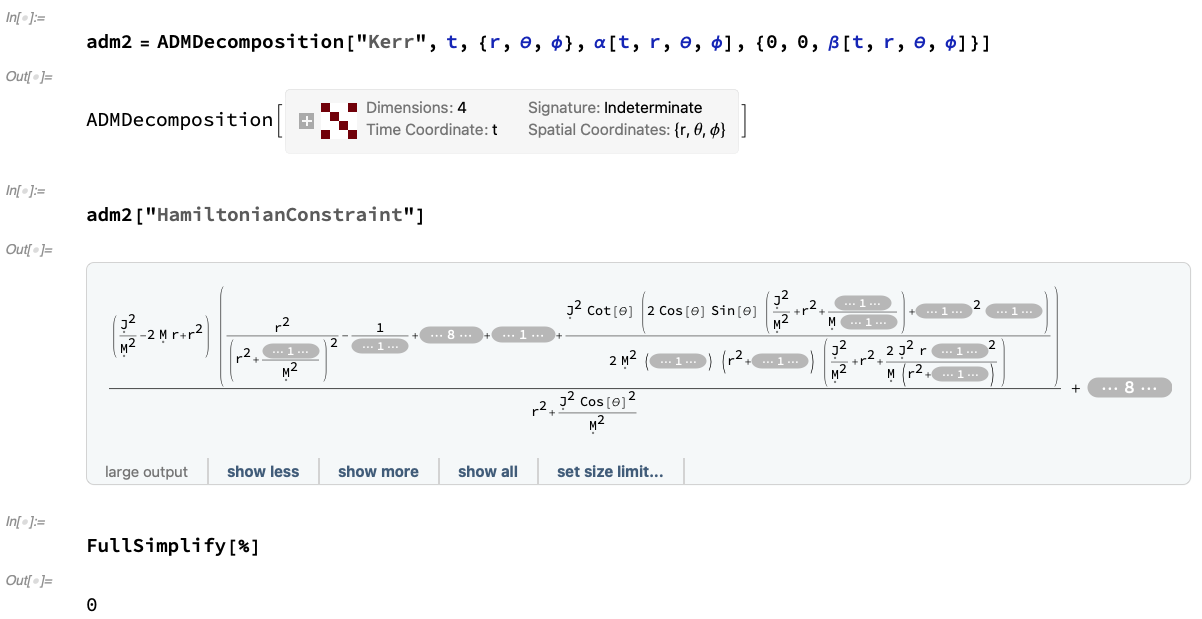}
\end{framed}
\caption{On the left, the value of the Hamiltonian constraint obtained from contracting the Gauss equations for the \texttt{ADMDecomposition} object for a Schwarzschild geometry (representing, for instance, an uncharged, non-rotating black hole with mass $M$ in Schwarzschild or spherical polar coordinates ${\left( t, r, \theta, \phi \right)}$) with lapse function ${\alpha \left( t, r, \theta, \phi \right)}$ and modified shift vector ${\left( \beta \left( t, r, \theta, \phi \right), 0, 0 \right)}$, together with a verification that it vanishes identically. On the right, the value of the Hamiltonian constraint obtained from contracting the Gauss equations for the \texttt{ADMDecomposition} object for a Kerr geometry (representing, for instance, an uncharged, spinning black hole with mass $M$ and angular momentum $J$ in Boyer-Lindquist or oblate spheroidal coordinates ${\left( t, r, \theta, \phi \right)}$) with lapse function ${\alpha \left( t, r, \theta, \phi \right)}$ and modified shift vector ${\left( 0, 0, \beta \left( t, r, \theta, \phi \right) \right)}$, together with a verification that it vanishes identically.}
\label{fig:Figure10}
\end{figure}

\begin{figure}[ht]
\centering
\begin{framed}
\includegraphics[width=0.545\textwidth]{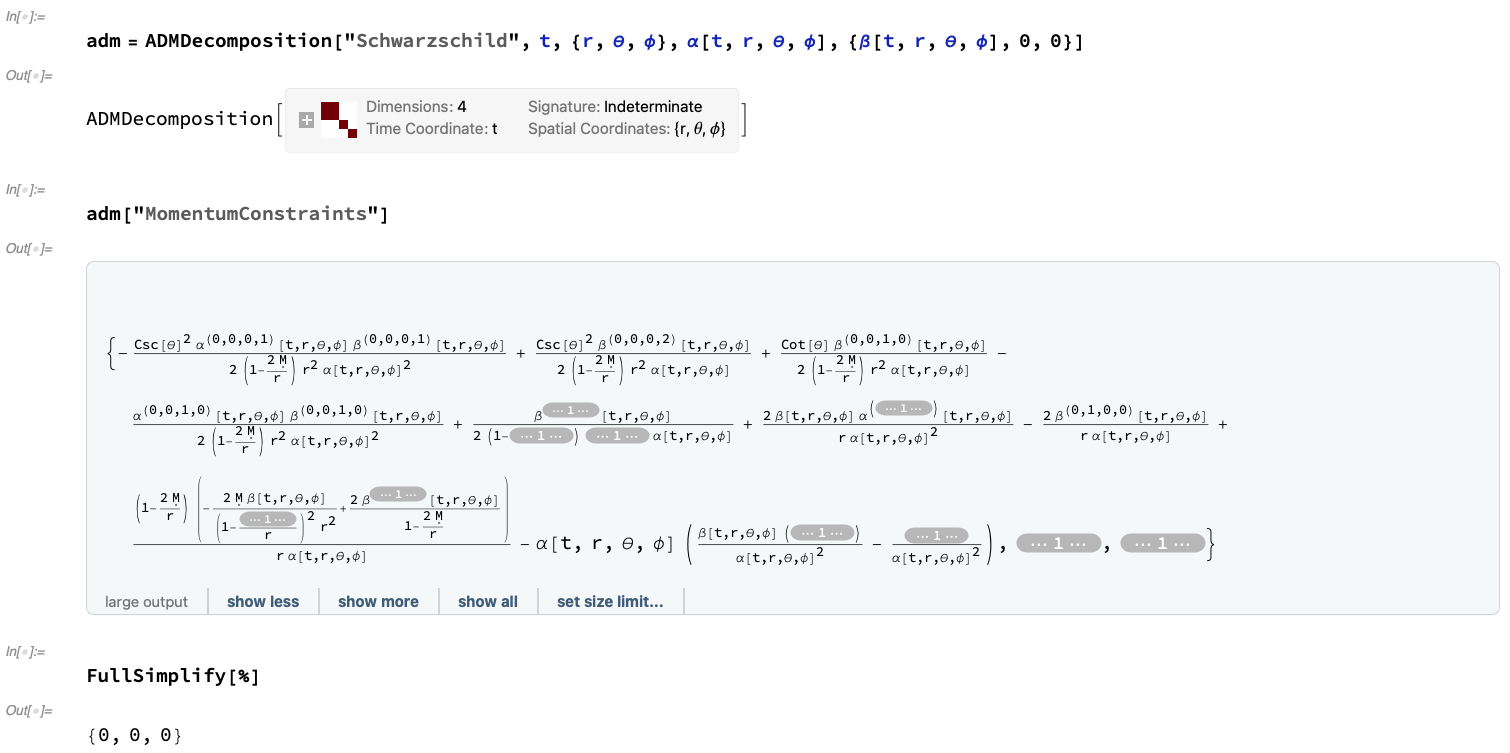}
\vrule
\includegraphics[width=0.445\textwidth]{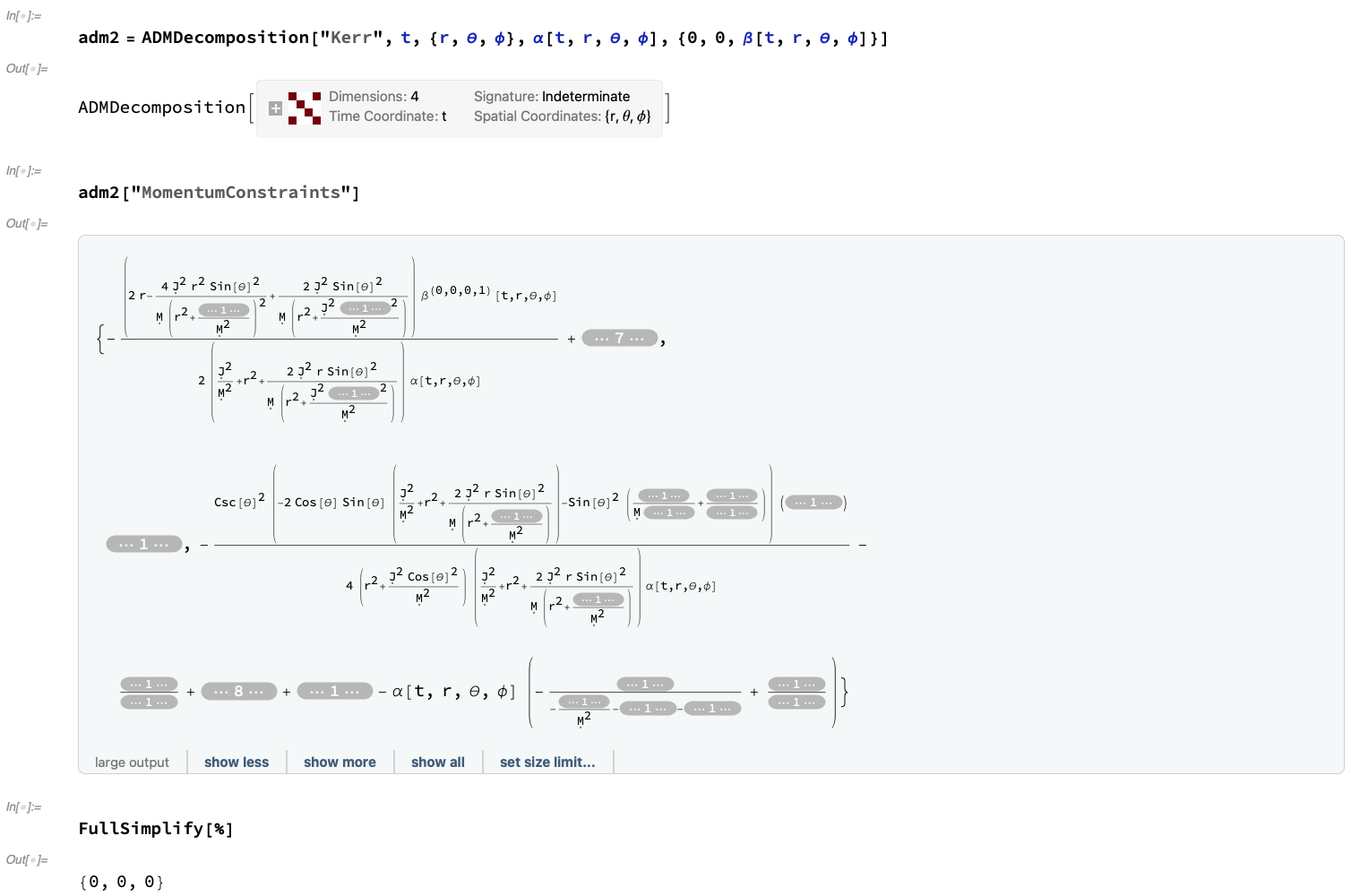}
\end{framed}
\caption{On the left, the values of the momentum constraints obtained from contracting the Codazzi-Mainardi equations for the \texttt{ADMDecomposition} object for a Schwarzschild geometry (representing, for instance, an uncharged, non-rotating black hole with mass $M$ in Schwarzschild or spherical polar coordinates ${\left( t, r, \theta, \phi \right)}$) with lapse function ${\alpha \left( t, r, \theta, \phi \right)}$ and modified shift vector ${\left( \beta \left( t, r, \theta, \phi \right), 0, 0 \right)}$, together with a verification that they all vanish identically. On the right, the values of the momentum constraints obtained from contracting the Codazzi-Mainardi equations for the \texttt{ADMDecomposition} object for a Kerr geometry (representing, for instance, an uncharged, spinning black hole with mass $M$ and angular momentum $J$ in Boyer-Lindquist or oblate spheroidal coordinates ${\left( t, r, \theta, \phi \right)}$) with lapse function ${\alpha \left( t, r, \theta, \phi \right)}$ and modified shift vector ${\left( 0, 0, \beta \left( t, r, \theta, \phi \right) \right)}$, together with a verification that they all vanish identically.}
\label{fig:Figure11}
\end{figure}

\begin{figure}[ht]
\centering
\begin{framed}
\includegraphics[width=0.495\textwidth]{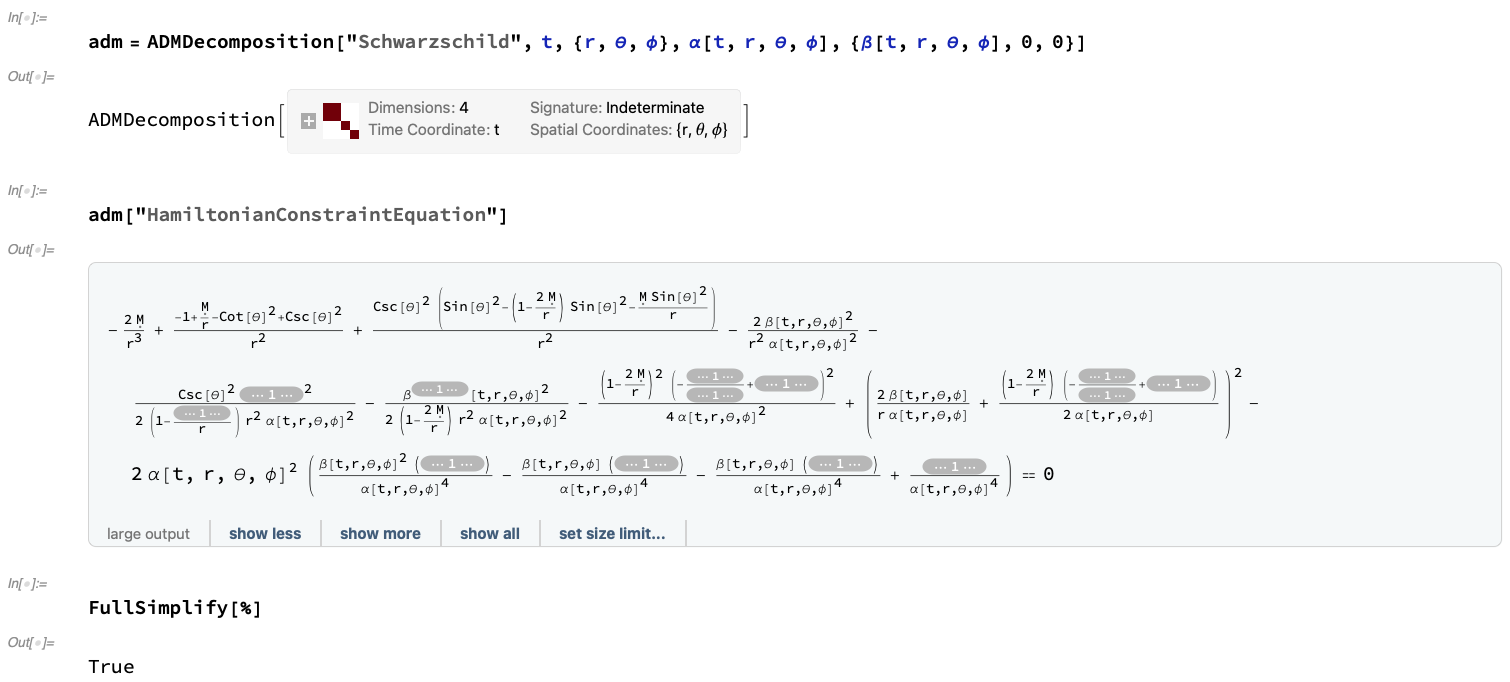}
\vrule
\includegraphics[width=0.495\textwidth]{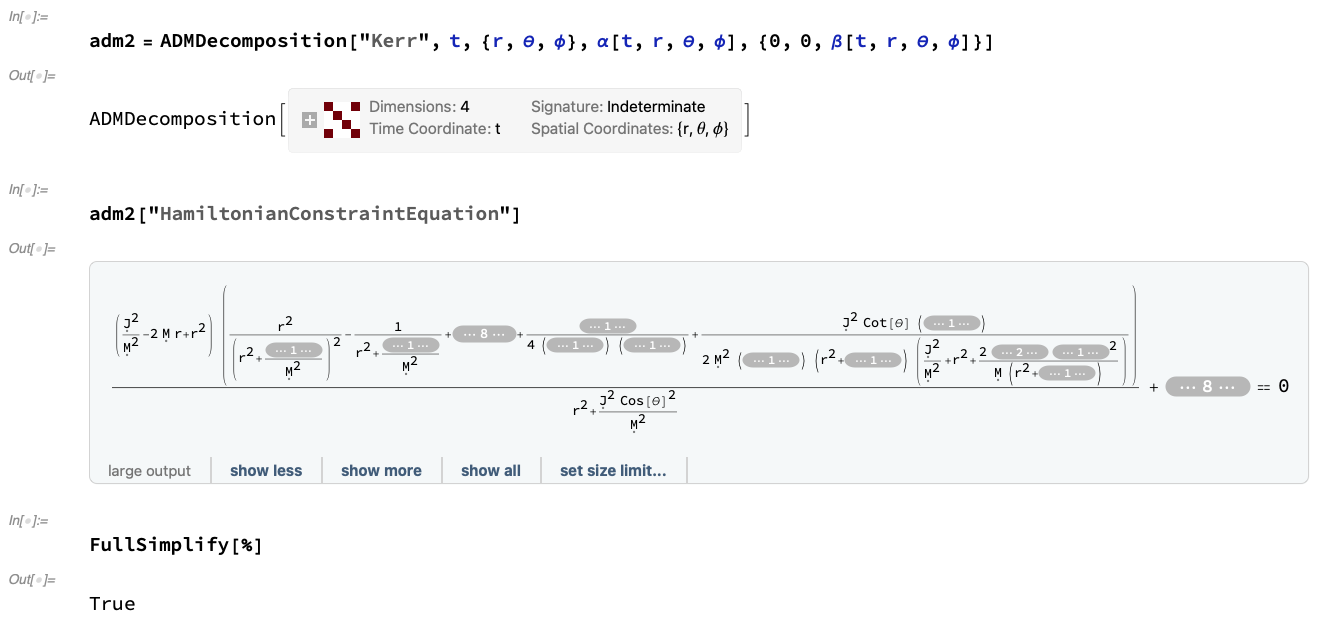}
\end{framed}
\caption{On the left, the condition required to guarantee that the Hamiltonian constraint obtained from contracting the Gauss equations for the \texttt{ADMDecomposition} object for a Schwarzschild geometry (representing, for instance, an uncharged, non-rotating black hole with mass $M$ in Schwarzschild or spherical polar coordinates ${\left( t, r, \theta, \phi \right)}$) with lapse function ${\alpha \left( t, r,\theta, \phi \right)}$ and modified shift vector ${\left( \beta \left( t, r, \theta, \phi \right), 0, 0 \right)}$ vanishes, together with a verification that it holds identically. On the right, the condition required to guarantee that the Hamiltonian constraint obtained from contracting the Gauss equations for the \texttt{ADMDecomposition} object for a Kerr geometry (representing, for instance, an uncharged, spinning black hole with mass $M$ and angular momentum $J$ in Boyer-Lindquist or oblate spheroidal coordinates ${\left( t, r, \theta, \phi \right)}$) with lapse function ${\alpha \left( t, r, \theta, \phi \right)}$ and modified shift vector ${\left( 0, 0, \beta \left( t, r, \theta, \phi \right) \right)}$ vanishes, together with a verification that it holds identically.}
\label{fig:Figure12}
\end{figure}

\begin{figure}[ht]
\centering
\begin{framed}
\includegraphics[width=0.495\textwidth]{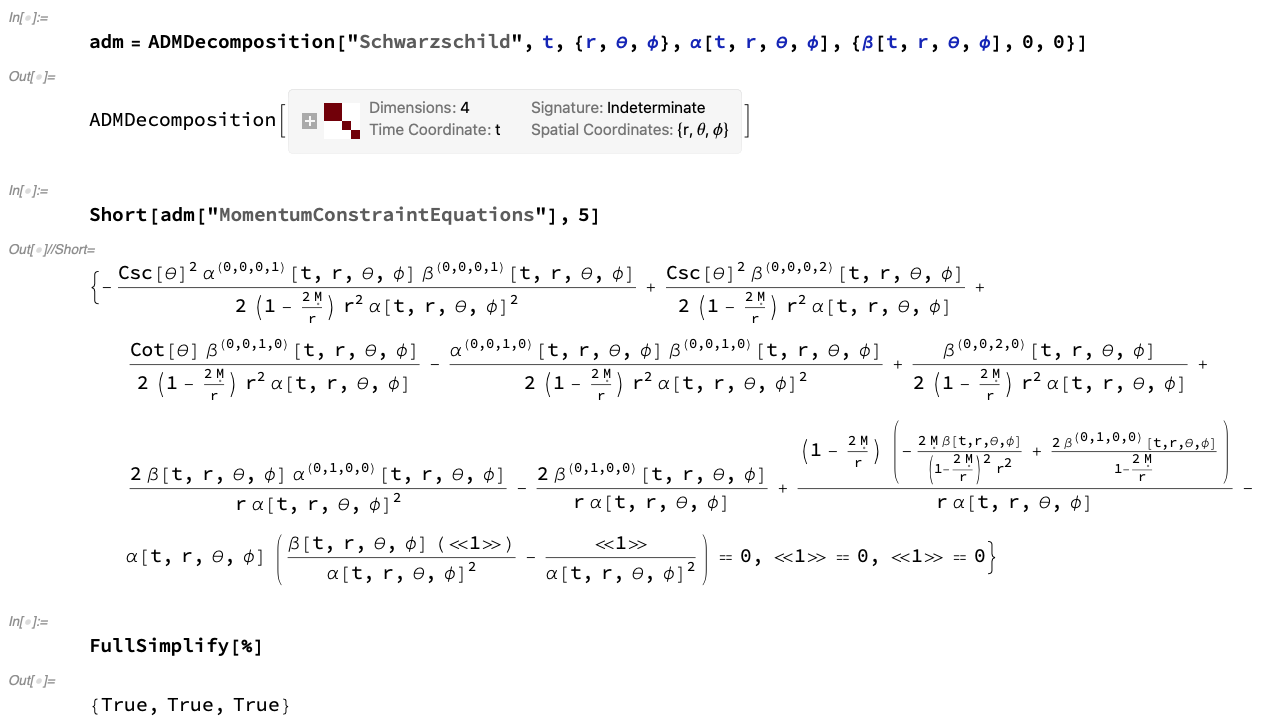}
\vrule
\includegraphics[width=0.495\textwidth]{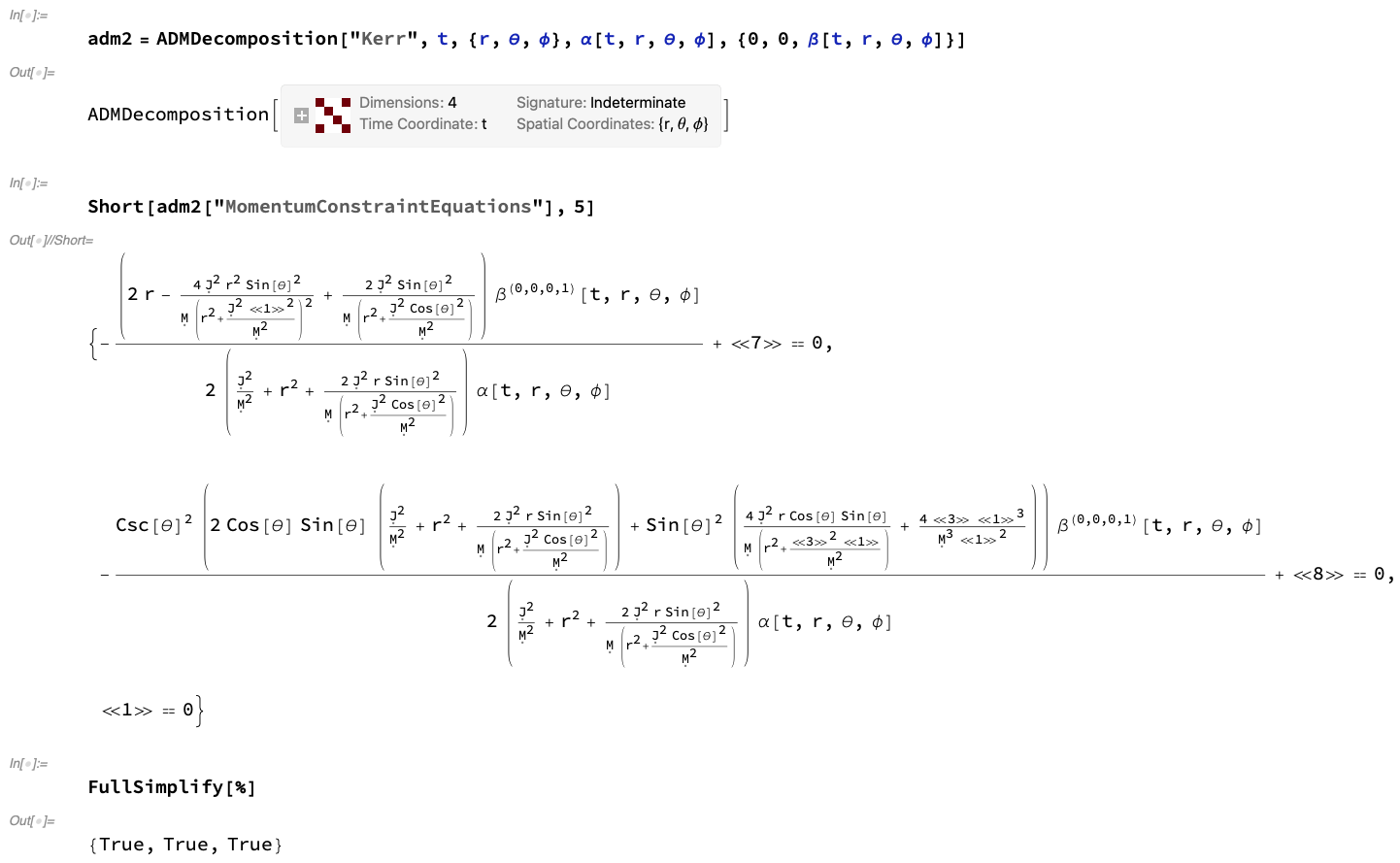}
\end{framed}
\caption{On the left, the conditions required to guarantee that the momentum constraints obtained from contracting the Codazzi-Mainardi equations for the \texttt{ADMDecomposition} object for a Schwarzschild geometry (representing, for instance, an uncharged, non-rotating black hole with mass $M$ in Schwarzschild or spherical polar coordinates ${\left( t, r, \theta, \phi \right)}$) with lapse function ${\alpha \left( t, r, \theta, \phi \right)}$ and modified shift vector ${\left( \beta \left( t, r, \theta, \phi \right), 0, 0 \right)}$ vanish, together with a verification that they all hold identically. On the right, the conditions required to guarantee that the momentum constraints obtained from contracting the Codazzi-Mainardi equations for the \texttt{ADMDecomposition} object for a Kerr geometry (representing, for instance, an uncharged, spinning black hole with mass $M$ and angular momentum $J$ in Boyer-Lindquist or oblate spheroidal coordinates ${\left( t, r, \theta, \phi \right)}$) with lapse function ${\alpha \left( t, r, \theta, \phi \right)}$ and modified shift vector ${\left( 0, 0, \beta \left( t, r, \theta, \phi \right) \right)}$ vanish, together with a verification that they all hold identically.}
\label{fig:Figure13}
\end{figure}

In much the same vein as the Gauss and Codazzi-Mainardi equations above, it is also possible to relate the components of the ambient/spacetime Ricci tensor ${{}^{\left( 4 \right)} R_{\mu \nu}}$ and the induced/spatial Ricci tensor ${{}^{\left( 3 \right)} R_{\mu \nu}}$; by making the time derivative of the extrinsic curvature tensor ${\frac{\partial}{\partial t} \left( K_{\nu}^{\mu} \right)}$ the subject of these equations, one can then proceed to formulate them as a system of time evolution equations for the components of the extrinsic curvature tensor ${K_{\mu \nu}}$ (regarded here as the conjugate momenta of the components of the spatial metric tensor ${\gamma_{\mu \nu}}$, which are in turn regarded here as the dynamical variables of the theory):

\begin{multline}
\frac{\partial}{\partial t} \left( K_{\nu}^{\mu} \right) = \alpha {}^{\left( 3 \right)} R_{\nu}^{\mu} - {}^{\left( 3 \right)} \nabla_{\rho} \left( {}^{\left( 3 \right)} \nabla_{\nu} \alpha \right) \gamma^{\rho \mu} + \alpha K K_{\nu}^{\mu} + \beta^{\rho} {}^{\left( 3 \right)} \nabla_{\rho} K_{\nu}^{\mu}\\
+ K_{\rho}^{\mu} {}^{\left( 3 \right)} \nabla_{\nu} \beta^{\rho} - K_{\nu}^{\rho} {}^{\left( 3 \right)} \nabla_{\rho} \beta^{\mu} - \alpha {}^{\left( 4 \right)} R_{\left( \rho + 1 \right) \left( \nu + 1 \right)} \gamma^{\rho \mu},
\end{multline}
i.e., in expanded form:

\begin{multline}
\frac{\partial}{\partial t} \left( K_{\nu}^{\mu} \right) = \alpha {}^{\left( 3 \right)} R_{\nu}^{\mu} - \left( \frac{\partial}{\partial x^{\rho}} \left( \frac{\partial}{\partial x^{\nu}} \left( \alpha \right) \right) - {}^{\left( 3 \right)} \Gamma_{\rho \nu}^{\sigma} \left( \frac{\partial}{\partial x^{\sigma}} \left( \alpha \right) \right) \right) \gamma^{\rho \mu} + \alpha K K_{\nu}^{\mu}\\
+ \beta^{\rho} \left( \frac{\partial}{\partial x^{\rho}} \left( K_{\nu}^{\mu} \right) + {}^{\left( 3 \right)} \Gamma_{\rho \sigma}^{\mu} K_{\nu}^{\sigma} - {}^{\left( 3 \right)} \Gamma_{\rho \nu}^{\sigma} K_{\sigma}^{\mu} \right) + K_{\rho}^{\mu} \left( \frac{\partial}{\partial x^{\nu}} \left( \beta^{\rho} \right) + {}^{\left( 3 \right)} \Gamma_{\nu \sigma}^{\rho} \beta^{\sigma} \right)\\
- K_{\nu}^{\rho} \left( \frac{\partial}{\partial x^{\rho}} \left( \beta^{\mu} \right) + {}^{\left( 3 \right)} \Gamma_{\rho \sigma}^{\mu} \beta^{\sigma} \right) - \alpha {}^{\left( 4 \right)} R_{\left( \rho + 1 \right) \left( \nu + 1 \right)} \gamma^{\rho \mu},
\end{multline}
with ${\mu, \nu, \rho, \sigma}$ ranging across all ${\left\lbrace 0, \dots, n - 2 \right\rbrace}$ (i.e. across spatial coordinate indices only). Since these are purely geometrical identities relating components of ${{}^{\left( 4 \right)} R_{\mu \nu}}$ to components of ${{}^{\left( 3 \right)} R_{\mu \nu}}$, these evolution equations necessarily hold identically for any \texttt{ADMDecomposition} object computed from an (initial) spatial \texttt{MetricTensor} object in the usual way, as shown in Figure \ref{fig:Figure9} for the cases of the Schwarzschild and Kerr metrics with the same restricted choice of gauge as before, consisting of the lapse function ${\alpha \left( t, r, \theta, \phi \right)}$ and the modified shift vectors ${\left( \beta \left( t, r, \theta, \phi \right), 0, 0 \right)}$ (for Schwarzschild) and ${\left( 0, 0, \beta \left( t, r, \theta, \phi \right) \right)}$ (for Kerr). However, as we shall see shortly, once the Einstein field equations are used to replace the ambient/spacetime Ricci tensor terms ${{}^{\left( 4 \right)} R_{\mu \nu}}$ with stress-energy tensor terms ${T_{\mu \nu}}$, these purely geometrical relations instead become bona fide evolution equations, describing how the components of the extrinsic curvature tensor ${K_{\mu \nu}}$ on spacelike hypersurfaces change as one moves forwards (or backwards) in time. By taking appropriate contractions of the Gauss equations, we also obtain a timelike projection of the contracted Bianchi identities, known as the Hamiltonian constraint ${\mathcal{H}}$:

\begin{equation}
\mathcal{H} = {}^{\left( 3 \right)} R + K^2 - K_{\nu}^{\mu} K_{\mu}^{\nu} - 2 \alpha^2 \left( {}^{\left( 4 \right)} R^{0 0} - \frac{1}{2} {}^{\left( 4 \right)} R g^{0 0} \right),
\end{equation}
or, more succinctly:

\begin{equation}
\mathcal{H} = {}^{\left( 3 \right)} R + K^2 - K_{\nu}^{\mu} K_{\mu}^{\nu} - 2 \alpha^2 {}^{\left( 4 \right)} G^{0 0},
\end{equation}
with ${\mu, \nu}$ ranging across all ${\left\lbrace 0, \dots, n - 2 \right\rbrace}$ (i.e. across spatial coordinate indices only), where we have used the definition of the ambient/spacetime Einstein tensor ${{}^{\left( 4 \right)} G_{\mu \nu}}$:

\begin{equation}
{}^{\left( 4 \right)} G_{\mu \nu} = {}^{\left( 4 \right)} R_{\mu \nu} - \frac{1}{2} {}^{\left( 4 \right)} R g_{\mu \nu},
\end{equation}
with ${\mu, \nu}$ here ranging across all ${\left\lbrace 0, \dots, n - 1 \right\rbrace}$ (i.e. across all spacetime coordinate indices), and where ${{}^{\left( 3 \right)} R}$ and ${{}^{\left( 4 \right)} R}$ denote the induced/spatial and ambient/spacetime Ricci scalars:

\begin{equation}
{}^{\left( 3 \right)} R = {}^{\left( 3 \right)} R_{\mu}^{\mu} = \gamma^{\mu \nu} {}^{\left( 3 \right)} R_{\mu \nu}, \qquad \text{ and } \qquad {}^{\left( 4 \right)} R = {}^{\left( 4 \right)} R_{\mu}^{\mu} = g^{\mu \nu} {}^{\left( 4 \right)} R_{\mu \nu},
\end{equation}
respectively. Likewise, one can obtain the corresponding spacelike projections of the contracted Bianchi identities by taking appropriate contractions of the Codazzi-Mainardi equations, thus yielding the so-called momentum constraints ${\mathcal{M}_{\mu}}$ (represented here in covector form):

\begin{equation}
\mathcal{M}_{\mu} = {}^{\left( 3 \right)} \nabla_{\nu} K_{\mu}^{\nu} - {}^{\left( 3 \right)} \nabla_{\mu} K - \alpha \left( {}^{\left( 4 \right)} R_{\left( \mu + 1 \right)}^{0} - \frac{1}{2} {}^{\left( 4 \right)} R \delta_{\left( \mu + 1 \right)}^{0} \right),
\end{equation}
i.e., in expanded form:

\begin{equation}
\mathcal{M}_{\mu} = \frac{\partial}{\partial x^{\nu}} \left( K_{\mu}^{\nu} \right) + {}^{\left( 3 \right)} \Gamma_{\nu \sigma}^{\nu} K_{\mu}^{\sigma} - {}^{\left( 3 \right)} \Gamma_{\nu \mu}^{\sigma} K_{\sigma}^{\nu} - \frac{\partial}{\partial x^{\mu}} \left( K \right) - \alpha \left( {}^{\left( 4 \right)} R_{\left( \mu + 1 \right)}^{0} - \frac{1}{2} {}^{\left( 4 \right)} R \delta_{\left( \mu + 1 \right)}^{0} \right),
\end{equation}
which, upon simplification again using the ambient/spacetime Einstein tensor ${{}^{\left( 4 \right)} G_{\mu \nu}}$, become:

\begin{equation}
\mathcal{M}_{\mu} = \frac{\partial}{\partial x^{\nu}} \left( K_{\mu}^{\nu} \right) + {}^{\left( 3 \right)} \Gamma_{\nu \sigma}^{\nu} K_{\mu}^{\sigma} - {}^{\left( 3 \right)} \Gamma_{\nu \mu}^{\sigma} K_{\sigma}^{\nu} - \frac{\partial}{\partial x^{\mu}} \left( K \right) - \alpha {}^{\left( 4 \right)} G_{\left( \mu + 1 \right)}^{0},
\end{equation}
with ${\mu, \nu, \sigma}$ here ranging across all ${\left\lbrace 0, \dots, n - 2 \right\rbrace}$ (i.e. across spatial coordinate indices only). Just as with the evolution equations above, these forms of the Hamiltonian ${\mathcal{H}}$ and momentum ${\mathcal{M}_{\mu}}$ constraints are purely geometrical identities, and therefore vanish identically for any \texttt{ADMDecomposition} object obtained from an (initial) spatial \texttt{MetricTensor} object in the same way, as illustrated in Figures \ref{fig:Figure10} and \ref{fig:Figure11}, again for the cases of the Schwarzschild and Kerr metrics with modified shift vectors ${\left( \beta \left( t, r, \theta, \phi \right), 0, 0 \right)}$ and ${\left( 0, 0, \beta \left( t, r, \theta, \phi \right) \right)}$, respectively. The equations asserting that the Hamiltonian and momentum constraints vanish identically, i.e. that ${\mathcal{H} = 0}$ and ${\mathcal{M}_{\mu} = 0}$, are consequently referred to as the Hamiltonian and momentum constraint equations, and they can be computed and verified directly within \textsc{Gravitas}, as shown in Figures \ref{fig:Figure12} and \ref{fig:Figure13}. Much like the evolution equations, once the Einstein field equations are applied and used to transform the ambient/spacetime Einstein tensor terms ${{}^{\left( 4 \right)} G_{\mu \nu}}$ into stress-energy tensor terms ${T_{\mu \nu}}$, the Hamiltonian and momentum constraint equations will cease to be purely geometrical identities and will instead become bona fide physical constraints on the evolution and choice of gauge. Indeed, in (for instance) the ${3 + 1}$-dimensional case, the ten independent Einstein field equations may be projected in the space-space directions (thus yielding the six evolution equations), the time-time direction (thus yielding the one Hamiltonian constraint equation) and the time-space/space-time directions (thus yielding the three momentum constraint equations), with the latter two corresponding to the four redundant degrees of freedom arising from the contracted Bianchi identities.

\begin{figure}[ht]
\centering
\begin{framed}
\includegraphics[width=0.495\textwidth]{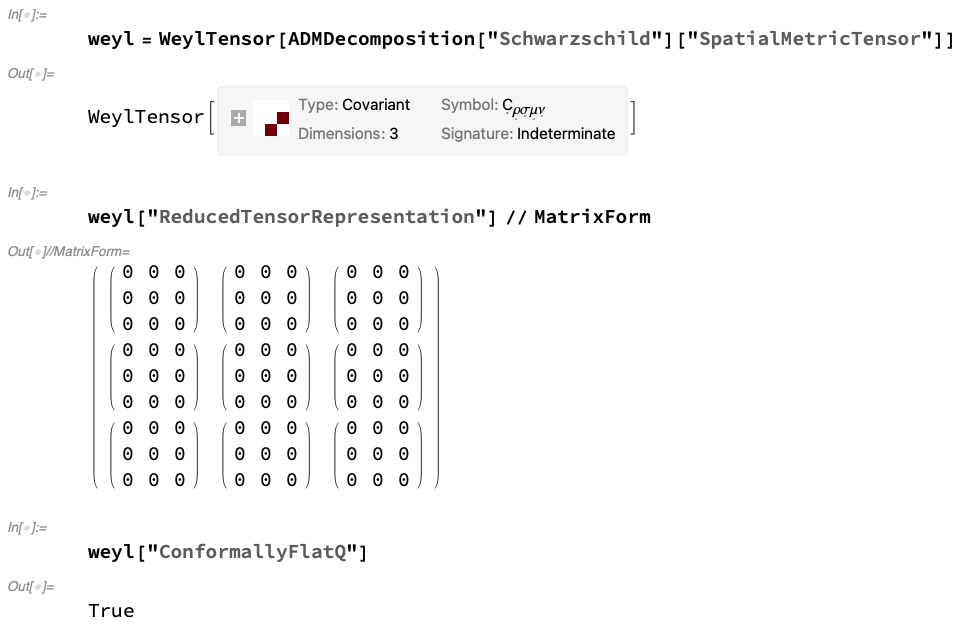}
\vrule
\includegraphics[width=0.495\textwidth]{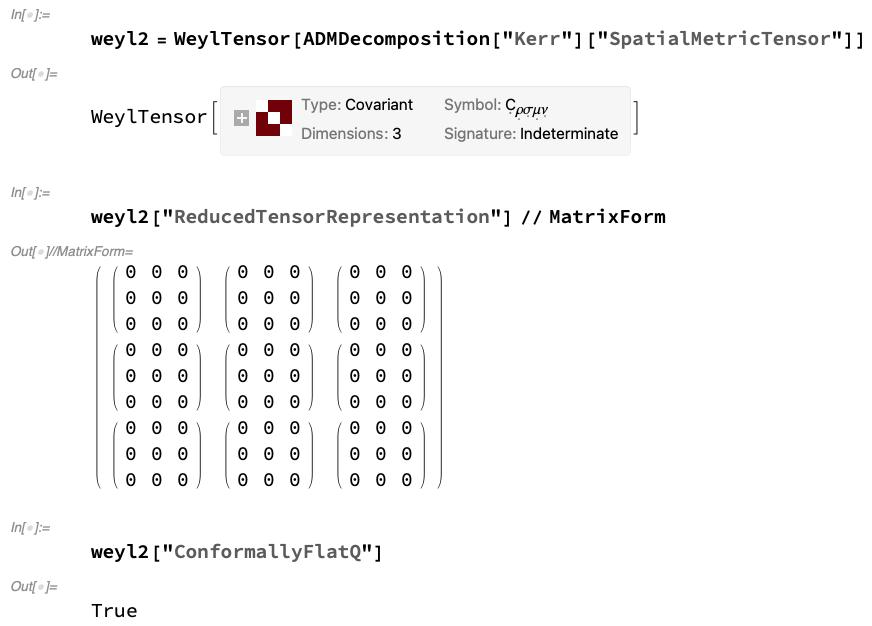}
\end{framed}
\caption{On the left, the \texttt{WeylTensor} object for the (initial) spacelike hypersurface of the \texttt{ADMDecomposition} object for a Schwarzschild geometry (representing, for instance, an uncharged, non-rotating black hole with mass $M$ in Schwarzschild or spherical polar coordinates ${\left( t, r, \theta, \phi \right)}$) with lapse function ${\alpha \left( t, r, \theta, \phi \right)}$ and shift vector ${\left( \beta^1 \left( t, r, \theta, \phi \right), \beta^2 \left( t, r, \theta, \phi \right), \beta^3 \left( t, r, \theta, \phi \right) \right)}$, illustrating that the Weyl tensor vanishes identically. On the right, the \texttt{WeylTensor} object for the (initial) spacelike hypersurface of the \texttt{ADMDecomposition} object for a Kerr geometry (representing, for instance, an uncharged, spinning black hole with mass $M$ and angular momentum $J$ in Boyer-Lindquist or oblate spheroidal coordinates ${\left( t, r, \theta, \phi \right)}$) with lapse function ${\alpha \left( t, r, \theta, \phi \right)}$ and shift vector ${\left( \beta^1 \left( t, r, \theta, \phi \right), \beta^2 \left( t, r, \theta, \phi \right), \beta^3 \left( t, r, \theta, \phi \right) \right)}$, illustrating that the Weyl tensor vanishes identically.}
\label{fig:Figure14}
\end{figure}

\begin{figure}[ht]
\centering
\begin{framed}
\includegraphics[width=0.445\textwidth]{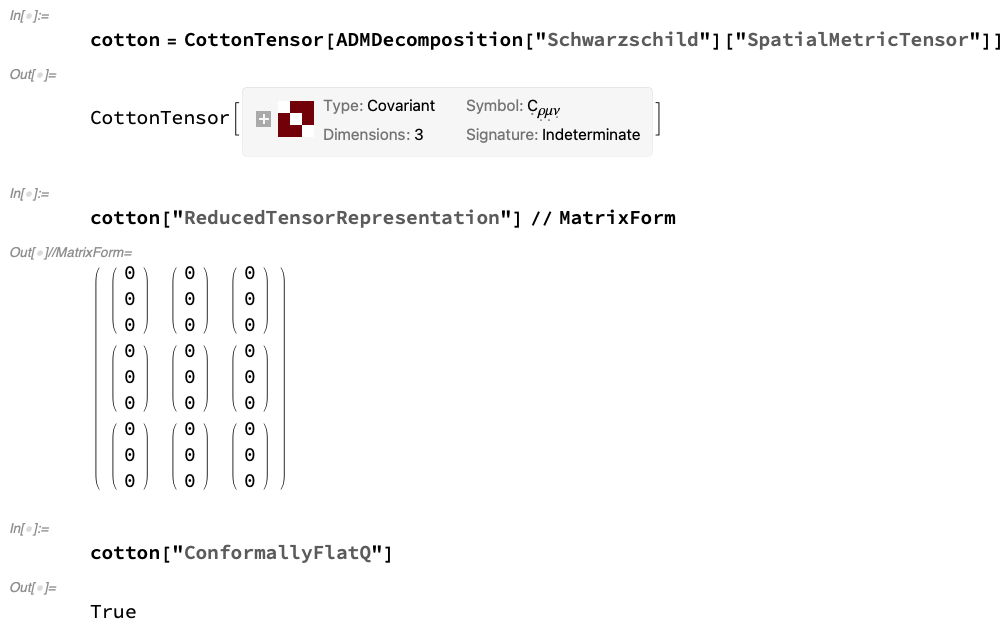}
\vrule
\includegraphics[width=0.545\textwidth]{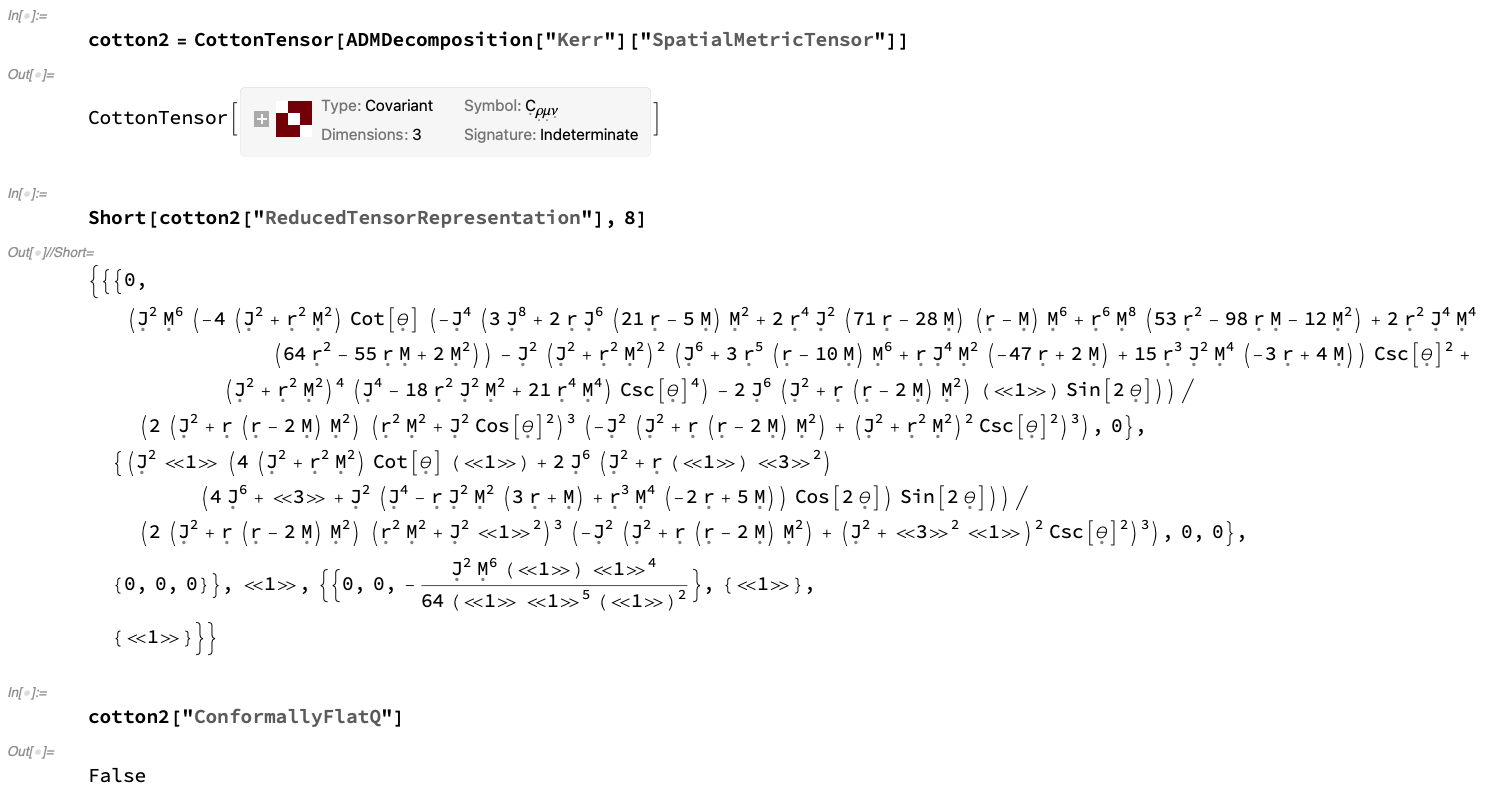}
\end{framed}
\caption{On the left, the \texttt{CottonTensor} object for the (initial) spacelike hypersurface of the \texttt{ADMDecomposition} object for a Schwarzschild geometry (representing, for instance, an uncharged, non-rotating black hole with mass $M$ in Schwarzschild or spherical polar coordinates ${\left( t, r, \theta, \phi \right)}$) with lapse function ${\alpha \left( t, r, \theta, \phi \right)}$ and shift vector ${\left( \beta^1 \left( t, r, \theta, \phi \right), \beta^2 \left( t, r, \theta, \phi \right), \beta^3 \left( t, r, \theta, \phi \right) \right)}$, illustrating that the hypersurface is conformally-flat. On the right, the \texttt{CottonTensor} object for the (initial) spacelike hypersurface of the \texttt{ADMDecomposition} object for a Kerr geometry (representing, for instance, an uncharged, spinning black hole with mass $M$ and angular momentum $J$ in Boyer-Lindquist or oblate spheroidal coordinates ${\left( t, r, \theta, \phi \right)}$) with lapse function ${\alpha \left( t, r, \theta, \phi \right)}$ and shift vector ${\left( \beta^1 \left( t, r, \theta, \phi \right), \beta^2 \left( t, r, \theta, \phi \right), \beta^3 \left( t, r, \theta, \phi \right) \right)}$, illustrating that the hypersurface is not conformally-flat.}
\label{fig:Figure15}
\end{figure}

\begin{figure}[ht]
\centering
\begin{framed}
\includegraphics[width=0.545\textwidth]{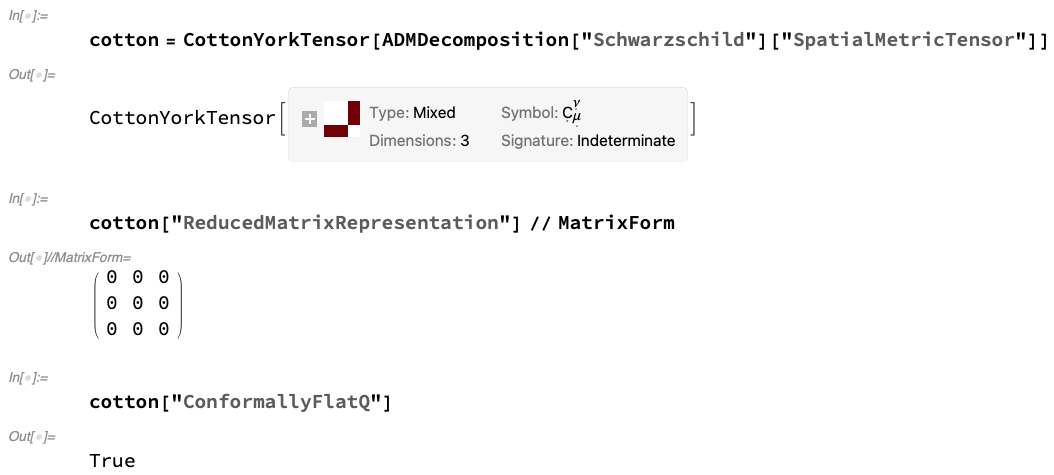}
\vrule
\includegraphics[width=0.445\textwidth]{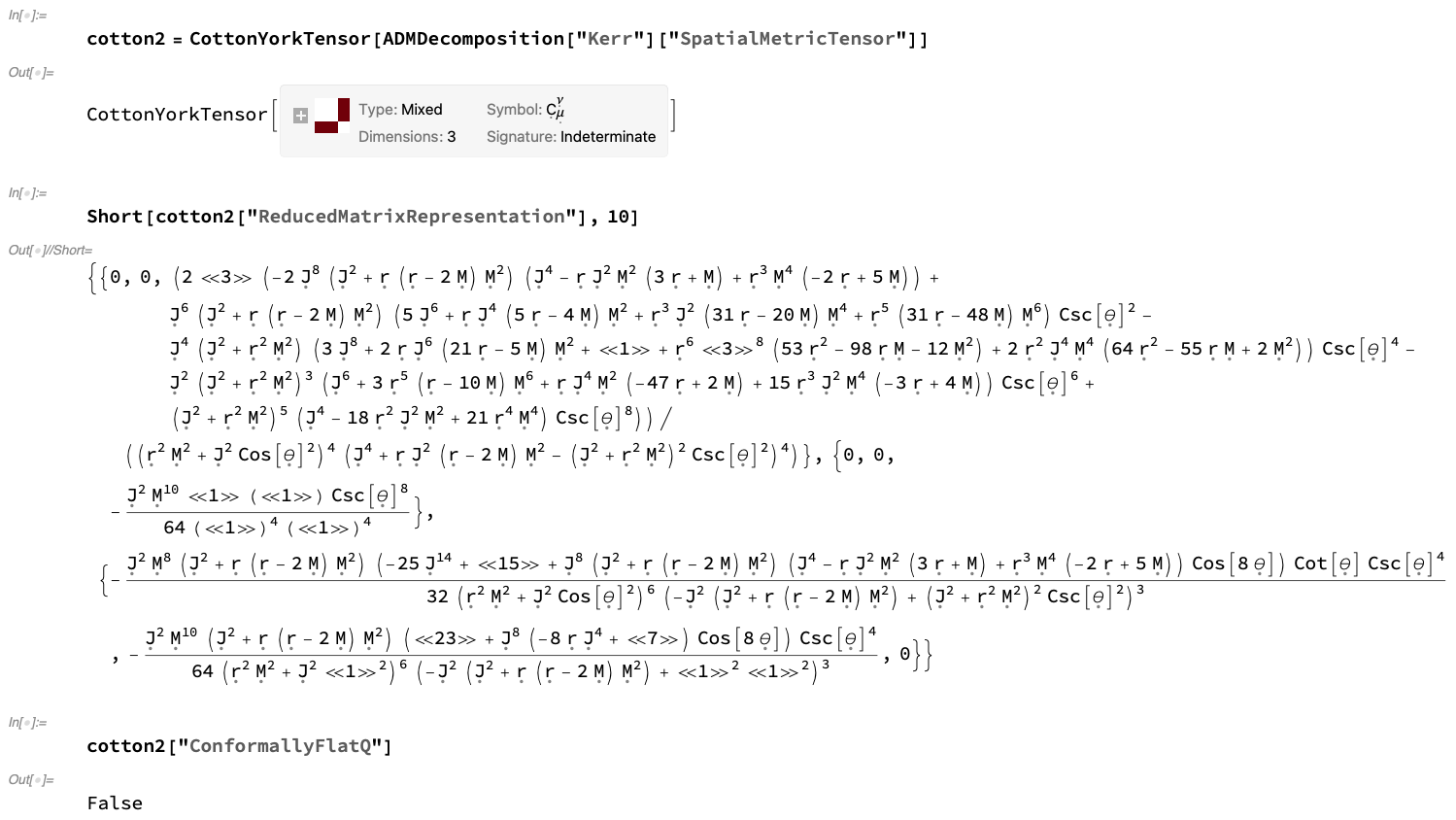}
\end{framed}
\caption{On the left, the \texttt{CottonYorkTensor} object for the (initial) spacelike hypersurface of the \texttt{ADMDecomposition} object for a Schwarzschild geometry (representing, for instance, an uncharged, non-rotating black hole with mass $M$ in Schwarzschild or spherical polar coordinates ${\left( t, r, \theta, \phi \right)}$) with lapse function ${\alpha \left( t, r, \theta, \phi \right)}$ and shift vector ${\left( \beta^1 \left( t, r, \theta, \phi \right), \beta^2 \left( t, r, \theta, \phi \right), \beta^3 \left( t, r, \theta, \phi \right) \right)}$, illustrating that the hypersurface is conformally-flat. On the right, the \texttt{CottonYorkTensor} object for the (initial) spacelike hypersurface of the \texttt{ADMDecomposition} object for a Kerr geometry (representing, for instance, an uncharged, spinning black hole with mass $M$ and angular momentum $J$ in Boyer-Lindquist or oblate spheroidal coordinates ${\left( t, r, \theta, \phi \right)}$) with lapse function ${\alpha \left( t, r, \theta, \phi \right)}$ and shift vector ${\left( \beta^1 \left( t, r, \theta, \phi \right), \beta^2 \left( t, r, \theta, \phi \right), \beta^3 \left( t, r, \theta, \phi \right) \right)}$, illustrating that the hypersurface is not conformally-flat.}
\label{fig:Figure16}
\end{figure}

In all dimensions ${n \geq 4}$, the vanishing of all components of the unique, trace-free, rank-4 Weyl tensor ${C_{\sigma \mu \nu}^{\rho}}$\cite{weyl2}, obtained by subtracting out all trace components (i.e. all Ricci tensor components ${R_{\mu \nu}}$) from the full Riemann tensor ${R_{\sigma \mu \nu}^{\rho}}$ via the Ricci decomposition theorem\cite{sharpe}:

\begin{multline}
C_{\rho \sigma \mu \nu} = R_{\rho \sigma \mu \nu} + \frac{1}{n - 1} \left( R_{\rho \nu} g_{\sigma \mu} - R_{\rho \mu} g_{\sigma \nu} + R_{\sigma \mu} g_{\rho \nu} - R_{\sigma \nu} g_{\rho \mu} \right)\\
+ \frac{R}{\left( n - 1 \right) \left( n - 2 \right)} \left( g_{\rho \mu} g_{\sigma \nu} - g_{\rho \nu} g_{\sigma \mu} \right),
\end{multline}
is both a necessary and sufficient condition for the underlying manifold ${\left( \mathcal{M}, g \right)}$ to be conformally-flat (by the Weyl-Schouten theorem\cite{eisenhart}), since the Weyl tensor ${C_{\sigma \mu \nu}^{\rho}}$ is invariant under all conformal transformations of the form ${\widetilde{g_{\mu \nu}} = \Omega^2 g_{\mu \nu}}$ with real conformal factor ${\Omega \in \mathbb{R}}$; the Weyl tensor ${C_{\sigma \mu \nu}^{\rho}}$ is also significant from a physical standpoint, since it governs the propagation of gravitational radiation within vacuum spacetimes. However, the fact that the Weyl tensor ${C_{\sigma \mu \nu}^{\rho}}$ vanishes identically in dimension ${n = 3}$, as demonstrated in Figure \ref{fig:Figure14} for the cases of spatial \texttt{MetricTensor} objects extracted from \texttt{ADMDecomposition} objects for the Schwarzschild and Kerr metrics (with the most general choice of gauge, although this makes no difference to the spacelike hypersurface geometry) presents problems in the case of ${3 + 1}$ decomposition and the ADM formalism, since it makes it more difficult to determine whether a given 3-dimensional spacelike hypersurface is conformally-flat (or, indeed, to extract gravitational wave data from 3-dimensional spacelike hypersurfaces directly); indeed, as seen here, the \texttt{WeylTensor} objects will (incorrectly) report both manifolds to be conformally-flat in these cases. To this end, it is instructive instead to consider the rank-3 Cotton tensor ${C_{\rho \mu \nu}}$\cite{cotton}, defined by:

\begin{equation}
C_{\rho \mu \nu} = \nabla_{\nu} R_{\rho \mu} - \nabla_{\mu} R_{\rho \nu} + \frac{1}{2 \left( n - 1 \right)} \left( \nabla_{\mu} \left( R g_{\rho \nu} \right) - \nabla_{\nu} \left( R g_{\rho \mu} \right) \right),
\end{equation}
i.e., in expanded form:

\begin{multline}
C_{\rho \mu \nu} = \left( \frac{\partial}{\partial x^{\nu}} \left( R_{\rho \mu} \right) - \Gamma_{\nu \rho}^{\sigma} R_{\sigma \mu} - \Gamma_{\nu \mu}^{\sigma} R_{\rho \sigma} \right) - \left( \frac{\partial}{\partial x^{\mu}} \left( R_{\rho \nu} \right) - \Gamma_{\mu \rho}^{\sigma} R_{\sigma \nu} - \Gamma_{\mu \nu}^{\sigma} R_{\rho \sigma} \right)\\
+ \frac{1}{2 \left( n - 1 \right)} \left( \left( \frac{\partial}{\partial x^{\mu}} \left( R g_{\rho \nu} \right) - \Gamma_{\mu \rho}^{\sigma} \left( R g_{\sigma \nu} \right) - \Gamma_{\mu \nu}^{\sigma} \left( R g_{\rho \sigma} \right) \right) \right.\\
\left. - \left( \frac{\partial}{\partial x^{\nu}} \left( R g_{\rho \mu} \right) - \Gamma_{\nu \rho}^{\sigma} \left( R g_{\sigma \mu} \right) - \Gamma_{\nu \mu}^{\sigma} \left( R g_{\rho \sigma} \right) \right) \right),
\end{multline}
for which vanishing of all components ${C_{\rho \mu \nu} = 0}$ is both a necessary and sufficient condition for conformal-flatness in dimension ${n = 3}$ (in dimensions ${n > 3}$, the vanishing of the Cotton tensor ${C_{\rho \mu \nu}}$ is a necessary but not sufficient condition). Figure \ref{fig:Figure15} illustrates the use of the \texttt{CottonTensor} function in \textsc{Gravitas} to demonstrate that the spatial \texttt{MetricTensor} object extracted from the \texttt{ADMDecomposition} object is conformally-flat for the case of the Schwarzschild metric, but not for the Kerr metric (since it is provable that there do not exist any conformally-flat spatial slices of the Kerr metric\cite{garat}). In fact, in the particular case of dimension ${n = 3}$, one can simplify the Cotton tensor ${C_{\rho \mu \nu}}$ even further by using the Hodge dual operation ${\star}$ in the associated Grassmann algebra to reduce the rank-3 Cotton tensor to the rank-2 Cotton-York tensor (density) ${C_{\mu}^{\nu}}$, defined by:

\begin{equation}
C_{\mu}^{\nu} = \nabla_{\rho} \left( R_{\sigma \mu} - \frac{1}{4} R g_{\sigma \mu} \right) \varepsilon^{\rho \sigma \nu},
\end{equation}
i.e., in expanded form:

\begin{equation}
C_{\mu}^{\nu} = \left( \frac{\partial}{\partial x^{\rho}} \left( R_{\sigma \mu} - \frac{1}{4} R g_{\sigma \mu} \right) - \Gamma_{\rho \sigma}^{\lambda} \left( R_{\lambda \mu} - \frac{1}{4} R g_{\lambda \mu} \right) - \Gamma_{\rho \mu}^{\lambda} \left( R_{\sigma \lambda} - \frac{1}{4} R g_{\sigma \lambda} \right) \right) \varepsilon^{\rho \sigma \nu},
\end{equation}
where ${\varepsilon_{\rho \sigma \nu}}$ designates the totally-antisymmetric Levi-Civita symbol of rank-3, and therefore the Cotton-York tensor (density) ${C_{\mu}^{\nu}}$ is definable \textit{only} in three dimensions (in which, like the full Cotton tensor ${C_{\rho \mu \nu}}$, the vanishing of all components of the Cotton-York tensor ${C_{\mu}^{\nu} = 0}$ is both a necessary and sufficient condition for conformal-flatness). Figure \ref{fig:Figure16} illustrates the use of the \texttt{CottonYorkTensor} function in \textsc{Gravitas} to demonstrate, once again, that the spatial \texttt{MetricTensor} object extracted from the \texttt{ADMDecomposition} object is conformally-flat in the Schwarzschild case but not in the Kerr case. The geometrical significance of the Cotton tensor ${C_{\rho \mu \nu}}$ lies in its very simple transformation law under conformal rescalings of the form ${\widetilde{g_{\\mu \nu}} = \Omega^2 g_{\mu \nu}}$ with ${\Omega \in \mathbb{R}}$, namely:

\begin{equation}
\widetilde{C_{\rho \mu \nu}} = C_{\rho \mu \nu} + \left( n - 2 \right) \frac{\partial}{\partial x^{\sigma}} \left( \log \left( \Omega \right) C_{\rho \mu \nu}^{\sigma} \right).
\end{equation}

\clearpage

\section{Gauge Conditions and Vacuum ADM Solutions}
\label{sec:Section3}

\begin{figure}[ht]
\centering
\begin{framed}
\includegraphics[width=0.545\textwidth]{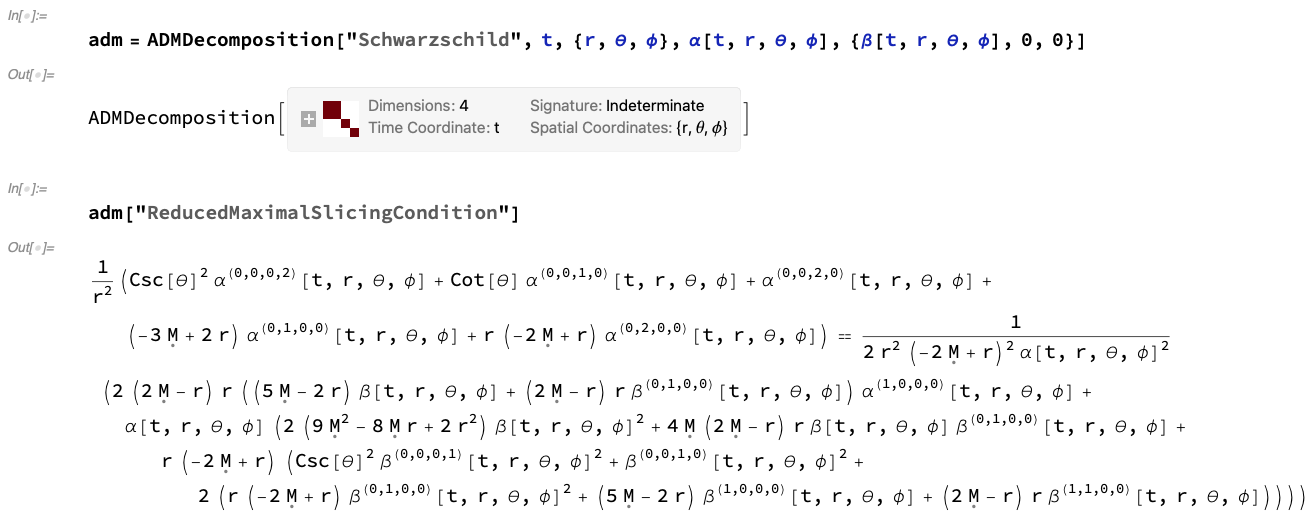}
\vrule
\includegraphics[width=0.445\textwidth]{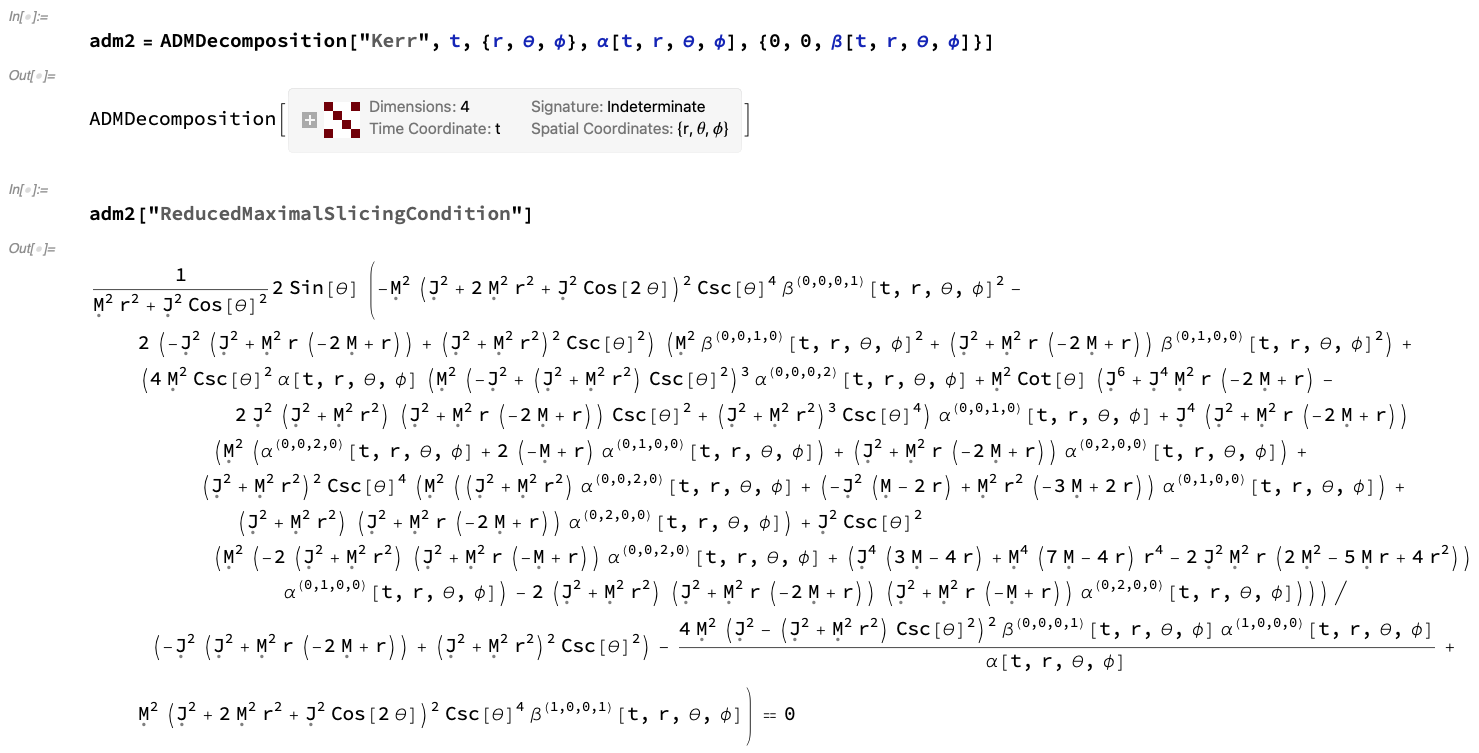}
\end{framed}
\caption{On the left, the maximal slicing gauge condition on the lapse function of the \texttt{ADMDecomposition} object for a Schwarzschild geometry (representing, for instance, an uncharged, non-rotating black hole with mass $M$ in Schwarzschild or spherical polar coordinates ${\left( t, r, \theta, \phi \right)}$) with lapse function ${\alpha \left( t, r, \theta, \phi \right)}$ and modified shift vector ${\left( \beta \left( t, r, \theta, \phi \right), 0, 0 \right)}$. On the right, the maximal slicing gauge condition on the lapse function of the \texttt{ADMDecomposition} object for a Kerr geometry (representing, for instance, an uncharged, spinning black hole with mass $M$ and angular momentum $J$ in Boyer-Lindquist or oblate spheroidal coordinates ${\left( t, r, \theta, \phi \right)}$) with lapse function ${\alpha \left( t, r, \theta, \phi \right)}$ and modified shift vector ${\left( 0, 0, \beta \left( t, r, \theta, \phi \right) \right)}$.}
\label{fig:Figure17}
\end{figure}

\begin{figure}[ht]
\centering
\begin{framed}
\includegraphics[width=0.495\textwidth]{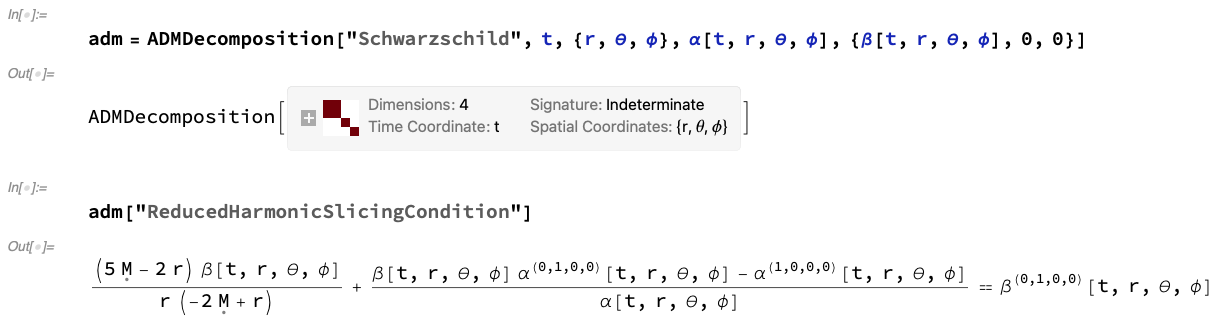}
\vrule
\includegraphics[width=0.495\textwidth]{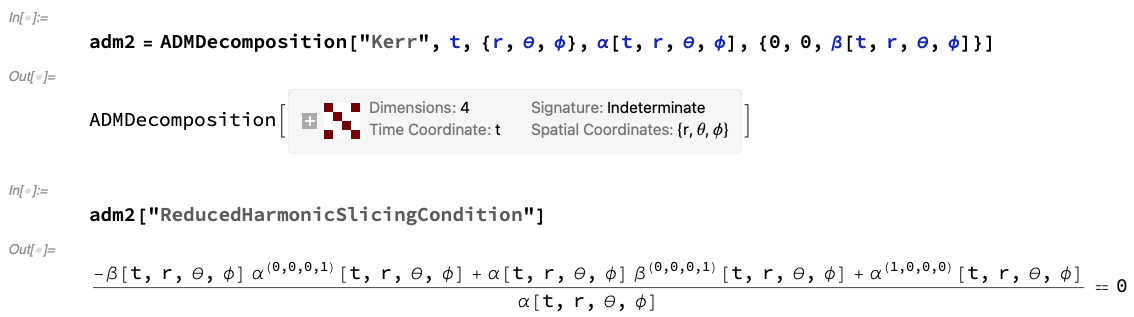}
\end{framed}
\caption{On the left, the harmonic slicing gauge condition on the lapse function of the \texttt{ADMDecomposition} object for a Schwarzschild geometry (representing, for instance, an uncharged, non-rotating black hole with mass $M$ in Schwarzschild or spherical polar coordinates ${\left( t, r, \theta, \phi \right)}$) with lapse function ${\alpha \left( t, r, \theta, \phi \right)}$ and modified shift vector ${\left( \beta \left( t, r, \theta, \phi \right), 0, 0 \right)}$. On the right, the harmonic slicing gauge condition on the lapse function of the \texttt{ADMDecomposition} object for a Kerr geometry (representing, for instance, an uncharged, spinning black hole of mass $M$ and angular momentum $J$ in Boyer-Lindquist or oblate spheroidal coordinates ${\left( t, r, \theta, \phi \right)}$) with lapse function ${\alpha \left( t, r, \theta, \phi \right)}$ and modified shift vector ${\left( 0, 0, \beta \left( t, r, \theta, \phi \right) \right)}$.}
\label{fig:Figure18}
\end{figure}

\begin{figure}[ht]
\centering
\begin{framed}
\includegraphics[width=0.495\textwidth]{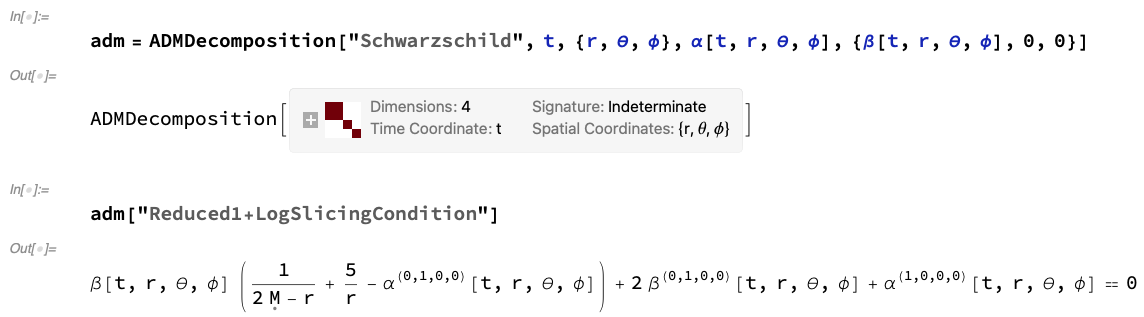}
\vrule
\includegraphics[width=0.495\textwidth]{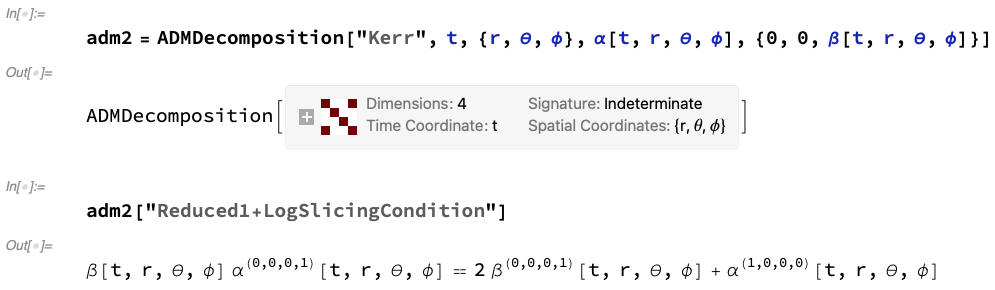}
\end{framed}
\caption{On the left, the ${1 + log}$ slicing gauge condition on the lapse function of the \texttt{ADMDecomposition} object for a Schwarzschild geometry (representing, for instance, an uncharged, non-rotating black hole with mass $M$ in Schwarzschild or spherical polar coordinates ${\left( t, r, \theta, \phi \right)}$) with lapse function ${\alpha \left( t, r, \theta, \phi \right)}$ and modified shift vector ${\left( \beta \left( t, r, \theta, \phi \right), 0, 0 \right)}$. On the right, the ${1 + log}$ slicing gauge condition on the lapse function of the \texttt{ADMDecomposition} object for a Kerr geometry (representing, for instance, an uncharged, spinning black hole of mass $M$ and angular momentum $J$ in Boyer-Lindquist or oblate spheroidal coordinates ${\left( t, r, \theta, \phi \right)}$) with lapse function ${\alpha \left( t, r, \theta, \phi \right)}$ and modified shift vector ${\left( 0, 0, \beta \left( t, r, \theta, \phi \right) \right)}$.}
\label{fig:Figure19}
\end{figure}

In order to be able to formulate the Einstein field equations as a (well-posed) initial value problem, it is necessary not only to specify the (initial) Cauchy data in the form of the (initial) spacelike hypersurface/spatial metric tensor ${\gamma_{\mu \nu}}$, but also to give an appropriate prescription for the \textit{gauge}, i.e. a sufficient set of mathematical conditions for one to be able to determine (uniquely) the values of the lapse function ${\alpha}$ and the shift vector field ${\beta^{\mu}}$ at every point in the ambient spacetime. The \textsc{Gravitas} framework makes the terminological distinction between \textit{slicing conditions} (which are constraints on the lapse function ${\alpha}$, and which therefore determine geometrically how the spacetime is foliated into a time-ordered sequence of spacelike hypersurfaces) and \textit{coordinate conditions} (which are constraints on the shift vector ${\beta^{\mu}}$, and which therefore determine how the spatial coordinates are relabeled as one moves between neighboring spacelike hypersurfaces). The most trivial choice of gauge would involve imposing the geodesic slicing condition, in which ${\alpha = 1}$ everywhere, along with the normal coordinate conditions, in which ${\beta^{\mu} = 0}$ everywhere; these conditions are extremely straightforward to impose, although they have a tendency to produce coordinate pathologies (and, in particular, make no effort to avoid singularities) and are therefore of limited utility for numerical simulations. In addition to geodesic slicing and normal coordinates, \texttt{ADMDecomposition} includes a small library of in-built gauge conditions (including most standard slicing and coordinate conditions), with many more planned for future inclusion. For example, a very commonly-used slicing condition in numerical relativity (due to its tendency to avoid singularities, and therefore its utility when simulating black hole spacetimes) is the maximal slicing condition, due originally to Lichnerowicz\cite{lichnerowicz} and subsequently developed further by York\cite{york}:

\begin{equation}
{}^{\left( 3 \right)} \Delta \alpha = \alpha K^{\mu \nu} K_{\mu \nu} - \frac{\partial}{\partial t} \left( K \right),
\end{equation}
where ${{}^{\left( 3 \right)} \Delta}$ designates the induced (connection) Laplacian on spacelike hypersurfaces, where the connection Laplacian ${\Delta}$ is defined abstractly (for arbitrary tensor fields $T$) as the trace of the second covariant derivative ${\nabla^2}$, i.e. ${\Delta T = \mathrm{tr} \left( \nabla^2 T \right)}$, where the second covariant derivative is itself given abstractly by:

\begin{equation}
\forall \mathbf{X}, \mathbf{Y} \in \Gamma \left( \bigsqcup_{\mathbf{x} \in \mathcal{M}} T_{\mathbf{x}} \mathcal{M} \right), \qquad \nabla_{\mathbf{X}, \mathbf{Y}}^{2} T = \nabla_{\mathbf{X}} \left( \nabla_{\mathbf{Y}} T \right) - \nabla_{\nabla_{\mathbf{X}} \mathbf{Y}} T.
\end{equation}
In explicit component form, the action of the induced (connection) Laplacian ${{}^{\left( 3 \right)} \Delta}$ on an arbitrary scalar field ${\phi}$ (i.e. a rank-0 tensor field) can therefore be written as:

\begin{equation}
{}^{\left( 3 \right)} \Delta \phi = {}^{\left( 3 \right)} \nabla^{\mu} \left( {}^{\left( 3 \right)} \nabla_{\mu} \phi \right) = \gamma^{\mu \sigma} {}^{\left( 3 \right)} \nabla_{\sigma} \left( {}^{\left( 3 \right)} \nabla_{\mu} \phi \right) = \gamma^{\mu \sigma} \left( \frac{\partial}{\partial x^{\sigma}} \left( \frac{\partial}{\partial x^{\mu}} \left( \phi \right) \right) - {}^{\left( 3 \right)} \Gamma_{\sigma \mu}^{\lambda} \left( \frac{\partial}{\partial x^{\lambda}} \left( \phi \right) \right) \right),
\end{equation}
or, equivalently:

\begin{equation}
{}^{\left( 3 \right)} \Delta \phi = \frac{1}{\sqrt{\det \left( \gamma_{\mu \nu} \right)}} \left( \frac{\partial}{\partial x^{\mu}} \left( \sqrt{\det \left( \gamma_{\mu \nu} \right)} \left( \gamma^{\mu \nu} \frac{\partial}{\partial x^{\nu}} \left( \phi \right) \right) \right) \right),
\end{equation}
hence allowing us to express the maximal slicing condition in the following (explicit, expanded) form, by treating the lapse function ${\alpha}$ as a pure scalar field defined over each spacelike hypersurface:

\begin{equation}
\frac{1}{\sqrt{\det \left( \gamma_{\mu \nu} \right)}} \left( \frac{\partial}{\partial x^{\mu}} \left( \sqrt{\det \left( \gamma_{\mu \nu} \right)} \left( \gamma^{\mu \nu} \frac{\partial}{\partial x^{\nu}} \left( \alpha \right) \right) \right) \right) = \alpha K^{\mu \nu} K_{\mu \nu} - \frac{\partial}{\partial t} \left( K \right),
\end{equation}
with ${\mu, \nu, \sigma, \lambda}$ in all of the above ranging across all ${\left\lbrace 0, \dots, n - 2 \right\rbrace}$ (i.e. across spatial coordinate indices only). Intuitively, the maximal slicing condition seeks to maximize the spatial volume of each hypersurface by essentially ``slowing'' the evolution in regions of high curvature and ``accelerating'' it in regions of low curvature. Figure \ref{fig:Figure17} shows the maximal slicing gauge condition, as computed directly from the \texttt{ADMDecomposition} objects for the Schwarzschild metric (representing, for instance, an uncharged, non-rotating black hole with mass $M$ in Schwarzschild or spherical polar coordinates ${\left( t, r, \theta, \phi \right)}$) and the Kerr metric (representing, for instance, an uncharged, spinning black hole with mass $M$ and angular momentum $J$ in Boyer-Lindquist or oblate spheroidal coordinates ${\left( t, r, \theta, \phi \right)}$), assuming a restricted choice of gauge consisting of the lapse function ${\alpha \left( t, r, \theta, \phi \right)}$ and the modified shift vectors ${\left( \beta \left( t, r, \theta, \phi \right), 0, 0 \right)}$ (for Schwarzschild) and ${\left( 0, 0, \beta \left( t, r, \theta, \phi \right) \right)}$ (for Kerr). Another slicing condition that is frequently used whenever one wishes to cast the Einstein field equations in a strongly hyperbolic form is the harmonic slicing condition first investigated by Bona and Mass\'o\cite{bona2}\cite{bona3}, though its singularity-avoidance properties were discovered by Alcubierre and Mass\'o\cite{alcubierre2} to be somewhat less favorable than those of maximal slicing, in which one simply asserts that the (spacetime) Laplacian of the time coordinate $t$ (considered now as a pure scalar field on the spacetime) vanishes identically:

\begin{equation}
{}^{\left( 4 \right)} \Delta t = {}^{\left( 4 \right)} \nabla^{\mu} \left( {}^{\left( 4 \right)} \nabla_{\mu} t \right) = g^{\mu \sigma} {}^{\left( 4 \right)} \nabla_{\sigma} \left( {}^{\left( 4 \right)} \nabla_{\mu} t \right) = 0,
\end{equation}
where ${{}^{\left( 4 \right)} \Delta}$ designates the ambient (connection) Laplacian on spacetime, i.e., in expanded form, one has:

\begin{equation}
g^{\mu \sigma} \left( \frac{\partial}{\partial x^{\sigma}} \left( \frac{\partial}{\partial x^{\mu}} \left( t \right) \right) - {}^{\left( 4 \right)} \Gamma_{\sigma \mu}^{\lambda} \left( \frac{\partial}{\partial x^{\lambda}} \left( t \right) \right) \right) = 0,
\end{equation}
with ${\mu, \sigma, \lambda}$ here ranging across all ${\left\lbrace 0, \dots, n - 1 \right\rbrace}$ (i.e. across all spacetime coordinate indices). Intuitively, the harmonic slicing condition replaces the typical elliptic slicing condition (as is the case with, say, maximal slicing) with a corresponding hyperbolic one. Figure \ref{fig:Figure18} demonstrates the harmonic slicing gauge condition, as computed directly from the \texttt{ADMDecomposition} objects for the Schwarzschild and Kerr metrics, with the same restricted choice of gauge (with lapse function ${\alpha \left( t, r, \theta, \phi \right)}$ and modified shift vectors ${\left( \beta \left( t, r, \theta, \phi \right), 0, 0 \right)}$ and ${\left( 0, 0, \beta \left( t, r, \theta, \phi \right) \right)}$, respectively). Finally, a slicing condition that is frequently used in the simulation of compact binary inspirals and mergers\cite{campanelli}\cite{baker} (due to its empirical stability properties and comparable singularity-avoidance properties to maximal slicing) is the ${1 + log}$ slicing condition:

\begin{equation}
\frac{\partial}{\partial t} \left( \alpha \right) = \mathcal{L}_{\boldsymbol{\beta}} \alpha - 2 K \alpha = \beta^{\sigma} \frac{\partial}{\partial x^{\sigma}} \left( \alpha \right) - 2 K \alpha,
\end{equation}
so-named because it can be rewritten in the slightly expanded form:

\begin{multline}
\frac{\partial}{\partial t} \left( \alpha \right) = \beta^{\sigma} \frac{\partial}{\partial x^{\sigma}} \left( \alpha \right) + \frac{\partial}{\partial t} \left( \log \left( \det \left( \gamma_{\mu \nu} \right) \right) \right) - 2 {}^{\left( 3 \right)} \nabla_{\sigma} \beta^{\sigma}\\
= \beta^{\sigma} \frac{\partial}{\partial x^{\sigma}} \left( \alpha \right) + \frac{\partial}{\partial t} \left( \log \left( \det \left( \gamma_{\mu \nu} \right) \right) \right) - 2 \frac{\partial}{\partial x^{\sigma}} \left( \beta^{\sigma} \right) - 2 {}^{\left( 3 \right)} \Gamma_{\sigma \rho}^{\sigma} \beta^{\rho},
\end{multline}
which, in the particular case where normal coordinate conditions are enforced and in which the shift vector therefore vanishes everywhere, i.e. ${\boldsymbol{\beta} = \mathbf{0}}$, reduces to:

\begin{equation}
\frac{\partial}{\partial t} \left( \alpha \right) = \frac{\partial}{\partial t} \left( \log \left( \det \left( \gamma_{\mu \nu} \right) \right) \right), \qquad \text{ with solution } \qquad \alpha = 1 + \log \left( \det \left( \gamma_{\mu \nu} \right) \right),
\end{equation}
with ${\sigma, \rho}$ in all of the above ranging across all ${\left\lbrace 0, \dots, n - 2 \right\rbrace}$ (i.e. across spatial coordinate indices only). ${1 + log}$ slicing is itself a special case of the general Bona-Mass\'o slicing condition presented in \cite{bona2}\cite{bona3}. Figure \ref{fig:Figure19} illustrates the ${1 + log}$ slicing gauge condition, as computed directly from the \texttt{ADMDecomposition} objects for the Schwarzschild and Kerr metrics, with the same restricted choice of gauge (with lapse function ${\alpha \left( t, r, \theta, \phi \right)}$ and modified shift vectors ${\left( \beta \left( t, r, \theta, \phi \right), 0, 0 \right)}$ and ${\left( 0, 0, \beta \left( t, r, \theta, \phi \right) \right)}$, respectively).

\begin{figure}[ht]
\centering
\begin{framed}
\includegraphics[width=0.545\textwidth]{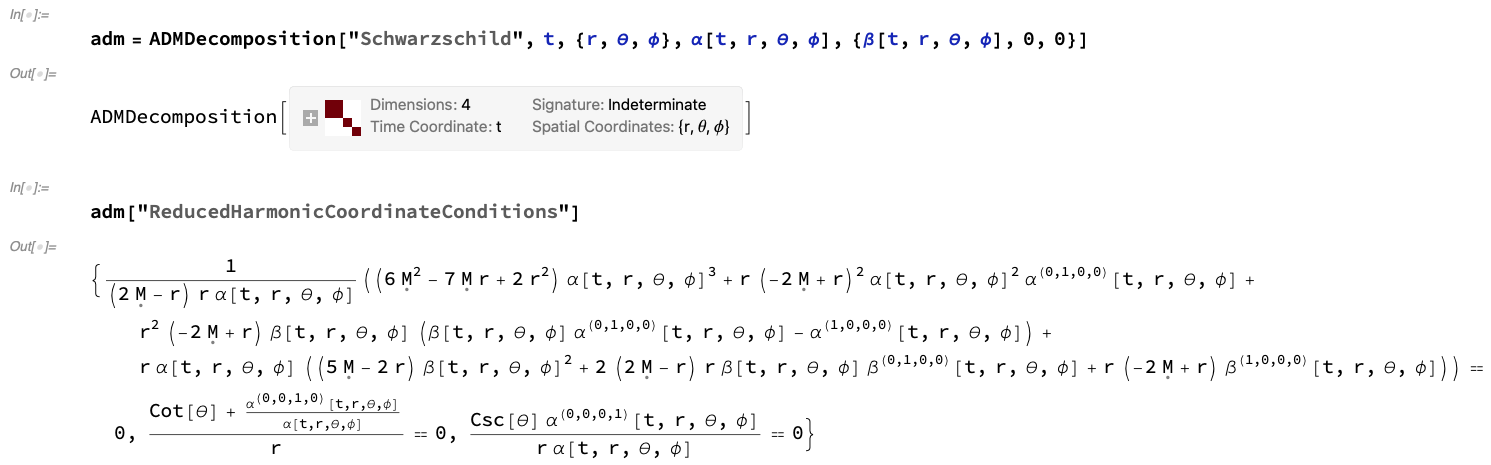}
\vrule
\includegraphics[width=0.445\textwidth]{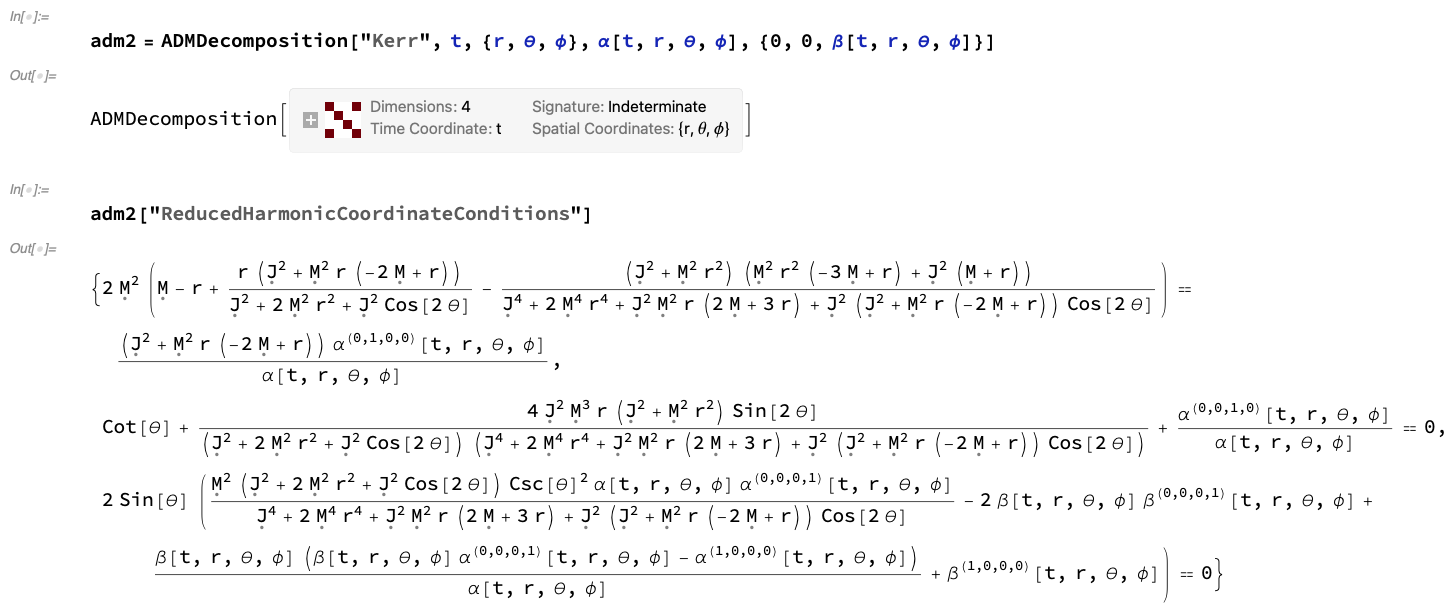}
\end{framed}
\caption{On the left, the harmonic coordinate gauge conditions on the shift vector of the \texttt{ADMDecomposition} object for a Schwarzschild geometry (representing, for instance, an uncharged, non-rotating black hole with mass $M$ in Schwarzschild or spherical polar coordinates ${\left( t, r, \theta, \phi \right)}$) with lapse function ${\alpha \left( t, r, \theta, \phi \right)}$ and modified shift vector ${\left( \beta \left( t, r, \theta, \phi \right), 0, 0 \right)}$. On the right, the harmonic coordinate gauge conditions on the shift vector of the \texttt{ADMDecomposition} object for a Kerr geometry (representing, for instance, an uncharged, spinning black hole with mass $M$ and angular momentum $J$ in Boyer-Lindquist or oblate spheroidal coordinates ${\left( t, r, \theta, \phi \right)}$) with lapse function ${\alpha \left( t, r, \theta, \phi \right)}$ and modified shift vector ${\left( 0, 0, \beta \left( t, r, \theta, \phi \right) \right)}$.}
\label{fig:Figure20}
\end{figure}

\begin{figure}[ht]
\centering
\begin{framed}
\includegraphics[width=0.495\textwidth]{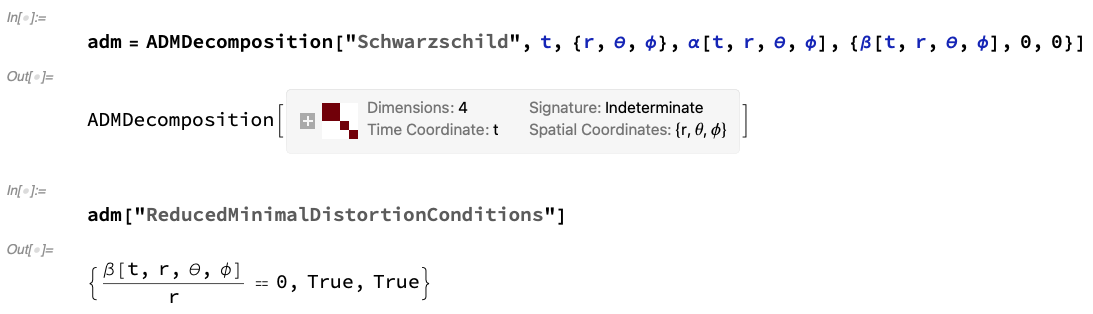}
\vrule
\includegraphics[width=0.495\textwidth]{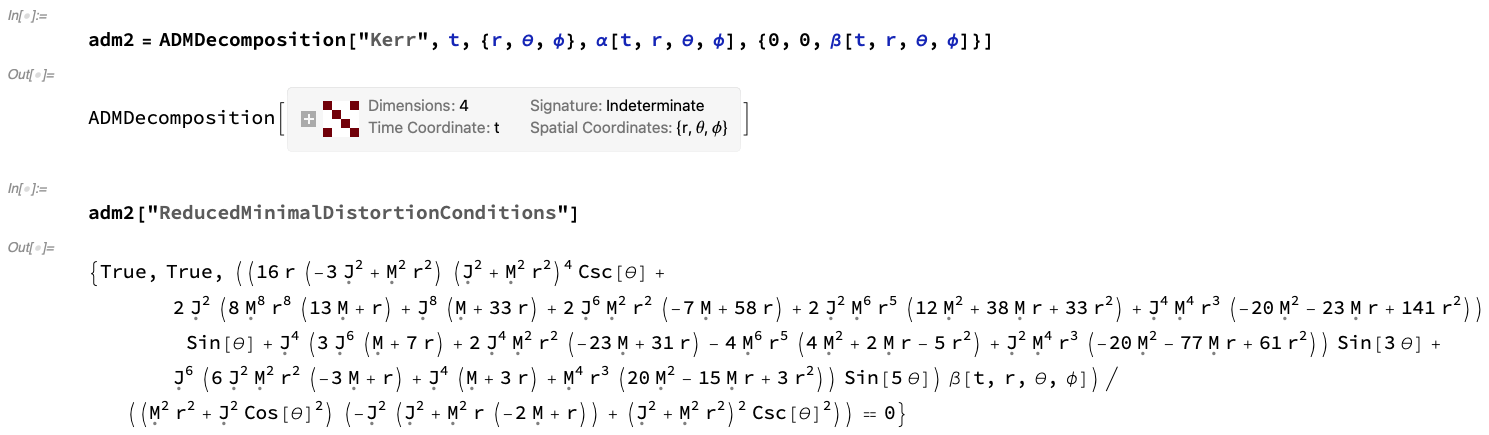}
\end{framed}
\caption{On the left, the minimal distortion coordinate gauge conditions on the shift vector of the \texttt{ADMDecomposition} object for a Schwarzschild geometry (representing, for instance, an uncharged, non-rotating black hole with mass $M$ in Schwarzschild or spherical polar coordinates ${\left( t, r, \theta, \phi \right)}$) with lapse function ${\alpha \left( t, r, \theta, \phi \right)}$ and modified shift vector ${\left( \beta \left( t, r, \theta, \phi \right), 0, 0 \right)}$. On the right, the minimal distortion coordinate gauge conditions on the shift vector of the \texttt{ADMDecomposition} object for a Kerr geometry (representing, for instance, an uncharged, spinning black hole with mass $M$ and angular momentum $J$ in Boyer-Lindquist or oblate spheroidal coordinates ${\left( t, r, \theta, \phi \right)}$) with lapse function ${\alpha \left( t, r, \theta, \phi \right)}$ and modified shift vector ${\left( 0, 0, \beta \left( t, r, \theta, \phi \right) \right)}$.}
\label{fig:Figure21}
\end{figure}

\begin{figure}[ht]
\centering
\begin{framed}
\includegraphics[width=0.495\textwidth]{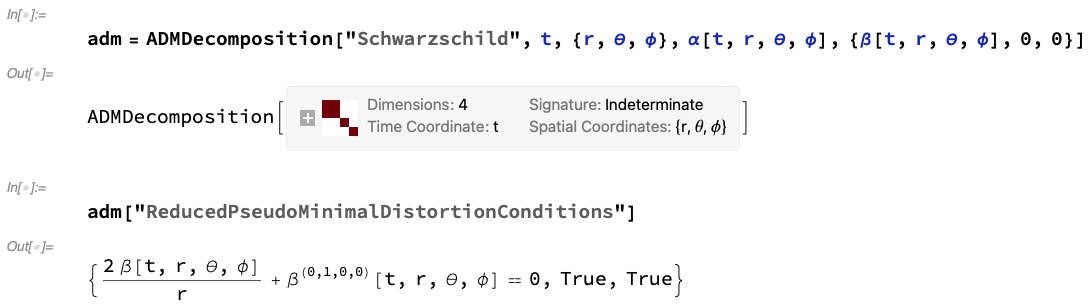}
\vrule
\includegraphics[width=0.495\textwidth]{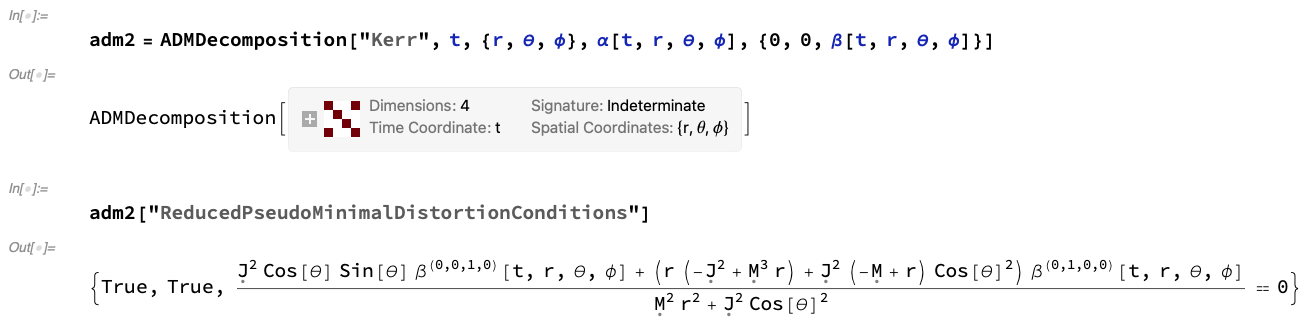}
\end{framed}
\caption{On the left, the pseudo-minimal distortion coordinate gauge conditions on the shift vector of the \texttt{ADMDecomposition} object for a Schwarzschild geometry (representing, for instance, an uncharged, non-rotating black hole with mass $M$ in Schwarzschild or spherical polar coordinates ${\left( t, r, \theta, \phi \right)}$) with lapse function ${\alpha \left( t, r, \theta, \phi \right)}$ and modified shift vector ${\left( \beta \left( t, r, \theta, \phi \right), 0, 0 \right)}$. On the right, the pseudo-minimal distortion coordinate gauge conditions on the shift vector of the \texttt{ADMDecomposition} object for a Kerr geometry (representing, for instance, an uncharged, spinning black hole with mass $M$ and angular momentum $J$ in Boyer-Lindquist or oblate spheroidal coordinates ${\left( t, r, \theta, \phi \right)}$) with lapse function ${\alpha \left( t, r, \theta, \phi \right)}$ and modified shift vector ${\left( 0, 0, \beta \left( t, r, \theta, \phi \right) \right)}$.}
\label{fig:Figure22}
\end{figure}

Dual to the harmonic slicing condition on the lapse function ${\alpha}$ are the harmonic coordinate conditions on the shift vector ${\beta^{\mu}}$\cite{bona2}\cite{bona3}, in which one treats each spatial coordinate ${x^{\rho}}$ as a pure scalar field on the spacetime, and asserts that the (spacetime) Laplacian of each such scalar field vanishes identically:

\begin{equation}
{}^{\left( 4 \right)} \Delta x^{\rho} = {}^{\left( 4 \right)} \nabla^{\mu} \left( {}^{\left( 4 \right)} \nabla_{\mu} x^{\rho} \right) = g^{\mu \sigma} {}^{\left( 4 \right)} \nabla_{\sigma} \left( {}^{\left( 4 \right)} \nabla_{\mu} x^{\rho} \right) = 0,
\end{equation}
i.e., in expanded form:

\begin{equation}
g^{\mu \sigma} \left( \frac{\partial}{\partial x^{\sigma}} \left( \frac{\partial}{\partial x^{\mu}} \left( x^{\rho} \right) \right) - {}^{\left( 4 \right)} \Gamma_{\sigma \mu}^{\lambda} \left( \frac{\partial}{\partial x^{\lambda}} \left( x^{\rho} \right) \right) \right) = 0,
\end{equation}
with ${\mu, \sigma, \lambda}$ here ranging across all ${\left\lbrace 0, \dots, n - 1 \right\rbrace}$ (i.e. across all spacetime coordinate indices), and with ${\rho}$ ranging across all ${\left\lbrace 0, \dots, n - 2 \right\rbrace}$ (i.e. across spatial coordinate indices only). The harmonic coordinate conditions enjoy the same hyperbolicity and (partial) singularity-avoidance properties as the harmonic slicing condition discussed above. Figure \ref{fig:Figure20} shows the harmonic coordinate gauge conditions, as computed directly from the \texttt{ADMDecomposition} object for the Schwarzschild and Kerr metrics, with the same restricted choice of gauge as above (with lapse function ${\alpha \left( t, r, \theta, \phi \right)}$ and modified shift vectors ${\left( \beta \left( t, r, \theta, \phi \right), 0, 0 \right)}$ and ${\left( 0, 0, \beta \left( t, r, \theta, \phi \right) \right)}$, respectively). Another common set of coordinate conditions, which seek to minimize the hypersurface ``strain'' (i.e. to minimize the distortion in the spatial coordinate systems between neighboring spacelike hypersurfaces), are the minimal distortion conditions, due originally to Smarr and York\cite{smarr}, and later adapted by Brady, Creighton and Thorne\cite{brady}:

\begin{multline}
{}^{\left( 3 \right)} \nabla^{\mu} \left( {}^{\left( 3 \right)} \nabla_{\mu} \beta^{\nu} \right) + {}^{\left( 3 \right)} \nabla^{\nu} \left( {}^{\left( 3 \right)} \nabla_{\mu} \beta^{\mu} \right) - 2 {}^{\left( 3 \right)} \nabla_{\mu} \left( \alpha K^{\mu \nu} \right)\\
= \gamma^{\mu \sigma} {}^{\left( 3 \right)} \nabla_{\sigma} \left( {}^{\left( 3 \right)} \nabla_{\mu} \beta^{\nu} \right) + \gamma^{\nu \sigma} {}^{\left( 3 \right)} \nabla_{\sigma} \left( {}^{\left( 3 \right)} \nabla_{\mu} \beta^{\mu} \right) - 2 {}^{\left( 3 \right)} \nabla_{\mu} \left( \alpha K^{\mu \nu} \right) = 0,
\end{multline}
i.e., in expanded form:

\begin{multline}
\gamma^{\mu \sigma} \left( \frac{\partial}{\partial x^{\sigma}} \left( D_{\mu}^{\nu} \right) + {}^{\left( 3 \right)} \Gamma_{\sigma \lambda}^{\nu} D_{\mu}^{\lambda} - {}^{\left( 3 \right)} \Gamma_{\sigma \mu}^{\lambda} D_{\lambda}^{\nu} \right) + \gamma^{\nu \sigma} \left( \frac{\partial}{\partial x^{\sigma}} \left( D_{\mu}^{\mu} \right) + {}^{\left( 3 \right)} \Gamma_{\sigma \lambda}^{\mu} D_{\mu}^{\lambda} - {}^{\left( 3 \right)} \Gamma_{\sigma \mu}^{\lambda} D_{\lambda}^{\mu} \right)\\
- 2 \left( \frac{\partial}{\partial x^{\mu}} \left( \alpha K^{\mu \nu} \right) + {}^{\left( 3 \right)} \Gamma_{\mu \sigma}^{\mu} \left( \alpha K^{\sigma \nu} \right) + {}^{\left( 3 \right)} \Gamma_{\mu \sigma}^{\nu} \left( \alpha K^{\mu \sigma} \right) \right) = 0,
\end{multline}
where we have introduced, for the sake of notational convenience, the rank-2 tensor ${D_{\mu}^{\nu}}$ consisting of (spatial) covariant derivatives of the shift vector ${\beta^{\nu}}$:

\begin{equation}
D_{\mu}^{\nu} = {}^{\left( 3 \right)} \nabla_{\mu} \beta^{\nu} = \frac{\partial}{\partial x^{\mu}} \left( \beta^{\nu} \right) + {}^{\left( 3 \right)} \Gamma_{\mu \sigma}^{\nu} \beta^{\sigma},
\end{equation}
and with ${\mu, \nu, \rho, \sigma, \lambda}$ here ranging across all ${\left\lbrace 0, \dots, n - 2 \right\rbrace}$ (i.e. across spatial coordinate indices only). Figure \ref{fig:Figure21} demonstrates the minimal distortion coordinate gauge conditions, as computed directly from the \texttt{ADMDecomposition} objects for the Schwarzschild and Kerr metrics, with the same restricted choice of gauge (with lapse function ${\alpha \left( t, r, \theta, \phi \right)}$ and modified shift vectors ${\left( \beta \left( t, r, \theta, \phi \right), 0, 0 \right)}$ and ${\left( 0, 0, \beta \left( t, r, \theta, \phi \right) \right)}$, respectively). In the case where the variations of the spatial metric tensor ${\gamma_{\mu \nu}}$ between neighboring hypersurfaces are relatively small, one can simplify the computation of the minimal distortion conditions substantially by approximating all covariant derivatives as partial derivatives (hence discarding all spatial Christoffel symbol terms), thus yielding the pseudo-minimal distortion conditions, as first proposed by Oohara and Nakamura\cite{oohara}:

\begin{equation}
\gamma^{\mu \sigma} \frac{\partial}{\partial x^{\sigma}} \left( \frac{\partial}{\partial x^{\mu}} \left( \beta^{\nu} \right) \right) + \gamma^{\nu \sigma} \frac{\partial}{\partial x^{\sigma}} \left( \frac{\partial}{\partial x^{\mu}} \left( \beta^{\mu} \right) \right) - 2 \frac{\partial}{\partial x^{\mu}} \left( \alpha K^{\mu \nu} \right) = 0.
\end{equation}
Figure \ref{fig:Figure22} illustrates the pseudo-minimal distortion coordinate gauge conditions, as computed directly from the \texttt{ADMDecomposition} objects for the Schwarzschild and Kerr metrics, with the same restricted choice of gauge (with lapse function ${\alpha \left( t, r, \theta, \phi \right)}$ and modified shift vectors ${\left( \beta \left( t, r, \theta, \phi \right), 0, 0 \right)}$ and ${\left( 0, 0, \beta \left( t, r, \theta, \phi \right) \right)}$, respectively).

\begin{figure}[ht]
\centering
\begin{framed}
\includegraphics[width=0.495\textwidth]{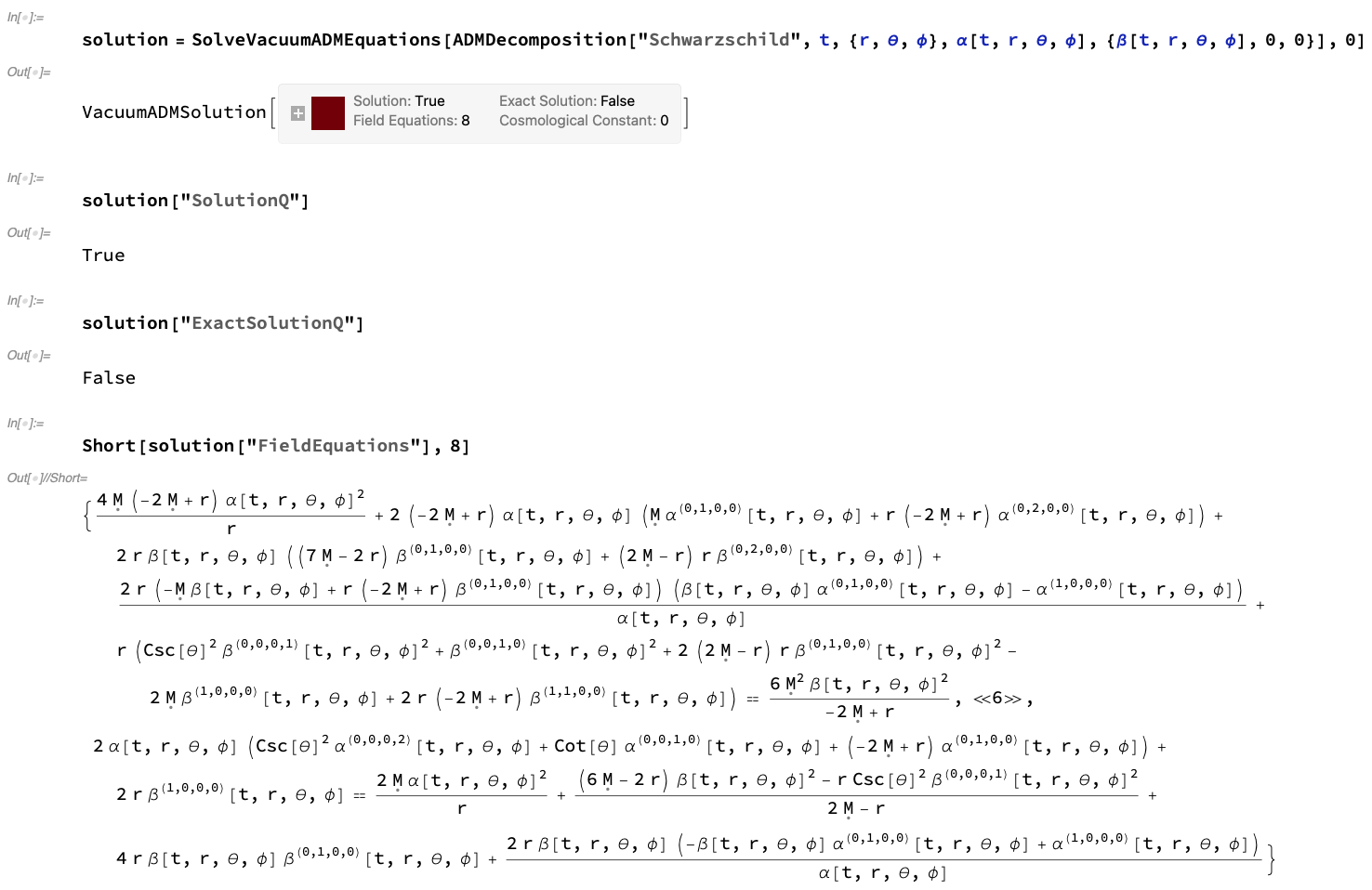}
\vrule
\includegraphics[width=0.495\textwidth]{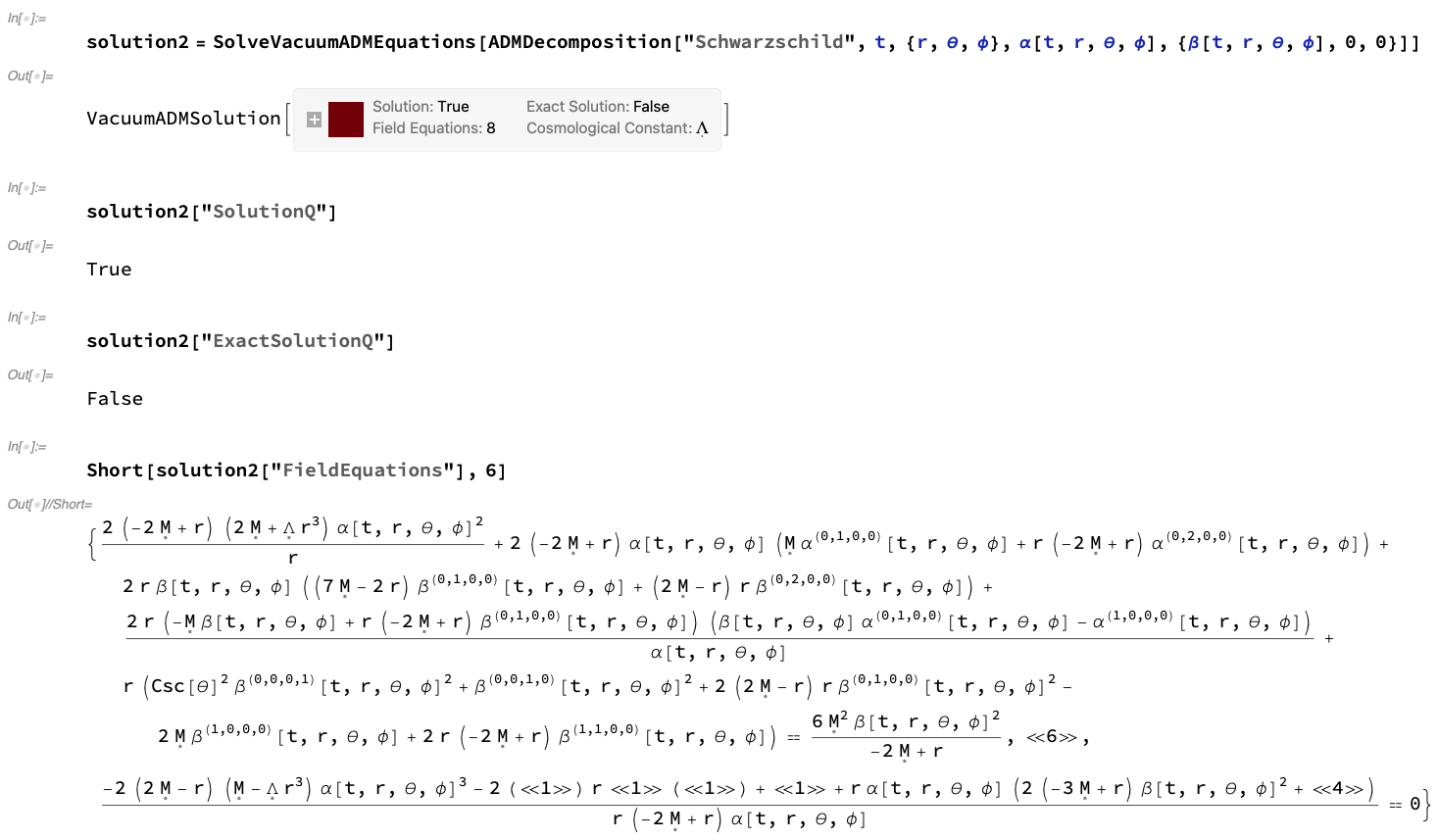}
\end{framed}
\caption{On the left, the \texttt{VacuumADMSolution} object for an ADM decomposition of a Schwarzschild geometry (representing, for instance, an uncharged, non-rotating black hole with mass $M$ in Schwarzschild or spherical polar coordinates ${\left( t, r, \theta, \phi \right)}$) with lapse function ${\alpha \left( t, r, \theta, \phi \right)}$ and modified shift vector ${\left( \beta \left( t, r, \theta, \phi \right), 0, 0 \right)}$, with zero cosmological constant ${\Lambda = 0}$, computed using \texttt{SolveVacuumADMEquations}, illustrating that this decomposition is a non-exact solution to the vacuum ADM evolution equations, with eight additional field equations required. On the right, the \texttt{VacuumADMSolution} object for an ADM decomposition of a Schwarzschild geometry (representing, for instance, an uncharged, non-rotating black hole with mass $M$ in Schwarzschild or spherical polar coordinates ${\left( t, r, \theta, \phi \right)}$) with lapse function ${\alpha \left( t, r, \theta, \phi \right)}$ and modified shift vector ${\left( \beta \left( t, r, \theta, \phi \right), 0, 0 \right)}$, with non-zero cosmological constant ${\Lambda \neq 0}$, computed using \texttt{SolveVacuumADMEquations}, illustrating that this decomposition is a non-exact solution to the vacuum ADM evolution equations, with eight additional field equations required.}
\label{fig:Figure23}
\end{figure}

\begin{figure}[ht]
\centering
\begin{framed}
\includegraphics[width=0.495\textwidth]{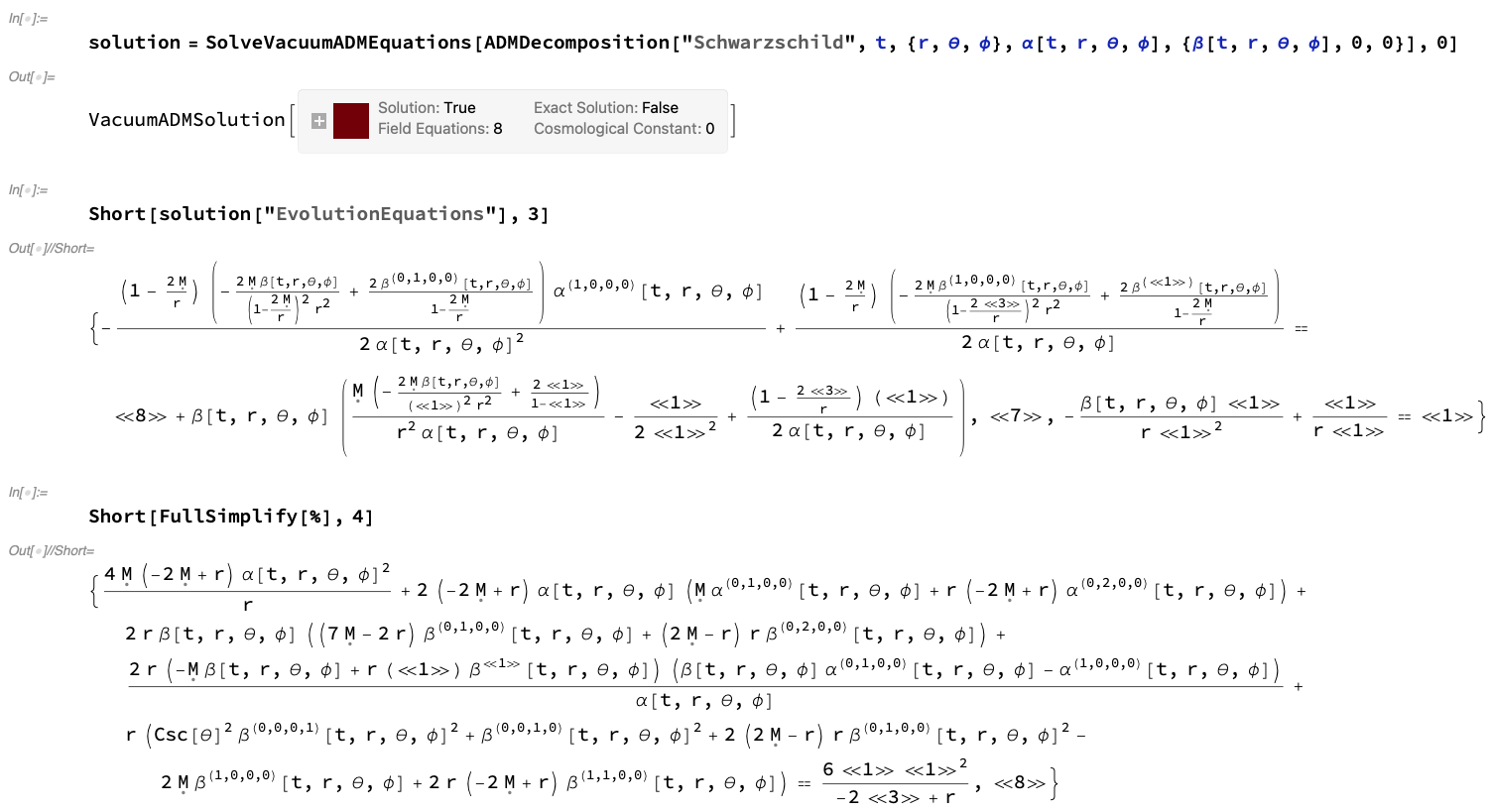}
\vrule
\includegraphics[width=0.495\textwidth]{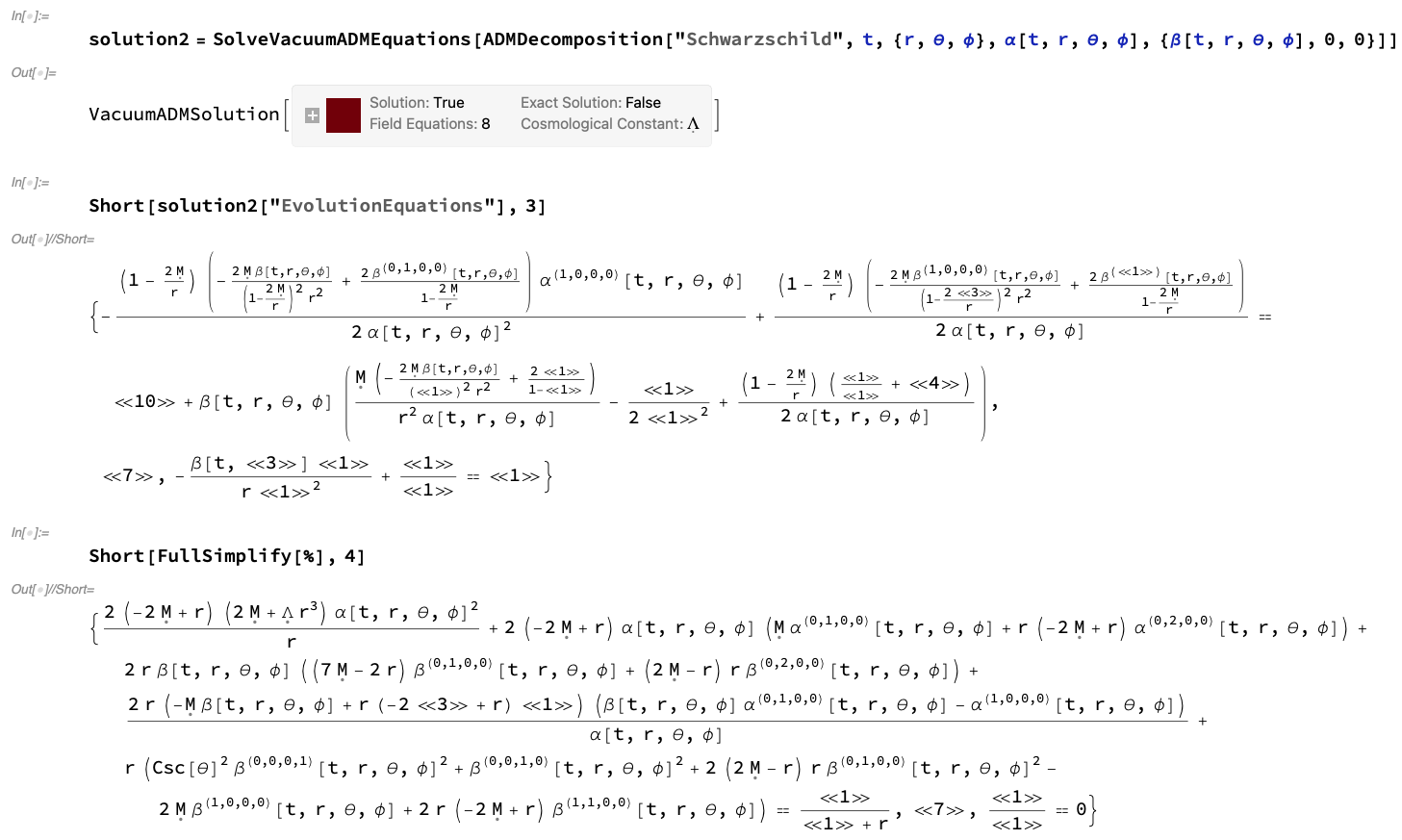}
\end{framed}
\caption{On the left, the list of vacuum ADM evolution equations, computed using the \texttt{VacuumADMSolution} object for an ADM decomposition of a Schwarzschild geometry (representing, for instance, an uncharged, non-rotating black hole with mass $M$ in Schwarzschild or spherical polar coordinates ${\left( t, r, \theta, \phi \right)}$) with lapse function ${\alpha \left( t, r, \theta, \phi \right)}$ and modified shift vector ${\left( \beta \left( t, r, \theta, \phi \right), 0, 0 \right)}$, with zero cosmological constant ${\Lambda = 0}$, together with a verification that they reduce down to a set of eight canonical field equations. On the right, the list of vacuum ADM evolution equations, computed using the \texttt{VacuumADMSolution} object for an ADM decomposition of a Schwarzschild geometry (representing, for instance, an uncharged, non-rotating black hole with mass $M$ in Schwarzschild or spherical polar coordinates ${\left( t, r, \theta, \phi \right)}$) with lapse function ${\alpha \left( t, r, \theta, \phi \right)}$ and modified shift vector ${\left( \beta \left( t, r, \theta, \phi \right), 0, 0 \right)}$, with non-zero cosmological constant ${\Lambda \neq 0}$, together with a verification that they reduce down to a set of eight canonical field equations.}
\label{fig:Figure24}
\end{figure}

\begin{figure}[ht]
\centering
\begin{framed}
\includegraphics[width=0.495\textwidth]{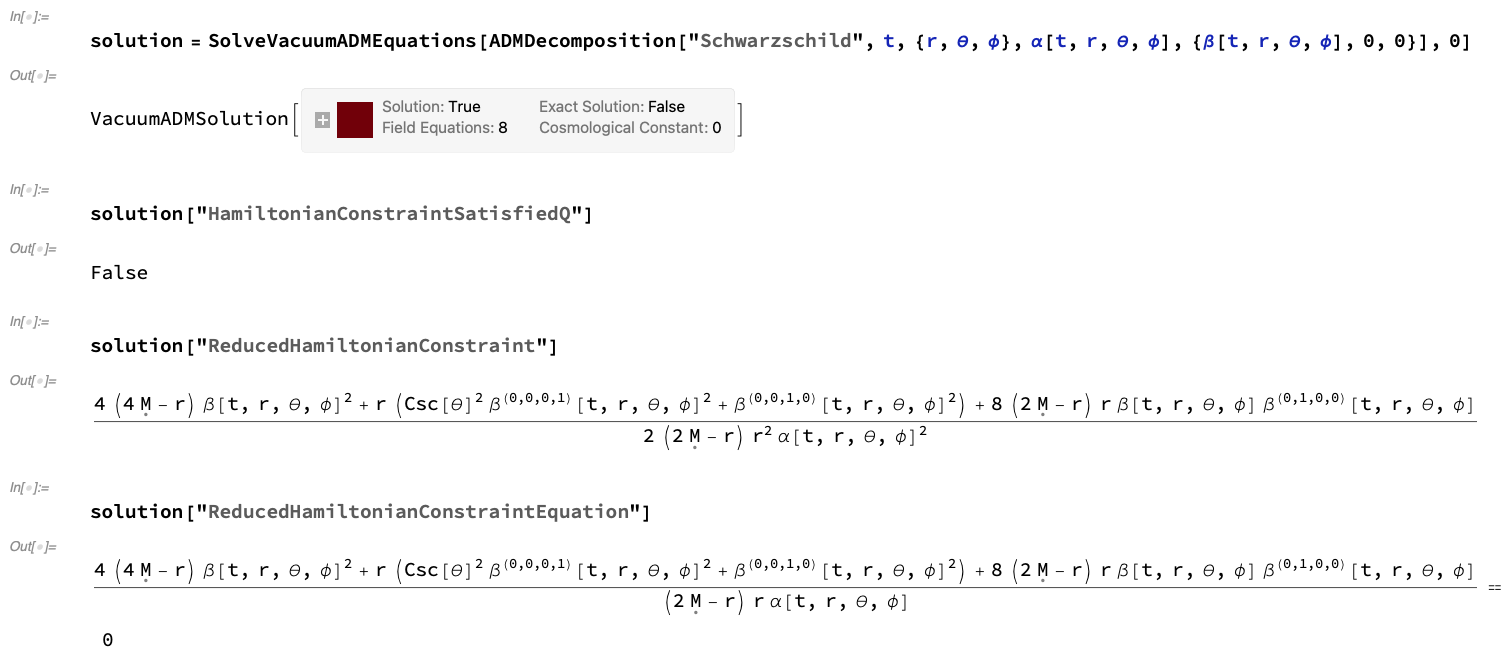}
\vrule
\includegraphics[width=0.495\textwidth]{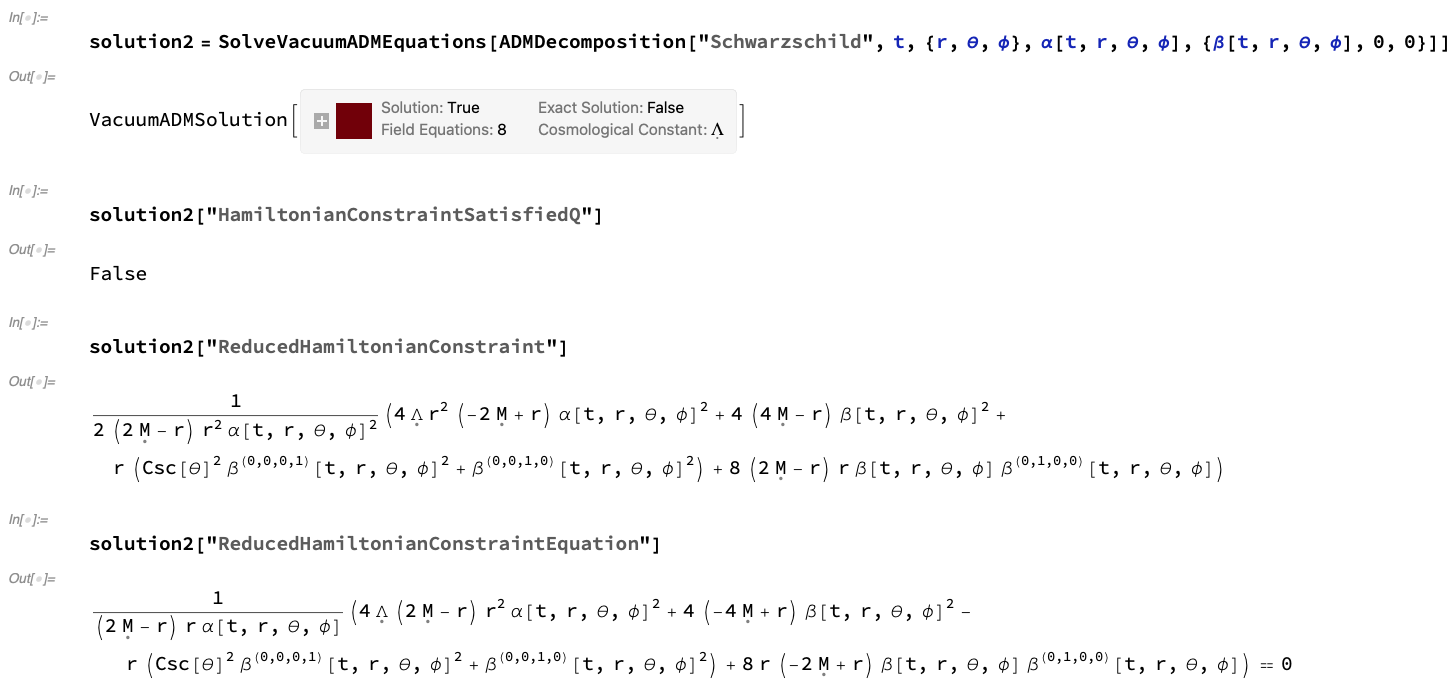}
\end{framed}
\caption{On the left, the vacuum ADM Hamiltonian constraint, computed using the \texttt{VacuumADMSolution} object for an ADM decomposition of a Schwarzschild geometry (representing, for instance, an uncharged, non-rotating black hole with mass $M$ in Schwarzschild or spherical polar coordinates ${\left( t, r, \theta, \phi \right)}$) with lapse function ${\alpha \left( t, r, \theta, \phi \right)}$ and modified shift vector ${\left( \beta \left( t, r, \theta, \phi \right), 0, 0 \right)}$, with zero cosmological constant ${\Lambda = 0}$, illustrating that it does not vanish identically. On the right, the vacuum ADM Hamiltonian constraint, computed using the \texttt{VacuumADMSolution} object for an ADM decomposition of a Schwarzschild geometry (representing, for instance, an uncharged, non-rotating black hole with mass $M$ in Schwarzschild or spherical polar coordinates ${\left( t, r, \theta, \phi \right)}$) with lapse function ${\alpha \left( t, r, \theta, \phi \right)}$ and modified shift vector ${\left( \beta \left( t, r, \theta, \phi \right), 0, 0 \right)}$, with non-zero cosmological constant ${\Lambda \neq 0}$, illustrating that it does not vanish identically.}
\label{fig:Figure25}
\end{figure}

\begin{figure}[ht]
\centering
\begin{framed}
\includegraphics[width=0.495\textwidth]{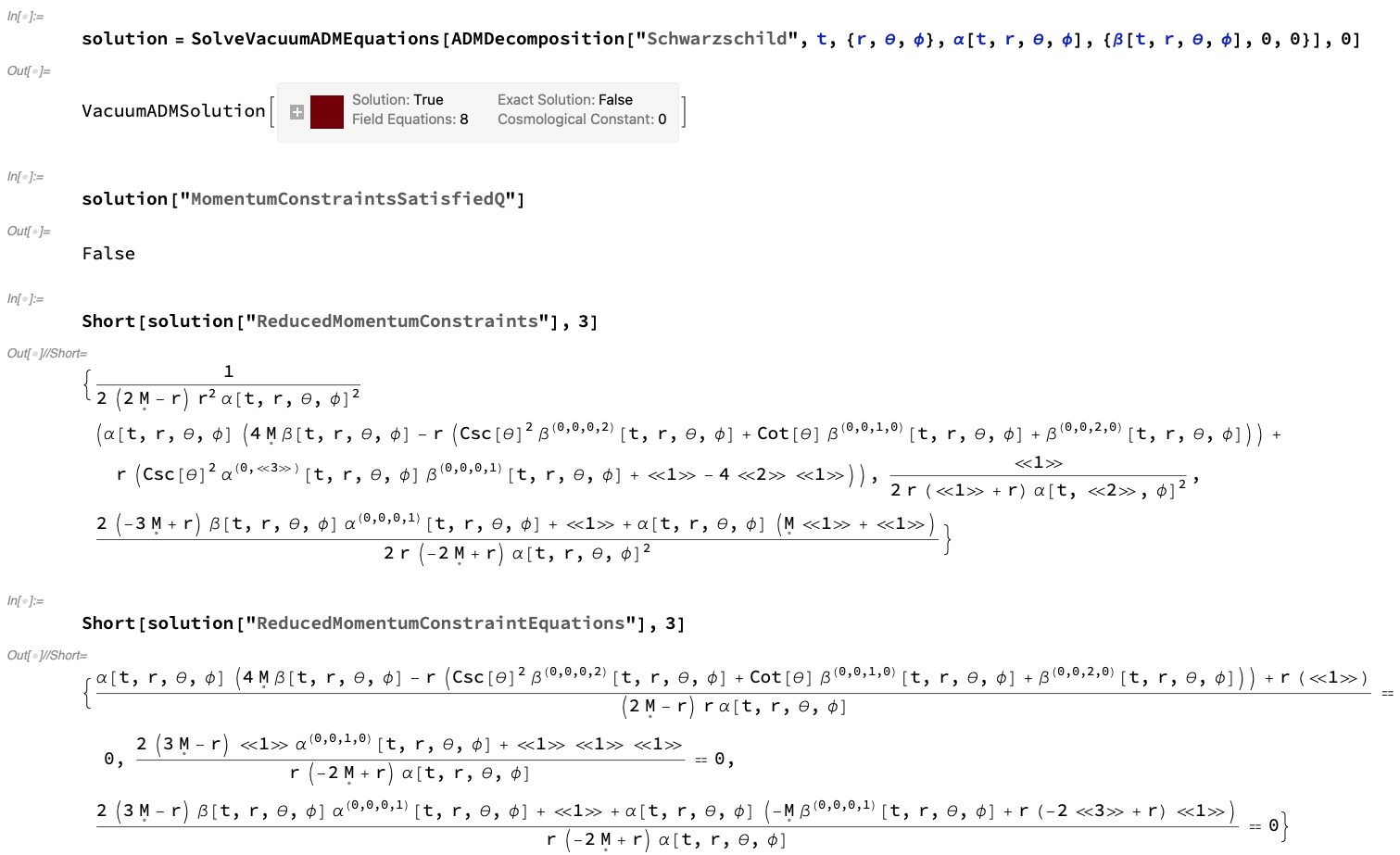}
\vrule
\includegraphics[width=0.495\textwidth]{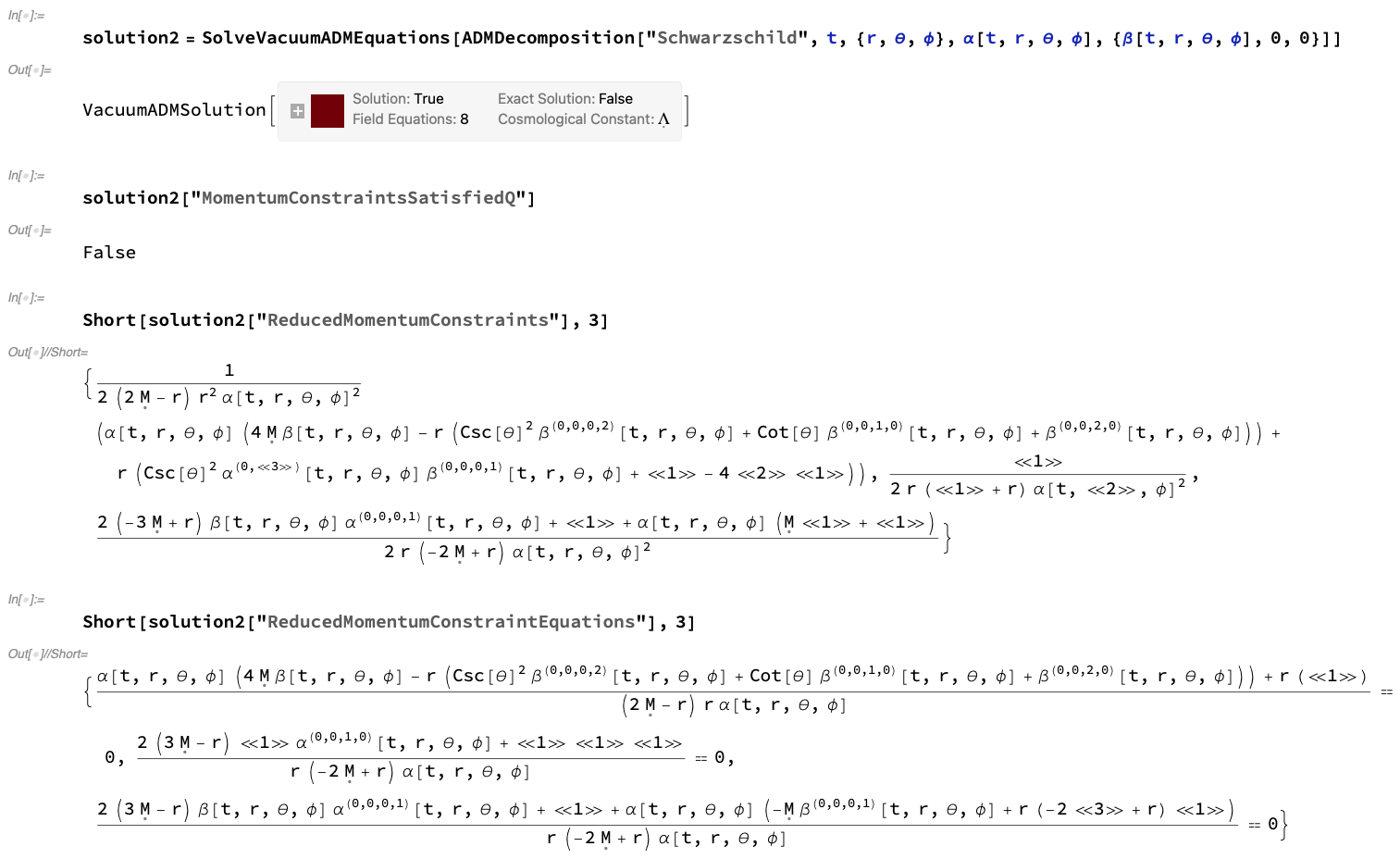}
\end{framed}
\caption{On the left, the list of vacuum ADM momentum constraints, computed using the \texttt{VacuumADMSolution} object for an ADM decomposition of a Schwarzschild geometry (representing, for instance, an uncharged, non-rotating black hole with mass $M$ in Schwarzschild or spherical polar coordinates ${\left( t, r, \theta, \phi \right)}$) with lapse function ${\alpha \left( t, r, \theta, \phi \right)}$ and modified shift vector ${\left( \beta \left( t, r, \theta, \phi \right), 0, 0 \right)}$, with zero cosmological constant ${\Lambda = 0}$, illustrating that they do not vanish identically. On the right, the list of vacuum ADM momentum constraints, computed using the \texttt{VacuumADMSolution} object for an ADM decomposition of a Schwarzschild geometry (representing, for instance, an uncharged, non-rotating black hole with mass $M$ in Schwarzschild or spherical polar coordinates ${\left( t, r, \theta, \phi \right)}$) with lapse function ${\alpha \left( t, r, \theta, \phi \right)}$ and modified shift vector ${\left( \beta \left( t, r, \theta, \phi \right), 0, 0 \right)}$, with non-zero cosmological constant ${\Lambda \neq 0}$, illustrating that they do not vanish identically.}
\label{fig:Figure26}
\end{figure}

We can now proceed to derive a system of vacuum ADM evolution equations by taking the aforementioned time evolution equations for the components of the extrinsic curvature tensor ${K_{\nu}^{\mu}}$, namely:

\begin{multline}
\frac{\partial}{\partial t} \left( K_{\nu}^{\mu} \right) = \alpha {}^{\left( 3 \right)} R_{\nu}^{\mu} - {}^{\left( 3 \right)} \nabla_{\rho} \left( {}^{\left( 3 \right)} \nabla_{\nu} \alpha \right) \gamma^{\rho \mu} + \alpha K K_{\nu}^{\mu} + \beta^{\rho} {}^{\left( 3 \right)} \nabla_{\rho} K_{\nu}^{\mu}\\
+ K_{\rho}^{\mu} {}^{\left( 3 \right)} \nabla_{\nu} \beta^{\rho} - K_{\nu}^{\rho} {}^{\left( 3 \right)} \nabla_{\rho} \beta^{\mu} - \alpha {}^{\left( 4 \right)} R_{\left( \rho + 1 \right) \left( \nu + 1 \right)} \gamma^{\rho \mu},
\end{multline}
or, equivalently:

\begin{multline}
\frac{\partial}{\partial t} \left( K_{\nu}^{\mu} \right) = \alpha {}^{\left( 3 \right)} R_{\nu}^{\mu} - \left( \frac{\partial}{\partial x^{\rho}} \left( \frac{\partial}{\partial x^{\nu}} \left( \alpha \right) \right) - {}^{\left( 3 \right)} \Gamma_{\rho \nu}^{\sigma} \left( \frac{\partial}{\partial x^{\sigma}} \left( \alpha \right) \right) \right) \gamma^{\rho \mu} + \alpha K K_{\nu}^{\mu}\\
+ \beta^{\rho} \left( \frac{\partial}{\partial x^{\rho}} \left( K_{\nu}^{\mu} \right) + {}^{\left( 3 \right)} \Gamma_{\rho \sigma}^{\mu} K_{\nu}^{\sigma} - {}^{\left( 3 \right)} \Gamma_{\rho \nu}^{\sigma} K_{\sigma}^{\mu} \right) + K_{\rho}^{\mu} \left( \frac{\partial}{\partial x^{\nu}} \left( \beta^{\rho} \right) + {}^{\left( 3 \right)} \Gamma_{\nu \sigma}^{\rho} \beta^{\sigma} \right)\\
- K_{\nu}^{\rho} \left( \frac{\partial}{\partial x^{\rho}} \left( \beta^{\mu} \right) + {}^{\left( 3 \right)} \Gamma_{\rho \sigma}^{\mu} \beta^{\sigma} \right) - \alpha {}^{\left( 4 \right)} R_{\left( \rho + 1 \right) \left( \nu + 1 \right)} \gamma^{\rho \mu},
\end{multline}
with ${\mu, \nu, \rho, \sigma}$ ranging across all ${\left\lbrace 0, \dots, n - 2 \right\rbrace}$ (i.e. across spatial coordinate indices only), and applying the vacuum Einstein field equations:

\begin{equation}
{}^{\left( 4 \right)} G_{\mu \nu} + \Lambda g_{\mu \nu} = {}^{\left( 4 \right)} R_{\mu \nu} - \frac{1}{2} {}^{\left( 4 \right)} R g_{\mu \nu} + \Lambda g_{\mu \nu} = 0,
\end{equation}
which, in particular, allow us to rewrite the spacetime Ricci tensor terms ${{}^{\left( 4 \right)} R_{\mu \nu}}$ appearing in these equations purely in terms of the cosmological constant ${\Lambda}$ and the spacetime metric tensor ${g_{\mu \nu}}$ as:

\begin{equation}
{}^{\left( 4 \right)} R_{\mu \nu} = \left( \frac{1}{2} {}^{\left( 4 \right)} R - \Lambda \right) g_{\mu \nu} = \frac{2 \Lambda}{n - 2} g_{\mu \nu},
\end{equation}
obtained by taking appropriate traces of the equations, and thereby deducing an explicit relationship between the spacetime Ricci scalar and the cosmological constant, namely ${{}^{\left( 4 \right)} R = \frac{2 \Lambda n}{n - 2}}$. This yields the vacuum ADM evolution equations:

\begin{multline}
\frac{\partial}{\partial t} \left( K_{\nu}^{\mu} \right) = \alpha {}^{\left( 3 \right)} R_{\nu}^{\mu} - {}^{\left( 3 \right)} \nabla_{\rho} \left( {}^{\left( 3 \right)} \nabla_{\nu} \alpha \right) \gamma^{\rho \mu} + \alpha K K_{\nu}^{\mu} + \beta^{\rho} {}^{\left( 3 \right)} \nabla_{\rho} K_{\nu}^{\mu}\\
+ K_{\rho}^{\mu} {}^{\left( 3 \right)} \nabla_{\nu} \beta^{\rho} - K_{\nu}^{\rho} {}^{\left( 3 \right)} \nabla_{\rho} \beta^{\mu} - \alpha \left( \frac{2 \Lambda}{n - 2} \gamma_{\rho \nu} \right) \gamma^{\rho \mu},
\end{multline}
i.e., in expanded form:

\begin{multline}
\frac{\partial}{\partial t} \left( K_{\nu}^{\mu} \right) = \alpha {}^{\left( 3 \right)} R_{\nu}^{\mu} - \left( \frac{\partial}{\partial x^{\rho}} \left( \frac{\partial}{\partial x^{\nu}} \left( \alpha \right) \right) - {}^{\left( 3 \right)} \Gamma_{\rho \nu}^{\sigma} \left( \frac{\partial}{\partial x^{\sigma}} \left( \alpha \right) \right) \right) \gamma^{\rho \mu} + \alpha K K_{\nu}^{\mu}\\
+ \beta^{\rho} \left( \frac{\partial}{\partial x^{\rho}} \left( K_{\nu}^{\mu} \right) + {}^{\left( 3 \right)} \Gamma_{\rho \sigma}^{\mu} K_{\nu}^{\sigma} - {}^{\left( 3 \right)} \Gamma_{\rho \nu}^{\sigma} K_{\sigma}^{\mu} \right) + K_{\rho}^{\mu} \left( \frac{\partial}{\partial x^{\nu}} \left( \beta^{\rho} \right) + {}^{\left( 3 \right)} \Gamma_{\nu \sigma}^{\rho} \beta^{\sigma} \right)\\
- K_{\nu}^{\rho} \left( \frac{\partial}{\partial x^{\rho}} \left( \beta^{\mu} \right) + {}^{\left( 3 \right)} \Gamma_{\rho \sigma}^{\mu} \beta^{\sigma} \right) - \alpha \left( \frac{2 \Lambda}{n - 2} \gamma_{\rho \nu} \right) \gamma^{\rho \mu},
\end{multline}
with ${\mu, \nu, \rho, \sigma}$ ranging across all ${\left\lbrace 0, \dots, n - 2 \right\rbrace}$ (i.e. across spatial coordinate indices only). Representations of the corresponding \texttt{VacuumADMSolution} objects for the ADM decomposition of the Schwarzschild metric (representing, for instance, an uncharged, non-rotating black hole with mass $M$ in Schwarzschild or spherical polar coordinates ${\left( t, r, \theta, \phi \right)}$), assuming the usual restricted choice of gauge consisting of the lapse function ${\alpha \left( t, r, \theta, \phi \right)}$ and the modified shift vector ${\left( \beta \left( t, r, \theta, \phi \right), 0, 0 \right)}$, both with vanishing (${\Lambda = 0}$) and non-vanishing (${\Lambda \neq 0}$) cosmological constant terms, computed using the \texttt{SolveVacuumADMEquations} function, are shown in Figure \ref{fig:Figure23}; these examples demonstrate that this ADM decomposition is a valid \textit{non-exact} solution of the vacuum ADM evolution equations, in the sense that eight additional field equations need to be assumed, in both the ${\Lambda = 0}$ and ${\Lambda \neq 0}$ cases. The complete lists of vacuum ADM evolution equations for both cases can be computed directly from the \texttt{VacuumADMSolution} object, and it can be verified in both cases that they do indeed reduce down to the eight canonical field equations previously mentioned, as illustrated in Figure \ref{fig:Figure24}. Similarly, applying the vacuum Einstein field equations to eliminate the spacetime Einstein tensor terms ${^{\left( 4 \right)} G_{\mu \nu}}$ appearing within the ADM Hamiltonian and momentum constraints:

\begin{equation}
\mathcal{H} = {}^{\left( 3 \right)} R + K^2 - K_{\nu}^{\mu} K_{\mu}^{\nu} - 2 \alpha^2 {}^{\left( 4 \right)} G^{0 0},
\end{equation}
and:

\begin{multline}
\mathcal{M}_{\mu} = {}^{\left( 3 \right)} \nabla_{\nu} K_{\mu}^{\nu} - {}^{\left( 3 \right)} \nabla_{\mu} K - \alpha G_{\left( \mu + 1 \right)}^{0}\\
= \frac{\partial}{\partial x^{\nu}} \left( K_{\mu}^{\nu} \right) + {}^{\left( 3 \right)} \Gamma_{\nu \sigma}^{\nu} K_{\mu}^{\sigma} - {}^{\left( 3 \right)} \Gamma_{\nu \mu}^{\sigma} K_{\sigma}^{\nu} - \frac{\partial}{\partial x^{\mu}} \left( K \right) - \alpha {}^{\left( 4 \right)} G_{\left( \mu + 1 \right)}^{0},
\end{multline}
respectively, yields the \textit{vacuum} forms of the ADM Hamiltonian and momentum constraints:

\begin{equation}
\mathcal{H} = {}^{\left( 3 \right)} R + K^2 - K_{\nu}^{\mu} K_{\mu}^{\nu} - 2 \Lambda,
\end{equation}
and:

\begin{equation}
\mathcal{M}_{\mu} = {}^{\left( 3 \right)} \nabla_{\nu} K_{\mu}^{\nu} - {}^{\left( 3 \right)} \nabla_{\mu} K = \frac{\partial}{\partial x^{\nu}} \left( K_{\mu}^{\nu} \right) + {}^{\left( 3 \right)} \Gamma_{\nu \sigma}^{\nu} K_{\mu}^{\sigma} - {}^{\left( 3 \right)} \Gamma_{\nu \mu}^{\sigma} K_{\sigma}^{\nu} - \frac{\partial}{\partial x^{\mu}} \left( K \right),
\end{equation}
respectively, with ${\mu, \nu, \sigma}$ as before ranging across all ${\left\lbrace 0, \dots, n - 2 \right\rbrace}$, which, as demonstrated in Figures \ref{fig:Figure25} and \ref{fig:Figure26}, do not vanish identically (i.e. the constraint equations ${\mathcal{H} = 0}$ and ${\mathcal{M}_{\mu} = 0}$ do not hold identically) for the case of the ADM decomposition of the Schwarzschild metric with the restricted choice of gauge described above, in either of the vanishing cosmological constant (${\Lambda = 0}$) or non-vanishing cosmological constant (${\Lambda \neq 0}$) cases.

\begin{figure}[ht]
\centering
\begin{framed}
\includegraphics[width=0.545\textwidth]{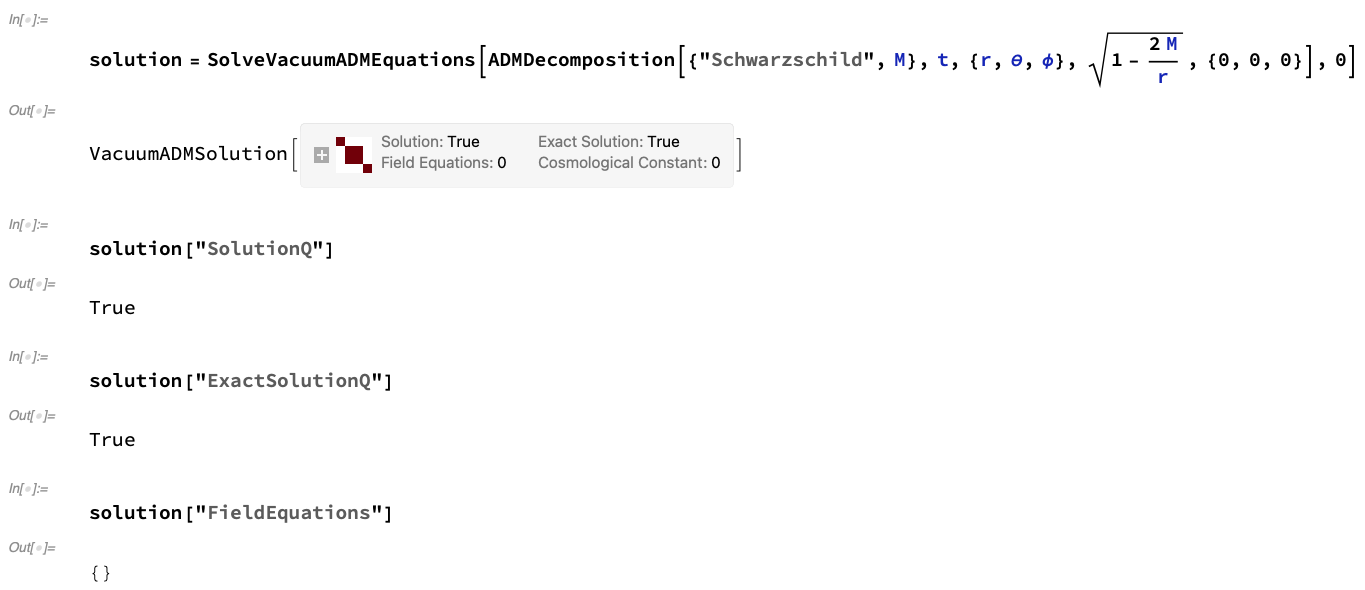}
\vrule
\includegraphics[width=0.445\textwidth]{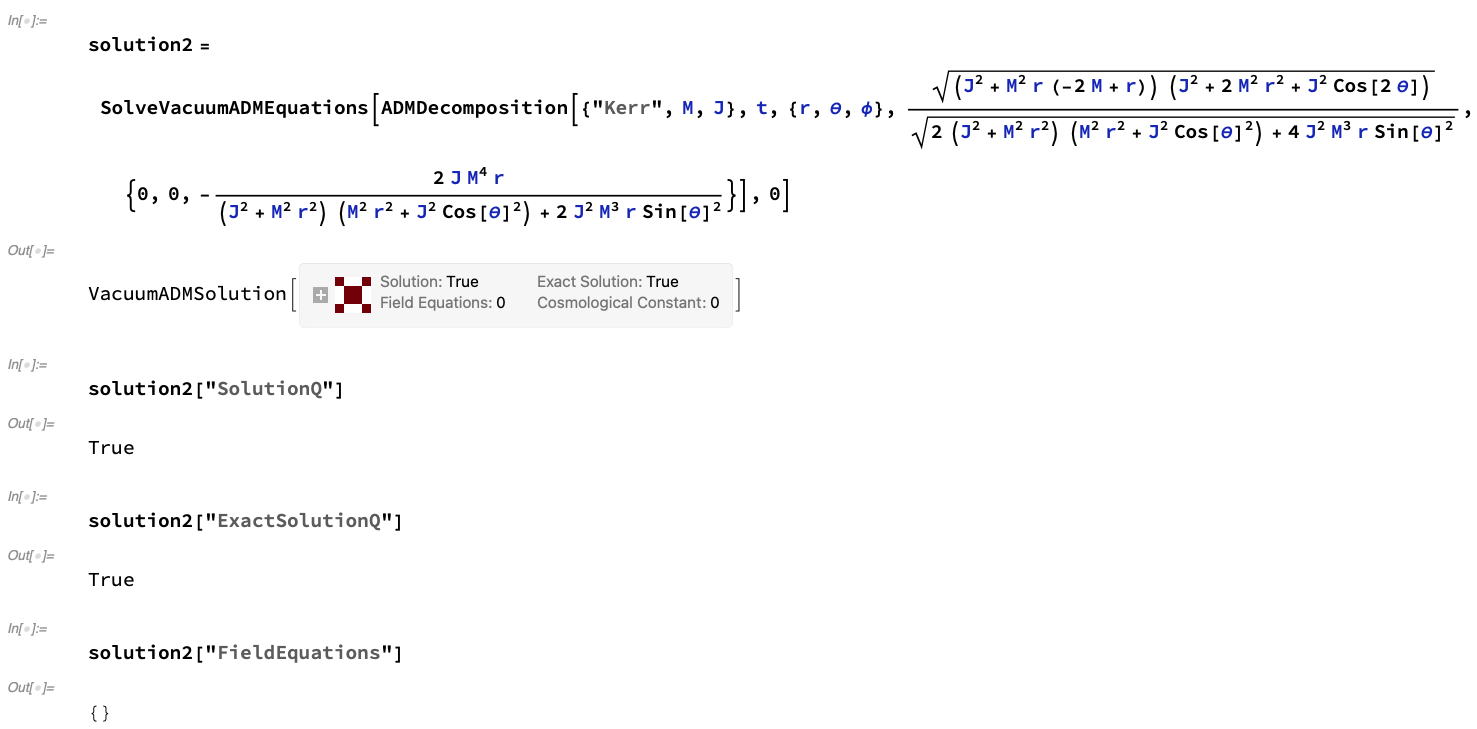}
\end{framed}
\caption{On the left, the \texttt{VacuumADMSolution} object for an ADM decomposition of a Schwarzschild geometry (representing, for instance, an uncharged, non-rotating black hole with mass $M$ in Schwarzschild or spherical polar coordinates ${\left( t, r, \theta, \phi \right)}$) with modified lapse function ${\alpha}$ and vanishing shift vector ${\beta^{\mu} = 0}$, with zero cosmological constant ${\Lambda = 0}$, computed using \texttt{SolveVacuumADMEquations}, illustrating that this decomposition is an exact solution to the vacuum ADM evolution equations. On the right, the \texttt{VacuumADMSolution} object for an ADM decomposition of a Kerr geometry (representing, for instance, an uncharged, spinning black hole with mass $M$ and angular momentum $J$ in Boyer-Lindquist or oblate spheroidal coordinates ${\left( t, r, \theta, \phi \right)}$) with modified lapse function ${\alpha}$ and modified shift vector ${\beta^{\mu}}$, with zero cosmological constant ${\Lambda = 0}$, computed using \texttt{SolveVacuumADMEquations}, illustrating that this decomposition is an exact solution to the vacuum ADM evolution equations.}
\label{fig:Figure27}
\end{figure}

\begin{figure}[ht]
\centering
\begin{framed}
\includegraphics[width=0.595\textwidth]{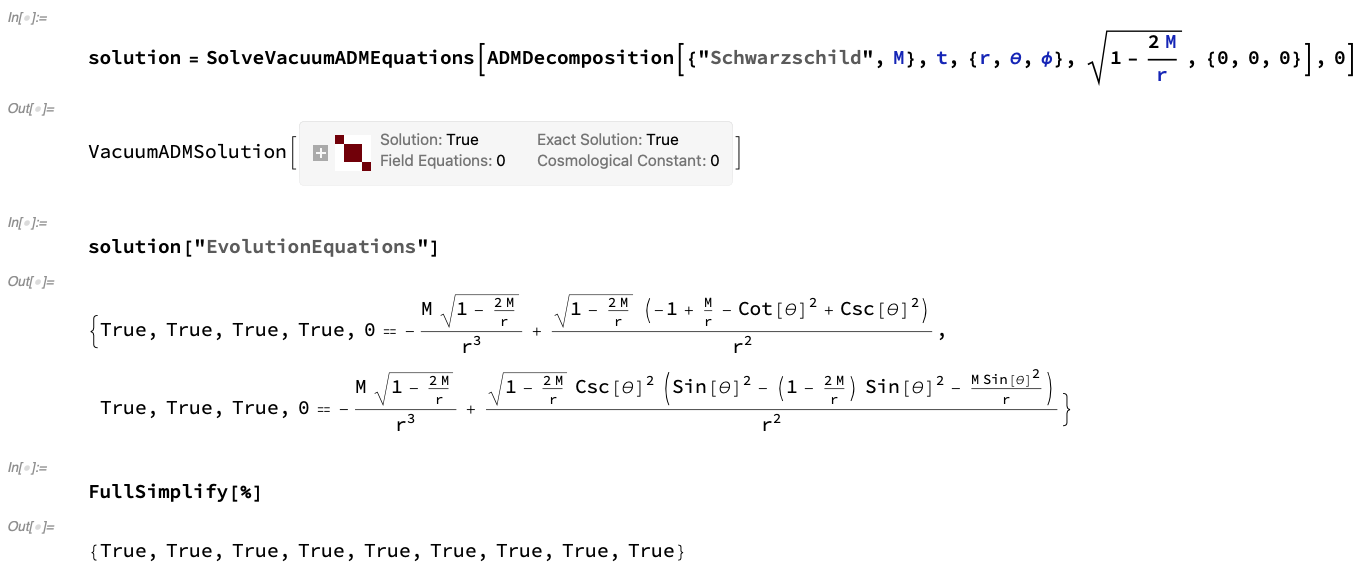}
\vrule
\includegraphics[width=0.395\textwidth]{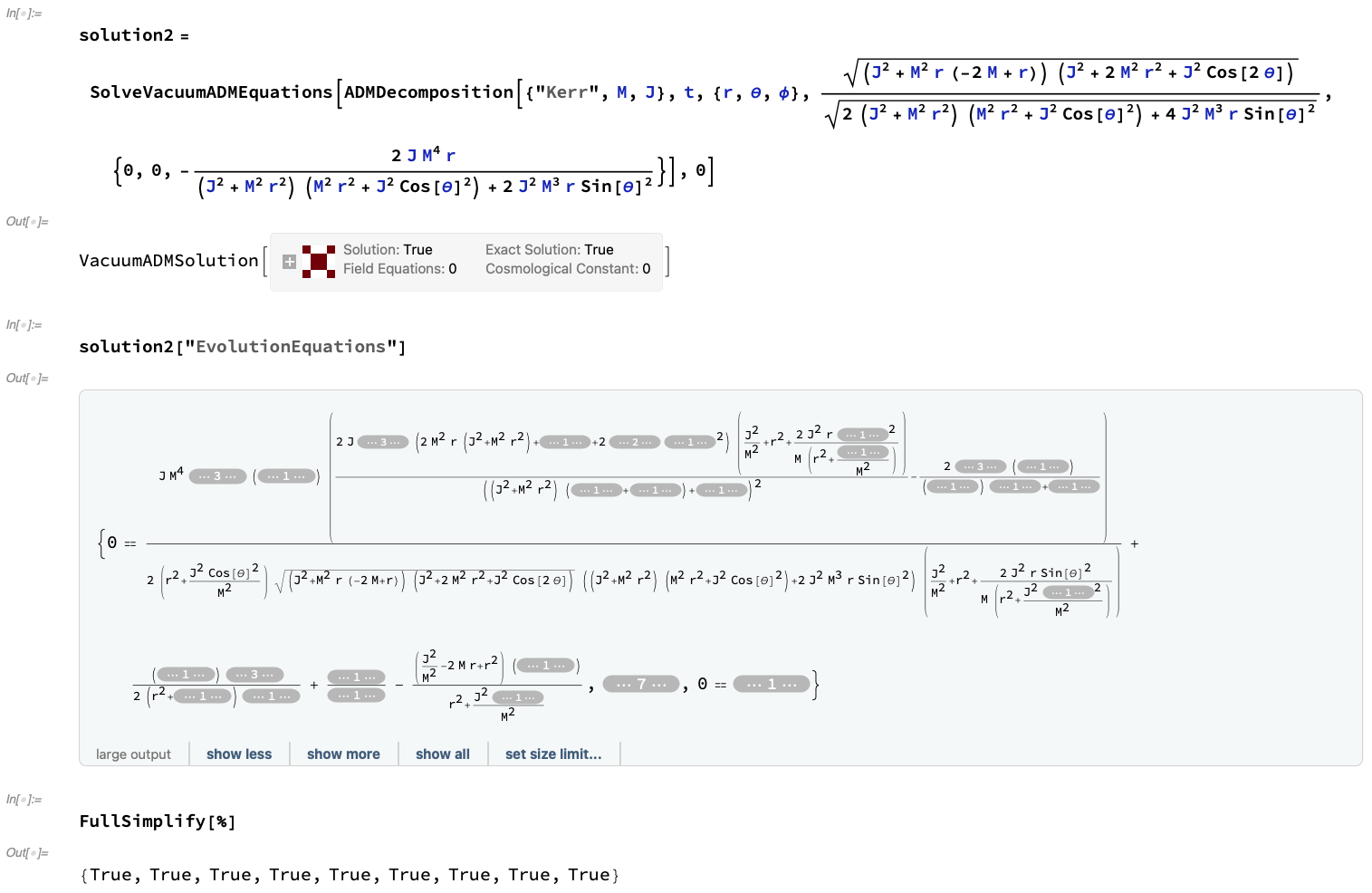}
\end{framed}
\caption{On the left, the list of vacuum ADM evolution equations, computed using the \texttt{VacuumADMSolution} object for an ADM decomposition of a Schwarzschild geometry (representing, for instance, an uncharged, non-rotating black hole with mass $M$ in Schwarzschild or spherical polar coordinates ${\left( t, r, \theta, \phi \right)}$) with modified lapse function ${\alpha}$ and vanishing shift vector ${\beta^{\mu} = 0}$, with zero cosmological constant ${\Lambda = 0}$, together with a verification that they all hold identically. On the right, the list of vacuum ADM evolution equations, computed using the \texttt{VacuumADMSolution} object for an ADM decomposition of a Kerr geometry (representing, for instance, an uncharged, spinning black hole with mass $M$ and angular momentum $J$ in Boyer-Lindquist or oblate spheroidal coordinates ${\left( t, r, \theta, \phi \right)}$) with modified lapse function ${\alpha}$ and modified shift vector ${\beta^{\mu}}$, with zero cosmological constant ${\Lambda = 0}$, together with a verification that they all hold identically.}
\label{fig:Figure28}
\end{figure}

\begin{figure}[ht]
\centering
\begin{framed}
\includegraphics[width=0.595\textwidth]{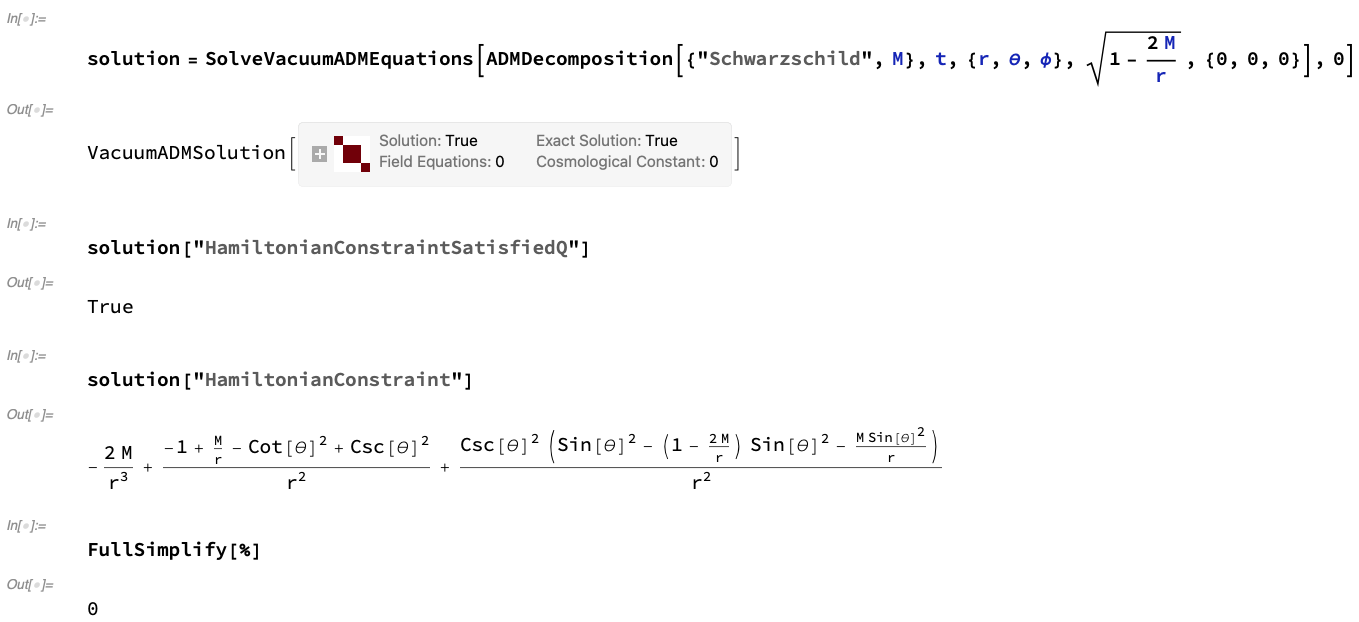}
\vrule
\includegraphics[width=0.395\textwidth]{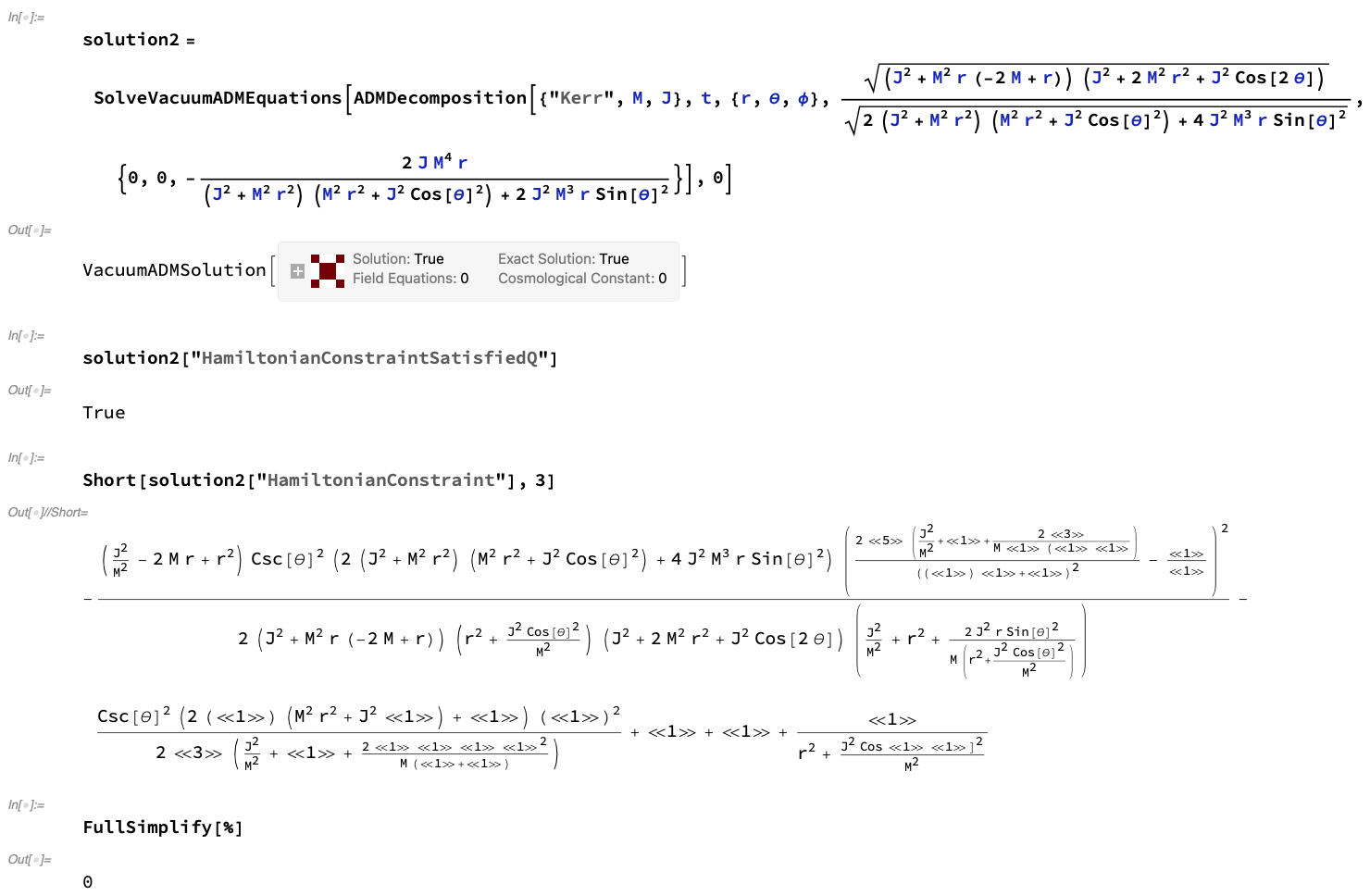}
\end{framed}
\caption{On the left, the vacuum ADM Hamiltonian constraint, computed using the \texttt{VacuumADMSolution} object for an ADM decomposition of a Schwarzschild geometry (representing, for instance, an uncharged, non-rotating black hole with mass $M$ in Schwarzschild or spherical polar coordinates ${\left( t, r, \theta, \phi \right)}$) with modified lapse function ${\alpha}$ and vanishing shift vector ${\beta^{\mu} = 0}$, with zero cosmological constant ${\Lambda = 0}$, together with a verification that it vanishes identically. On the right, the vacuum ADM Hamiltonian constraint, computed using the \texttt{VacuumADMSolution} object for an ADM decomposition of a Kerr geometry (representing, for instance, an uncharged, spinning black hole with mass $M$ and angular momentum $J$ in Boyer-Lindquist or oblate spheroidal coordinates ${\left( t, r, \theta, \phi \right)}$) with modified lapse function ${\alpha}$ and modified shift vector ${\beta^{\mu}}$, with zero cosmological constant ${\Lambda = 0}$, together with a verification that it vanishes identically.}
\label{fig:Figure29}
\end{figure}

\begin{figure}[ht]
\centering
\begin{framed}
\includegraphics[width=0.645\textwidth]{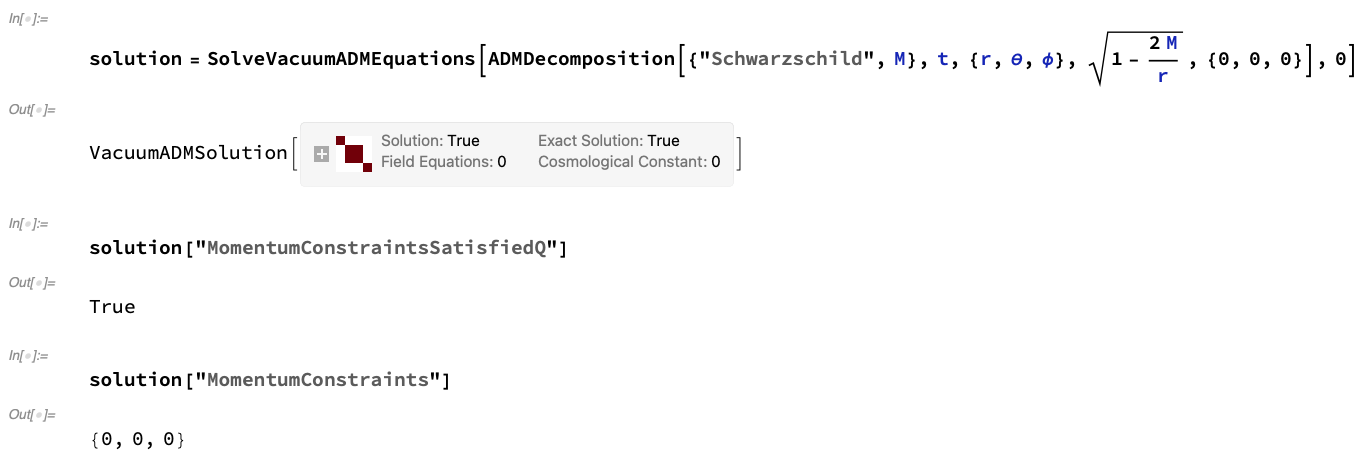}
\vrule
\includegraphics[width=0.345\textwidth]{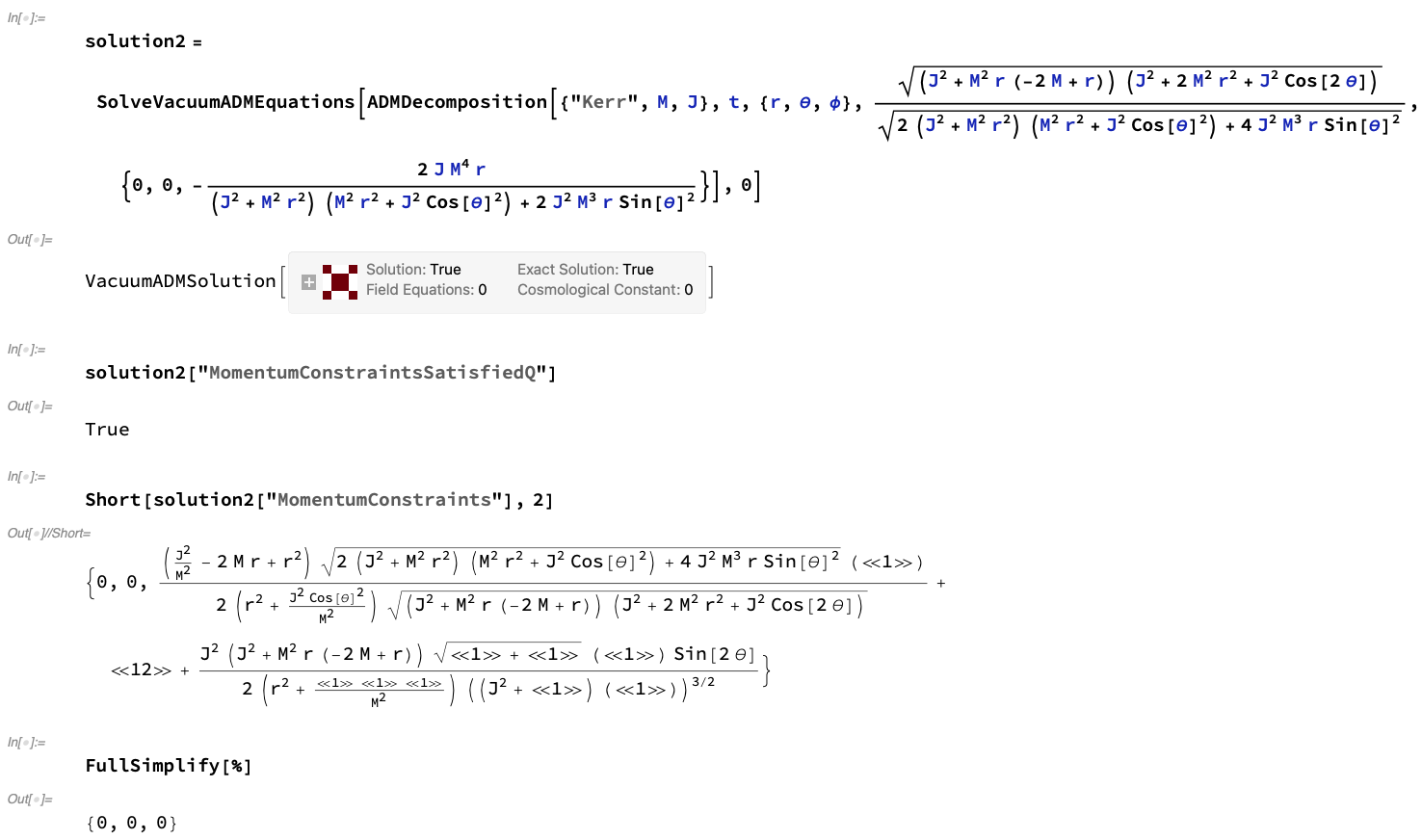}
\end{framed}
\caption{On the left, the list of vacuum ADM momentum constraints, computed using the \texttt{VacuumADMSolution} object for an ADM decomposition of a Schwarzschild geometry (representing, for instance, an uncharged, non-rotating black hole with mass $M$ in Schwarzschild or spherical polar coordinates ${\left( t, r, \theta, \phi \right)}$) with modified lapse function ${\alpha}$ and vanishing shift vector ${\beta^{\mu} = 0}$, with zero cosmological constant ${\Lambda = 0}$, together with a verification that they all vanish identically. On the right, the list of vacuum ADM momentum constraints, computed using the \texttt{VacuumADMSolution} object for an ADM decomposition of a Kerr geometry (representing, for instance, an uncharged, spinning black hole with mass $M$ and angular momentum $J$ in Boyer-Lindquist or oblate spheroidal coordinates ${\left( t, r, \theta, \phi \right)}$) with modified lapse function ${\alpha}$ and modified shift vector ${\beta^{\mu}}$, with zero cosmological constant ${\Lambda = 0}$, together with a verification that they all vanish identically.}
\label{fig:Figure30}
\end{figure}

We can force these vacuum ADM solutions to become exact by assuming a more restricted form of the gauge variables ${\alpha}$ and ${\beta^{\mu}}$; more specifically, we can adopt the following form of the lapse function ${\alpha \left( t, r, \theta, \phi \right)}$:

\begin{equation}
\alpha \left( t, r, \theta, \phi \right) = \sqrt{1 - \frac{2 M}{r}},
\end{equation}
with normal coordinate conditions (i.e. vanishing shift vector ${\beta^{\mu} = 0}$ everywhere), in the case of the ADM decomposition of the Schwarzschild metric, or otherwise the following, somewhat more complicated, form of the lapse function ${\alpha \left( t, r, \theta, \phi \right)}$:

\begin{equation}
\alpha \left( t, r, \theta, \phi \right) = \frac{\sqrt{\left( J^2 + M^2 r \left( r - 2 M \right) \right) \left( J^2 + 2 M^2 r^2 + J^2 \cos \left( 2 \theta \right) \right)}}{\sqrt{2 \left( J^2 + M^2 r^2 \right) \left( M^2 r^2 + J^2 \cos^2 \left( \theta \right) \right) + 4 J^2 M^3 r \sin^2 \left( \theta \right)}},
\end{equation}
and corresponding form of the shift vector ${\beta^{\mu} \left( t, r, \theta, \phi \right)}$:

\begin{equation}
\beta^{\mu} \left( t, r, \theta, \phi \right) = \left( 0, 0, - \frac{2 J M^4 r}{\left( J^2 + M^2 r^2 \right) \left( M^2 r^2 + J^2 \cos^2 \left( \theta \right) \right) + 2J^2 M^3 r \sin^2 \left( \theta \right)} \right),
\end{equation}
in the case of the ADM decomposition of the Kerr metric (representing, for instance, an uncharged, spinning black hole with mass $M$ and angular momentum $J$ in Boyer-Lindquist or oblate spheroidal coordinates ${\left( t, r, \theta, \phi \right)}$). Representations of the corresponding \texttt{VacuumADMSolution} objects for the Schwarzschild and Kerr metrics, assuming these modified forms of the gauge variables ${\alpha}$ and ${\beta^{\mu}}$, and with vanishing cosmological constant ${\Lambda = 0}$, are shown in Figure \ref{fig:Figure27}; these examples demonstrate that both the Schwarzschild and Kerr metrics represent \textit{exact} solutions to the (vacuum) ADM evolution equations if one adopts these restricted forms of the gauge variables ${\alpha}$ and ${\beta^{\mu}}$, in the sense that no additional field equations need to be assumed. The complete lists of (vacuum) ADM evolution equations for the Schwarzschild and Kerr metrics, with vanishing cosmological constant ${\Lambda = 0}$, can be computed directly from the respective \texttt{VacuumADMSolution} objects, and it can be verified in both cases that they all indeed hold identically, as illustrated in Figure \ref{fig:Figure28}. Figures \ref{fig:Figure29} and \ref{fig:Figure30} show the (vacuum) ADM Hamiltonian constraint ${\mathcal{H}}$ and the (vacuum) ADM momentum constraints ${\mathcal{M}_{\mu}}$, computed directly from the \texttt{VacuumADMSolution} objects for both a Schwarzschild and a Kerr metric, demonstrating that they all vanish identically (i.e. that the constraint equations ${\mathcal{H} = 0}$ and ${\mathcal{M}_{\mu} = 0}$ hold identically). Note that the necessary existence of a non-zero angular component ${\beta^{\phi}}$ of the shift vector in the case of the Kerr metric corresponds physically to the effect of frame-dragging, which is clearly absent in the case of Schwarzschild black holes.

\clearpage

\section{Stress-Energy Decomposition and Full ADM Solutions}
\label{sec:Section4}

In order to be able to solve the full (non-vacuum) Einstein field equations numerically using the ADM formalism, we must proceed to construct a corresponding ADM-type decomposition (i.e. a ${3 + 1}$ decomposition) of the full spacetime stress-energy tensor ${T^{\mu \nu}}$, by projecting it both onto, and normal to, the spacelike hypersurfaces of our foliation, analogous to the decomposition of the spacetime metric tensor ${g_{\mu \nu}}$ described previously. The analog of the lapse function ${\alpha}$ is the energy density $E$ perceived by an observer who is moving normal to the spacelike hypersurfaces (i.e. in the direction of ${\mathbf{n}}$), which we can obtain via a straightforward projection of ${T_{\mu \nu}}$ in the normal direction to the spacelike hypersurfaces\cite{gourgoulhon}:

\begin{equation}
E = T \left( \mathbf{n}, \mathbf{n} \right), \qquad \text{ i.e., in component form } \qquad E = T_{\mu \nu} n^{\mu} n^{\nu}.
\end{equation}
The analog of the shift vector ${\boldsymbol{\beta}}$ is the (spatial) momentum density ${\mathbf{p}}$ perceived by an observer moving in the direction of ${\mathbf{n}}$, which we can obtain via  a projection of ${T_{\mu \nu}}$ that is, instead, parallel to the spacelike hypersurfaces (and therefore orthogonal to the normal vector ${\mathbf{n}}$):

\begin{equation}
\forall \mathbf{x} \in \mathcal{M}, \qquad \forall \mathbf{v} \in T_{\mathbf{x}} \mathcal{M}, \qquad g_{\mathcal{M}} \left( \mathbf{p}, \mathbf{v} \right) = - T \left( \mathbf{n}, \bot \left( \mathbf{v} \right) \right),
\end{equation}
i.e., in component form:

\begin{equation}
p_{\alpha} = - T_{\mu \nu} n^{\mu} \bot_{\alpha}^{\nu}, \qquad \text{such that } \qquad p^{\alpha} = \gamma^{\alpha \sigma} p_{\sigma}.
\end{equation}
Finally, the analog of the induced/spatial metric tensor ${\gamma_{\mu \nu}}$ (i.e. the value of the spacetime stress-energy tensor ${T_{\mu \nu}}$ localized to each spacelike hypersurface) is the (Cauchy) stress tensor perceived by an observer moving in the direction of ${\mathbf{n}}$, which we can obtain via a projection of ${T_{\mu \nu}}$ in a mixture of parallel directions to the spacelike hypersurfaces (i.e. a mixture of orthogonal directions to the normal vector ${\mathbf{n}}$):

\begin{equation}
S = \bot^{\star} T, \qquad \text{ i.e., in component form } \qquad S_{\alpha \beta} = T_{\mu \nu} \bot_{\alpha}^{\mu} \bot_{\beta}^{\nu},
\end{equation}
such that its trace, $S$, is simply given by ${S = \gamma^{\alpha \beta} S_{\alpha \beta}}$. In all of the above, ${\alpha, \beta, \sigma}$ range across all ${\left\lbrace 0, \dots, n - 2 \right\rbrace}$ (i.e. across spatial coordinate indices only), whereas ${\mu, \nu}$ range across all ${\left\lbrace 0, \dots, n - 1 \right\rbrace}$ (i.e. across all spacetime coordinate indices). If we now define spacetime-extended versions of the momentum density covector/one-form ${p_{\alpha}}$ and Cauchy stress tensor ${S_{\alpha \beta}}$, which we shall denote ${\overline{p_{\alpha}}}$ and ${\overline{S_{\alpha \beta}}}$, respectively, namely:

\begin{equation}
\overline{p_{\mu}} = \begin{cases}
p_{\left( \mu - 1 \right)}, \qquad & \text{ if } \mu > 1,\\
0, \qquad & \text{ otherwise},
\end{cases}
\end{equation}
and:

\begin{equation}
\overline{S_{\mu \nu}} = \begin{cases}
S_{\left( \mu - 1 \right) \left( \nu - 1 \right)}, \qquad & \text{ if } \mu > 1 \text{ and } \nu > 1,\\
0, \qquad & \text{ otherwise},
\end{cases}
\end{equation}
then we are able to represent the full ADM decomposition of the spacetime stress-energy tensor ${T_{\mu \nu}}$ as:

\begin{equation}
T_{\mu \nu} = \overline{S_{\mu \nu}} + n_{\mu} \overline{p_{\nu}} + \overline{p_{\mu}} n_{\nu} + E n_{\mu} n_{\nu}.
\end{equation}

\begin{figure}[ht]
\centering
\begin{framed}
\includegraphics[width=0.545\textwidth]{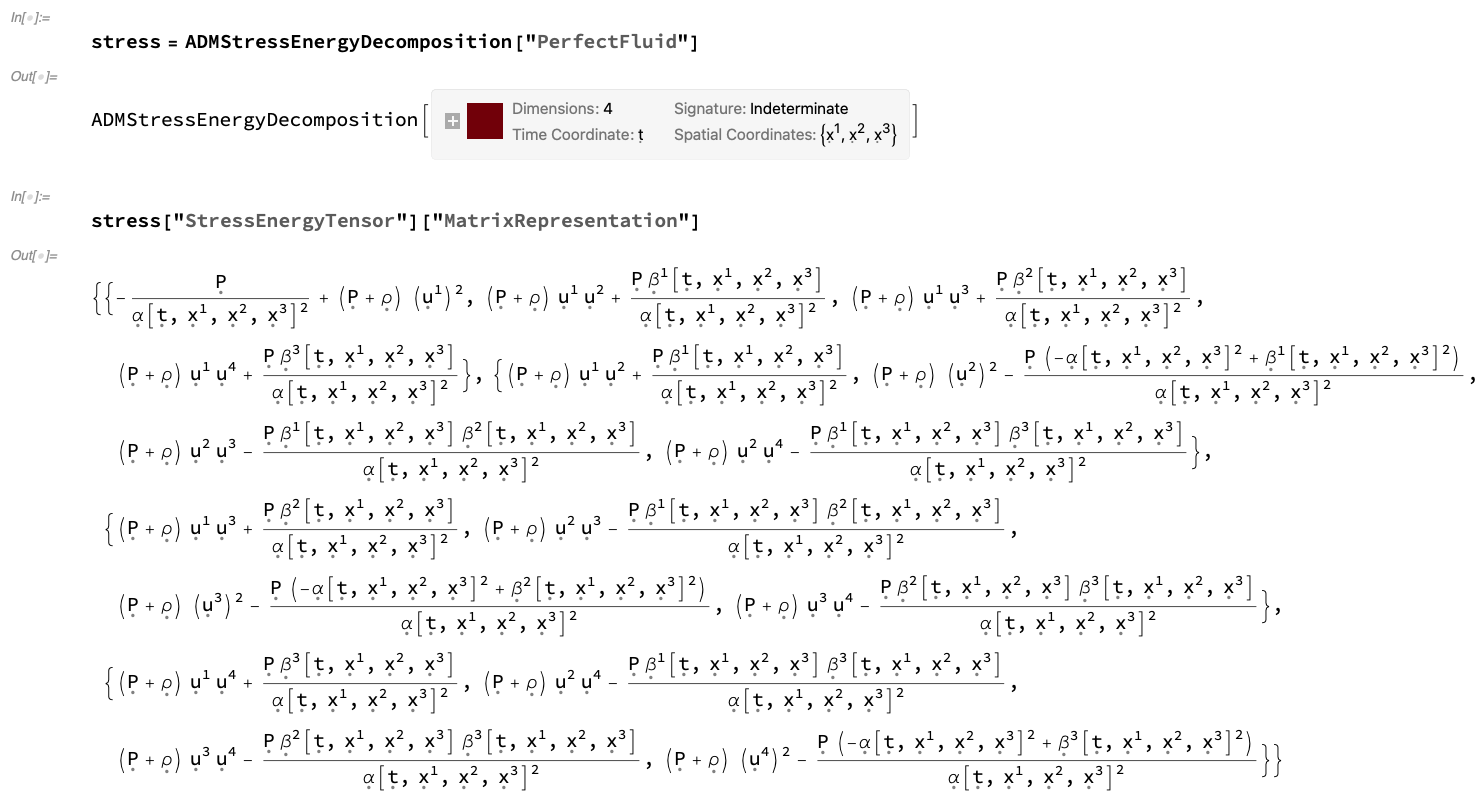}
\vrule
\includegraphics[width=0.445\textwidth]{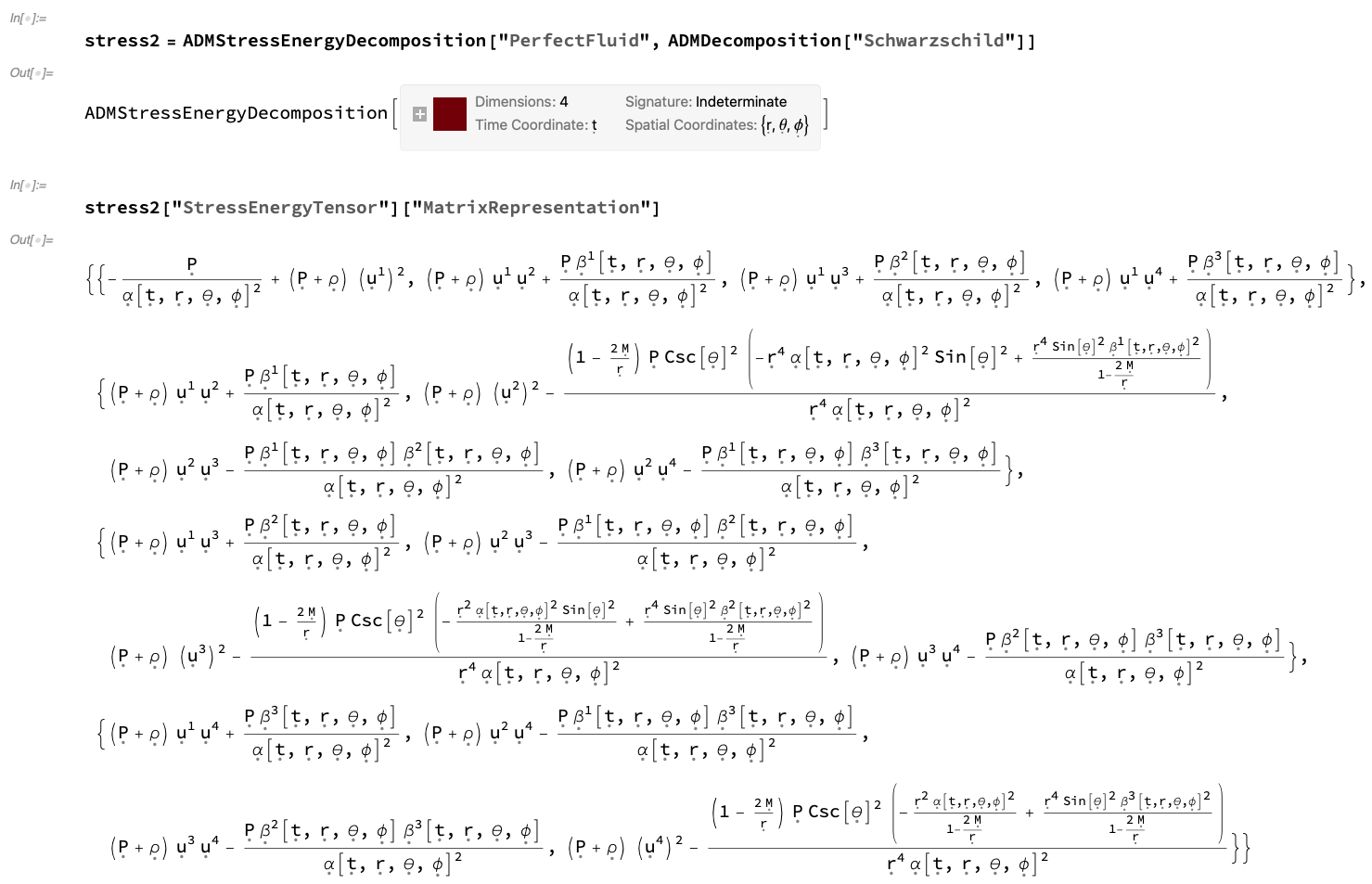}
\end{framed}
\caption{On the left, the \texttt{ADMStressEnergyDecomposition} object for a perfect relativistic fluid (representing an idealized fluid with density ${\rho}$, pressure $P$ and spacetime velocity ${u^{\mu}}$) embedded within an ADM decomposition of a Minkowski geometry with lapse function ${\alpha \left( t, r, \theta, \phi \right)}$ and shift vector ${\left( \beta^1 \left( t, r, \theta, \phi \right), \beta^2 \left( t, r, \theta, \phi \right), \beta^3 \left( t, r, \theta, \phi \right) \right)}$, in explicit contravariant matrix form. On the right, the \texttt{ADMStressEnergyDecomposition} object for a perfect relativistic fluid (representing an idealized fluid with density ${\rho}$, pressure $P$ and spacetime velocity ${u^{\mu}}$) embedded within an ADM decomposition of a Schwarzschild geometry with lapse function ${\alpha \left( t, r, \theta, \phi \right)}$ and shift vector ${\left( \beta^1 \left( t, r, \theta, \phi \right), \beta^2 \left( t, r, \theta, \phi \right), \beta^3 \left( t, r, \theta, \phi \right) \right)}$, in explicit contravariant matrix form.}
\label{fig:Figure31}
\end{figure}

\begin{figure}[ht]
\centering
\begin{framed}
\includegraphics[width=0.645\textwidth]{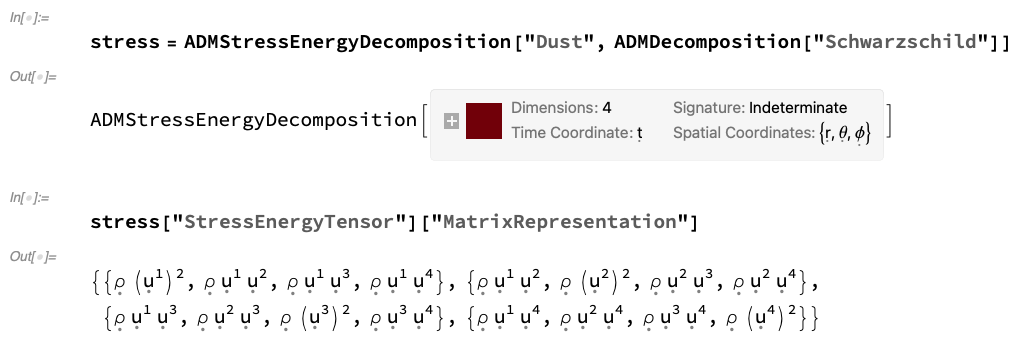}
\vrule
\includegraphics[width=0.345\textwidth]{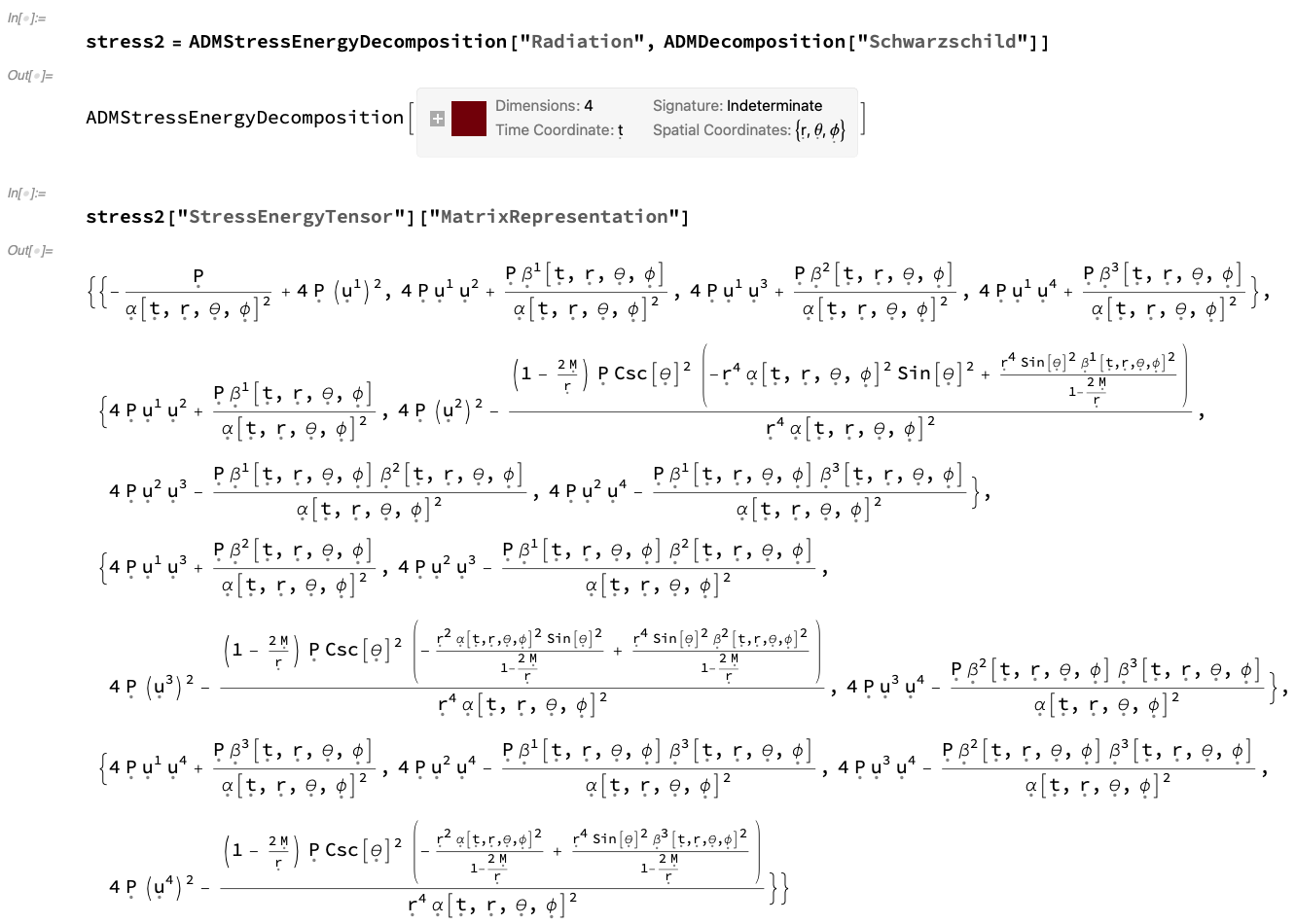}
\end{framed}
\caption{On the left, the \texttt{ADMStressEnergyDecomposition} object for a perfect relativistic dust (representing an idealized dust distribution with density ${\rho}$ and spacetime velocity ${u^{\mu}}$) embedded within an ADM decomposition of a Schwarzschild geometry with lapse function ${\alpha \left( t, r, \theta, \phi \right)}$ and shift vector ${\left( \beta^1 \left( t, r, \theta, \phi \right), \beta^2 \left( t, r, \theta, \phi \right), \beta^3 \left( t, r, \theta, \phi \right) \right)}$, in explicit contravariant matrix form. On the right, the \texttt{ADMStressEnergyDecomposition} object for a perfect relativistic radiation distribution (representing an idealized distribution of radiation with pressure $P$ and spacetime velocity ${u^{\mu}}$) embedded within an ADM decomposition of a Schwarzschild geometry with lapse function ${\alpha \left( t, r, \theta, \phi \right)}$ and shift vector ${\left( \beta^1 \left( t, r, \theta, \phi \right), \beta^2 \left( t, r, \theta, \phi \right), \beta^3 \left( t, r, \theta, \phi \right) \right)}$, in explicit contravariant matrix form.}
\label{fig:Figure32}
\end{figure}

\begin{figure}[ht]
\centering
\begin{framed}
\includegraphics[width=0.545\textwidth]{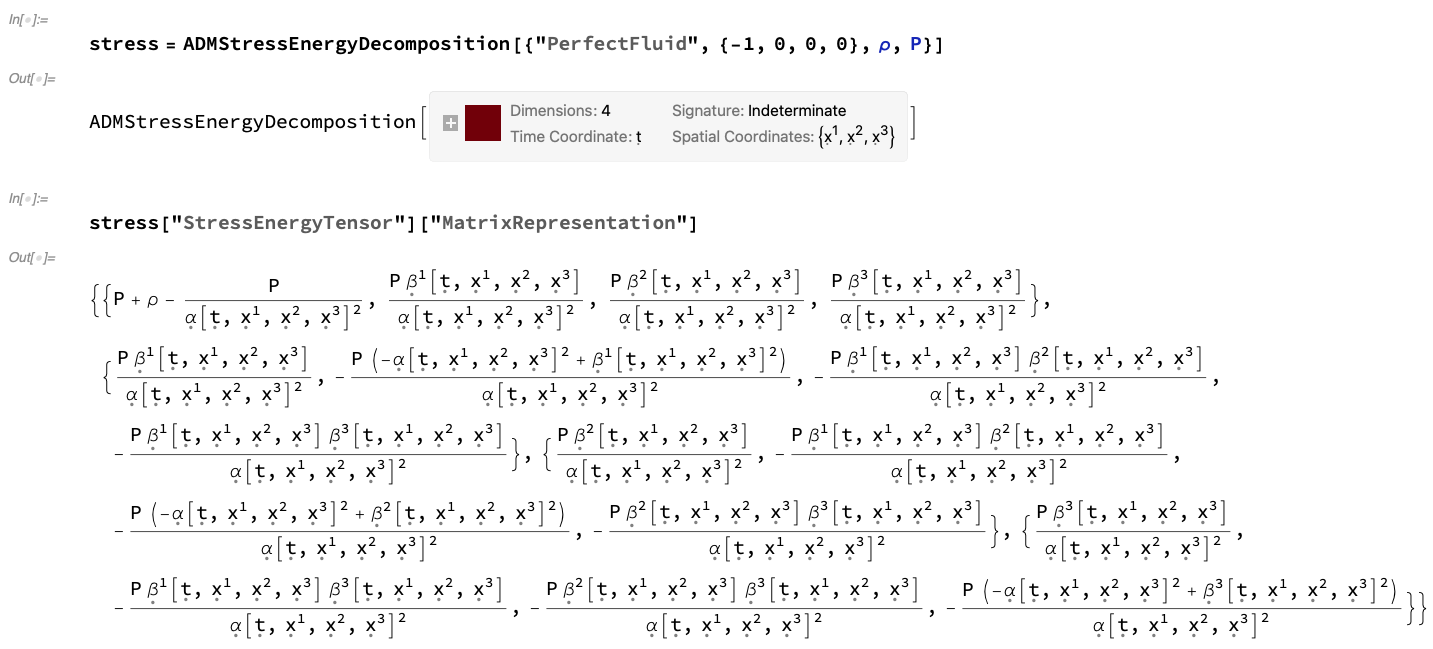}
\vrule
\includegraphics[width=0.445\textwidth]{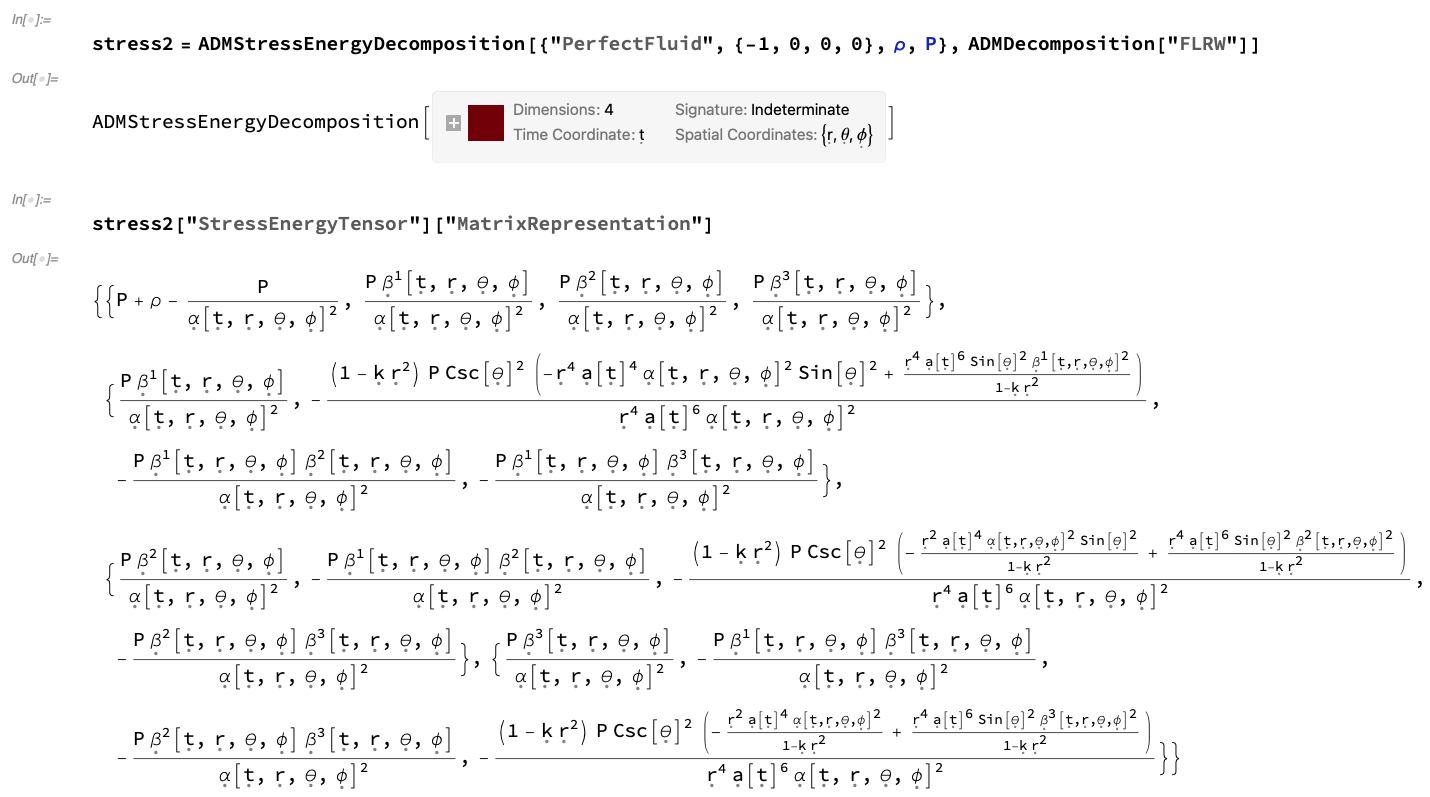}
\end{framed}
\caption{On the left, the \texttt{ADMStressEnergyDecomposition} object for a perfect relativistic fluid (representing an idealized fluid with density ${\rho}$, pressure $P$ and spacetime velocity ${\left( -1, 0, 0, 0 \right)}$) embedded within an ADM decomposition of a Minkowski geometry with lapse function ${\alpha \left( t, r, \theta, \phi \right)}$ and shift vector ${\left( \beta^1 \left( t, r, \theta, \phi \right), \beta^2 \left( t, r, \theta, \phi \right), \beta^3 \left( t, r, \theta, \phi \right) \right)}$, in explicit contravariant matrix form. On the right, the \texttt{ADMStressEnergyDecomposition} object for a perfect relativistic fluid (representing an idealized fluid with density ${\rho}$, pressure $P$ and spacetime velocity ${\left( -1, 0, 0, 0 \right)}$) embedded within an ADM decomposition of an FLRW geometry with lapse function ${\alpha \left( t, r, \theta, \phi \right)}$ and shift vector ${\left( \beta^1 \left( t, r, \theta, \phi \right), \beta^2 \left( t, r, \theta, \phi \right), \beta^3 \left( t, r, \theta, \phi \right) \right)}$, in explicit contravariant matrix form.}
\label{fig:Figure33}
\end{figure}

\begin{figure}[ht]
\centering
\begin{framed}
\includegraphics[width=0.495\textwidth]{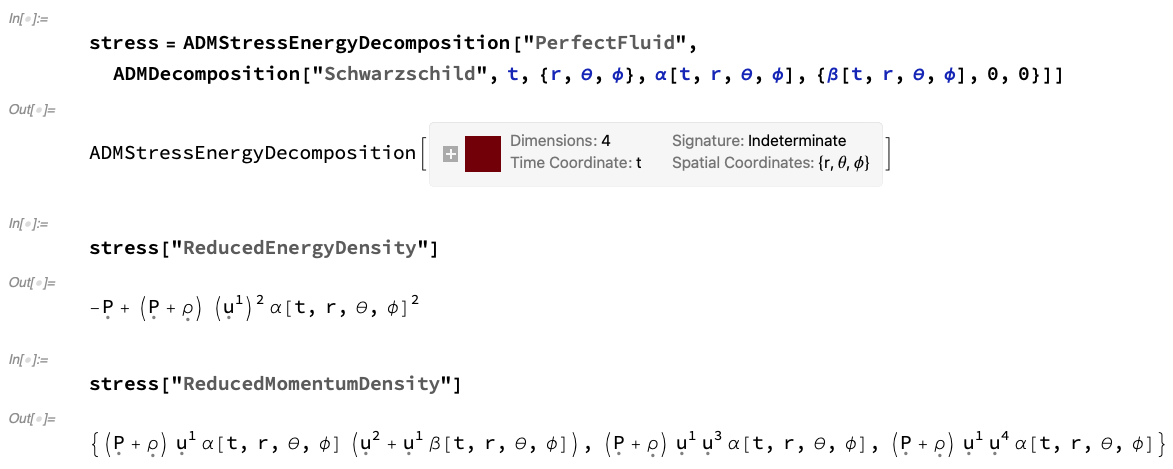}
\vrule
\includegraphics[width=0.495\textwidth]{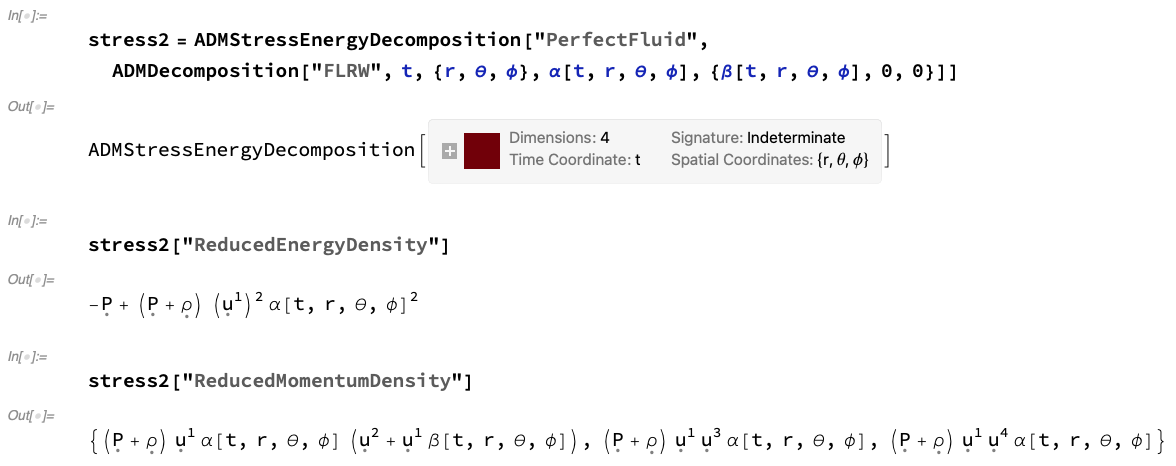}
\end{framed}
\caption{On the left, the relativistic energy density and relativistic momentum density of the \texttt{ADMStressEnergyDecomposition} object for a perfect relativistic fluid (representing an idealized fluid with density ${\rho}$, pressure $P$ and spacetime velocity ${u^{\mu}}$) embedded within an ADM decomposition of a Schwarzschild geometry with lapse function ${\alpha \left( t, r, \theta, \phi \right)}$ and modified shift vector ${\left( \beta \left( t, r, \theta, \phi \right), 0, 0 \right)}$. On the right, the relativistic energy density and relativistic momentum density of the \texttt{ADMStressEnergyDecomposition} object for a perfect relativistic fluid (representing an idealized fluid with density ${\rho}$, pressure $P$ and spacetime velocity ${u^{\mu}}$) embedded within an ADM decomposition of an FLRW geometry with lapse function ${\alpha \left( t, r, \theta, \phi \right)}$ and modified shift vector ${\left( \beta \left( t, r, \theta, \phi \right), 0, 0 \right)}$.}
\label{fig:Figure34}
\end{figure}

\begin{figure}[ht]
\centering
\begin{framed}
\includegraphics[width=0.495\textwidth]{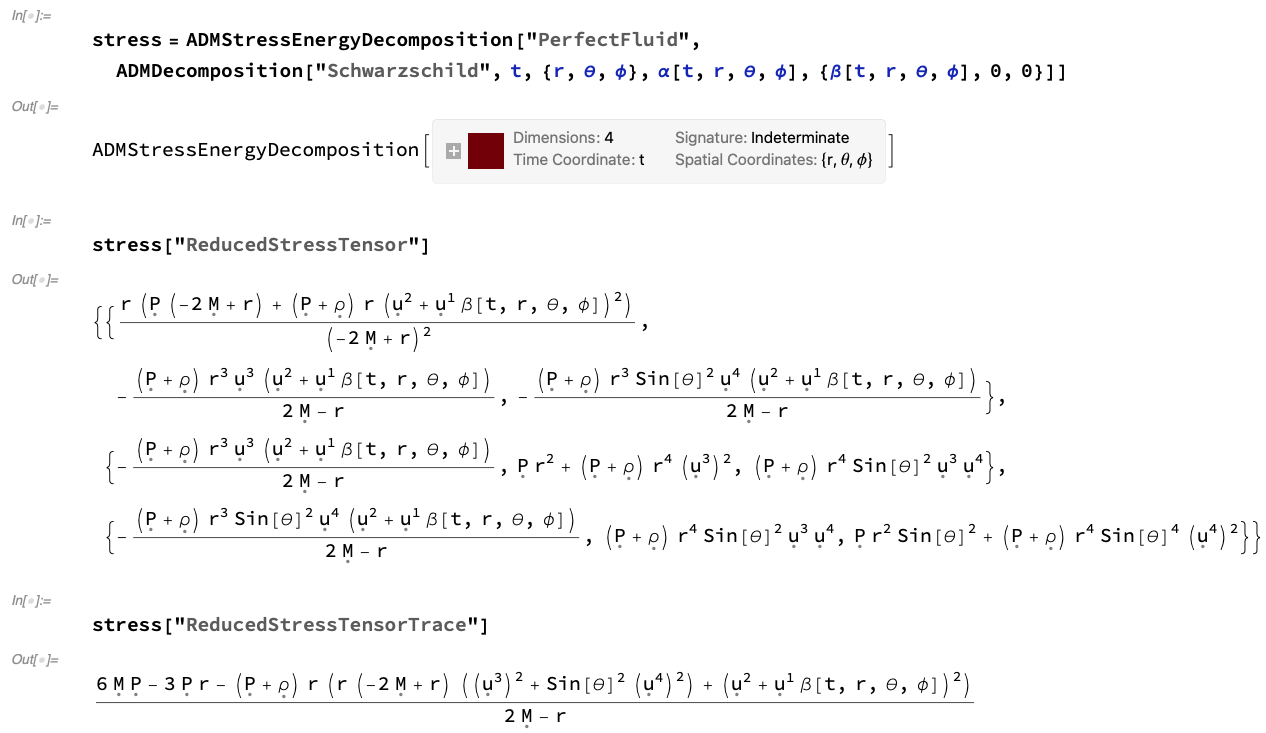}
\vrule
\includegraphics[width=0.495\textwidth]{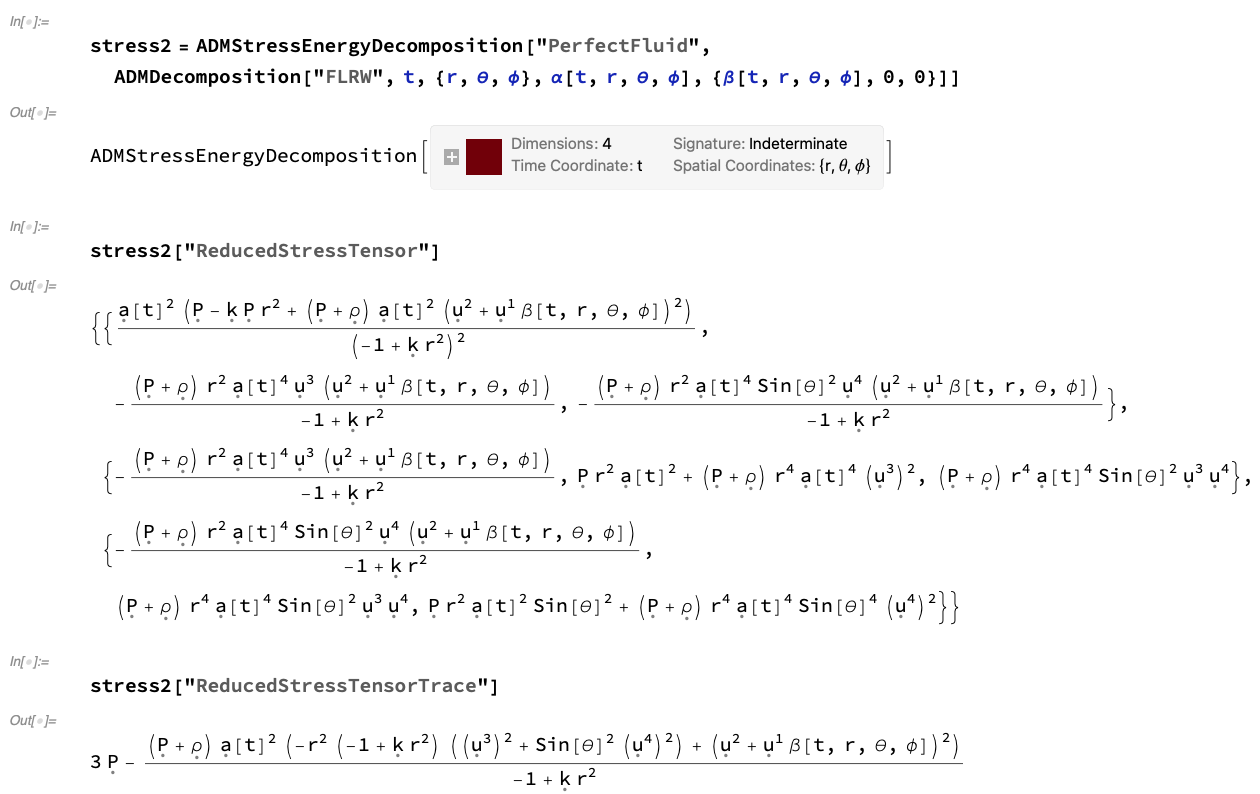}
\end{framed}
\caption{On the left, the relativistic Cauchy stress tensor and the corresponding stress tensor trace of the \texttt{ADMStressEnergyDecomposition} object for a perfect relativistic fluid (representing an idealized fluid with density ${\rho}$, pressure $P$ and spacetime velocity ${u^{\mu}}$) embedded within an ADM decomposition of a Schwarzschild geometry with lapse function ${\alpha \left( t, r, \theta, \phi \right)}$ and modified shift vector ${\left( \beta \left( t, r, \theta, \phi \right), 0, 0 \right)}$. On the right, the relativistic Cauchy stress tensor and the corresponding stress tensor trace of the \texttt{ADMStressEnergyDecomposition} object for a perfect relativistic fluid (representing an idealized fluid with density ${\rho}$, pressure $P$ and spacetime velocity ${u^{\mu}}$) embedded within an ADM decomposition of an FLRW geometry with lapse function ${\alpha \left( t, r, \theta, \phi \right)}$ and modified shift vector ${\left( \beta \left( t, r, \theta, \phi \right), 0, 0 \right)}$.}
\label{fig:Figure35}
\end{figure}

Representations of the ADM decompositions of the stress-energy tensors for a perfect relativistic fluid (representing an idealized fluid with density ${\rho}$, pressure $P$ and spacetime velocity ${u^{\mu}}$):

\begin{equation}
T^{\mu \nu} = \left( \rho + P \right) u^{\mu} u^{\nu} + P g^{\mu \nu},
\end{equation}
embedded within both an ADM decomposition of a Minkowski metric (representing a flat spacetime in Minkowski/Cartesian coordinates ${\left( t, x^1, x^2, x^3 \right)}$) and an ADM decomposition of a Schwarzschild metric (representing, for instance, an uncharged, non-rotating black hole with mass $M$ in Schwarzschild or spherical polar coordinates ${\left( t, r, \theta, \phi \right)}$), assuming the most general choice of gauge (i.e. lapse function ${\alpha \left( t, r, \theta, \phi \right)}$ and shift vector ${\left( \beta^1 \left( t, r, \theta, \phi \right), \beta^2 \left( t, r, \theta, \phi \right), \beta^3 \left( t, r, \theta, \phi \right) \right)}$), using the \texttt{ADMStressEnergyDecomposition} function, are shown in Figure \ref{fig:Figure31}. In addition to perfect relativistic fluids, \texttt{ADMStressEnergyDecomposition} also includes a small library of other in-built ADM decompositions of common relativistic energy-matter distributions (with, just as for \texttt{ADMDecomposition}, many more planned for future inclusion), including, but not limited to, perfect relativistic dust (with density ${\rho}$ and spacetime velocity ${u^{\mu}}$) and perfect relativistic radiation (with pressure $P$ and spacetime velocity ${u^{\mu}}$), namely:

\begin{equation}
T^{\mu \nu} = \rho u^{\mu} u^{\nu}, \qquad \text{ and } \qquad T^{\mu \nu} = n P u^{\mu} u^{\nu} + P g^{\mu \nu},
\end{equation}
respectively, as shown in Figure \ref{fig:Figure32}, in both cases embedded within an ADM decomposition of a Schwarzschild metric with the same generic choice of gauge (with lapse function ${\alpha \left( t, r, \theta, \phi \right)}$ and shift vector ${\left( \beta^1 \left( t, r, \theta, \phi \right), \beta^2 \left( t, r, \theta, \phi \right), \beta^3 \left( t, r, \theta, \phi \right) \right)}$). By default, appropriate formal symbols are chosen for the various parameters of the energy-matter distribution (e.g. mass-energy density ${\rho}$ or spacetime velocity ${u^{\mu}}$); however, these defaults can be easily overridden by passing additional arguments to \texttt{ADMStressEnergyDecomposition}, as shown in Figure \ref{fig:Figure33}. Figure \ref{fig:Figure34} shows the relativistic energy density $E$ (obtained from normal projection of ${T_{\mu \nu}}$) and the relativistic momentum density ${p^{\mu}}$ (obtained from orthogonal projection of ${T_{\mu \nu}}$), computed directly from the \texttt{ADMStressEnergyDecomposition} object for a perfect relativistic fluid (representing an idealized fluid with density ${\rho}$, pressure $P$ and spacetime velocity ${u^{\mu}}$) embedded within both an ADM decomposition of a Schwarzschild metric and an ADM decomposition of a Friedmann-Lema\^itre-Robertson-Walker or FLRW metric (representing, for instance, a perfectly homogeneous and isotropic universe with scale factor ${a \left( t \right)}$ and curvature $k$ in spherical polar coordinates ${\left( t, r, \theta, \phi \right)}$), assuming in both cases a restricted choice of gauge with lapse function ${\alpha \left( t, r, \theta, \phi \right)}$ and modified shift vector ${\left( \beta \left( t, r, \theta, \phi \right), 0, 0 \right)}$. Likewise, Figure \ref{fig:Figure35} shows the relativistic Cauchy stress tensor ${S_{\mu \nu}}$ (obtained from a mixture of orthogonal projections of ${T_{\mu \nu}}$) and its corresponding trace ${S = \gamma^{\mu \nu} S_{\mu \nu}}$, also computed directly from the \texttt{ADMStressEnergyDecomposition} object for a perfect relativistic fluid, and again embedded within both an ADM decomposition of a Schwarzschild metric and an ADM decomposition of an FLRW metric, with the same restricted choice of gauge.

\begin{figure}[ht]
\centering
\begin{framed}
\includegraphics[width=0.495\textwidth]{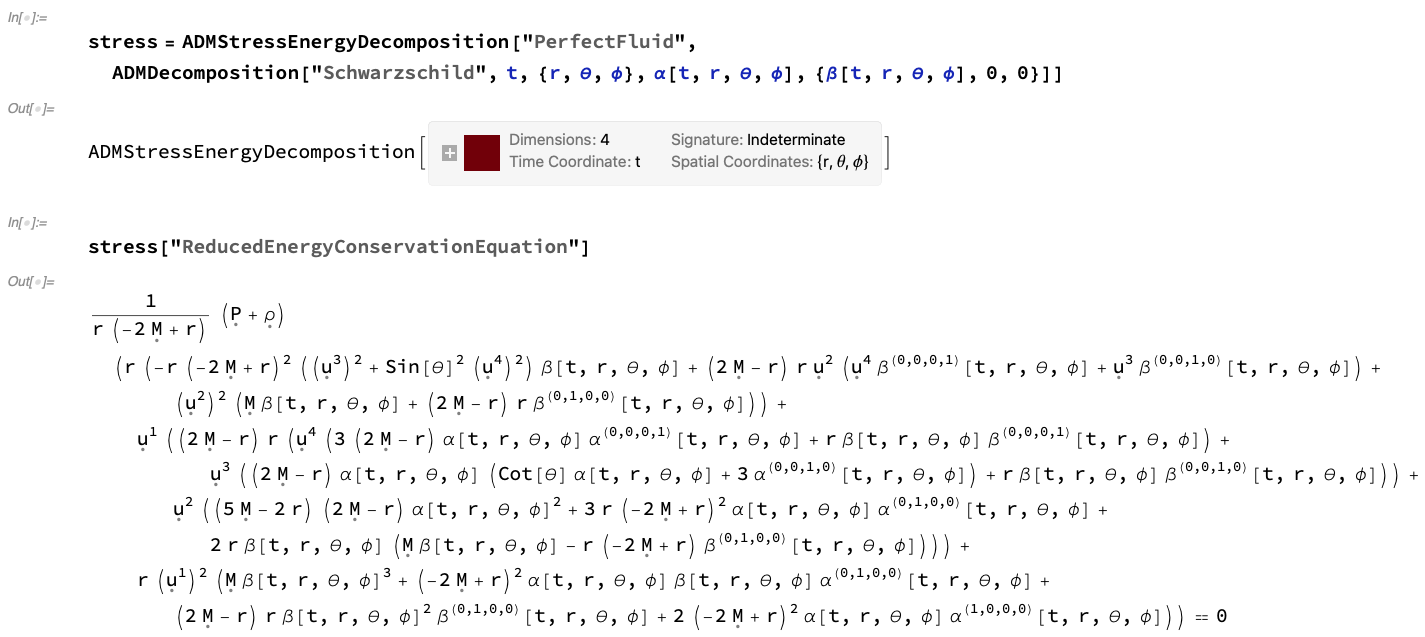}
\vrule
\includegraphics[width=0.495\textwidth]{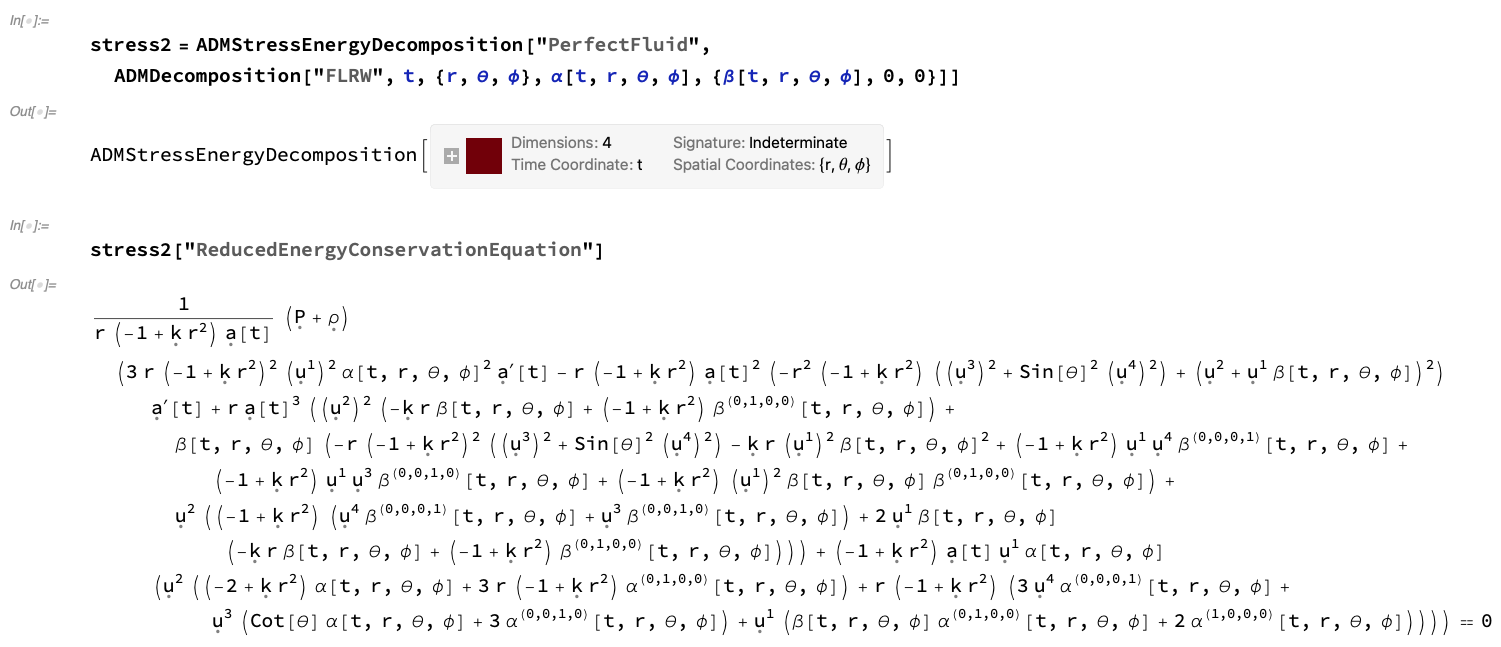}
\end{framed}
\caption{On the left, the energy conservation equation for the \texttt{ADMStressEnergyDecomposition} object for a perfect relativistic fluid (representing an idealized fluid with density ${\rho}$, pressure $P$ and spacetime velocity ${u^{\mu}}$) embedded within an ADM decomposition of a Schwarzschild geometry with lapse function ${\alpha \left( t, r, \theta, \phi \right)}$ and modified shift vector ${\left( \beta \left( t, r, \theta, \phi \right), 0, 0 \right)}$. On the right, the energy conservation equation for the \texttt{ADMStressEnergyDecomposition} object for a perfect relativistic fluid (representing an idealized fluid with density ${\rho}$, pressure $P$ and spacetime velocity ${u^{\mu}}$) embedded within an ADM decomposition of an FLRW geometry with lapse function ${\alpha \left( t, r, \theta, \phi \right)}$ and modified shift vector ${\left( \beta \left( t, r, \theta, \phi \right), 0, 0 \right)}$.}
\label{fig:Figure36}
\end{figure}

\begin{figure}[ht]
\centering
\begin{framed}
\includegraphics[width=0.495\textwidth]{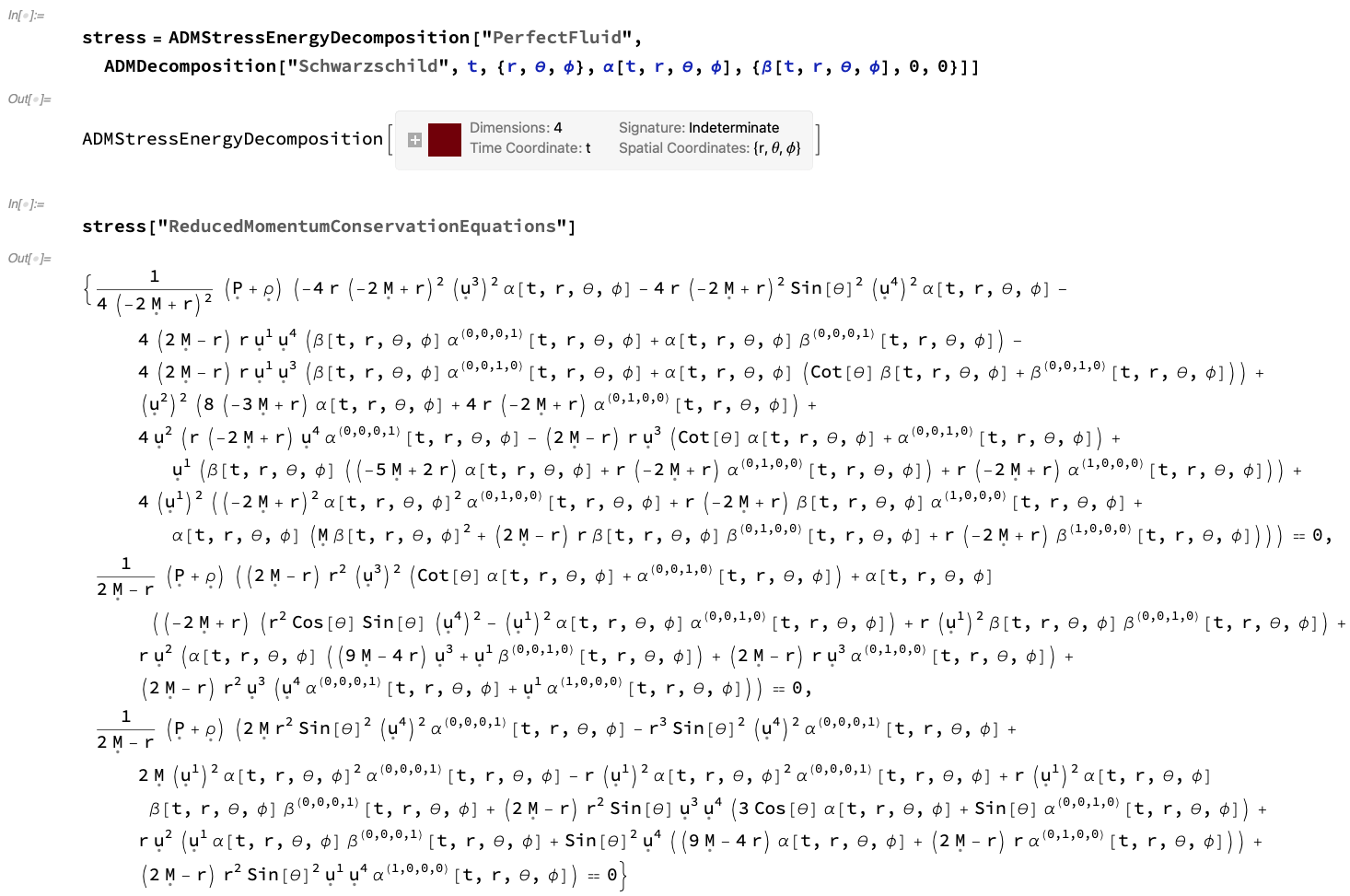}
\vrule
\includegraphics[width=0.495\textwidth]{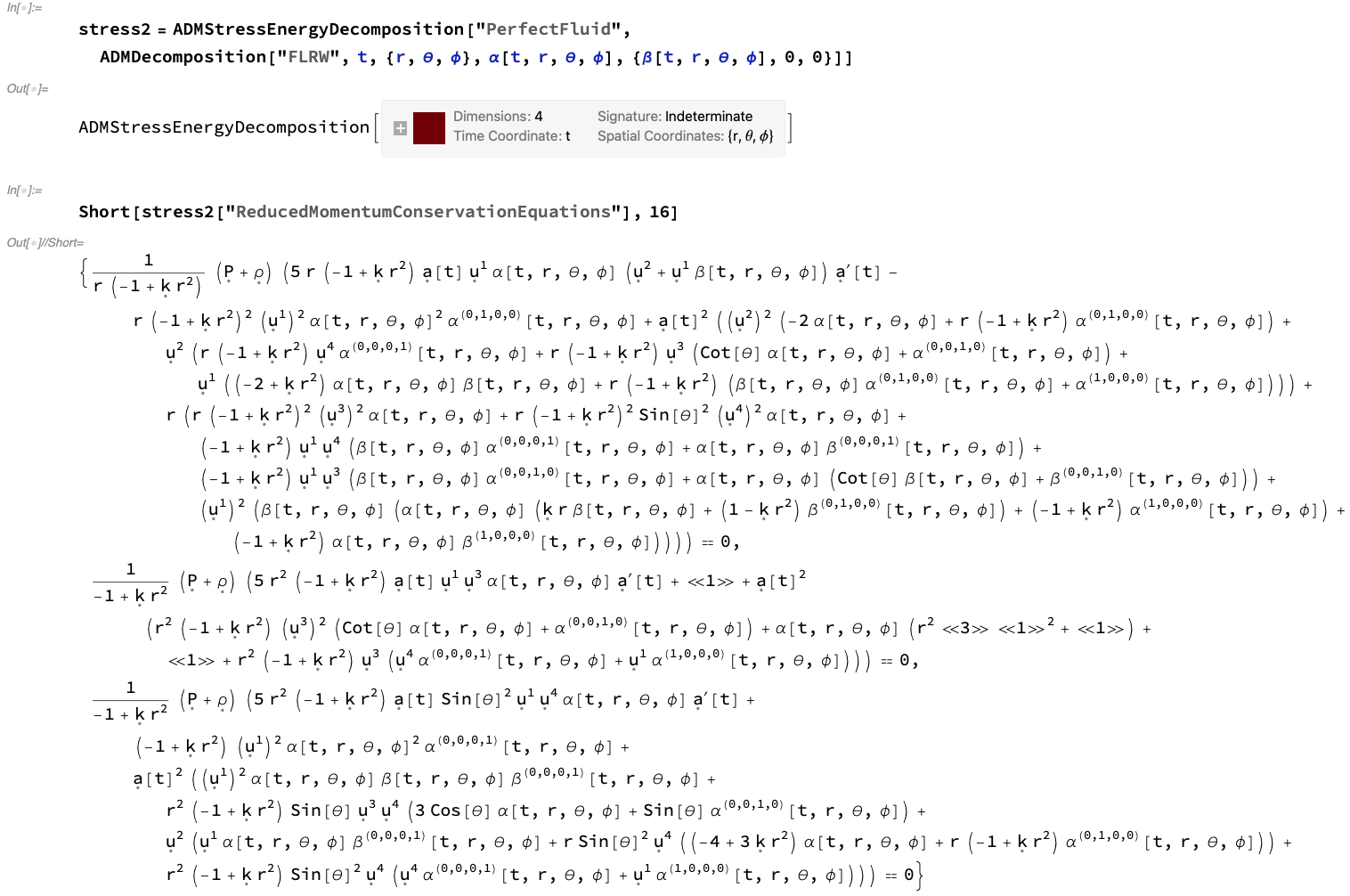}
\end{framed}
\caption{On the left, the list of momentum conservation equations for the \texttt{ADMStressEnergyDecomposition} object for a perfect relativistic fluid (representing an idealized fluid with density ${\rho}$, pressure $P$ and spacetime velocity ${u^{\mu}}$) embedded within an ADM decomposition of a Schwarzschild geometry with lapse function ${\alpha \left( t, r, \theta, \phi \right)}$ and modified shift vector ${\left( \beta \left( t, r, \theta, \phi \right), 0, 0 \right)}$. On the right, the list of momentum conservation equations for the \texttt{ADMStressEnergyDecomposition} object for a perfect relativistic fluid (representing an idealized fluid with density ${\rho}$, pressure $P$ and spacetime velocity ${u^{\mu}}$) embedded within an ADM decomposition of an FLRW geometry with lapse function ${\alpha \left( t, r, \theta, \phi \right)}$ and modified shift vector ${\left( \beta \left( t, r, \theta, \phi \right), 0, 0 \right)}$.}
\label{fig:Figure37}
\end{figure}

This decomposition now also enables us to decompose the continuity equations for the full (spacetime) stress-energy tensor ${T_{\mu \nu}}$, which assert that the spacetime covariant divergence of the stress-energy tensor ${{}^{\left( 4 \right)} \nabla_{\nu} T^{\mu \nu}}$ must vanish identically, and which follow from a combination of the Einstein field equations and the Bianchi identities (which assert that the spacetime covariant divergence of the Einstein tensor ${{}^{\left( 4 \right)} \nabla_{\nu} {}^{\left( 4 \right)} G^{\mu \nu}}$ must also vanish identically, for purely geometrical reasons):

\begin{multline}
{}^{\left( 4 \right)} \nabla_{\nu} {}^{\left( 4 \right)} G^{\mu \nu} = \frac{\partial}{\partial x^{\nu}} \left( {}^{\left( 4 \right)} G^{\mu \nu} \right) + {}^{\left( 4 \right)} \Gamma_{\nu \sigma}^{\mu} {}^{\left( 4 \right)} G^{\sigma \nu} + {}^{\left( 4 \right)} \Gamma_{\nu \sigma}^{\nu} {}^{\left( 4 \right)} G^{\mu \sigma} = 0,\\
\implies {}^{\left( 4 \right)} \nabla_{\nu} T^{\mu \nu} = \frac{\partial}{\partial x^{\nu}} \left( T^{\mu \nu} \right) + {}^{\left( 4 \right)} \Gamma_{\nu \sigma}^{\mu} T^{\sigma \nu} + {}^{\left( 4 \right)} \Gamma_{\nu \sigma}^{\nu} T^{\mu \sigma} = 0,
\end{multline}
and which guarantee the conservation of relativistic energy and momentum, into purely timelike and purely spacelike components, just as we did previously with the decomposition of the Bianchi identities themselves into the timelike Hamiltonian constraint ${\mathcal{H}}$ and the spacelike momentum constraints ${\mathcal{M}_{\mu}}$. In the above, ${\mu, \nu, \sigma}$ range across all ${\left\lbrace 0, \dots, n - 1 \right\rbrace}$ (i.e. across all spacetime coordinate indices). The purely timelike projection of the continuity equations yields the energy conservation equation:

\begin{equation}
\frac{\partial}{\partial t} \left( E \right) - \mathcal{L}_{\boldsymbol{\beta}} E + \alpha \left( {}^{\left( 3 \right)} \nabla_{\mu} p^{\mu} - K E - K_{\mu \nu} S^{\mu \nu} \right) + 2 p^{\mu} {}^{\left( 3 \right)} \nabla_{\mu} \alpha = 0,
\end{equation}
where the Lie derivative term ${\mathcal{L}_{\boldsymbol{\beta}} E}$ expands out to give:

\begin{equation}
\frac{\partial}{\partial t} \left( E \right) - \beta^{\mu} \frac{\partial}{\partial x^{\mu}} \left( E \right) + \alpha \left( {}^{\left( 3 \right)} \nabla_{\mu} p^{\mu} - K E - K_{\mu \nu} S^{\mu \nu} \right) + 2 p^{\mu} {}^{\left( 3 \right)} \nabla_{\mu} \alpha = 0,
\end{equation}
i.e., in fully-expanded form:

\begin{equation}
\frac{\partial}{\partial t} \left( E \right) - \beta^{\mu} \frac{\partial}{\partial x^{\mu}} \left( E \right) + \alpha \left( \frac{\partial}{\partial x^{\mu}} \left( p^{\mu} \right) + {}^{\left( 3 \right)} \Gamma_{\mu \sigma}^{\mu} p^{\sigma} - K E - K_{\mu \nu} S^{\mu \nu} \right) + 2 p^{\mu} \frac{\partial}{\partial x^{\mu}} \left( \alpha \right) = 0,
\end{equation}
which holds identically by virtue of the vanishing of the ADM Hamiltonian constraint ${\mathcal{H} = 0}$. Here, ${\mu, \nu, \sigma}$ range across all ${\left\lbrace 0, \dots, n - 2 \right\rbrace}$ (i.e. across spatial coordinate indices only). The purely spacelike projections of the continuity equations yield the momentum conservation equations:

\begin{equation}
\frac{\partial}{\partial t} \left( p_{\mu} \right) - \mathcal{L}_{\boldsymbol{\beta}}  p_{\mu} + \alpha {}^{\left( 3 \right)} \nabla_{\nu} S_{\mu}^{\nu} + S_{\mu \nu} {}^{\left( 3 \right)} \nabla^{\nu} \alpha - \alpha K p_{\mu} + E {}^{\left( 3 \right)} \nabla_{\mu} \alpha = 0,
\end{equation}
which, after expansion of the Lie derivative term ${\mathcal{L}_{\boldsymbol{\beta}} p_{\mu}}$ and replacement of the contravariant derivative operator ${{}^{\left( 3 \right)} \nabla^{\nu}}$ with a corresponding covariant derivative operator ${{}^{\left( 3 \right)} \nabla_{\sigma}}$, become:

\begin{equation}
\frac{\partial}{\partial t} \left( p_{\mu} \right) - \beta^{\sigma} \frac{\partial}{\partial x^{\sigma}} \left( p_{\mu} \right) - p_{\sigma} \frac{\partial}{\partial x^{\mu}} \left( \beta^{\sigma} \right) + \alpha {}^{\left( 3 \right)} \nabla_{\nu} S_{\mu}^{\nu} + S_{\mu \nu} \gamma^{\nu \sigma} {}^{\left( 3 \right)} \nabla_{\sigma} \alpha - \alpha K p_{\mu} + E {}^{\left( 3 \right)} \nabla_{\mu} \alpha = 0,
\end{equation}
i.e., in fully-expanded form:

\begin{multline}
\frac{\partial}{\partial t} \left( p_{\mu} \right) - \beta^{\sigma} \frac{\partial}{\partial x^{\sigma}} \left( p_{\mu} \right) - p_{\sigma} \frac{\partial}{\partial x^{\mu}} \left( \beta^{\sigma} \right) + \alpha \left( \frac{\partial}{\partial x^{\nu}} \left( S_{\mu}^{\nu} \right) + {}^{\left( 3 \right)} \Gamma_{\nu \sigma}^{\nu} S_{\mu}^{\sigma} - {}^{\left( 3 \right)} \Gamma_{\nu \mu}^{\sigma} S_{\sigma}^{\nu} \right)\\
+ S_{\mu \nu} \gamma^{\nu \sigma} \frac{\partial}{\partial x^{\sigma}} \left( \alpha \right) - \alpha K p_{\mu} + E \frac{\partial}{\partial x^{\mu}} \left( \alpha \right) = 0,
\end{multline}
which hold identically by virtue of the vanishing of the ADM momentum constraints ${\mathcal{M}_{\mu} = 0}$. Figures \ref{fig:Figure36} and \ref{fig:Figure37} show how the energy and momentum conservation equations may be derived directly from the \texttt{ADMStressEnergyDecomposition} object for a perfect relativistic fluid embedded within both an ADM decomposition of a Schwarzschild metric and an ADM decomposition of an FLRW metric, with the same restricted choice of gauge as above (i.e. lapse function ${\alpha \left( t, r, \theta, \phi \right)}$ and modified shift vector ${\left( \beta \left( t, r, \theta, \phi \right), 0, 0 \right)}$).

\begin{figure}[ht]
\centering
\begin{framed}
\includegraphics[width=0.495\textwidth]{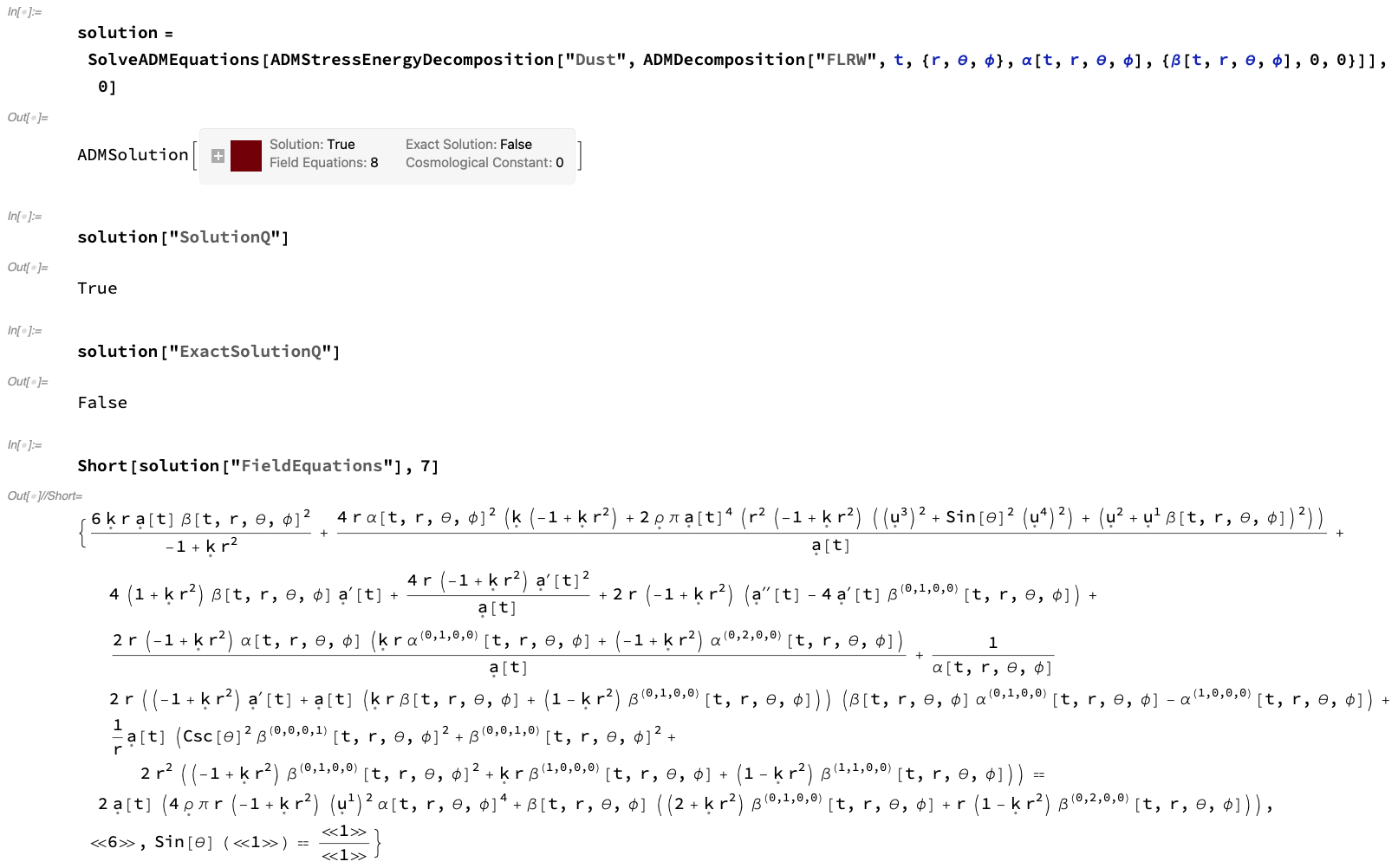}
\vrule
\includegraphics[width=0.495\textwidth]{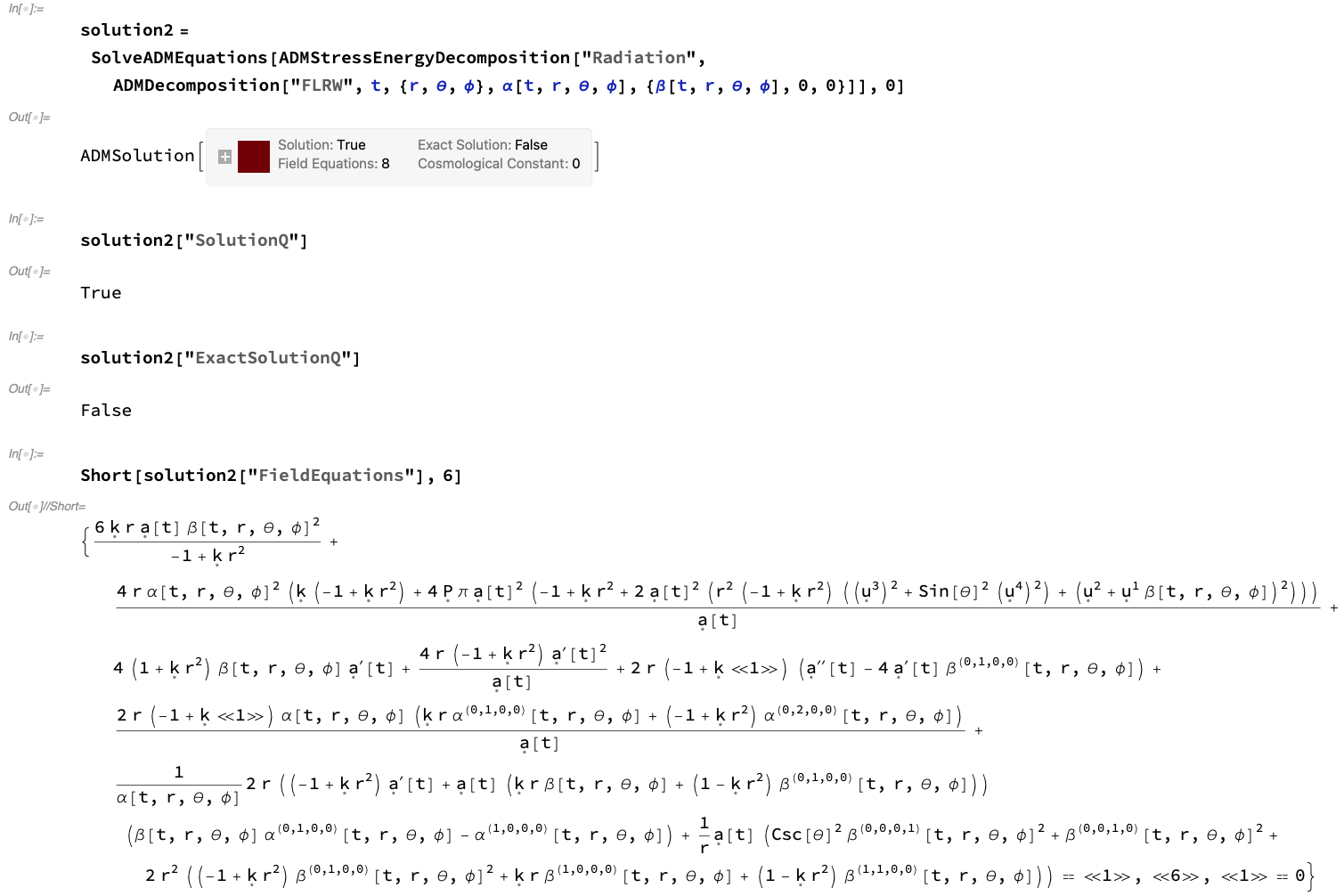}
\end{framed}
\caption{On the left, the \texttt{ADMSolution} object for an ADM decomposition of an FLRW geometry equipped with a perfect relativistic dust (representing a perfectly homogeneous and isotropic universe filled with an idealized distribution of dust) with lapse function ${\alpha \left( t, r, \theta, \phi \right)}$ and modified shift vector ${\left( \beta \left( t, r, \theta, \phi \right), 0, 0 \right)}$, with zero cosmological constant ${\Lambda = 0}$, computed using \texttt{SolveADMEquations}, illustrating that this decomposition is a non-exact solution to the ADM evolution equations, with eight additional field equations required. On the right, the \texttt{ADMSolution} object for an ADM decomposition of an FLRW geometry equipped with a perfect relativistic radiation distribution (representing a perfectly homogeneous and isotropic universe filled with an idealized distribution of radiation) with lapse function ${\alpha \left( t, r, \theta, \phi \right)}$ and modified shift vector ${\left( \beta \left( t, r, \theta, \phi \right), 0, 0 \right)}$, with zero cosmological constant ${\Lambda = 0}$, computed using \texttt{SolveADMEquations}, illustrating that this decomposition is a non-exact solution to the ADM evolution equations, with eight additional field equations required.}
\label{fig:Figure38}
\end{figure}

\begin{figure}[ht]
\centering
\begin{framed}
\includegraphics[width=0.495\textwidth]{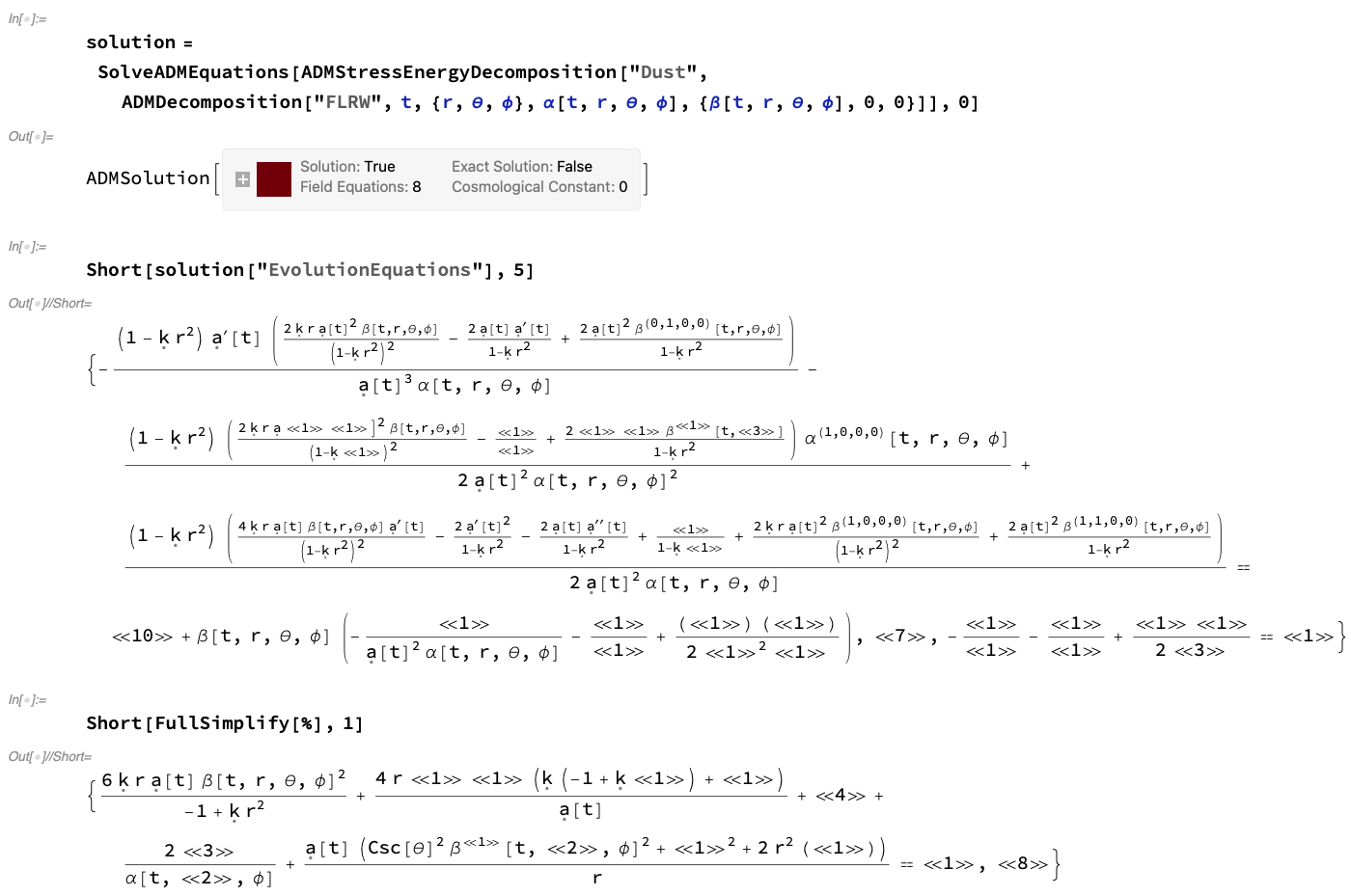}
\vrule
\includegraphics[width=0.495\textwidth]{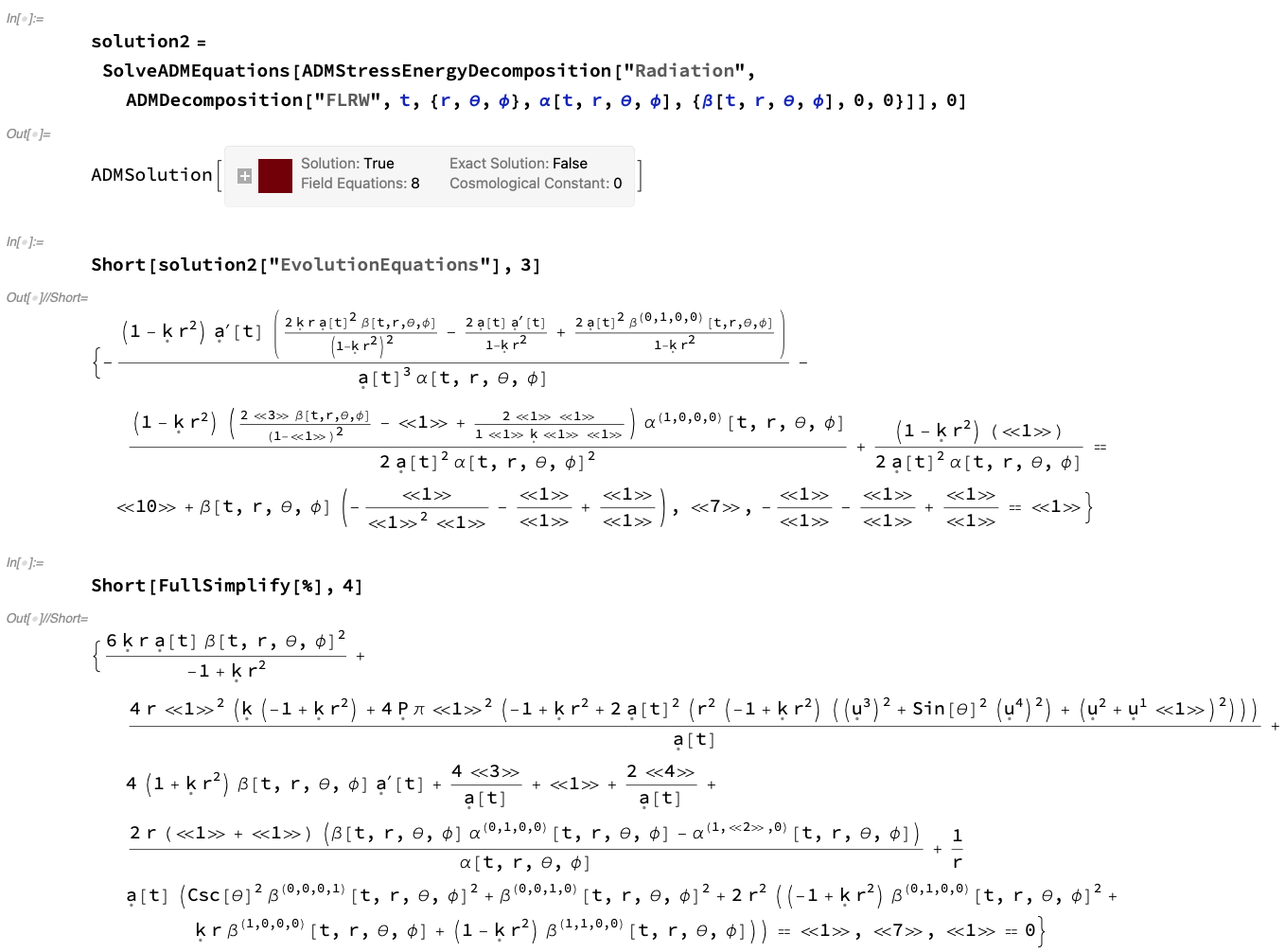}
\end{framed}
\caption{On the left, the list of ADM evolution equations, computed using the \texttt{ADMSolution} object for an ADM decomposition of an FLRW geometry equipped with a perfect relativistic dust (representing a perfectly homogeneous and isotropic universe filled with an idealized distribution of dust) with lapse function ${\alpha \left( t, r, \theta, \phi \right)}$ and modified shift vector ${\left( \beta \left( t, r, \theta, \phi \right), 0, 0 \right)}$, with zero cosmological constant ${\Lambda = 0}$, together with a verification that they reduce down to a set of eight canonical field equations. On the right, the list of ADM evolution equations, computed using the \texttt{ADMSolution} object for an ADM decomposition of an FLRW geometry equipped with a perfect relativistic radiation distribution (representing a perfectly homogeneous and isotropic universe filled with an idealized distribution of radiation) with lapse function ${\alpha \left( t, r, \theta, \phi \right)}$ and modified shift vector ${\left( \beta \left( t, r, \theta, \phi \right), 0, 0 \right)}$, with zero cosmological constant ${\Lambda = 0}$, together with a verification that they reduce down to a set of eight canonical field equations.}
\label{fig:Figure39}
\end{figure}

\begin{figure}[ht]
\centering
\begin{framed}
\includegraphics[width=0.495\textwidth]{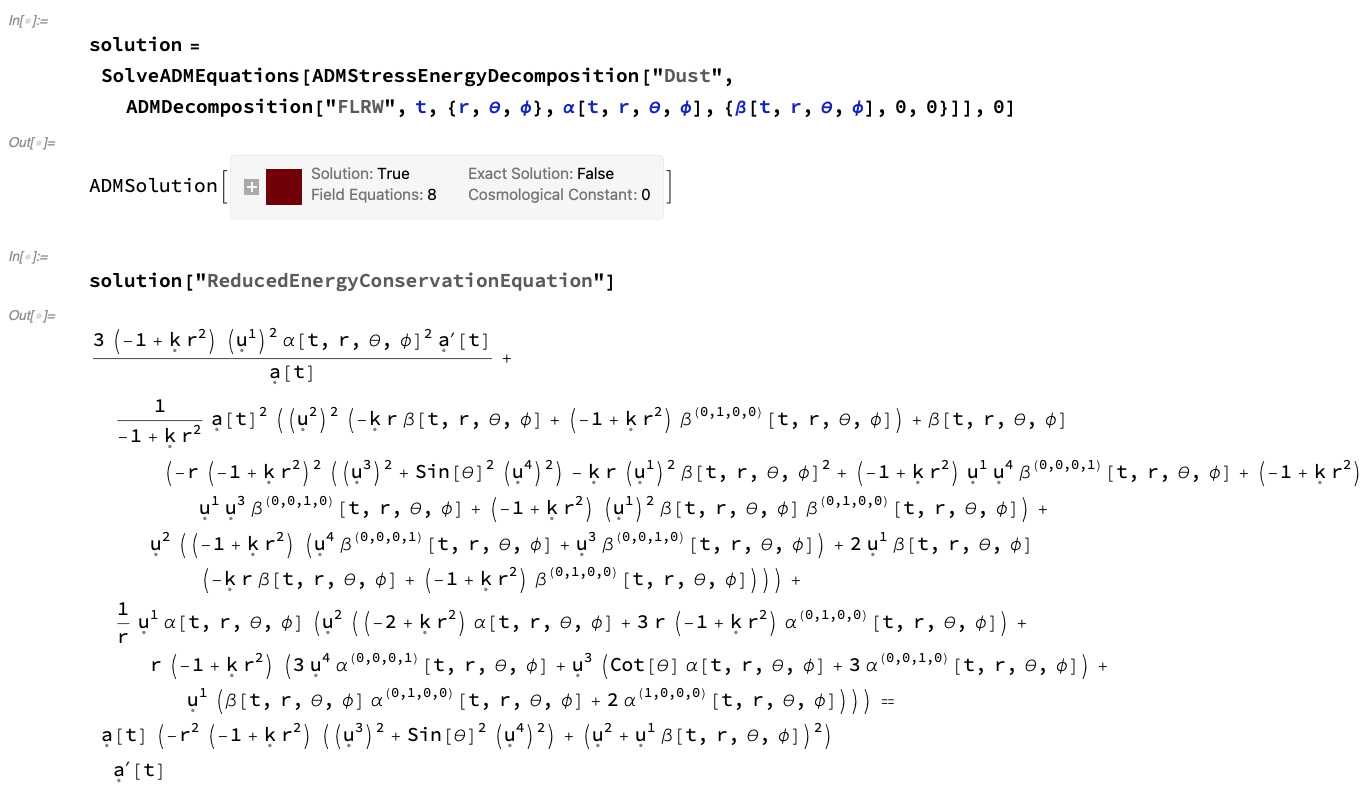}
\vrule
\includegraphics[width=0.495\textwidth]{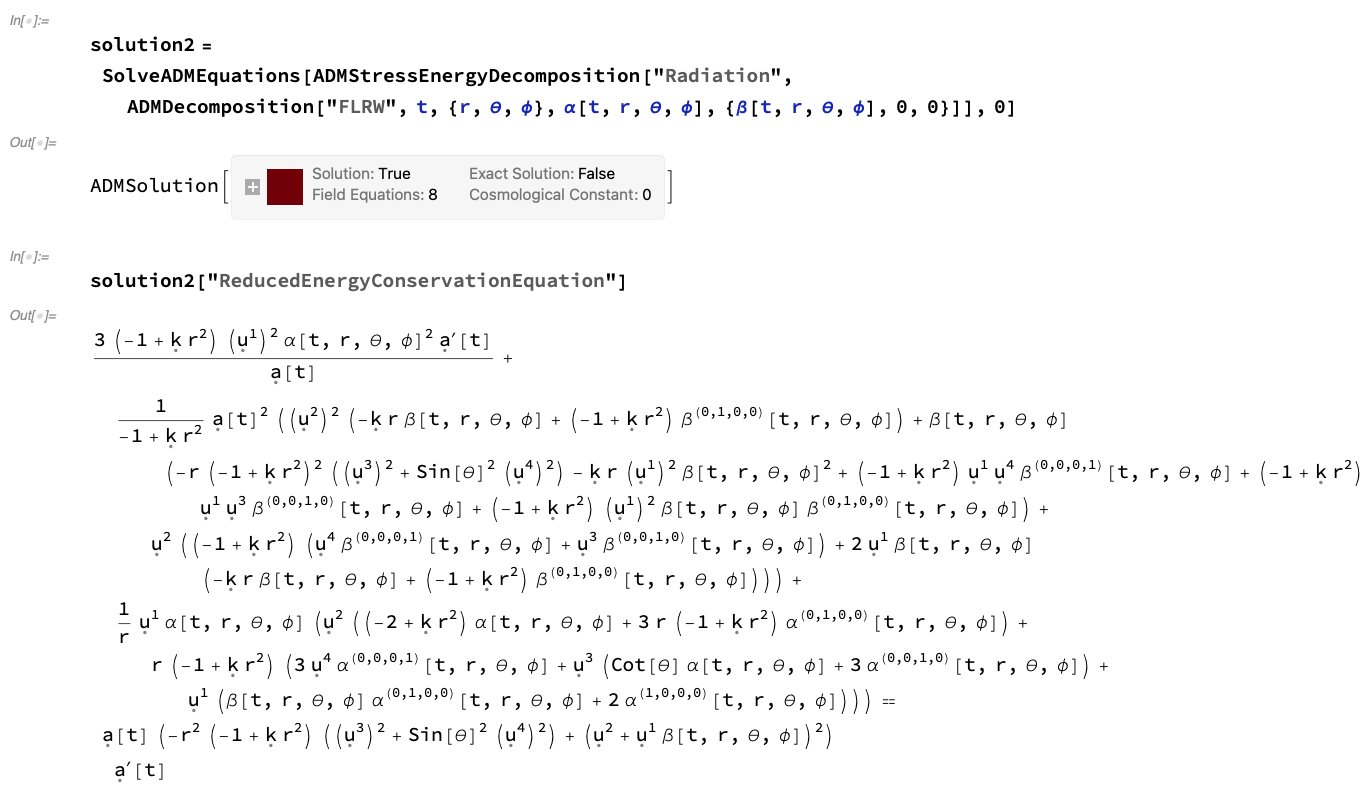}
\end{framed}
\caption{On the left, the energy conservation equation obtained from the purely timelike projection of the full continuity equations, computed using the \texttt{ADMSolution} object for an ADM decomposition of an FLRW geometry equipped with a perfect relativistic dust (representing a perfectly homogeneous and isotropic universe filled with an idealized distribution of dust) with lapse function ${\alpha \left( t, r, \theta, \phi \right)}$ and modified shift vector ${\left( \beta \left( t, r, \theta, \phi \right), 0, 0 \right)}$, with zero cosmological constant ${\Lambda = 0}$. On the right, the energy conservation equation obtained from the purely timelike projection of the full continuity equations, computed using the \texttt{ADMSolution} object for an ADM decomposition of an FLRW geometry equipped with a perfect relativistic radiation distribution (representing a perfectly homogeneous and isotropic universe filled with an idealized distribution of radiation) with lapse function ${\alpha \left( t, r, \theta, \phi \right)}$ and modified shift vector ${\left( \beta \left( t, r, \theta, \phi \right), 0, 0 \right)}$, with zero cosmological constant ${\Lambda = 0}$.}
\label{fig:Figure40}
\end{figure}

\begin{figure}[ht]
\centering
\begin{framed}
\includegraphics[width=0.495\textwidth]{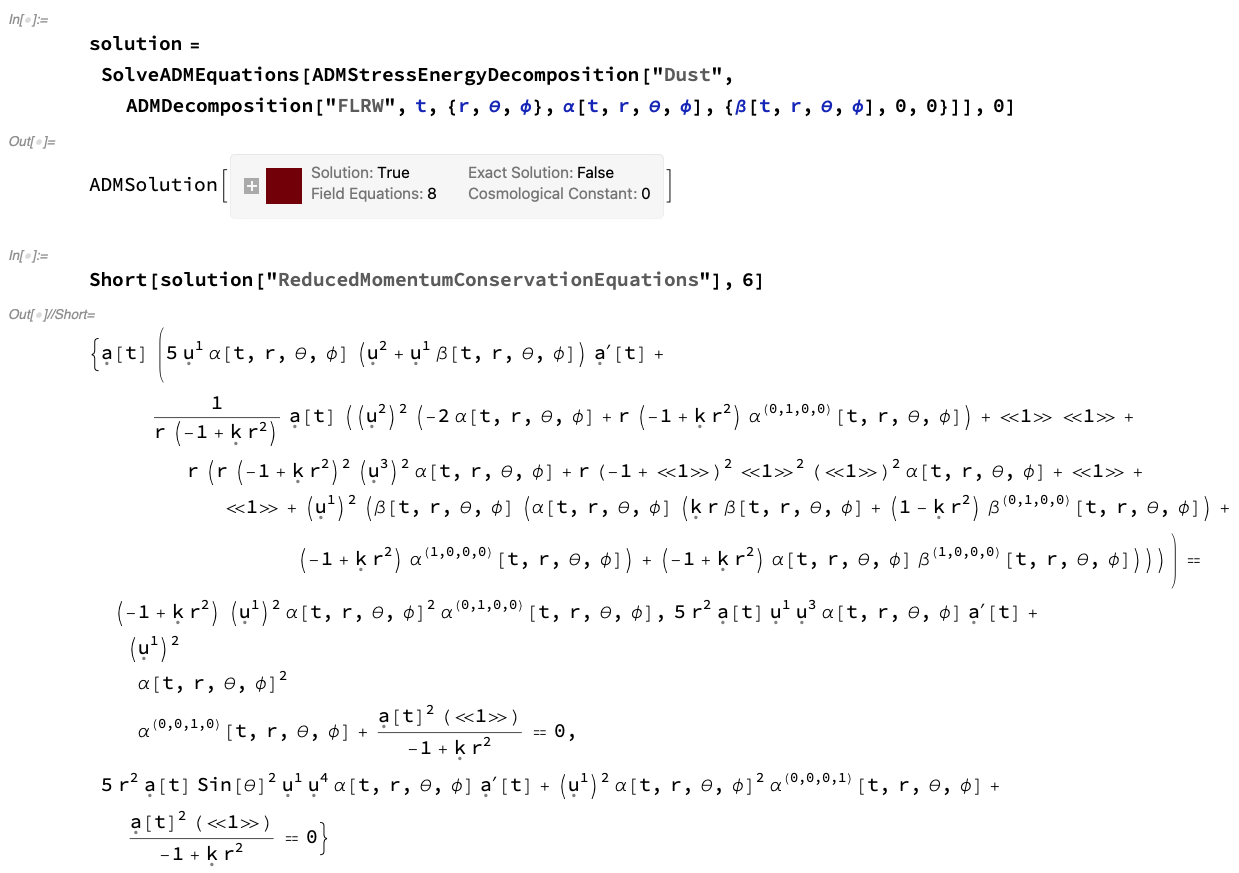}
\vrule
\includegraphics[width=0.495\textwidth]{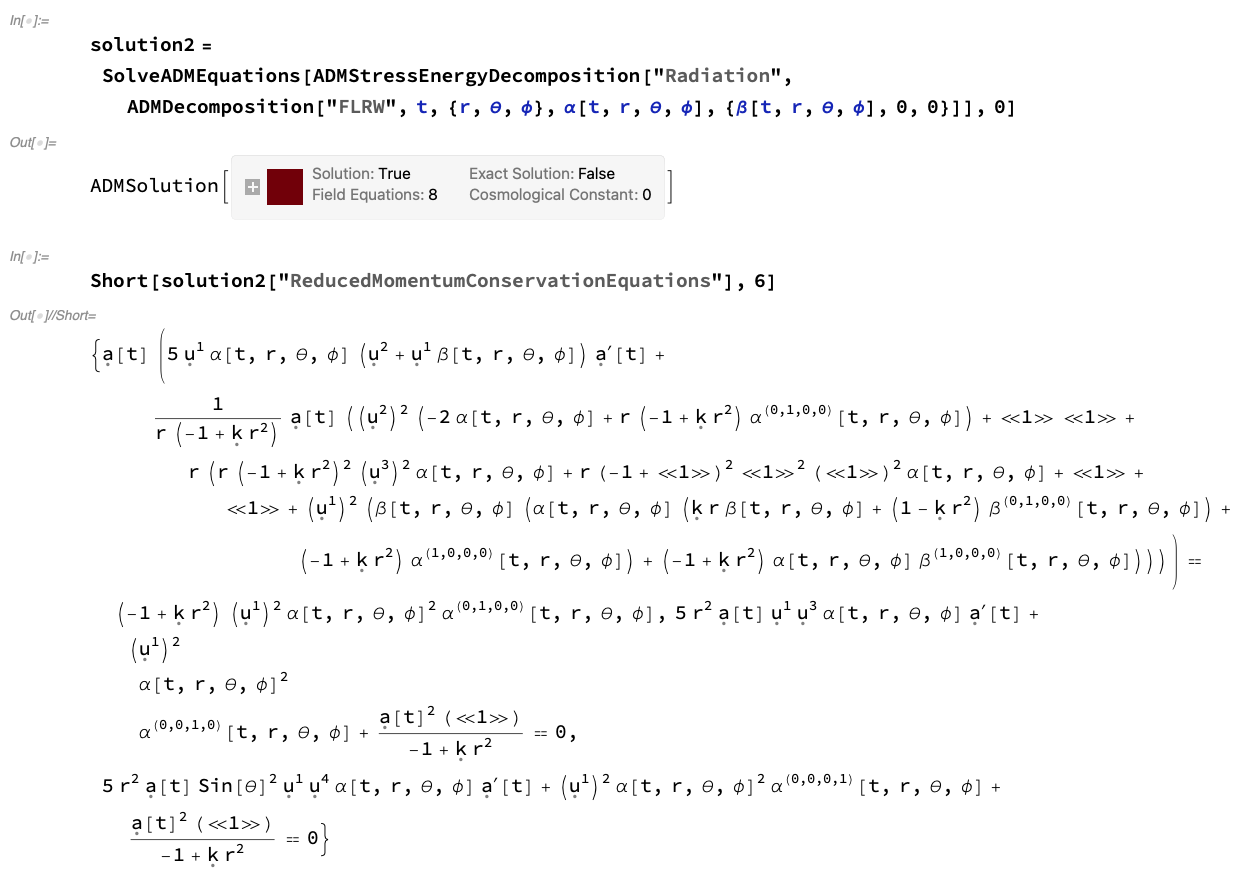}
\end{framed}
\caption{On the left, the list of momentum conservation equations obtained from the purely spacelike projections of the full continuity equations, computed using the \texttt{ADMSolution} object for an ADM decomposition of an FLRW geometry equipped with a perfect relativistic dust (representing a perfectly homogeneous and isotropic universe filled with an idealized distribution of dust) with lapse function ${\alpha \left( t, r, \theta, \phi \right)}$ and modified shift vector ${\left( \beta \left( t, r, \theta, \phi \right), 0, 0 \right)}$, with zero cosmological constant ${\Lambda = 0}$. On the right, the list of momentum conservation equations obtained from the purely spacelike projections of the full continuity equations, computed using the \texttt{ADMSolution} object for an ADM decomposition of an FLRW geometry equipped with a perfect relativistic radiation distribution (representing a prefectly homogeneous and isotropic universe filled with an idealized distribution of radiation) with lapse function ${\alpha \left( t, r, \theta, \phi \right)}$ and modified shift vector ${\left( \beta \left( t, r, \theta, \phi \right), 0, 0 \right)}$, with zero cosmological constant ${\Lambda = 0}$.}
\label{fig:Figure41}
\end{figure}

\begin{figure}[ht]
\centering
\begin{framed}
\includegraphics[width=0.495\textwidth]{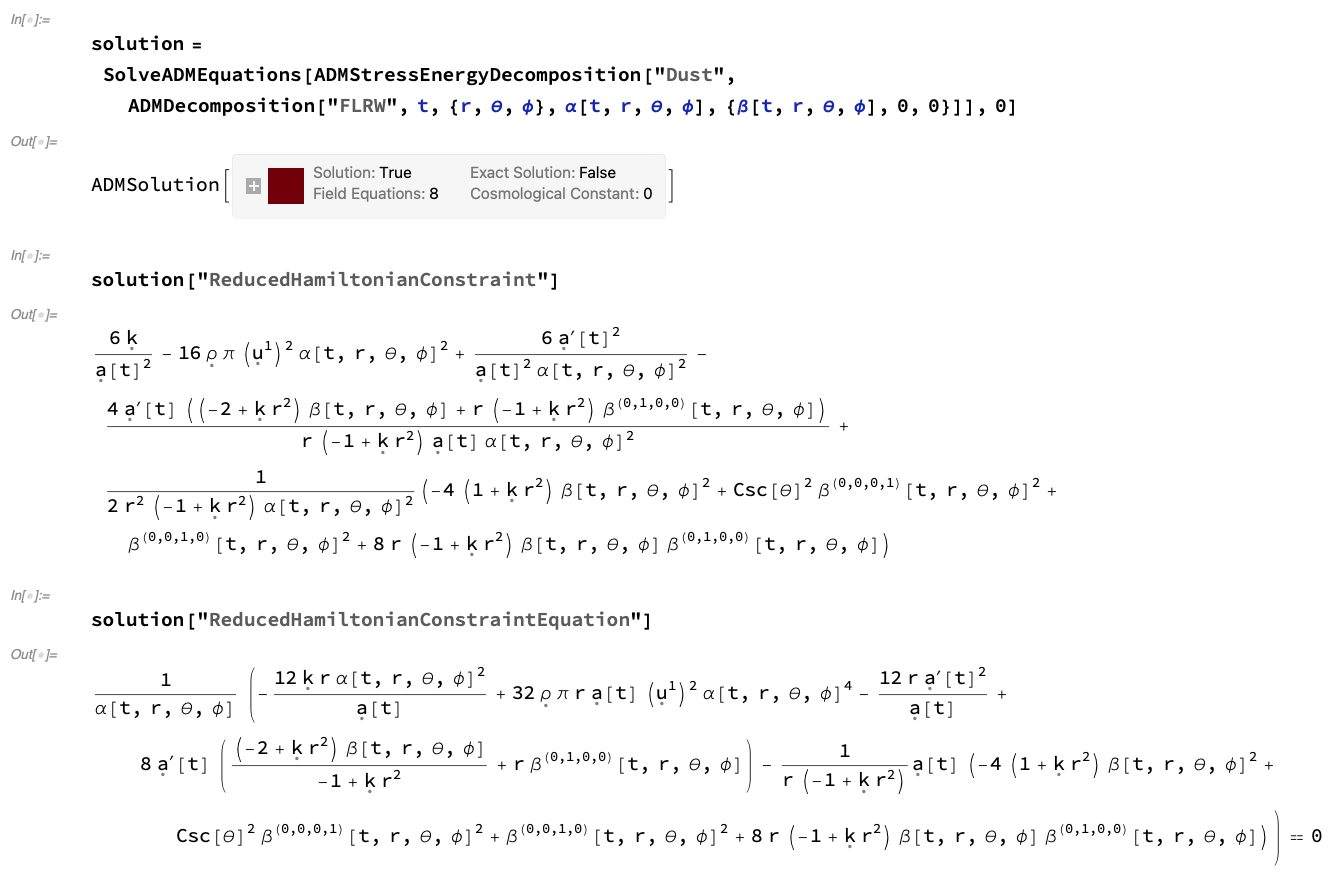}
\vrule
\includegraphics[width=0.495\textwidth]{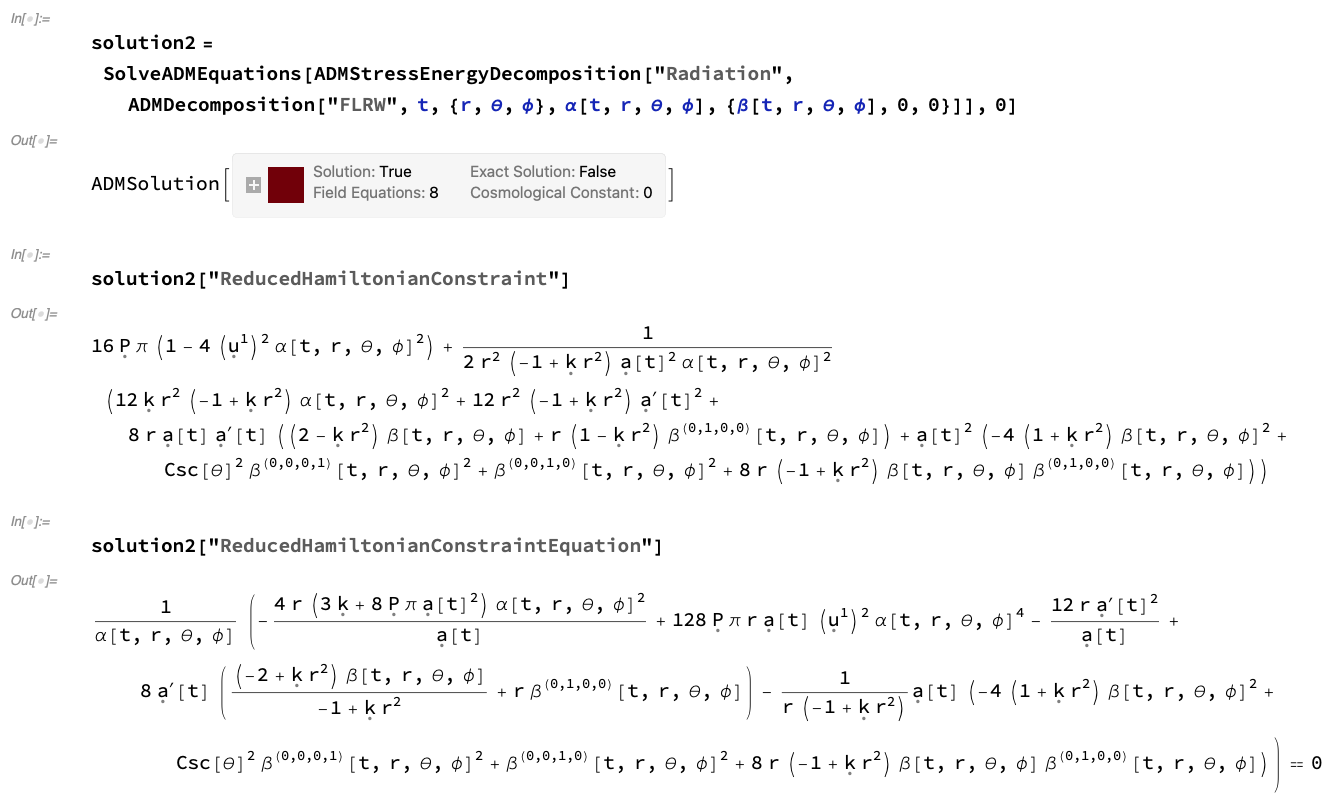}
\end{framed}
\caption{On the left, the ADM Hamiltonian constraint, computed using the \texttt{ADMSolution} object for an ADM decomposition of an FLRW geometry equipped with a perfect relativistic dust (representing a perfectly homogeneous and isotropic universe filled with an idealized distribution of dust) with lapse function ${\alpha \left( t, r, \theta, \phi \right)}$ and modified shift vector ${\left( \beta \left( t, r, \theta, \phi \right), 0, 0 \right)}$, with zero cosmological constant ${\Lambda = 0}$, illustrating that it does not vanish identically. On the right, the ADM Hamiltonian constraint, computed using the \texttt{ADMSolution} object for an ADM decomposition of an FLRW geometry equipped with a perfect relativistic radiation distribution (representing a perfectly homogeneous and isotropic universe filled with an idealized distribution of radiation) with lapse function ${\alpha \left( t, r, \theta, \phi \right)}$ and modified shift vector ${\left( \beta \left( t, r, \theta, \phi \right), 0, 0 \right)}$, with zero cosmological constant ${\Lambda = 0}$, illustrating that it does not vanish identically.}
\label{fig:Figure42}
\end{figure}

\begin{figure}[ht]
\centering
\begin{framed}
\includegraphics[width=0.495\textwidth]{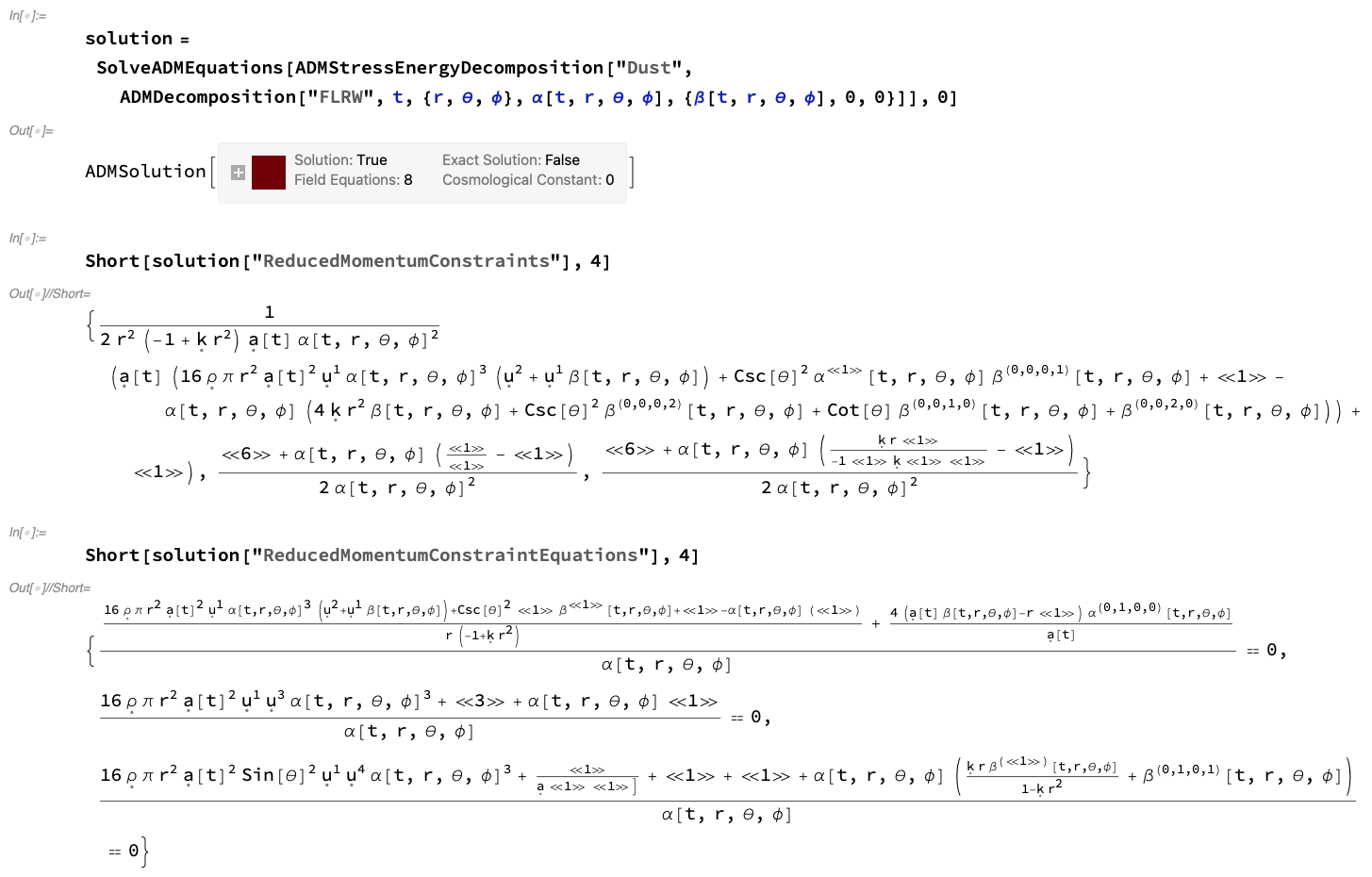}
\vrule
\includegraphics[width=0.495\textwidth]{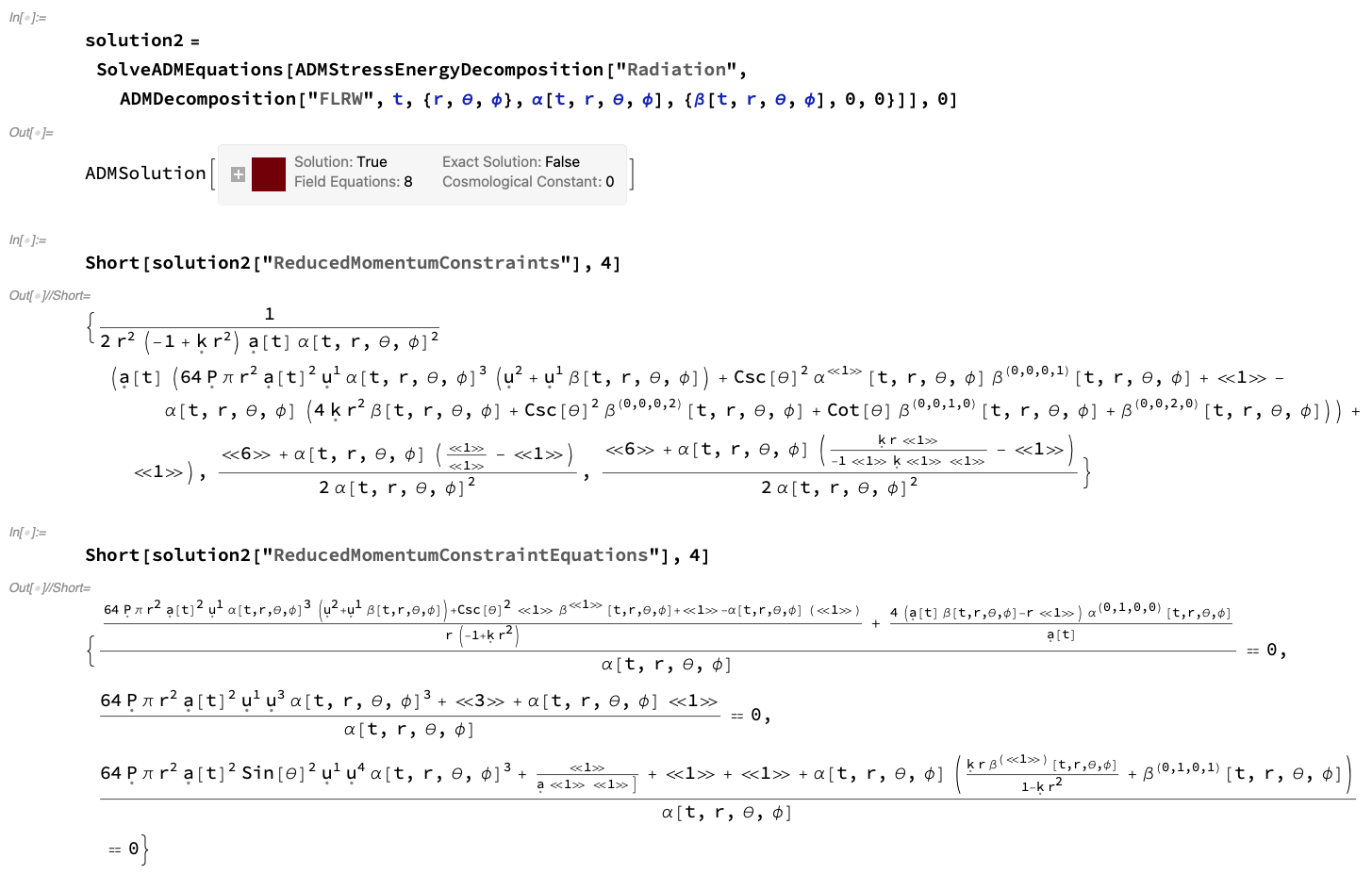}
\end{framed}
\caption{On the left, the list of ADM momentum constraints, computed using the \texttt{ADMSolution} object for an ADM decomposition of an FLRW geometry equipped with a perfect relativistic dust (representing a perfectly homogeneous and isotropic universe filled with an idealized distribution of dust) with lapse function ${\alpha \left( t, r, \theta, \phi \right)}$ and modified shift vector ${\left( \beta \left( t, r, \theta, \phi \right), 0, 0 \right)}$, with zero cosmological constant ${\Lambda = 0}$, illustrating that they do not vanish identically. On the right, the list of ADM momentum constraints, computed using the \texttt{ADMSolution} object for an ADM decomposition of an FLRW geometry equipped with a perfect relativistic radiation distribution (representing a perfectly homogeneous and isotropic universe filled with an idealized distribution of radiation) with lapse function ${\alpha \left( t, r, \theta, \phi \right)}$ and modified shift vector ${\left( \beta \left( t, r, \theta, \phi \right), 0, 0 \right)}$, with zero cosmological constant ${\Lambda = 0}$, illustrating that they do not vanish identically.}
\label{fig:Figure43}
\end{figure}

\begin{figure}[ht]
\centering
\begin{framed}
\includegraphics[width=0.495\textwidth]{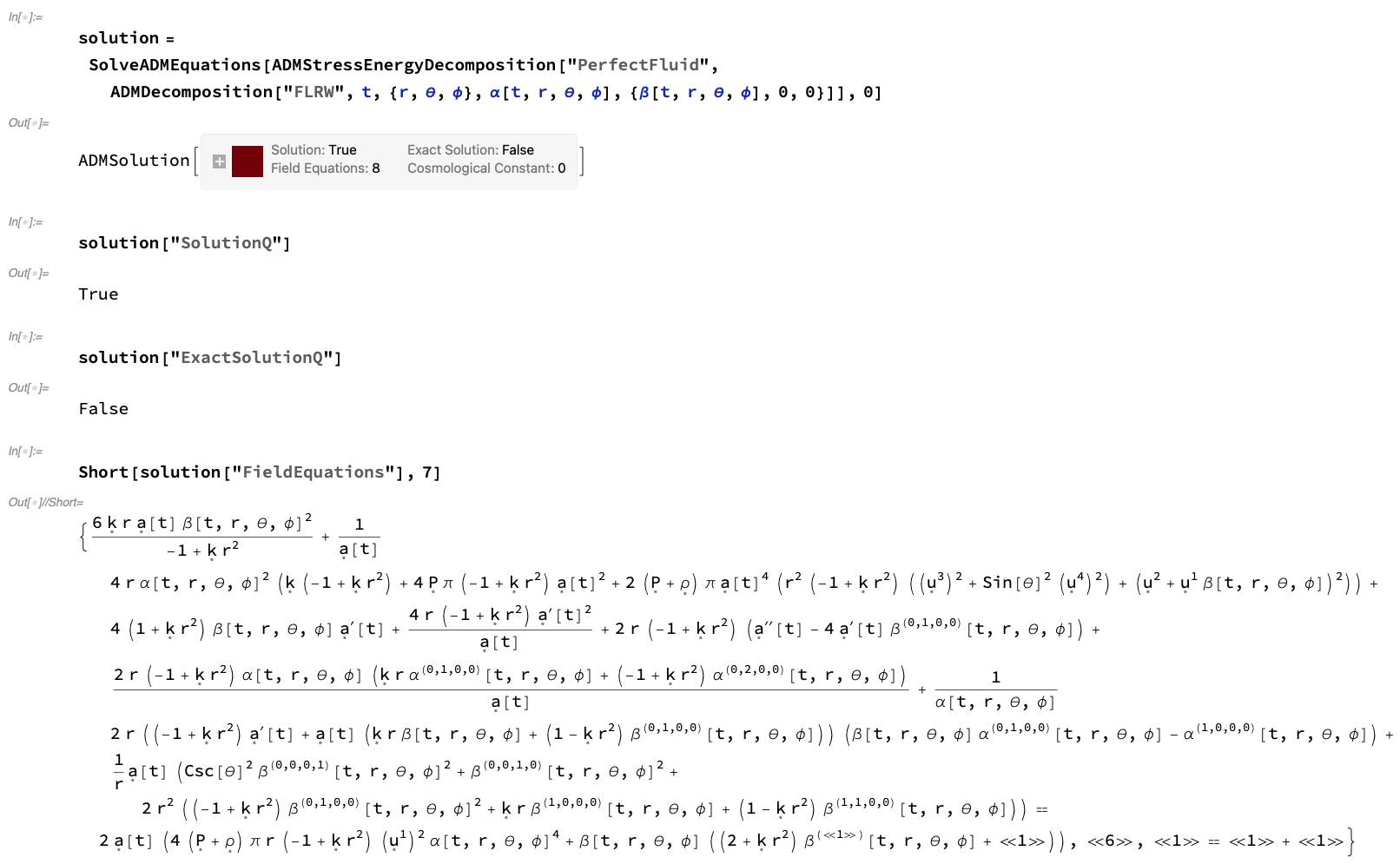}
\vrule
\includegraphics[width=0.495\textwidth]{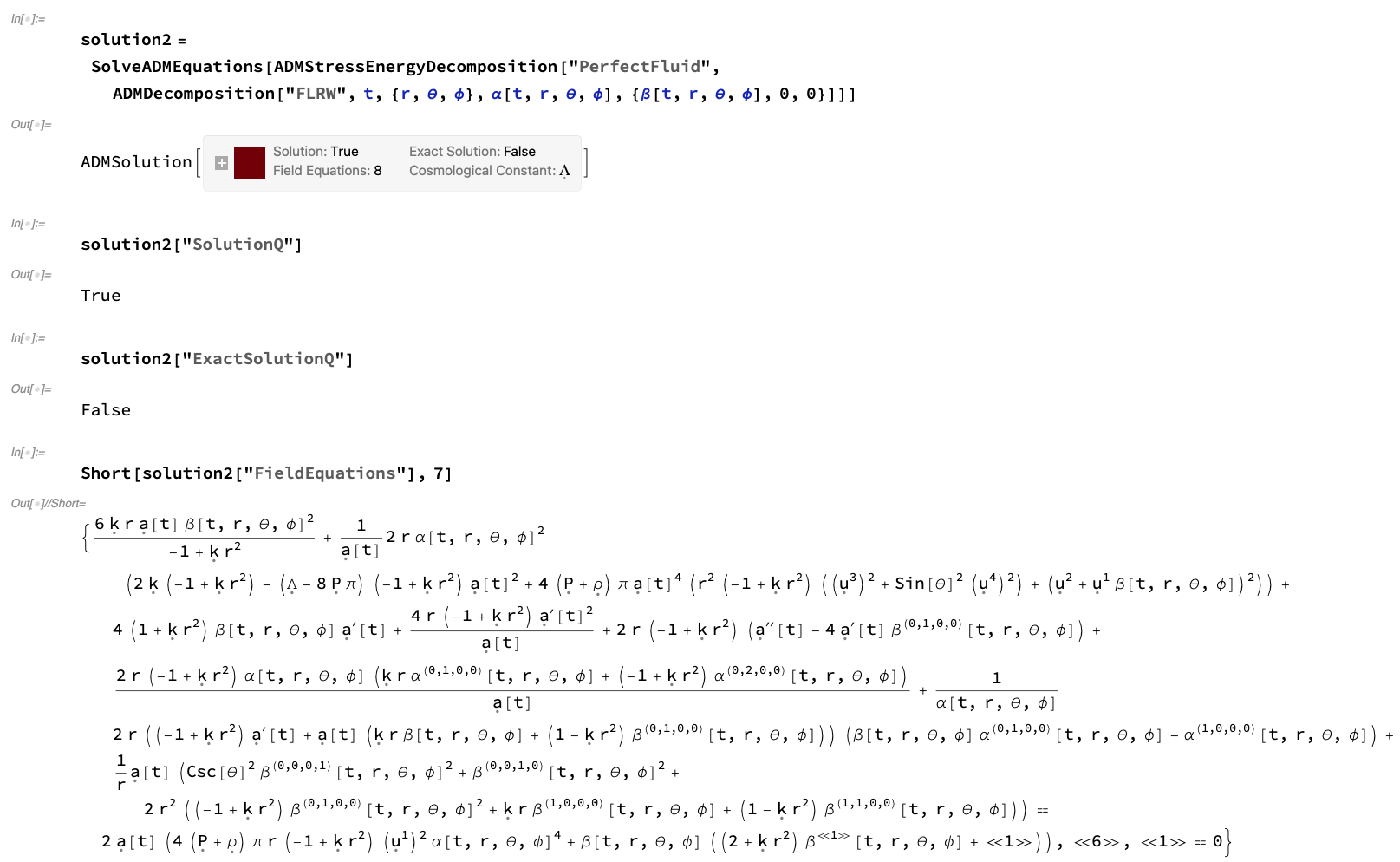}
\end{framed}
\caption{On the left, the \texttt{ADMSolution} object for an ADM decomposition of an FLRW geometry equipped with a perfect relativistic fluid (representing a perfectly homogeneous and isotropic universe filled with an idealized fluid) with lapse function ${\alpha \left( t, r, \theta, \phi \right)}$ and modified shift vector ${\left( \beta \left( t, r, \theta, \phi \right), 0, 0 \right)}$, with zero cosmological constant ${\Lambda = 0}$, computed using \texttt{SolveADMEquations}, illustrating that this decomposition is a non-exact solution to the ADM evolution equations, with eight additional field equations required. On the right, the \texttt{ADMSolution} object for an ADM decomposition of an FLRW geometry equipped with a perfect relativistic fluid (representing a perfectly homogeneous and isotropic universe filled with an idealized fluid) with lapse function ${\alpha \left( t, r, \theta, \phi \right)}$ and modified shift vector ${\left( \beta \left( t, r, \theta, \phi \right), 0, 0 \right)}$, with non-zero cosmological constant ${\Lambda \neq 0}$, computed using \texttt{SolveADMEquations}, illustrating that this decomposition is a non-exact solution to the ADM evolution equations, with eight additional field equations required.}
\label{fig:Figure46}
\end{figure}

We are now in a position to derive the \textit{full} system of ADM evolution equations by again taking the time evolution equations for the components of the extrinsic curvature tensor ${K_{\nu}^{\mu}}$, namely:

\begin{multline}
\frac{\partial}{\partial t} \left( K_{\nu}^{\mu} \right) = \alpha {}^{\left( 3 \right)} R_{\nu}^{\mu} - {}^{\left( 3 \right)} \nabla_{\rho} \left( {}^{\left( 3 \right)} \nabla_{\nu} \alpha \right) \gamma^{\rho \mu} + \alpha K K_{\nu}^{\mu} + \beta^{\rho} {}^{\left( 3 \right)} \nabla_{\rho} K_{\nu}^{\mu}\\
+ K_{\rho}^{\mu} {}^{\left( 3 \right)} \nabla_{\nu} \beta^{\rho} - K_{\nu}^{\rho} {}^{\left( 3 \right)} \nabla_{\rho} \beta^{\mu} - \alpha {}^{\left( 4 \right)} R_{\left( \rho + 1 \right) \left( \nu + 1 \right)} \gamma^{\rho \mu},
\end{multline}
or, equivalently:

\begin{multline}
\frac{\partial}{\partial t} \left( K_{\nu}^{\mu} \right) = \alpha {}^{\left( 3 \right)} R_{\nu}^{\mu} - \left( \frac{\partial}{\partial x^{\rho}} \left( \frac{\partial}{\partial x^{\nu}} \left( \alpha \right) \right) - {}^{\left( 3 \right)} \Gamma_{\rho \nu}^{\sigma} \left( \frac{\partial}{\partial x^{\sigma}} \left( \alpha \right) \right) \right) \gamma^{\rho \mu} + \alpha K K_{\nu}^{\mu}\\
+ \beta^{\rho} \left( \frac{\partial}{\partial x^{\rho}} \left( K_{\nu}^{\mu} \right) + {}^{\left( 3 \right)} \Gamma_{\rho \sigma}^{\mu} K_{\nu}^{\sigma} - {}^{\left( 3 \right)} \Gamma_{\rho \nu}^{\sigma} K_{\sigma}^{\mu} \right) + K_{\rho}^{\mu} \left( \frac{\partial}{\partial x^{\nu}} \left( \beta^{\rho} \right) + {}^{\left( 3 \right)} \Gamma_{\nu \sigma}^{\rho} \beta^{\sigma} \right)\\
- K_{\nu}^{\rho} \left( \frac{\partial}{\partial x^{\rho}} \left( \beta^{\mu} \right) + {}^{\left( 3 \right)} \Gamma_{\rho \sigma}^{\mu} \beta^{\sigma} \right) - \alpha {}^{\left( 4 \right)} R_{\left( \rho + 1 \right) \left( \nu + 1 \right)} \gamma^{\rho \mu},
\end{multline}
with ${\mu, \nu, \rho, \sigma}$ ranging across all ${\left\lbrace 0, \dots, n - 2 \right\rbrace}$ (i.e. across spatial coordinate indices only), and imposing the \textit{full} Einstein field equations (including stress-energy source terms):

\begin{equation}
{}^{\left( 4 \right)} G_{\mu \nu} + \Lambda g_{\mu \nu} = {}^{\left( 4 \right)} R_{\mu \nu} - \frac{1}{2} {}^{\left( 4 \right)} R g_{\mu \nu} + \Lambda g_{\mu \nu} = 8 \pi T_{\mu \nu},
\end{equation}
thus allowing us to replace the spacetime Ricci tensor terms ${{}^{\left( 4 \right)} R_{\mu \nu}}$ with terms involving the stress-energy tensor ${T_{\mu \nu}}$, the cosmological constant ${\Lambda}$ and the spacetime metric tensor ${g_{\mu \nu}}$, since one has:

\begin{equation}
{}^{\left( 4 \right)} R_{\mu \nu} = 8 \pi T_{\mu \nu} - 4 \pi T g_{\mu \nu} + \frac{2 \Lambda}{n - 2} g_{\mu \nu},
\end{equation}
where $T$ denotes the trace of the stress-energy tensor, i.e. ${T = g^{\mu \nu} T_{\mu \nu}}$. We thereby deduce the full ADM evolution equations:

\begin{multline}
\frac{\partial}{\partial t} \left( K_{\nu}^{\mu} \right) = \alpha {}^{\left( 3 \right)} R_{\nu}^{\mu} - {}^{\left( 3 \right)} \nabla_{\rho} \left( {}^{\left( 3 \right)} \nabla_{\nu} \alpha \right) \gamma^{\rho \mu} + \alpha K K_{\nu}^{\mu} + \beta^{\rho} {}^{\left( 3 \right)} \nabla_{\rho} K_{\nu}^{\mu}\\
+ K_{\rho}^{\mu} {}^{\left( 3 \right)} \nabla_{\nu} \beta^{\rho} - K_{\nu}^{\rho} {}^{\left( 3 \right)} \nabla_{\rho} \beta^{\mu} - \alpha \left( 8 \pi T_{\rho \nu} \gamma^{\rho \mu} - 4 \pi T \delta_{\nu}^{\mu} \right) - \alpha \left( \frac{2 \Lambda}{n - 2} \gamma_{\rho \nu} \right) \gamma^{\rho \mu},
\end{multline}
i.e., in expanded form:

\begin{multline}
\frac{\partial}{\partial t} \left( K_{\nu}^{\mu} \right) = \alpha {}^{\left( 3 \right)} R_{\nu}^{\mu} - \left( \frac{\partial}{\partial x^{\rho}} \left( \frac{\partial}{\partial x^{\nu}} \left( \alpha \right) \right) - {}^{\left( 3 \right)} \Gamma_{\rho \nu}^{\sigma} \left( \frac{\partial}{\partial x^{\sigma}} \left( \alpha \right) \right) \right) \gamma^{\rho \mu} + \alpha K _{\nu}^{\mu}\\
+ \beta^{\rho} \left( \frac{\partial}{\partial x^{\rho}} \left( K_{\nu}^{\mu} \right) + {}^{\left( 3 \right)} \Gamma_{\rho \sigma}^{\mu} K_{\nu}^{\sigma} - {}^{\left( 3 \right)} \Gamma_{\rho \nu}^{\sigma} K_{\sigma}^{\mu} \right) + K_{\rho}^{\mu} \left( \frac{\partial}{\partial x^{\nu}} \left( \beta^{\rho} \right) + {}^{\left( 3 \right)} \Gamma_{\nu \sigma}^{\rho} \beta^{\sigma} \right)\\
- K_{\nu}^{\rho} \left( \frac{\partial}{\partial x^{\rho}} \left( \beta^{\mu} \right) + {}^{\left( 3 \right)} \Gamma_{\rho \sigma}^{\mu} \beta^{\sigma} \right) - \alpha \left( 8 \pi T_{\left( \rho + 1 \right) \left( \nu + 1 \right)} \gamma^{\rho \mu} - 4 \pi T \delta_{\nu}^{\mu} \right) - \alpha \left( \frac{2 \Lambda}{n - 2} \gamma_{\rho \nu} \right) \gamma^{\rho \mu},
\end{multline}
with ${\mu, \nu, \rho, \sigma}$ again ranging across all ${\left\lbrace 0, \dots, n - 2 \right\rbrace}$ (i.e. across spatial coordinate indices only). Representations of the corresponding \texttt{ADMSolution} objects for the ADM decompositions of an FLRW metric (representing, for instance, a perfectly homogeneous and isotropic universe with scale factor ${a \left( t \right)}$ and curvature $k$ in spherical polar coordinates ${\left( t, r, \theta, \phi \right)}$) equipped with a perfect relativistic dust (with density ${\rho}$ and spacetime velocity ${u^{\mu}}$) and a perfect relativistic radiation distribution (with pressure $P$ and spacetime velocity ${u^{\mu}}$), assuming in both cases the same restricted choice of gauge with lapse function ${\alpha \left( t, r, \theta, \phi \right)}$ and modified shift vector ${\left( \beta \left( t, r, \theta, \phi \right), 0, 0 \right)}$, are shown in Figure \ref{fig:Figure38}; these examples demonstrate that these ADM decompositions are both valid \textit{non-exact} solutions of the full ADM evolution equations for these particular energy-matter distributions, in the sense that eight additional field equations need to be assumed in each case. The complete lists of ADM evolution equations for both ADM decomposition/stress-energy combinations can be computed directly from the \texttt{ADMSolution} object, and it can be verified in both cases that they do indeed reduce down to the eight canonical field equations previously mentioned, as illustrated in Figure \ref{fig:Figure39}. Figures \ref{fig:Figure40} and \ref{fig:Figure41} show the energy and momentum conservation equations, obtained from the purely timelike and purely spacelike projections of the full continuity equations, respectively, computed directly from the \texttt{ADMSolution} objects for a perfect relativistic dust and a perfect relativistic radiation distribution embedded within an ADM decomposition of an FLRW metric with this same restricted choice of gauge. Similarly, applying the full Einstein field equations to replace the spacetime Einstein tensor terms ${{}^{\left( 4 \right)} G_{\mu \nu}}$ appearing within the ADM Hamiltonian and momentum constraints:

\begin{equation}
\mathcal{H} = {}^{\left( 3 \right)} R + K^2 - K_{\nu}^{\mu} K_{\mu}^{\nu} - 2 \alpha^2 {}^{\left( 4 \right)} G^{0 0},
\end{equation}
and:

\begin{multline}
\mathcal{M}_{\mu} = {}^{\left( 3 \right)} \nabla_{\nu} K_{\mu}^{\nu} - {}^{\left( 3 \right)} \nabla_{\mu} K - \alpha {}^{\left( 4 \right)} G_{\left( \mu + 1 \right)}^{0}\\
= \frac{\partial}{\partial x^{\nu}} \left( K_{\mu}^{\nu} \right) + {}^{\left( 3 \right)} \Gamma_{\nu \sigma}^{\nu} K_{\mu}^{\sigma} - {}^{\left( 3 \right)} \Gamma_{\nu \mu}^{\sigma} K_{\sigma}^{\nu} - \frac{\partial}{\partial x^{\mu}} \left( K \right) - \alpha {}^{\left( 4 \right)} G_{\left( \mu + 1 \right)}^{0},
\end{multline}
respectively, yields the full forms of the ADM Hamiltonian and momentum constraints:

\begin{equation}
\mathcal{H} = {}^{\left( 3 \right)} R + K^2 - K_{\nu}^{\mu} K_{\mu}^{\nu} - 16 \pi \alpha^2 T^{0 0} - 2 \Lambda,
\end{equation}
and:

\begin{equation}
\mathcal{M}_{\mu} = {}^{\left( 3 \right)} \nabla_{\nu} K_{\mu}^{\nu} - {}^{\left( 3 \right)} \nabla_{\mu} K - 8 \pi T_{\left( \mu + 1 \right)}^{0} = \frac{\partial}{\partial x^{\nu}} \left( K_{\mu}^{\nu} \right) + {}^{\left( 3 \right)} \Gamma_{\nu \sigma}^{\nu} K_{\mu}^{\sigma} - {}^{\left( 3 \right)} \Gamma_{\nu \mu}^{\sigma} K_{\sigma}^{\nu} - \frac{\partial}{\partial x^{\mu}} \left( K \right) - 8 \pi T_{\left( \mu + 1 \right)}^{0},
\end{equation}
respectively, with ${\mu, \nu, \sigma}$ ranging as usual across all ${\left\lbrace 0, \dots, n - 2 \right\rbrace}$, as computed and shown in Figures \ref{fig:Figure42} and \ref{fig:Figure43}. Although these examples have all assumed a vanishing value of the cosmological constant ${\Lambda = 0}$, the \texttt{SolveADMEquations} and \texttt{ADMSolution} functionality in \textsc{Gravitas} works equally well in the presence of a non-vanishing cosmological constant ${\Lambda \neq 0}$, as shown in Figure \ref{fig:Figure46} for the case of a perfect relativistic fluid embedded within an ADM decomposition of an FLRW metric with the usual restricted gauge choice.

\clearpage

\section{Adaptive Hypergraph Refinement and Numerical Examples}
\label{sec:Section5}

\begin{figure}[ht]
\centering
\begin{framed}
\includegraphics[width=0.495\textwidth]{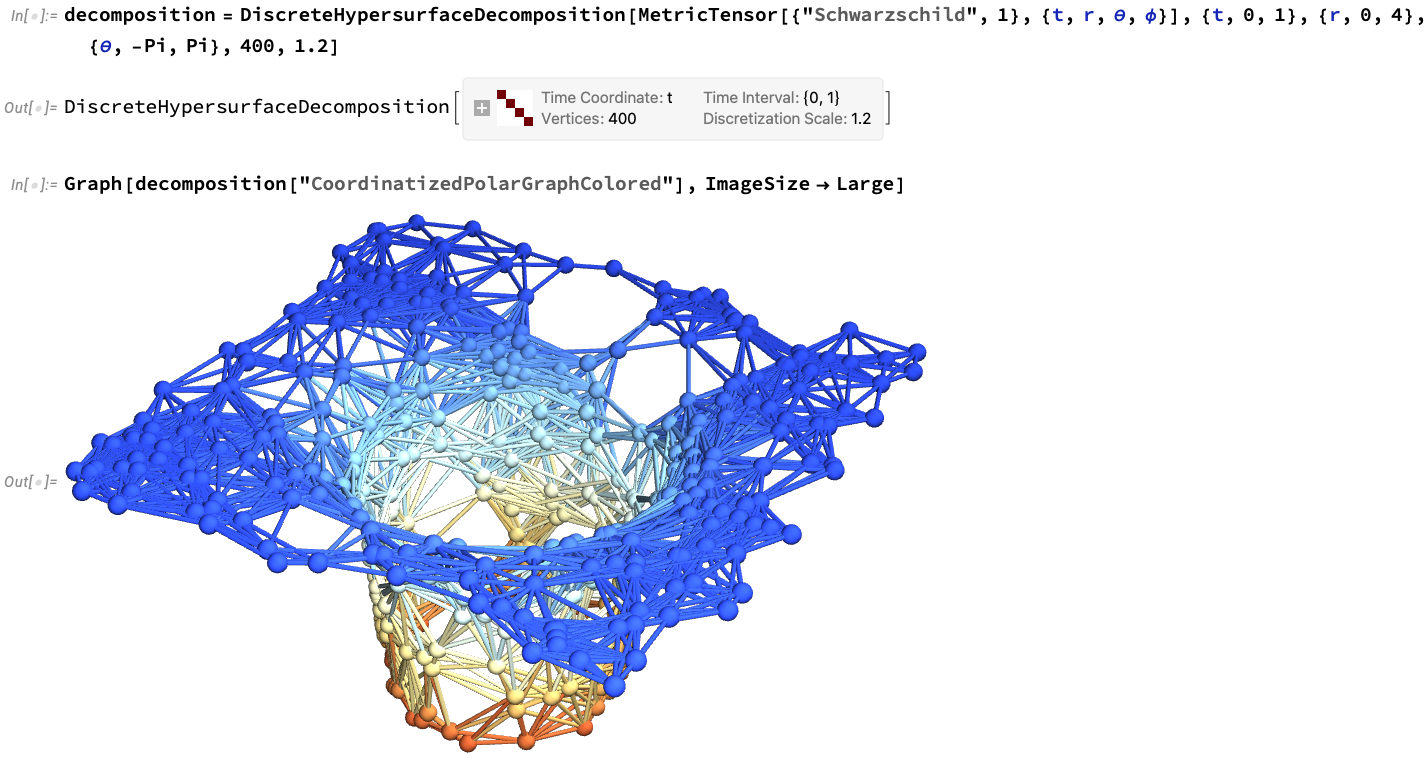}
\vrule
\includegraphics[width=0.495\textwidth]{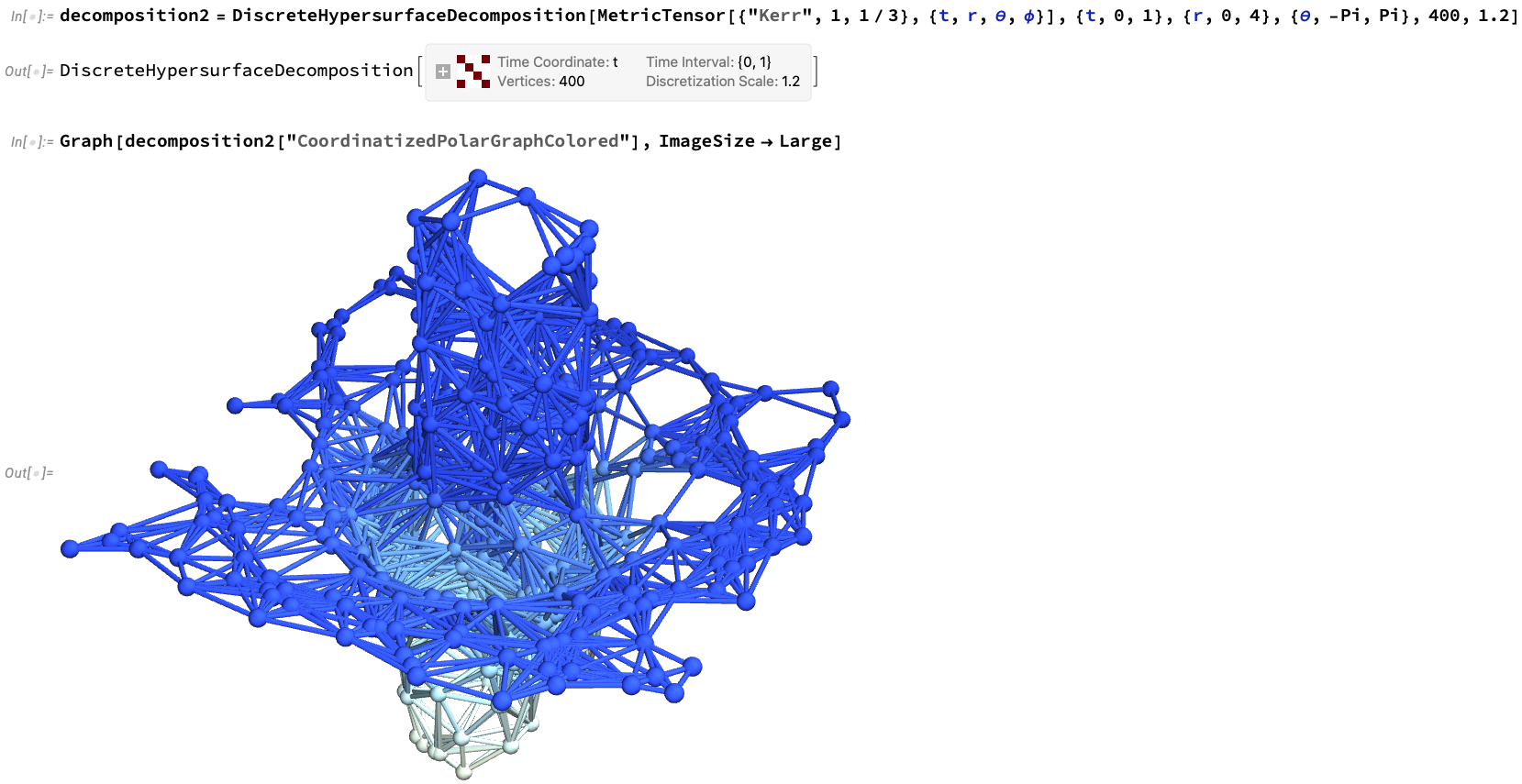}
\end{framed}
\caption{On the left, a \texttt{DiscreteHypersurfaceDecomposition} object for a Schwarzschild geometry (representing, for instance, an uncharged, non-rotating black hole with numerical mass 1 in Schwarzschild or spherical polar coordinates ${\left( t, r, \theta, \phi \right)}$ with ${t \in \left[ 0, 1 \right]}$, ${r \in \left[ 0, 4 \right]}$ and ${\theta \in \left[ - \pi, \pi \right]}$), with a resolution of 400 vertices and a discretization scale of 1.2, represented by its final spatial hypergraph at coordinate time ${t = 1}$ (colored based on extrinsic curvature) with spatial coordinate information assigned to the vertices. On the right, a \texttt{DiscreteHypersurfaceDecomposition} object for a Kerr geometry (representing, for instance, an uncharged, spinning black hole with numerical mass 1 and numerical angular momentum ${\frac{1}{3}}$ in Boyer-Lindquist or oblate spheroidal coordinates ${\left( t, r, \theta, \phi \right)}$ with ${t \in \left[ 0, 1 \right]}$, ${r \in \left[ 0, 4 \right]}$ and ${\theta \in \left[ - \pi, \pi \right]}$), with a resolution of 400 vertices and a discretization scale of 1.2, represented by its final spatial hypergraph at coordinate time ${t = 1}$ (colored based on extrinsic curvature) with spatial coordinate information assigned to the vertices.}
\label{fig:Figure47}
\end{figure}

\begin{figure}[ht]
\centering
\begin{framed}
\includegraphics[width=0.495\textwidth]{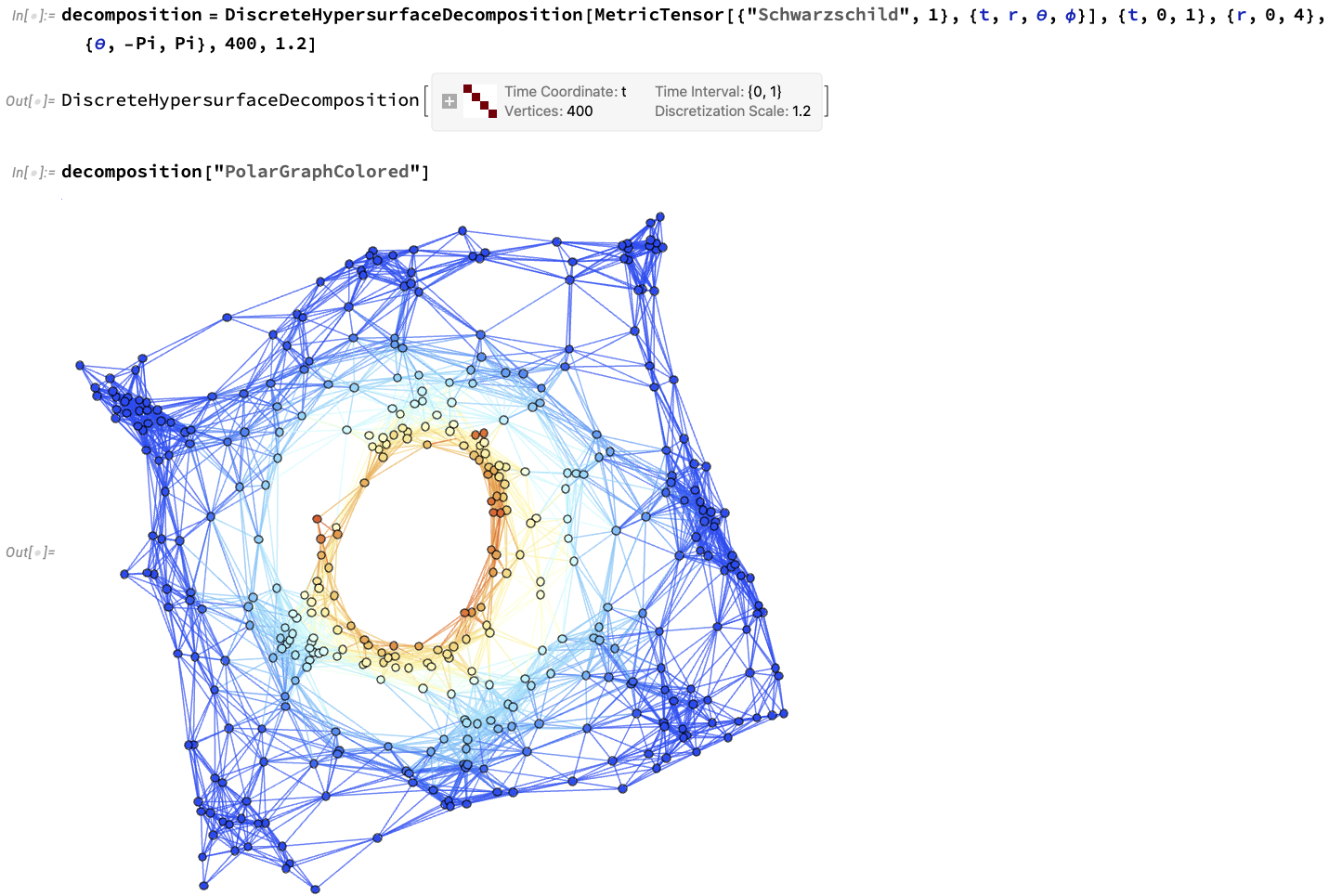}
\vrule
\includegraphics[width=0.495\textwidth]{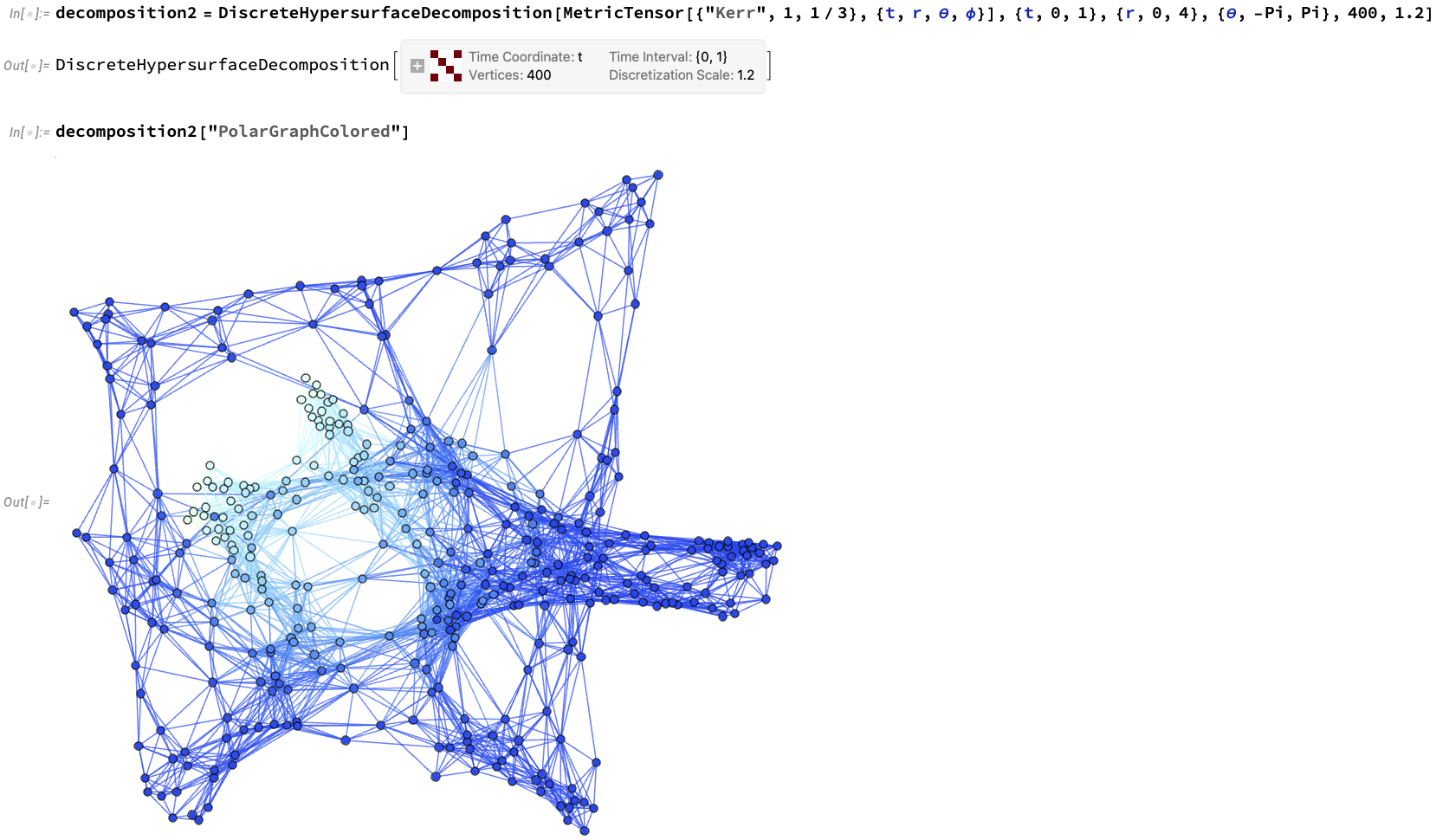}
\end{framed}
\caption{On the left, a \texttt{DiscreteHypersurfaceDecomposition} object for a Schwarzschild geometry (representing, for instance, an uncharged, non-rotating black hole with numerical mass 1 in Schwarzschild or spherical polar coordinates ${\left( t, r, \theta, \phi \right)}$ with ${t \in \left[ 0, 1 \right]}$, ${r \in \left[ 0, 4 \right]}$ and ${\theta \in \left[ - \pi, \pi \right]}$), with a resolution of 400 vertices and a discretization scale of 1.2, represented by its final spatial hypergraph at coordinate time ${t = 1}$ (colored based on extrinsic curvature) without any vertex coordinate information assigned. On the right, a \texttt{DiscreteHypersurfaceDecomposition} object for a Kerr geometry (representing, for instance, an uncharged, spinning black hole with numerical mass 1 and numerical angular momentum ${\frac{1}{3}}$ in Boyer-Lindquist or oblate spheroidal coordinates ${\left( t, r, \theta, \phi \right)}$ with ${t \in \left[ 0, 1 \right]}$, ${r \in \left[ 0, 4 \right]}$ and ${\theta \in \left[ - \pi, \pi \right]}$), with a resolution of 400 vertices and a discretization scale of 1.2, represented by its final spatial hypergraph at coordinate time ${t = 1}$ (colored based on extrinsic curvature) without any vertex coordinate information assigned.}
\label{fig:Figure48}
\end{figure}

\begin{figure}[ht]
\centering
\begin{framed}
\includegraphics[width=0.495\textwidth]{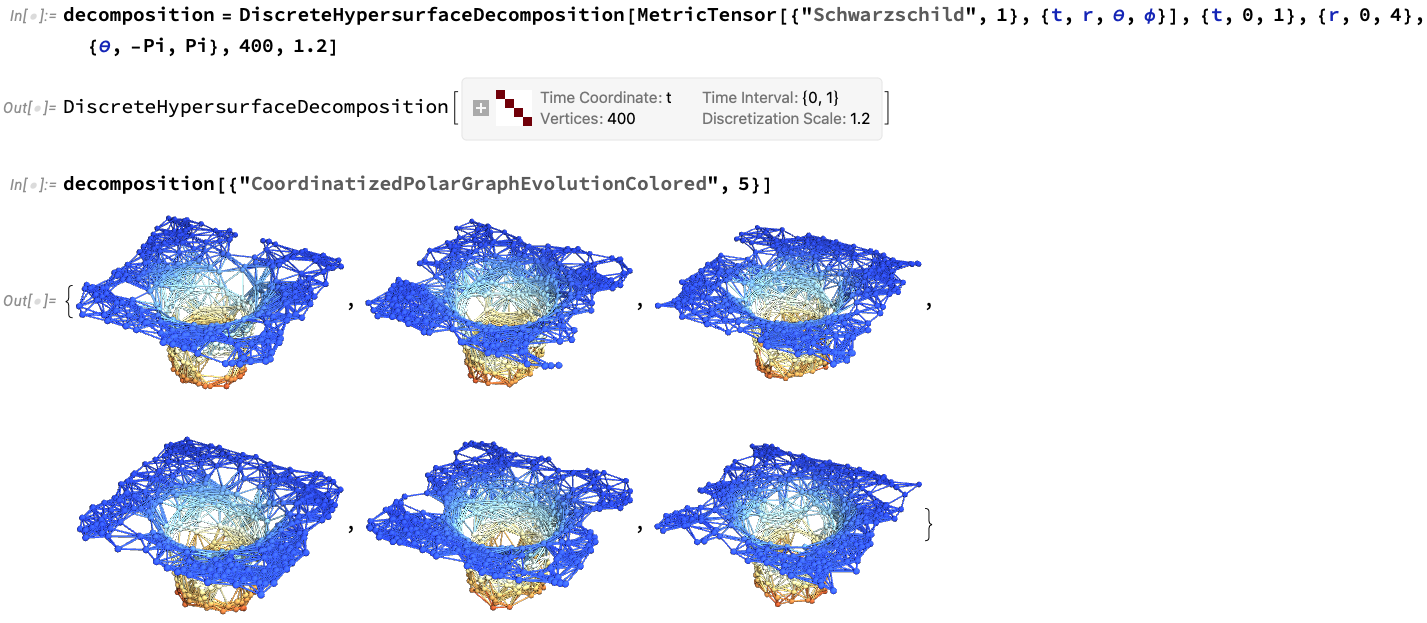}
\vrule
\includegraphics[width=0.495\textwidth]{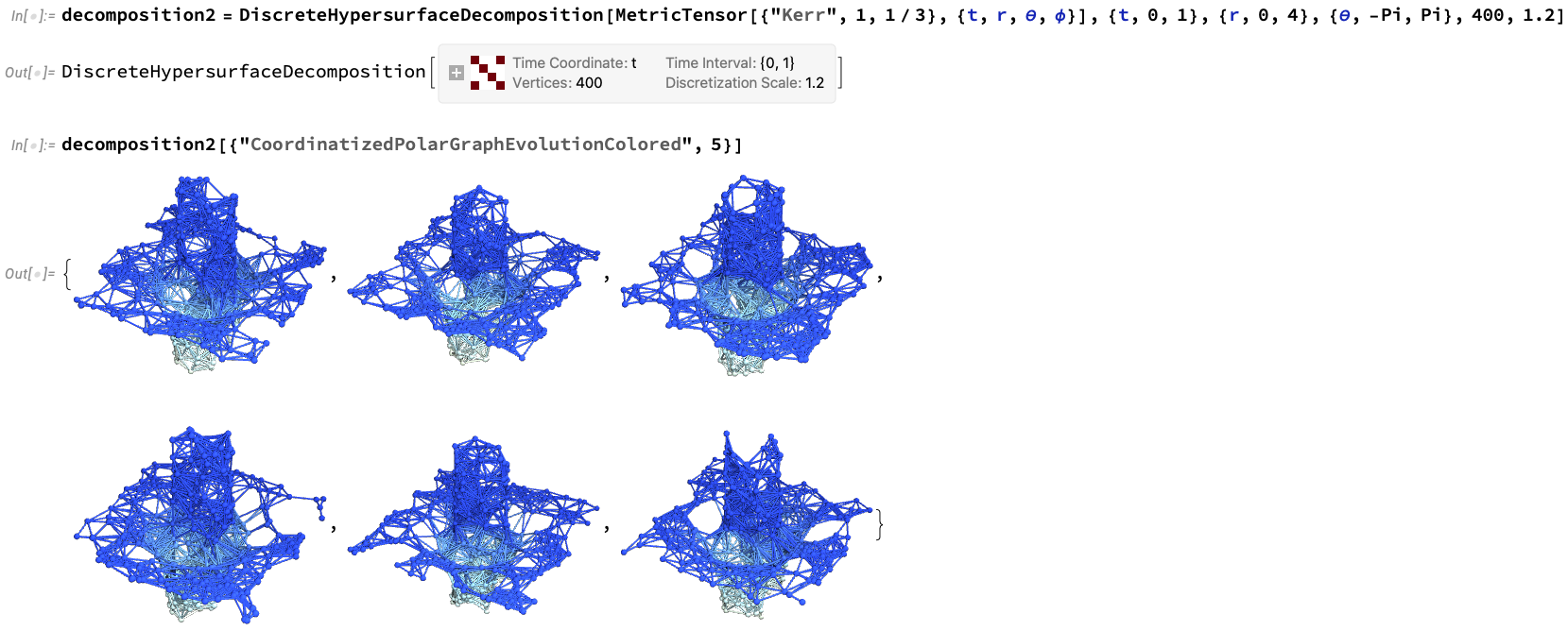}
\end{framed}
\caption{On the left, a \texttt{DiscreteHypersurfaceDecomposition} object for a Schwarzschild geometry (representing, for instance, an uncharged, non-rotating black hole with numerical mass 1 in Schwarzschild or spherical polar coordinates ${\left( t, r, \theta, \phi \right)}$ with ${t \in \left[ 0, 1 \right]}$, ${r \in \left[ 0, 4 \right]}$ and ${\theta \in \left[ - \pi, \pi \right]}$), with a resolution of 400 vertices and a discretization scale of 1.2, represented by the evolution of its spatial hypergraphs for five time steps from coordinate time ${t = 0}$ to coordinate time ${t = 1}$ (colored based on extrinsic curvature) with spatial coordinate information assigned to the vertices. On the right, a \texttt{DiscreteHypersurfaceDecomposition} object for a Kerr geometry (representing, for instance, an uncharged, spinning black hole with numerical mass 1 and numerical angular momentum ${\frac{1}{3}}$ in Boyer-Lindquist or oblate spheroidal coordinates ${\left( t, r, \theta, \phi \right)}$ with ${t \in \left[ 0, 1 \right]}$, ${r \in \left[ 0, 4 \right]}$ and ${\theta \in \left[ - \pi, \pi \right]}$), with a resolution of 400 vertices and a discretization scale of 1.2, represented by the evolution of its spatial hypergraphs for five time steps from coordinate time ${t = 0}$ to coordinate time ${t = 1}$ (colored based on extrinsic curvature) with spatial coordinate information assigned to the vertices.}
\label{fig:Figure49}
\end{figure}

\begin{figure}[ht]
\centering
\begin{framed}
\includegraphics[width=0.495\textwidth]{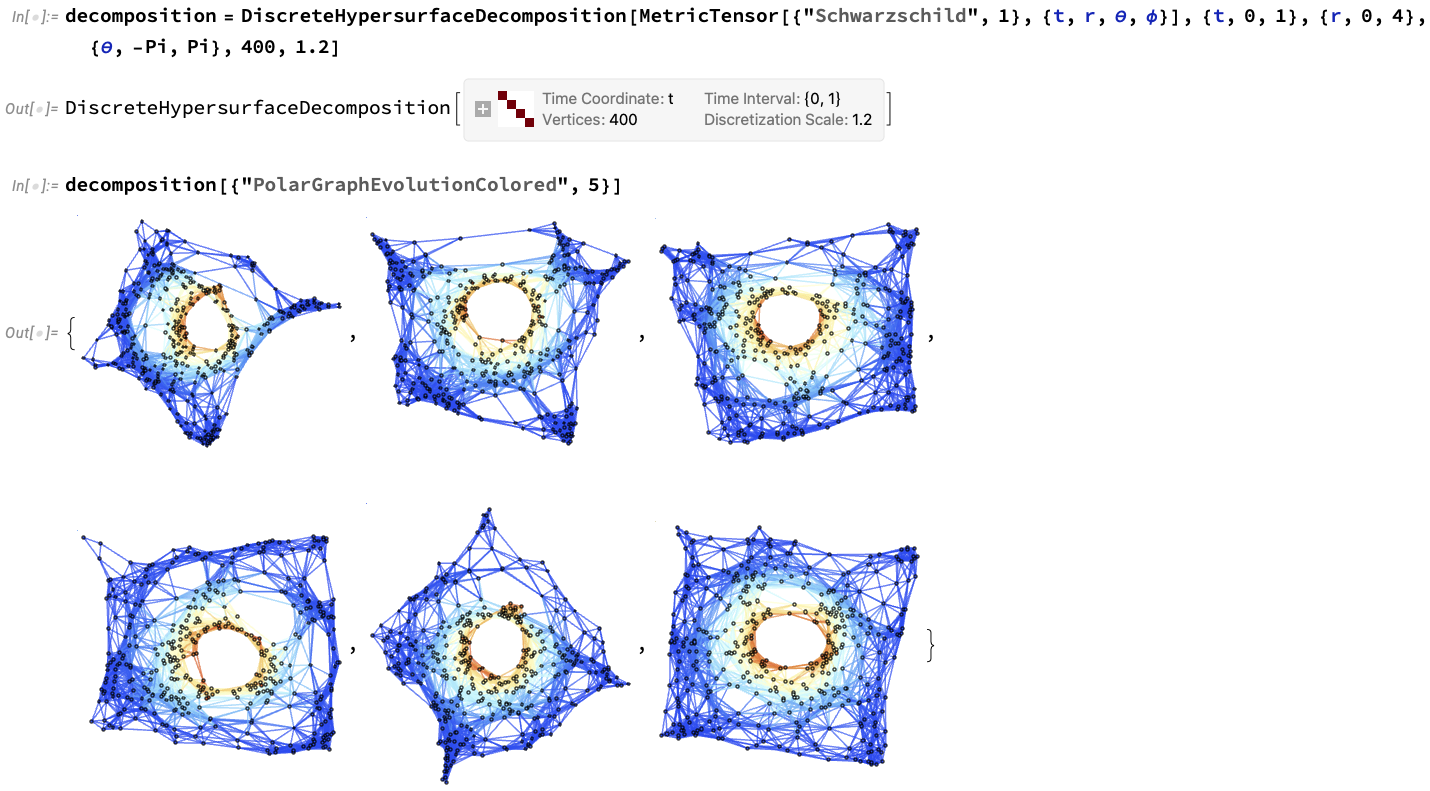}
\vrule
\includegraphics[width=0.495\textwidth]{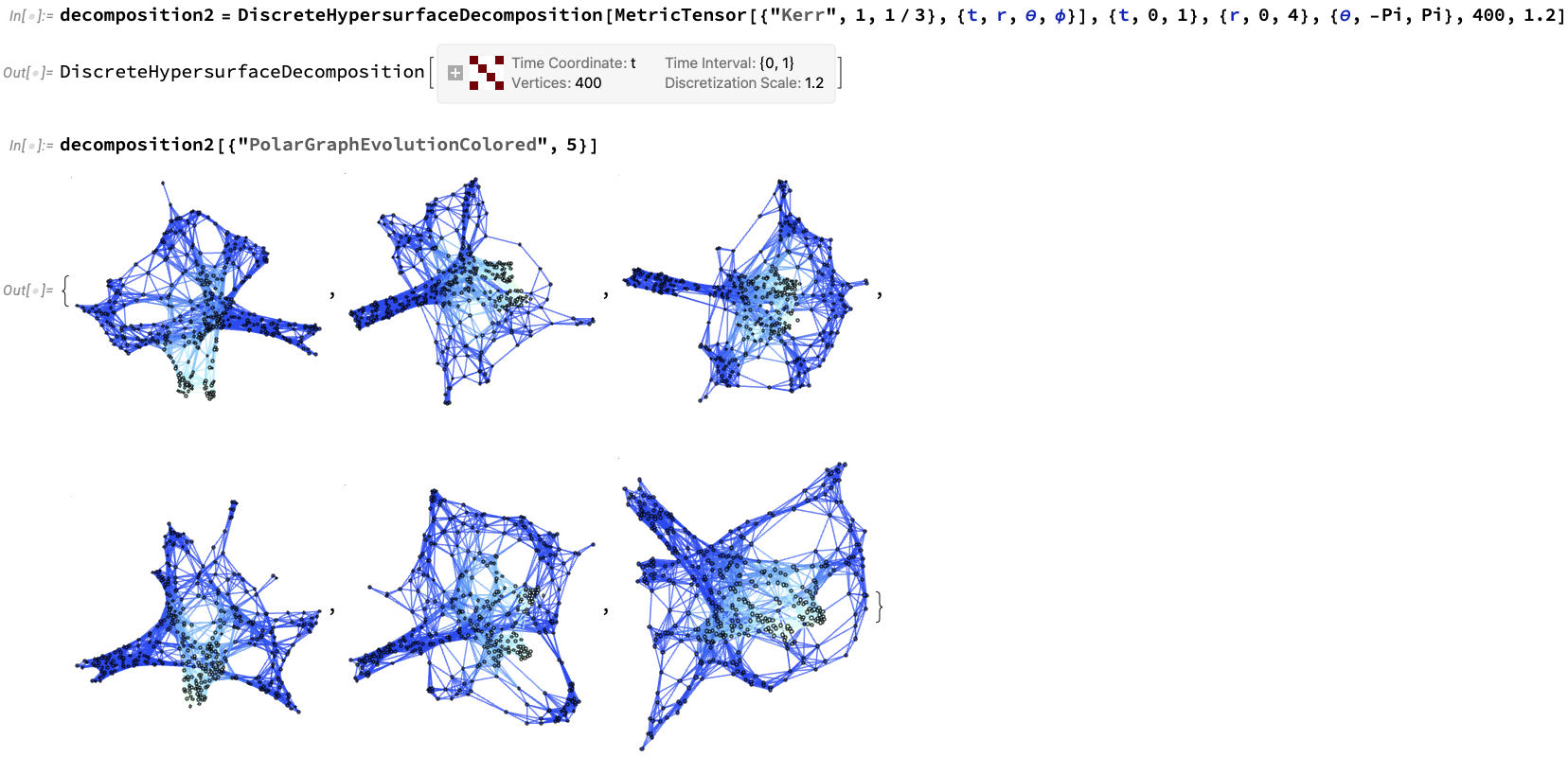}
\end{framed}
\caption{On the left, a \texttt{DiscreteHypersurfaceDecomposition} object for a Schwarzschild geometry (representing, for instance, an uncharged, non-rotating black hole with numerical mass 1 in Schwarzschild or spherical polar coordinates ${\left( t, r, \theta, \phi \right)}$ with ${t \in \left[ 0, 1 \right]}$, ${r \in \left[ 0, 4 \right]}$ and ${\theta \in \left[ - \pi, \pi \right]}$), with a resolution of 400 vertices and a discretization scale of 1.2, represented by the evolution of its spatial hypergraphs for five time steps from coordinate time ${t = 0}$ to coordinate time ${t = 1}$ (colored based on extrinsic curvature) without any vertex coordinate information assigned. On the right, a \texttt{DiscreteHypersurfaceDecomposition} object for a Kerr geometry (representing, for instance, an uncharged, spinning black hole with numerical mass 1 and numerical angular momentum ${\frac{1}{3}}$ in Boyer-Lindquist or oblate spheroidal coordinates ${\left( t, r, \theta, \phi \right)}$ with ${t \in \left[ 0, 1 \right]}$, ${r \in \left[ 0, 4 \right]}$ and ${\theta \in \left[ - \pi, \pi \right]}$), with a resolution of 400 vertices and a discretization scale of 1.2, represented by the evolution of its spatial hypergraphs for five time steps from coordinate time ${t = 0}$ to coordinate time ${t = 1}$ (colored based on extrinsic curvature) without any vertex coordinate information assigned.}
\label{fig:Figure50}
\end{figure}

Thus far, we have neglected to mention any of the internal algorithmic details of how \textsc{Gravitas} actually obtains its non-exact solutions to the vacuum and non-vacuum ADM evolution equations, and of how the ADM Hamiltonian and momentum constraints are numerically, if not analytically, enforced in such cases. This is intentional, since the \textsc{Gravitas} framework has been purposefully designed so as to abstract the complexities of numerical simulation away from the user to the greatest extent possible, thereby making it feasible to configure, run and analyze sophisticated relativistic simulations without specialized numerical relativity knowledge, using only the high-level functionality demonstrated within the previous sections. Internally, \textsc{Gravitas} uses a standard fourth-order explicit Runge-Kutta method\cite{kreiss}\cite{levy} to integrate the relevant evolution equations (and the hyperbolic forms of the gauge/constraint equations) in time, making use of the fourth-order centered finite-difference stencils proposed by Zlochower, Baker, Campanelli and Lousto\cite{zlochower} in the default case, or fourth-order upwind finite-difference stencils in the presence of non-zero advection terms (i.e. terms involving spatial derivatives multiplied by the shift vector ${\boldsymbol\beta}$). In order to eliminate the possibility of numerical instabilities propagating due to the appearance of spurious high-frequency modes as a consequence of our choice of finite-difference stencils, we also modify the right-hand sides of the time evolution equations to incorporate a Kreiss-Oliger dissipation term\cite{kreiss2}. However, in contrast to other similar numerical relativity codes, the unique feature of \textsc{Gravitas} in this respect lies in its employment of generalized forms of these numerical algorithms, defined over hypergraphs with potentially highly non-trivial (and, in particular, highly irregular and non-grid-like) topologies without any a priori coordinate system, which in turn automatically coarsen and refine via hypergraph rewriting in order to adapt to the topology and dynamics of the problem being solved, using a hypergraph generalization of the local adaptive mesh refinement algorithm developed by Gropp, Berger, Oliger and Colella\cite{gropp}\cite{berger}\cite{berger2}. This generalization involves introducing a binary function which determines whether a given vertex or subhypergraph should be tagged for refinement, then partitioning the overall hypergraph into subhypergraphs, integrating this binary tagging function over each subhypergraph (thus yielding a \textit{signature} for the subhypergraph), and then determining the local coordinate direction of refinement using the discrete Laplacians of these signatures; the refinement criterion, as in standard adaptive mesh refinement codes, is typically based upon the values of certain projections of either spatial or spacetime curvature tensors, or indeed of the total stress-energy tensor. This algorithm requires appropriate modifications to be made to the finite-difference stencils, in order to accommodate the presence of boundaries between coarser and finer subhypergraphs. The boundary-extrapolated values of the state vector along each hyperedge (necessary for the computation of the inter-cell flux functions in the time evolution step) are computed using a third-order weighted essentially non-oscillatory (WENO) reconstruction step\cite{jiang}\cite{jiang2}, with three linearly-independent nodal basis polynomials, each of degree two (taken, by default, to be the Lagrange interpolating polynomials), so as to ensure that the overall accuracy of the numerical integration remains fourth-order. The full algorithmic details are contained in \cite{gorard9} (with an abridged summary also present in \cite{gorard11}). The \texttt{DiscreteHypersurfaceDecomposition} object in \textsc{Gravitas} allows one to visualize explicitly the decomposition of an arbitrary spacetime into a sequence of discrete spacelike hypersurfaces, whose topologies are represented using hypergraphs. Figures \ref{fig:Figure47} and \ref{fig:Figure48} show the final spatial hypergraph configuration at coordinate time ${t = 1}$ of such a decomposition, for both the Schwarzschild metric (representing, for instance, an uncharged, non-rotating black hole with mass $M$ in Schwarzschild or spherical polar coordinates ${\left( t, r, \theta, \phi \right)}$):

\begin{equation}
d s^2 = g_{\mu \nu} d x^{\mu} d x^{\nu} = - \left( 1 - \frac{2 M}{r} \right) d t^2 + \left( 1 - \frac{2 M}{r} \right)^{-1} d r^2 + r^2 \left( d \theta^2 + \sin^2 \left( \theta \right) d \phi^2 \right),
\end{equation}
and the Kerr metric (representing, for instance, an uncharged, spinning black hole with mass $M$ and angular momentum $J$ in Boyer-Lindquist or oblate spheroidal coordinates ${\left( t, r, \theta, \phi \right)}$):

\begin{multline}
d s^2 = g_{\mu \nu} d x^{\mu} d x^{\nu} = - \left( 1 - \frac{2 M}{\left(  r^2 + \left( \frac{J}{M} \right)^2 \cos^2 \left( \theta \right) \right)} \right) d t^2 + \left( \frac{r^2 + \left( \frac{J}{M} \right)^2 \cos^2 \left( \theta \right)}{r^2 - 2 M + \left( \frac{J}{M} \right)^2} \right) d r^2\\
+ \left( r^2 + \left( \frac{J}{M} \right)^2 \cos^2 \left( \theta \right) \right) d \theta^2 + \left( r^2 + \left( \frac{J}{M} \right)^2 + \frac{2 J^2 \sin^2 \left( \theta \right)}{M \left( r^2 + \left( \frac{J}{M} \right)^2 \cos^2 \left( \theta \right) \right)} \right) \sin^2 \left( \theta \right) d \phi^2\\
- \left( \frac{4 J \sin^2 \left( \theta \right)}{r^2 + \left( \frac{J}{M} \right)^2 \cos^2 \left( \theta \right)} \right) d t d \phi,
\end{multline}
both with and without spatial coordinates assigned to the vertices, respectively. Figures \ref{fig:Figure49} and \ref{fig:Figure50} instead illustrate the complete evolution of spatial hypergraphs from coordinate time ${t = 0}$ to coordinate time ${t = 1}$ via a sequence of five discrete time steps, again for discrete decompositions of both the Schwarzschild and Kerr metrics, both with and without spatial coordinate information included in the visualization. In each of these examples, the numerical mass of the black hole is 1, the numerical angular momentum (in the case of Kerr black holes) is ${\frac{1}{3}}$, the evolution takes place within coordinate time ${t \in \left[ 0, 1 \right]}$ and the spatial domain is bounded by ${r \in \left[ 0, 4 \right]}$ and ${\theta \in \left[ - \pi, \pi \right]}$; all simulations are run with a resolution of 400 vertices and a discretization scale (which determines the hypergraph density, as described in \cite{gorard4} and \cite{gorard9}) of 1.2, with vertices and edges colored based on the extrinsic curvature of the spacelike hypersurface at the corresponding point.

\begin{figure}[ht]
\centering
\begin{framed}
\includegraphics[width=0.895\textwidth]{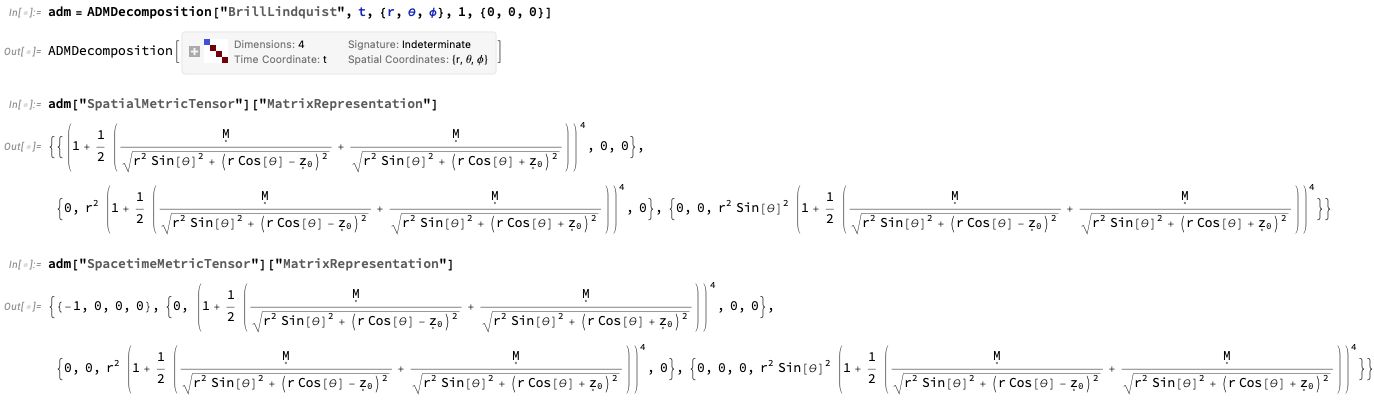}
\end{framed}
\caption{The \texttt{ADMDecomposition} object for a Brill-Lindquist geometry (representing, for instance, a pair of uncharged, non-rotating black holes, initially at rest, each of mass $M$ in Schwarzschild or spherical polar coordinates ${\left( t, r, \theta, \phi \right)}$) with trivial lapse function $1$ and trivial shift vector ${\left( 0, 0, 0 \right)}$, represented by its induced (spatial) and ambient (spacetime) \texttt{MetricTensor} objects in explicit covariant matrix form.}
\label{fig:Figure51}
\end{figure}

\begin{figure}[ht]
\centering
\begin{framed}
\includegraphics[width=0.895\textwidth]{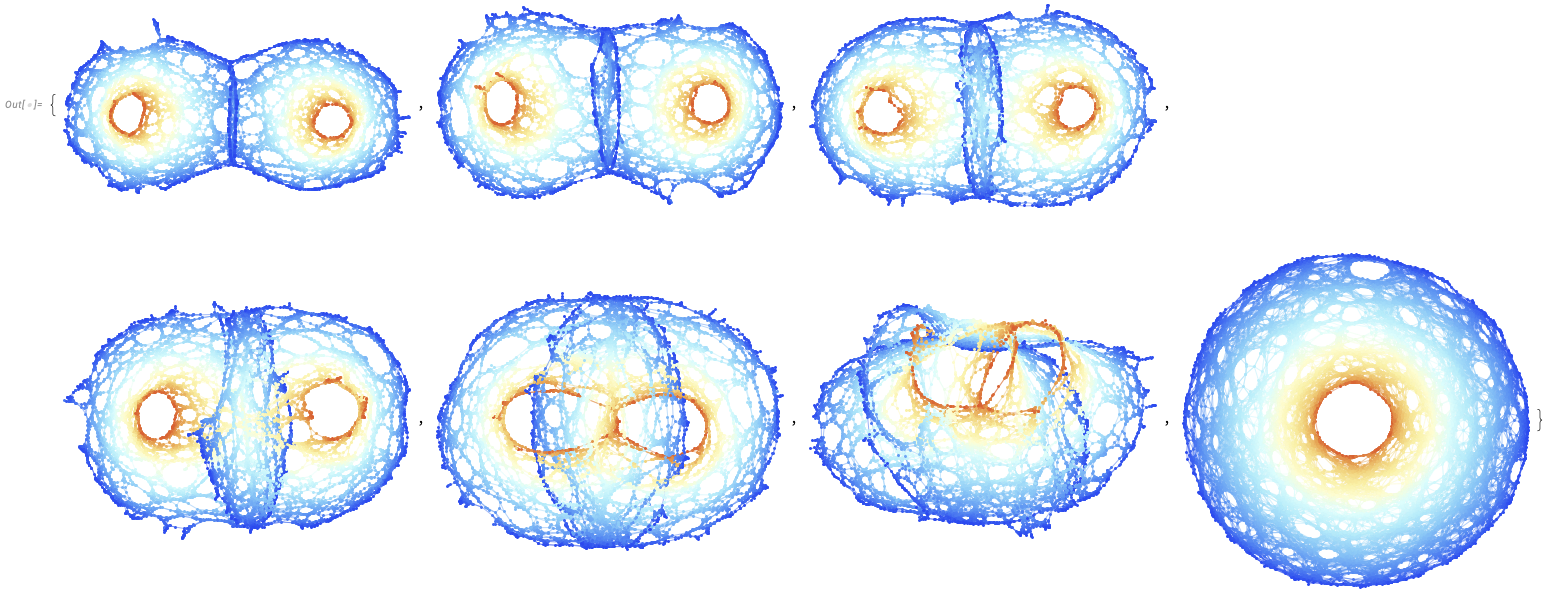}
\end{framed}
\caption{The \texttt{DiscreteHypersurfaceDecomposition} object for a Brill-Lindquist geometry (representing, for instance, a pair of uncharged, non-rotating black holes, initially at rest, each of mass $M$ in Schwarzschild or spherical polar coordinates ${\left( t, r, \theta, \phi \right)}$ with the radial boundary $R$ of the domain set to be comparable to the Schwarzschild radii of the black holes, i.e. ${R \sim 2 M}$) with the ${1 + log}$ slicing condition and the minimal distortion coordinate conditions for the gauge, represented by the evolution of its spatial hypergraphs for six time steps from coordinate time ${t = 0 M}$ to coordinate time ${t = 24 M}$ (colored based on extrinsic curvature) without any vertex coordinate information assigned.}
\label{fig:Figure52}
\end{figure}

\begin{figure}[ht]
\centering
\begin{framed}
\includegraphics[width=0.895\textwidth]{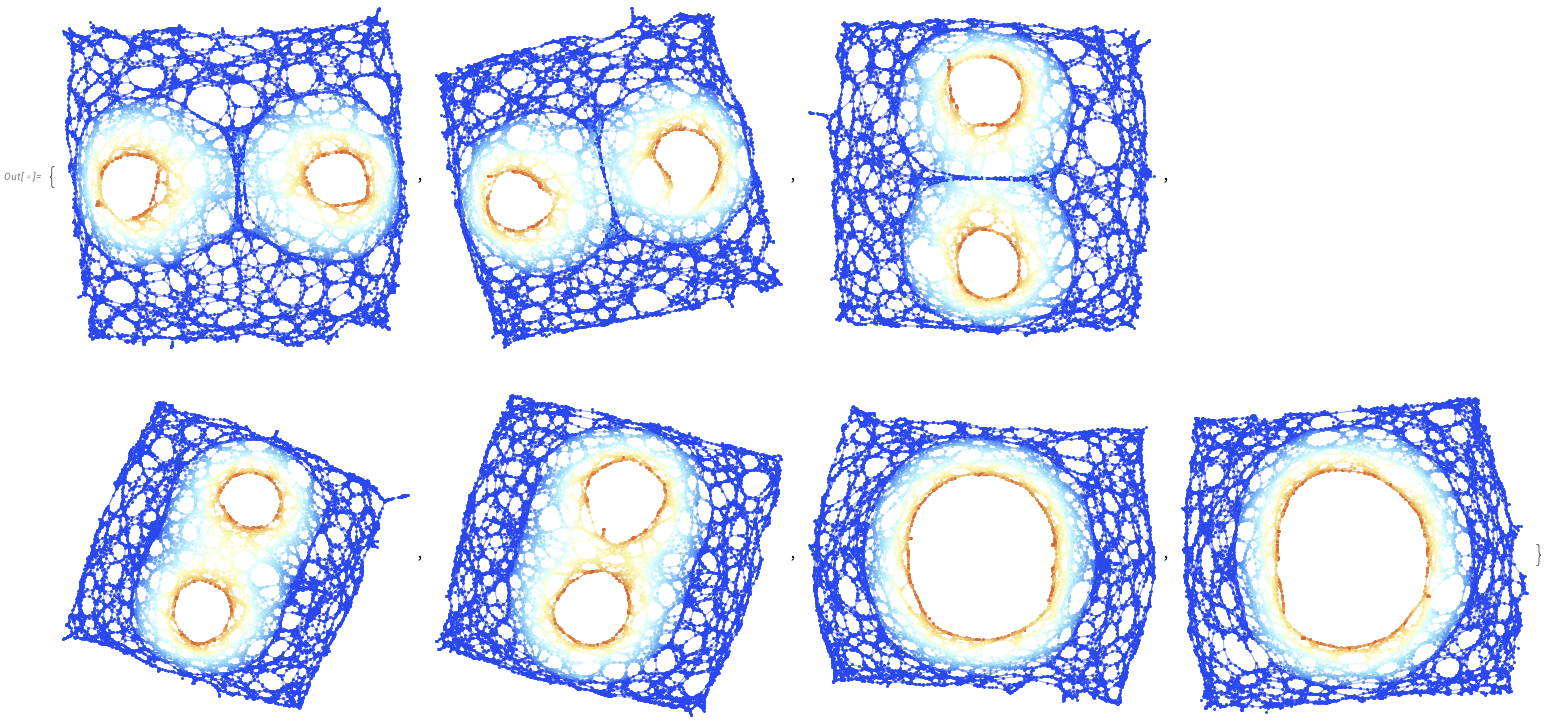}
\end{framed}
\caption{The \texttt{DiscreteHypersurfaceDecomposition} object for a Brill-Lindquist geometry (representing, for instance, a pair of uncharged, non-rotating black holes, initially at rest, each of mass $M$ in Schwarzschild or spherical polar coordinates ${\left( t, r, \theta, \phi \right)}$ with the radial boundary $R$ of the domain set to be much larger than the Schwarzschild radii of the black holes, i.e. ${R \gg 2 M}$) with the ${1 + log}$ slicing condition and the minimal distortion coordinate conditions for the gauge, represented by the evolution of its spatial hypergraphs for six time steps from coordinate time ${t = 0 M}$ to coordinate time ${t = 24 M}$ (colored based on extrinsic curvature) without any vertex coordinate information assigned.}
\label{fig:Figure53}
\end{figure}

Finally, we conclude by showcasing an example of a fairly typical but also reasonably challenging numerical relativity test case, namely the head-on collision and merger of a binary black hole system within a vacuum spacetime. We begin by setting up Brill-Lindquist initial data\cite{misner}\cite{brill} (representing, for instance, a pair of uncharged, non-rotating black holes, initially at rest, each of mass $M$ and separated by a distance of ${2 z_0}$ in Schwarzschild or spherical polar coordinates ${\left( t, r, \theta, \phi \right)}$), with the spatial line element on the initial spacelike hypersurface given by:

\begin{equation}
d l^2 = \gamma_{\mu \nu} d x^{\mu} d x^{\nu} = \left( 1 + \frac{1}{2} \left( \frac{M}{\sqrt{r^2 \sin^2 \left( \theta \right) + \left( r \cos \left( \theta \right) - z_0 \right)^2}} + \frac{M}{\sqrt{r^2 \sin^2 \left( \theta \right) + \left( r \cos \left( \theta \right) + z_0 \right)^2}} \right) \right)^4 \delta,
\end{equation}
where ${\delta}$ represents the form of the default spatial metric in spherical symmetry:

\begin{equation}
\delta = d r^2 + r^2 \left( d \theta^2 + \sin^2 \left( \theta \right) d \phi^2 \right).
\end{equation}
Being a purely spatial metric tensor, ${\mu, \nu}$ here range across all ${\left\lbrace 0, \dots, n - 2 \right\rbrace}$ (i.e. across spatial coordinate indices only). Note that the Brill-Lindquist initial geometry is built-in to \textsc{Gravitas}'s standard library of Cauchy initial data; the corresponding \texttt{ADMDecomposition} object, including representations of both the induced/spatial and ambient/spacetime \texttt{MetricTensor} objects, for the Brill-Lindquist initial metric is shown in Figure \ref{fig:Figure51}, assuming a trivial choice of gauge with lapse function identically equal to $1$ (i.e. ${\alpha \left( t, r, \theta, \phi \right) = 1}$) and shift vector (field) identically equal to zero (i.e. ${\boldsymbol\beta \left( t, r, \theta, \phi \right) = \left( 0, 0, 0 \right)}$). However, solving the vacuum ADM evolution equations:

\begin{multline}
\frac{\partial}{\partial t} \left( K_{\nu}^{\mu} \right) = \alpha {}^{\left( 3 \right)} R_{\nu}^{\mu} - {}^{\left( 3 \right)} \nabla_{\rho} \left( {}^{\left( 3 \right)} \nabla_{\nu} \alpha \right) \gamma^{\rho \mu} + \alpha K K_{\nu}^{\mu} + \beta^{\rho} {}^{\left( 3 \right)} \nabla_{\rho} K_{\nu}^{\mu}\\
+ K_{\rho}^{\mu} {}^{\left( 3 \right)} \nabla_{\nu} \beta^{\rho} - K_{\nu}^{\rho} {}^{\left( 3 \right)} \nabla_{\rho} \beta^{\mu} - \alpha \left( \frac{2 \Lambda}{n - 2} \gamma_{\rho \nu} \right) \gamma^{\rho \mu},
\end{multline}
or, equivalently:

\begin{multline}
\frac{\partial}{\partial t} \left( K_{\nu}^{\mu} \right) = \alpha {}^{\left( 3 \right)} R_{\nu}^{\mu} - \left( \frac{\partial}{\partial x^{\rho}} \left( \frac{\partial}{\partial x^{\nu}} \left( \alpha \right) \right) - {}^{\left( 3 \right)} \Gamma_{\rho \nu}^{\sigma} \left( \frac{\partial}{\partial x^{\sigma}} \left( \alpha \right) \right) \right) \gamma^{\rho \mu} + \alpha K K_{\nu}^{\mu}\\
+ \beta^{\rho} \left( \frac{\partial}{\partial x^{\rho}} \left( K_{\nu}^{\mu} \right) + {}^{\left( 3 \right)} \Gamma_{\rho \sigma}^{\mu} K_{\nu}^{\sigma} - {}^{\left( 3 \right)} \Gamma_{\rho \nu}^{\sigma} K_{\sigma}^{\mu} \right) K_{\rho}^{\mu} \left( \frac{\partial}{\partial x^{\nu}} \left( \beta^{\rho} \right) + {}^{\left( 3 \right)} \Gamma_{\nu \sigma}^{\rho} \beta^{\sigma} \right)\\
- K_{\nu}^{\rho} \left( \frac{\partial}{\partial x^{\rho}} \left( \beta^{\mu} \right) + {}^{\left( 3 \right)} \Gamma_{\rho \sigma}^{\mu} \beta^{\sigma} \right) - \alpha \left( \frac{2 \Lambda}{n - 2} \gamma_{\rho \nu} \right) \gamma^{\rho \mu},
\end{multline}
with ${\mu, \nu, \rho, \sigma}$ ranging across all ${\left\lbrace 0, \dots, n - 2 \right\rbrace}$ (i.e. across spatial coordinate indices only), using the \texttt{SolveVacuumADMEquations} function requires first imposing a more appropriate (or at least more physically reasonable) choice of gauge. For this particular class of initial-value problem, involving the head-on collision of two or more black holes, we typically choose to employ the ${1 + log}$ slicing condition\cite{campanelli}\cite{baker} for the lapse function:

\begin{equation}
\frac{\partial}{\partial t} \left( \alpha \right) = \mathcal{L}_{\boldsymbol\beta} \alpha - 2 K \alpha = \beta^{\sigma} \frac{\partial}{\partial x^{\sigma}} \left( \alpha \right) - 2 K \alpha,
\end{equation}
or, equivalently, in slightly expanded form:

\begin{multline}
\frac{\partial}{\partial t} \left( \alpha \right) = \beta^{\sigma} \frac{\partial}{\partial x^{\sigma}} \left( \alpha \right) + \frac{\partial}{\partial t} \left( \log \left( \det \left( \gamma_{\mu \nu} \right) \right) \right) - 2 {}^{\left( 3 \right)} \nabla_{\sigma} \beta^{\sigma}\\
= \beta^{\sigma} \frac{\partial}{\partial x^{\sigma}} \left( \alpha \right) + \frac{\partial}{\partial t} \left( \log \left( \det \left( \gamma_{\mu \nu} \right) \right) \right) - 2 \frac{\partial}{\partial x^{\sigma}} \left( \beta^{\sigma} \right) - 2 {}^{\left( 3 \right)} \Gamma_{\sigma \rho}^{\sigma} \beta^{\rho},
\end{multline}
with ${\sigma, \rho}$ ranging across all ${\left\lbrace 0, \dots, n - 2 \right\rbrace}$ (i.e. across spatial coordinate indices only), coupled with the minimal distortion coordinate conditions\cite{smarr}\cite{brady} for the shift vector:

\begin{multline}
{}^{\left( 3 \right)} \nabla^{\mu} \left( {}^{\left( 3 \right)} \nabla_{\mu} \beta^{\nu} \right) + {}^{\left( 3 \right)} \nabla^{\nu} \left( {}^{\left( 3 \right)} \nabla_{\mu} \beta^{\mu} \right) - 2 {}^{\left( 3 \right)} \nabla_{\mu} \left( \alpha K^{\mu \nu} \right)\\
= \gamma^{\mu \sigma} {}^{\left( 3 \right)} \nabla_{\sigma} \left( {}^{\left( 3 \right)} \nabla_{\mu} \beta^{\nu} \right) + \gamma^{\nu \sigma} {}^{\left( 3 \right)} \nabla_{\sigma} \left( {}^{\left( 3 \right)} \nabla_{\mu} \beta^{\mu} \right) - 2 {}^{\left( 3 \right)} \nabla_{\mu} \left( \alpha K^{\mu \nu} \right) = 0,
\end{multline}
or, equivalently:

\begin{multline}
\gamma^{\mu \sigma} \left( \frac{\partial}{\partial x^{\sigma}} \left( D_{\mu}^{\nu} \right) + {}^{\left( 3 \right)} \Gamma_{\sigma \lambda}^{\nu} D_{\mu}^{\lambda} - {}^{\left( 3 \right)} \Gamma_{\sigma \mu}^{\lambda} D_{\lambda}^{\nu} \right) + \gamma^{\nu \sigma} \left( \frac{\partial}{\partial x^{\sigma}} \left( D_{\mu}^{\mu} \right) + {}^{\left( 3 \right)} \Gamma_{\sigma \lambda}^{\mu} D_{\mu}^{\lambda} - {}^{\left( 3 \right)} \Gamma_{\sigma \mu}^{\lambda} D_{\lambda}^{\mu} \right)\\
- 2 \left( \frac{\partial}{\partial x^{\mu}} \left( \alpha K^{\mu \nu} \right) + {}^{\left( 3 \right)} \Gamma_{\mu \sigma}^{\mu} \left( \alpha K^{\sigma \nu} \right) + {}^{\left( 3 \right)} \Gamma_{\mu \sigma}^{\nu} \left( \alpha K^{\mu \sigma} \right) \right) = 0,
\end{multline}
with the rank-2 tensor ${D_{\mu}^{\nu}}$ of spatial derivatives of the shift vector ${\beta^{\nu}}$ given by:

\begin{equation}
D_{\mu}^{\nu} = {}^{\left( 3 \right)} \nabla_{\mu} \beta^{\nu} = \frac{\partial}{\partial x^{\mu}} \left( \beta^{\nu} \right) + {}^{\left( 3 \right)} \Gamma_{\mu \sigma}^{\nu} \beta^{\sigma},
\end{equation}
and with ${\mu, \nu, \rho, \sigma, \lambda}$ again ranging across all ${\left\lbrace 0, \dots, n - 2 \right\rbrace}$ (i.e. across spatial coordinate indices only). Figures \ref{fig:Figure52} and \ref{fig:Figure53} use the \texttt{DiscreteHypersurfaceDecomposition} function to show the full evolution of spatial hypergraphs from the initial coordinate time ${t = 0 M}$ to the final coordinate time ${t = 24 M}$ via a sequence of six discrete time steps, for decompositions of the resulting Brill-Lindquist geometry obtained from the simulated collision, without spatial coordinate information included in the visualization, with the radial boundary $R$ of the domain set to be comparable to the Schwarzschild radius ${R \sim 2 M}$ and much larger than the Schwarzschild radius ${R \gg 2 M}$, respectively. In both of these examples, the numerical masses of the black holes are both $M$, their initial separation is ${2 M}$ (i.e. ${z_0 = M}$), and all simulations are run with a resolution of 10,000 vertices and a discretization scale of ${0.2 M}$, with vertices and edges colored, as before, based on the extrinsic curvature of the spacelike hypersurface at the corresponding point. Although all other examples shown within this article may be reproduced in under a few seconds on a typical laptop computer, producing such comparatively high-resolution simulations and visualizations was possible in this case only with the aid of a supercomputer.

\begin{figure}[ht]
\centering
\begin{framed}
\includegraphics[width=0.595\textwidth]{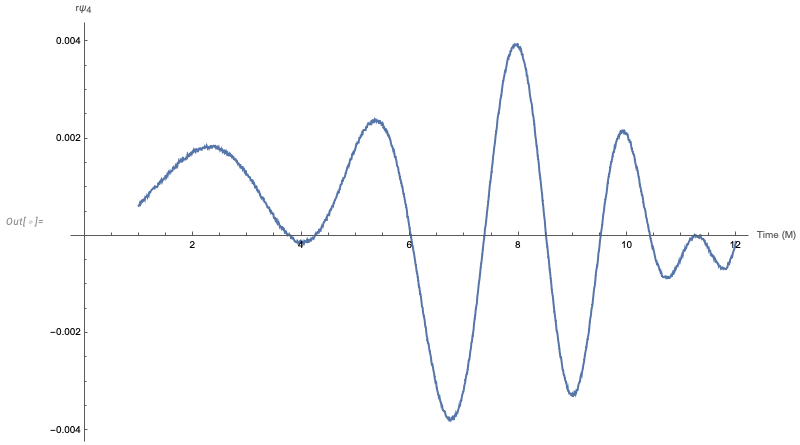}
\end{framed}
\caption{The outgoing gravitational wave strain for a Brill-Lindquist geometry (representing, for instance, a pair of uncharged, non-rotating black holes, initially at rest, each of mass $M$ in Schwarzschild or spherical polar coordinates ${\left(t, r, \theta, \phi \right)}$) with the ${1 + \log}$ slicing condition and the minimal distortion coordinate conditions for the gauge, calculated from the real part of the ${\ell = 2}$, ${m = 0}$ mode of the radial Weyl scalar ${r \Psi_4}$, projected onto a sphere of radius ${6 M}$.}
\label{fig:Figure54}
\end{figure}

Once the collision simulation has been completed and the resulting (discrete) spacetime extracted, it is possible to use \textsc{Gravitas}'s in-built \texttt{WeylTensor} functionality to extract the outgoing gravitational radiation from this binary black hole merger; within the Newman-Penrose formalism\cite{newman3}, the Weyl scalar ${\Psi_4}$ is defined as:

\begin{equation}
C_{\mu \nu \rho \sigma} n^{\mu} \bar{m}^{\nu} n^{\rho} \bar{m}^{\sigma},
\end{equation}
where ${C_{\mu \nu \rho \sigma}}$ is the (covariant) Weyl tensor and ${n^{\mu}, \bar{m}^{\mu}}$ are elements of a complex null tetrad ${\left\lbrace l^{\mu}, n^{\mu}, m^{\mu}, \bar{m}^{\mu} \right\rbrace}$. Once such a tetrad has been selected, the Weyl scalar ${\Psi_4}$ allows one to compute the outgoing gravitational wave strain $h$ in the spacetime as:

\begin{equation}
h = h_{+} - i h_{\times} = - \int_{- \infty}^{t} \int_{- \infty}^{t^{\prime}} \Psi_4 d t^{\prime \prime} d t^{\prime},
\end{equation}
where we have decomposed the total gravitational wave strain $h$ into plus polarization and cross polarization modes ${h_{+}}$ and ${h_{\times}}$, respectively. The total Weyl scalar field ${\Psi_4 \left( t, r, \theta, \phi \right)}$ is then interpolated onto a coordinate sphere, and the outgoing gravitational wave strain through that sphere is calculated by projecting the Weyl scalar field ${\Psi_4 \left( t, r, \theta, \phi \right)}$ onto the spin-weighted spherical harmonic functions ${{}_{s} Y_{\ell m} \left( \theta, \phi \right)}$\cite{dray}, yielding the coefficients:

\begin{equation}
C^{\ell m} \left( t, r \right) = \int {}_{s} \bar{Y}_{\ell m} \left( \theta, \phi \right) \Psi_4 \left( t, r, \theta, \phi \right) r^2 d \Omega.
\end{equation}
Figure \ref{fig:Figure54} shows the outgoing gravitational wave strain obtained by projecting the radial Weyl scalar ${r \Psi_4}$ onto a sphere of radius ${6 M}$, and then computing the real part of its ${\ell = 2}$, ${m = 0}$ mode.

\clearpage

\section{Future Research and Development Directions}
\label{sec:Section6}

Both the analytical and numerical subsystems of the \textsc{Gravitas} framework are currently being used, or are currently planned for use in the very near-term, in a variety of different research projects spanning mathematical relativity, astrophysics, cosmology and quantum gravity, including the investigation of black hole thermodynamics\cite{shah} and quantum computational complexity theory\cite{gorard12} in discrete spacetimes, the study of discrete spacetime structure via methods of global topology and homotopy theory\cite{arsiwalla}\cite{arsiwalla2}, the study of general relativistic Bondi-Hoyle accretion of perfect fluids onto spinning black holes\cite{papadopoulos}\cite{font}, and the simulation of binary neutron star collisions\cite{baiotti}\cite{bernuzzi}. In the latter two cases, the principal objective is to determine whether there might be observable signatures (for instance, those detectable within either the electromagnetic or gravitational wave spectra) of spacetime discreteness in certain high-energy astrophysical events, especially ones that might conceivably remain detectable even at high-redshift. In order to be able to achieve these ambitious research objectives, much further algorithmic and software development work on the \textsc{Gravitas} framework must be performed, including the implementation of alternative initial-value formulations of the Einstein field equations (moving beyond the ADM formalism), especially conformal formulations such as the BSSN formulation\cite{nakamura}\cite{shibata}\cite{baumgarte} and the conformally-covariant Z4/CCZ4 formulation\cite{alic}\cite{bona}, which typically have more favorable stability numerical properties when simulating more complex relativistic and longer-timescale phenomena such as compact binary inspirals. (Note that both the BSSN and CCZ4 formalisms have previously been implemented in an experimental form within \cite{gorard9}, although this experimental implementation has not yet been fully integrated with the design of the rest of the \textsc{Gravitas} framework). Relatedly, there are plans to implement a much wider class of algorithms for enforcing the Hamiltonian and momentum constraint equations (which, within the current \textsc{Gravitas} implementation, reduce to a slow, iterative numerical solution method for elliptic equations in the worst case), such as constraint-violation damping, as first developed by Gundlach, Mart\'in-Garc\'ia, Calabrese and Hinder\cite{gundlach}, based on the ${\lambda}$-system formalism previously proposed by Brodbeck, Frittelli, H\"ubner and Reula\cite{brodbeck}; once again, a prototype implementation of this, not yet fully integrated with the \textsc{Gravitas} design, is found in \cite{gorard9}. This, when combined with the planned extensions to \textsc{Gravitas}'s symbolic and analytic subsystems described within the concluding sections of \cite{gorard} (such as fully-integrated support for the tetrad formalism, frame fields, spin coefficients, connections on spinor bundles, scalar-tensor gravity theories, etc.), should help to ensure that \textsc{Gravitas} remains a robust, powerful, flexible and scalable computational tool for numerical and mathematical relativity research, across a variety of distinct domains, well into the future.

\end{document}